\pdfoutput=1

\input{aipcheck}

\documentclass[
    ,final            
  ]
  {aipproc}
\layoutstyle{8x11single}
\def\ga{\mathrel{\raise.3ex\hbox{$>$\kern-.75em\lower1ex\hbox{$\sim$}}}}
\def\la{\mathrel{\raise.3ex\hbox{$<$\kern-.75em\lower1ex\hbox{$\sim$}}}}
\def\beq{\begin{equation}}
\def\eeq{\end{equation}}

\begin{document}

\title{Non-Baryonic Dark Matter in Cosmology}


\author{A. Del Popolo}{
  address={Dipartimento di Fisica e Astronomia, Universit� di Catania, Viale Andrea Doria 6, 95125 Catania, Italy\\
Instituto de Astronomia, Geof\'{\i}sica e Ci\^{e}ncias Atmosf\'{e}ricas, Universidade de S\~ao Paulo, Rua do Mat\~ao 1226, 05508-900, S\~ao Paulo, SP, Brazil
}
}




\begin{abstract}

This paper is based on lectures given at the ninth Mexican School on Gravitation and Mathematical Physics. The lectures (as the paper) were 
a broad-band review of the current status of non-baryonic dark matter research. I start with a historical overview
of the evidences of dark matter existence, then I discuss how dark matter is distributed from small scale to large scale, and I then verge the attention to dark matter nature: dark matter candidates and their detection. I finally discuss some of the limits of the $\Lambda$CDM model, with particular emphasis on the small scale problems of the paradigm. 

\end{abstract}

\maketitle


\section{Introduction}

There are many reasons, discussed in the following, to believe that the universe is full of non-luminous and non-baryonic matter, not seen directly in our present observations, but influencing the evolution of the universe gravitationally\footnote{To date there is no conclusive
evidence of dark matter from electroweak or other kinds of non-gravitational interactions.}. This mass that would fill the Universe but is not observed is usually termed "dark matter", and the term, or a similar one, was introduced by Fritz Zwicky when in 1933\cite{zwicky} he studied the Coma cluster and observed that galaxies relative speeds in the cluster were much too great for them to be held together by the gravitational attraction of the visible matter alone, and that therefore, there must have been something else holding them together. 

Evidently if a component of the universe is non-visible one must have strong evidences, though indirect, to assume its existence.  
Plenty of indirect evidences for the existence of dark matter (DM) have been accumulated from the last century to now. Starting from smaller scales, the flatness of the rotation curves of spiral galaxies further the optical radius\cite{bosma}, and then the absence of observation of the "Keplerian fall",  are a strong evidence that some mass is non-visible in these objects, and we are led to similar conclusions by studying the velocity dispersion of stars in dwarf galaxies, that in some cases lead us to conclude that in some of them there is $10^2-10^3$ times more mass than can be attributed to their luminosity \cite{munoz}.
X-ray emission in Ellipticals and Clusters\cite{sarazin} of galaxies are another evidence of DM existence. Large elliptical galaxies have extended atmospheres of hot gas which appear to be in equilibrium, and similarly the intergalactic space in clusters is filled with a hot gas with temperature of tens of millions of degrees, emitting in X-rays by brehmstrahlung. By studying the distribution and temperature of the hot gas it is possible to measure 
the gravitational potential of the cluster.
This allows the determination of the total mass contained in the quoted objects. Remarkably, it turns out there is five times more material in clusters of galaxies than we would expect from the galaxies and hot gas we can see. 
Another evidence of the existence of DM in galaxies and clusters come form gravitational lensing\cite{kneib}, and
by combining lensing and the X-ray technique.
 The study of baryons location (through X-ray observations) and gravitational potential location (through weak lensing) in clusters which suffer collisions (e.g. the Bullet Cluster\cite{clowe})
is a strong evidence of the DM existence and of its dissipationless nature. These kind of studies favor the DM paradigm over modified gravity models. 
Spiral galaxies satellites have high velocity dispersion suggesting the existence of DM halos a distances > 200 kpc from the galactic center\cite{zaritsky}.
At large scales, the study of the anisotropies in the cosmic microwave background radiation (CMBR), and comparison 
of the CMBR with observations (e.g., WMAP), and other data allows to determine the DM content which is largely superior to that of baryonic matter 
(\cite{jarosik}).
Better constraints to the DM content of the Universe can be obtained combining different techniques (e.g. Baryonic Acoustic Oscillations (BAO), CMBR, and Supernovae (\cite{amanullah})). Structure formation requires DM, and can be used to put constraints on it (\cite{springel}).

In this paper, I will discuss in section 1 the evidences for DM existence, treated following an historical perspective. In section 2, I will describe how DM is distributed in structures. In section 3, I discuss the nature of dark matter (DM), the reasons why DM cannot be all baryonic, why it should be in particles, the characteristics of these particles, and the detection methods and experiments set up to find it. In section 4, I will discuss the limits and problems of the paradigm at small scale and how they can be overcome.

\section{DM evidences}

Before starting to discuss the evidences of DM existence, I will introduce the luminosity-mass ratio, $L/M$, and its connection with the matter density
parameter, $\Omega_m$.

As usual, we define the critical density $\rho_c  = 3H^2 / 8 \pi G_N	=  1.88\times 10^{-29} {h_o}^2$ g cm$^{-3}$,
where $H_o = 100 h_o$ km Mpc$^{-1}$ s$^{-1}$ is the present value of the Hubble, and $G_N$ the gravitational constant. 
If $\phi(L)$ is the number density of galaxies having a total luminosity $L$, the galaxies mean luminosity density is given by:
\beq
{\mathcal L}  = \int L \phi(L) dL  \simeq 2 \pm 0.2 \times 10^8  h_o L_\odot Mpc^{-3} 	
\eeq
where  $L_\odot = 3.8 \times 10^{33}$   erg s$^{-1}$ is the solar 
luminosity.  

A connection between $\Omega_m$ and $M/L$ can be obtained as: 
\beq
\Omega_m =  {\rho \over
\rho_c} = (M/L)/(M/L)_c 	
\eeq
if we define a critical mass-to-light ratio as:
\beq
(M/L)_c = \rho_c/{\mathcal L}  \simeq  1390 h_o (M_\odot/L_\odot)	
\eeq

The mass-luminosity ratio is a useful measure of mass and a good indicator of DM presence in a given region.
If the typical star in a given region was equal in mass to the sun, the ratio of total mass to total light would be unity, larger than unity if the typical star was less massive, and smaller than unity for more massive stars. 

For nearby stars the resulting ratio is found to be $\simeq 2$ (in solar units of solar mass (M) to solar luminosity (L)) then we conclude that the average star near the sun is slightly less massive than the sun. In the solar vicinity, there is little necessity for any dark matter other than 
"dark baryonic matter" (as we will see in the rest of the discussion).


Typical values of $M/L$ are: $M/L=(10-20) h_o M_\odot/L_\odot$ ($\Omega_m \simeq 0.01$), in the bright central part of galaxies; $(60-180) h_o M_\odot/L_\odot$ ($\Omega_m \simeq 0.1$) for small galaxies groups; $(200-500) h_o M_\odot/L_\odot$ ($\Omega_m \simeq 0.3$) for clusters of galaxies\cite{mihalas}). The quoted values depend on the wavelength. 


As was told in the introduction, DM is not directly observed and its existence is inferred from its influence on the evolution of the universe through gravitation. From this point of view, the "DM problem" is similar to the old problem of invisible stars or planets (Procyon B; Sirius B; Uranus, Neptune, etc.), which were not visible but their existence was deduced through their perturbation on "neighboring" objects. 
In 1844 Bessel deduced the existence of the unseen companions
for Sirius and Procyon and the invisible companions were detected in 1862 by Alvan Graham Clark and his father Alvan Clark and by John Schaeberle in 1896, respectively. Similarly Neptune was detected in 1846 by J. G. Galle using J. C. Adams (1845) and U. Leverrier (1846) calculations linked to the anomalous
motion of Uranus. After Neptune's discovery, Lowell proposed the Planet X hypothesis, namely the existence of a ninth planet whose gravity could have perturbed Uranus orbit, in order to explain apparent discrepancies in the orbits of the gas giants, particularly Uranus and Neptune, . The ninth planet, was discovered in 1930 by C. Tombaugh and named Pluto. It was later understood that Lowell's prediction was wrong. The mass of Pluto is so small ($0.002 M_{\bigoplus}$
\cite{christy}) that is cannot produce the observed anomalies. Uranus anomalies disappeared when Neptune mass was corrected in 1989 using Voyager 2's fly by of the planet. Another story of a blunder was that of 
the planet Vulcan, hypothetical planet in the orbit between Mercury and the Sun which should have been the responsible of the anomalies in Mercury motion. 
Mercury's orbit has now been explained by Albert Einstein's theory of general relativity.

The "DM problem" is somehow similar to the problem of unseen planets. Observing astrophysical systems (galaxies, clusters of galaxies) we notice some "anomalies" (e.g., rotation curves different than what expected) that can be explained assuming that the Universe contains a large amount of invisible matter, or assuming that our knowledge of the laws of gravitation is not correct. 
 
We will start our trip in search of the quoted irregularities (i.e., DM) starting from the solar's neighborhood. In 1922 Kapteyn (\cite{kapteyn}) studied the vertical motions of all known stars near
the Galactic plane to determine the acceleration of matter. He found that the spatial density is sufficient to explain the vertical
motions. Jeans (1923)\cite{jeans} reanalyzing Kapteyn's data reached the conclusion that some mass deficit should exist to explain stars motion. 
This result was confirmed by several other authors. Oort (1932)\cite{oort}: reanalyzed the vertical motions and came to the same conclusion as Jeans. 
The value of DM that Oort found is called the "Oort limit".

Combining Poisson's equation with the first moment of the Boltzmann Equation in z for an infinite disk
one obtains:
\begin{eqnarray}
\frac{d}{dz} [\frac{1}{n(z)} \frac{d(n(z) v_z^2)}{dz}  ] &=&  4 \pi G \rho_o
\nonumber \\
\frac{1}{n(z)} \frac{d(n(z) v_z^2)}{dz} &=& 2 \pi \Sigma(z)
\end{eqnarray}
Solving numerically the coupled equations for a detailed Galaxy model gives the density, $\rho_0$. 

In Table 1, I summarize the results comparing observations and theory, starting from the first attempts of Kapteyn. 

\begin{table}
\begin{tabular}{lrrrr}
\hline
  & \tablehead{1}{r}{b}{density(Theory)}
  & \tablehead{1}{r}{b}{density(Observed)}   \\
\hline
Katpeyn(1922) & No DM&\\
Jeans(1923) & Two dark stars to each bright star-> DM needed&  \\
Oort(1932) & 0.09 $M_{\odot}/pc^3$ & 0.03 $M_{\odot}/pc^3$\\
Bahcall(1984) & 0.19 $M_{\odot}/pc^3$ & 0.14 $M_{\odot}/pc^3$\\
Kuijken \& Gilmore(1988) & 50 $M_{\odot}/pc^2$ & 50 $M_{\odot}/pc^2$\\
\hline
\end{tabular}
\caption{Local density expectations from theory and observations.}
\label{tab:a}
\end{table}

\begin{table}
\begin{tabular}{lrrrr}
\hline
  & \tablehead{1}{r}{b}{$\rho_0$} \\
\hline
Caldwell \& Ostriker (1981)\cite{caldwell1} & $0.23 \pm 0.2 Gev/cm^3$ & \\
Gates, Gyuk \& Turner (1995)\cite{gates} & $0.30^{+0.12}_{-0.11} Gev/cm^3$&\\
Crez\' et al. (1998)\cite{creze} & $0.076 \pm 0.015 M_{\odot}/pc^3$ & \\
Moore et al. (2001)\cite{moore_ett} & $0.18-0.30 Gev/cm^3$ & \\
Belli et al. (2002)\cite{belli} & $0.18-0.71 Gev/cm^3$ & \\
Strigari \& Trotta(2009)\cite{striga_tro} & $0.32 \pm 0.07 Gev/cm^3$ &\\
Weber \& de Boer (2009)\cite{weber} & $0.4^{+0.005}_{-0.01} Gev/cm^3$ & \\
Catena \& Ulio (2009)\cite{catena} & $0.39 \pm 0.03 Gev/cm^3$ &  \\
Salucci et al. (2010)\cite{salucci} & $0.43 \pm 0.21 Gev/cm^3$ &\\
\hline
\end{tabular}
\caption{Local density obtained from dynamics and gravitational lensing.}
\label{tab:a}
\end{table}

In Table 2, I summarize the results obtained just through observations from dynamics and microlensing (see also Pato et al. (2010)\cite{pato} and Iocco et al. (2011)\cite{iocco}, also for results from simulations). It is important to stress that a correct evaluation of $\rho_0$ is of fundamental importance in direct and indirect detection of DM (see the section on DM detection).


On galaxy clusters scale, 
Zwicky (1933)\cite{zwicky}
measured the radial velocities for eight galaxies in the Coma cluster finding an unexpectedly large velocity dispersion of $\simeq 1200$ km/s. Applying 
the virial theorem he observed that much more mass than the observed was needed, in order the cluster conserved its structure. He suggested this light deficit was due to "dark matter" (Dunkle Materie).
He found a value for the mass-light ratio $M/L \simeq 400$ much larger than the value measured more recently (e.g., $M/L \simeq 160$\cite{fusco})
because he assumed a Hubble parameter value of $\simeq 500$ km/s/Mpc, at that time considered the correct value. Some years later, Smith (1936)\cite{smith}
confirmed Zwicky's earlier result using galaxies in
the Virgo cluster and speculated that "a great mass of internebular material" existed "within the cluster".
Differently from Zwicky and Smith, Babcock (1939)\cite{babcock}
obtained long slit spectra of M31 and noted that the outer parts of the disk were rotating with unexpectedly high velocities.
Babcock's study of the rotation curve of M31 predated the work of Rubin \& Ford (1970)\cite{rubin}
and Roberts \& Whitehurst (1975)\cite{roberts}, 
which provided the first widely recognized observational evidence in favor of DM in galaxies. 
Before discussing the studies of rotation curves of spiral galaxies, I want to recall another evidence of DM evidence, namely the "timing argument of M31" of Kahn \& Woltjer (1959)\cite{kahn}. 
\begin{figure}
\resizebox{8.65cm}{!}{\includegraphics{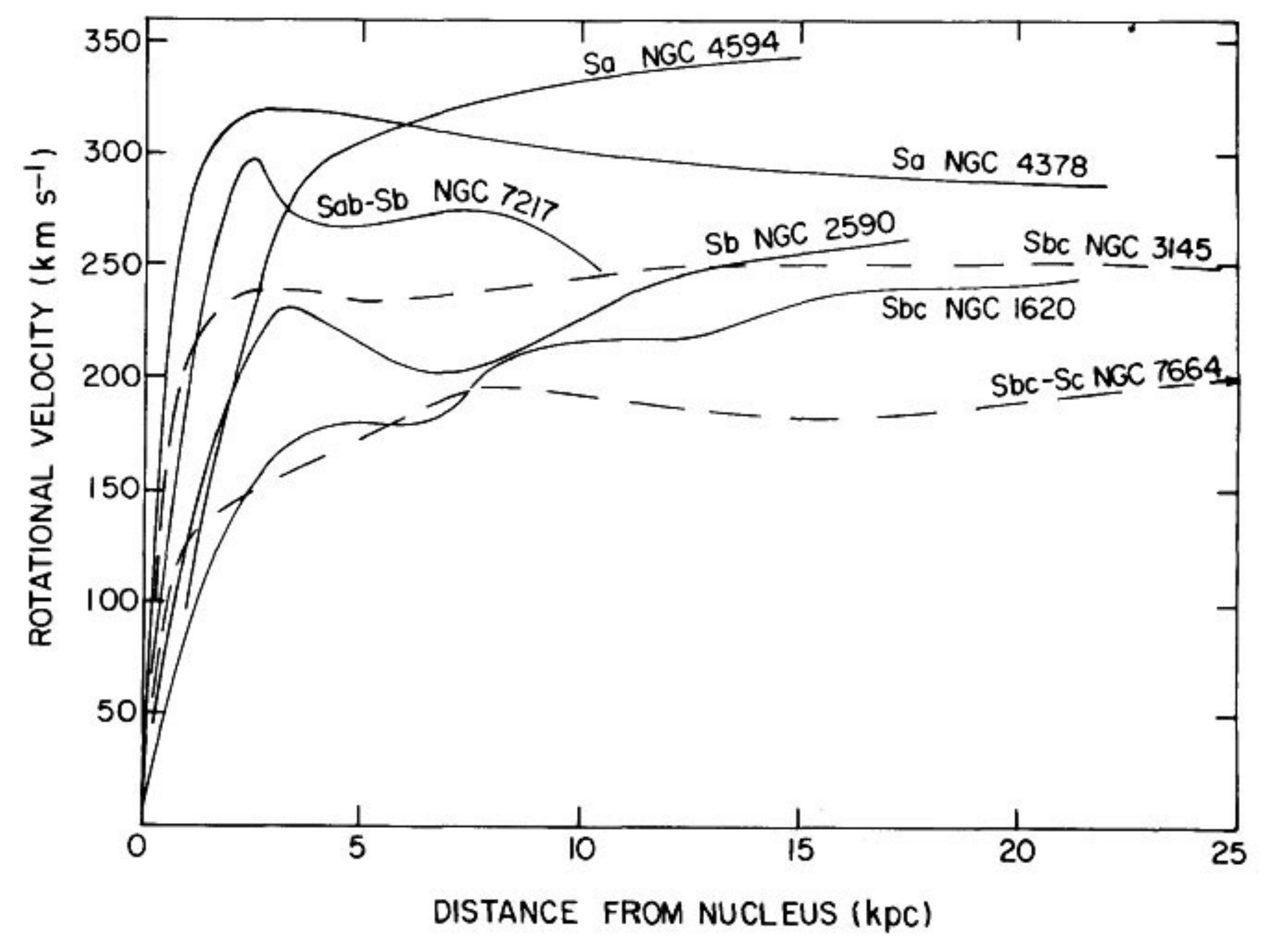}}
\resizebox{8.65cm}{!}{\includegraphics{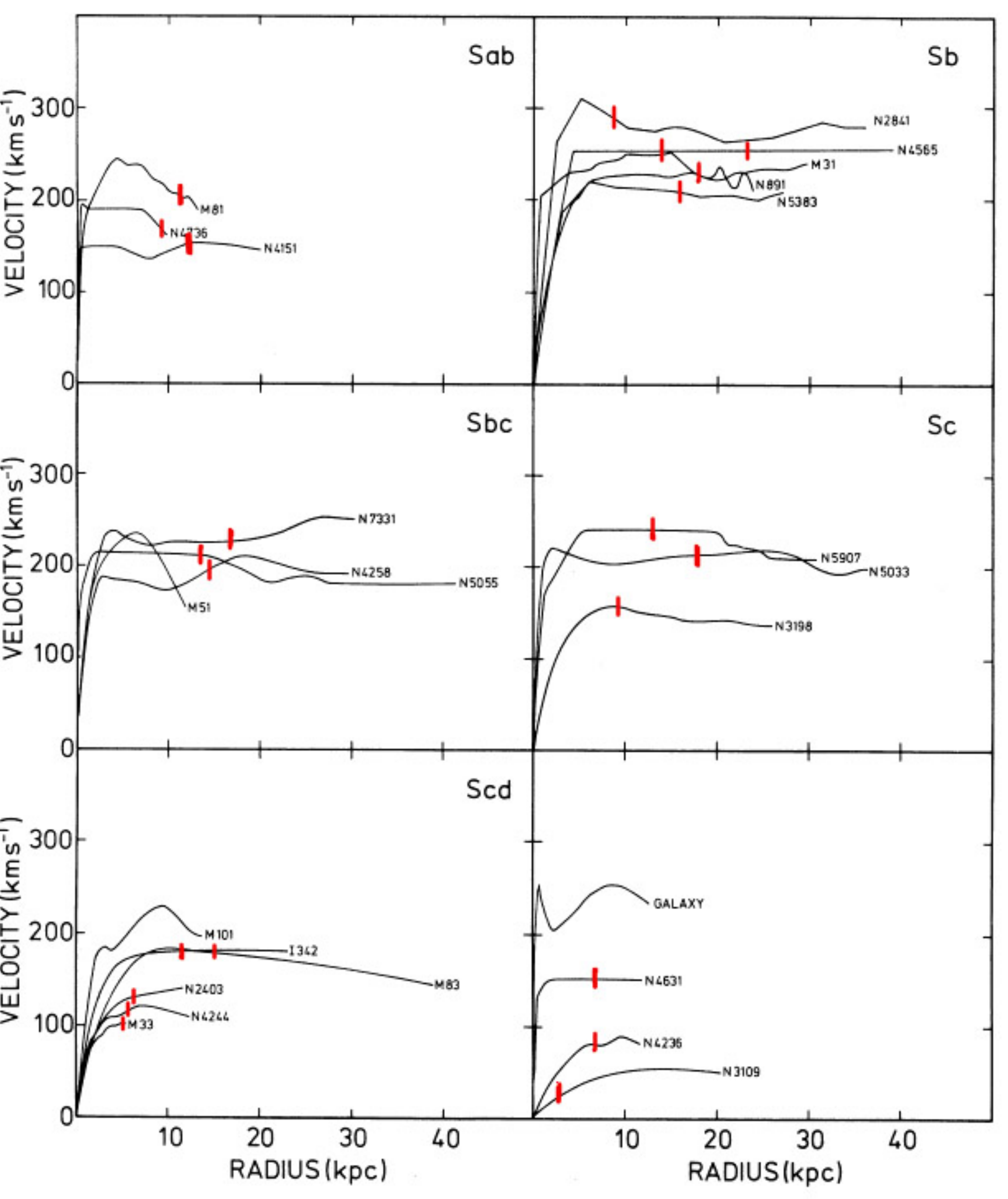}}
\caption{Left Panel: Rotation curves of seven galaxies of different Hubble type. At some tens of kpc the Keplerian fall is not observed. Early type galaxies have higher peaks than later types (from Rubin, Ford, and Thonnard (1978)\cite{rubin1}). Right panel: rotation curves in HI\cite{bosma}. The vertical lines on the curves represent the optical radius 
$R_{opt}=3.2 R_{D}$. It is evident the absence of the Keplerian fall further the optical radius.}
\end{figure}

As it is known, M31 is approaching the Milky Way (MW). It is possible to calculate the mass needed to have 
$t_{orbit} < t_{Universe}$. Kahn \& Woltjer (1959)\cite{kahn} from this argument deduced that the effective mass was larger than 
$1.8 \times 10^{12} M_{\odot}$, namely several times larger than the sum of M31 and MW mass. A more precise description was presented by Lynden-Bell (1983)\cite{lynden} and reproduced in 
Battaner \& Florido (2000)\cite{battaner}.
A simple deduction of the mass of the system is the following. Let's use a parametric representation of the orbit
\begin{eqnarray}
r&=&a(1-e cos \eta) 
\nonumber \\
t&=& T/(2 \pi) (\eta- e sin \eta ) 
\label{param}
\end{eqnarray}
that can combined in the equation

\begin{equation}
\frac{d log r}{d log t}= \frac{t}{r}\frac{dr}{dt}=\frac{e sin \eta (\eta- e sin \eta)}{(1-e cos \eta)^2}
\label{para}
\end{equation}
with 
\begin{equation}
T=2 \pi \sqrt{a^3/(GM)}
\label{temp}
\end{equation}
where $M$ is the system mass, $e$ the eccentricity, $a$ the semi-major axis.
Eq. (\ref{para})  for $e=1$, $t=13$ Gyr (age of the universe), $r=740$ kpc (distance between the two galaxies), $v=-125$km/s (approaching velocity) gives $\eta=4.26$. Introducing the previous values in 
Eq. (\ref{param}), and Eq. (\ref{temp}) one gets a system of equations for a, and M. Solving them one gets: $a=515$ kpc, $M_{Tot}=4.8 \times 10^{12} M_{\odot}$.

More recent evaluations (with other techniques) of the masses of M31 and the MW give: $M31(total)=1.23^{+1.8}_{-0.6} 10^{12}M_{\odot}$
(Evans \& Wilkinson 2000\cite{evans}), $MW(total)=(1.26 \pm 0.24) \times 10^{12}M_{\odot}$
(McMillan 2011\cite{mcmillan}).

The flatness of the rotation curves of galaxies is a strong evidence of DM existence. Rubin \& Ford (1970)\cite{rubin} observed  
the rotational velocities of OB associations for M31 up to 20 kpc. If a system is in virial equilibrium its circular velocity is given by $v^2_c=GM/r$, being M the mass in a given radius r. The previous expression implies that $v_c \propto 1/r$, named Keplerian fall. They did not observe any Keplerian fall, expected if all the system mass is associated with light, but instead observed a flat rotation curve (see Fig. 1). If $v_c=constant$ this imply that $M \propto r $ beyond the point where no light is visible, and this implies that there is a great quantity of matter (DM) which we do not observe and responsible of the flatness of the rotation curves (Freeman 1970\cite{freeman}).
Supposing DM is spherically distributed the $M(r) \propto r$ implies that the DM distribution has a singular isothermal sphere (SIS) density profile $\rho \propto 1/r^2$ at distances far beyond the visible region of the galaxy, which, at those distances coincides with the pseudo isothermal profile (ISO)
: $\rho= \frac{\rho_0}{1+(r/r_0)^2}$.
Other observations (e.g., \cite{rubin1})
in optical light and in the HI 21 cm emission lines (e.g., \cite{bosma}), showed that the flat behavior continued at several optical radii, $R_{opt}$ (defined as $R_{opt}=3.2 R_d$, where $R_D$ is the disc scale-length). Combining the Oort limit and the measured values of $M/L$ suggest
that DM is necessary but it cannot be all in the disc. At around this time, there were a number of influential theoretical studies that examined the implications of DM in galaxies. Ostriker \& Peebles (1973)\cite{ostriker} suggested that cold disks cannot survive, since they would be prone to bar-instability. According to them, the stability of galactic disks requires the presence of a massive halo of DM around galaxies. Ostriker-Peebles criterion for stability put a limit to the kinetic to gravitational energy: $T/W <0.14 \pm 0.02$ (see Fig. 2). Kalnajs 1983\cite{kalnajs} and Sellwood 1985\cite{sellwood} were not convinced from the previous argument. According to them the bulge was equally efficient to stabilize discs. 

\begin{figure}
\resizebox{8.65cm}{!}{\includegraphics{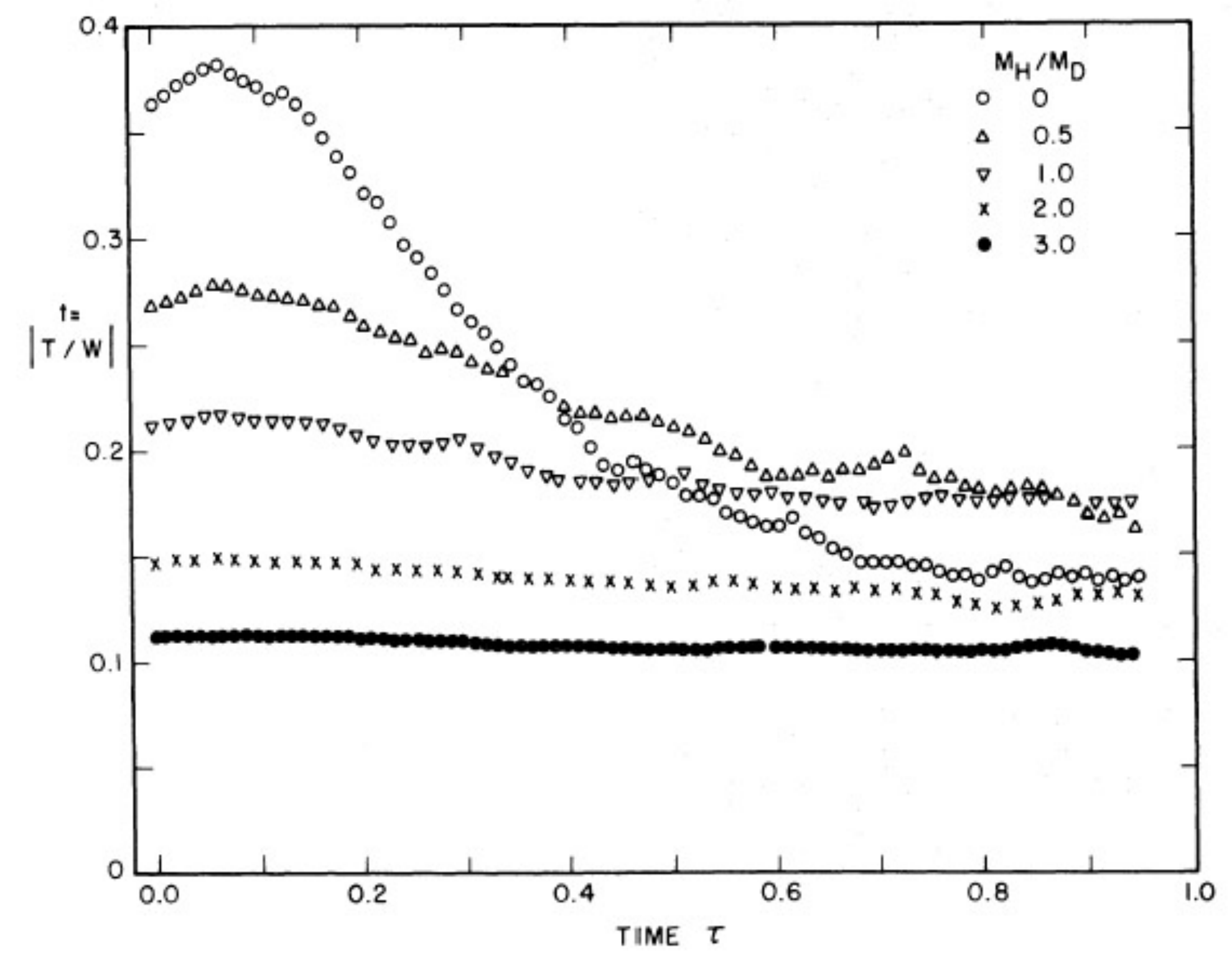}}
\caption{Effect oh the halo mass, $M_H$, on the evolution of the galaxy model. $M_D$ is the mass of the disk. From Ostriker and 
Peebles (1973)\cite{ostriker} }
\end{figure}

\begin{figure}
\resizebox{10.65cm}{!}{\includegraphics{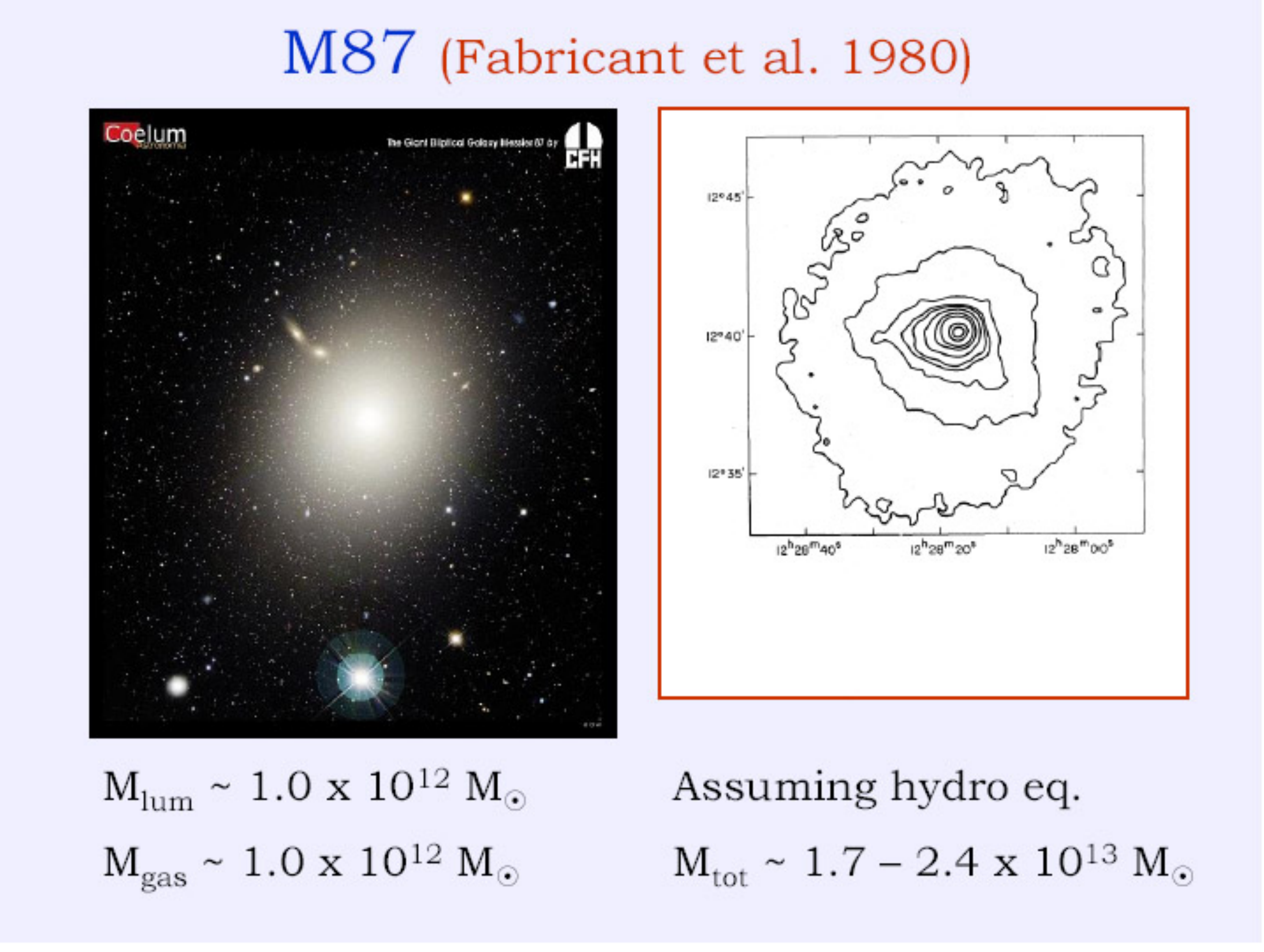}}
\resizebox{7.65cm}{!}{\includegraphics{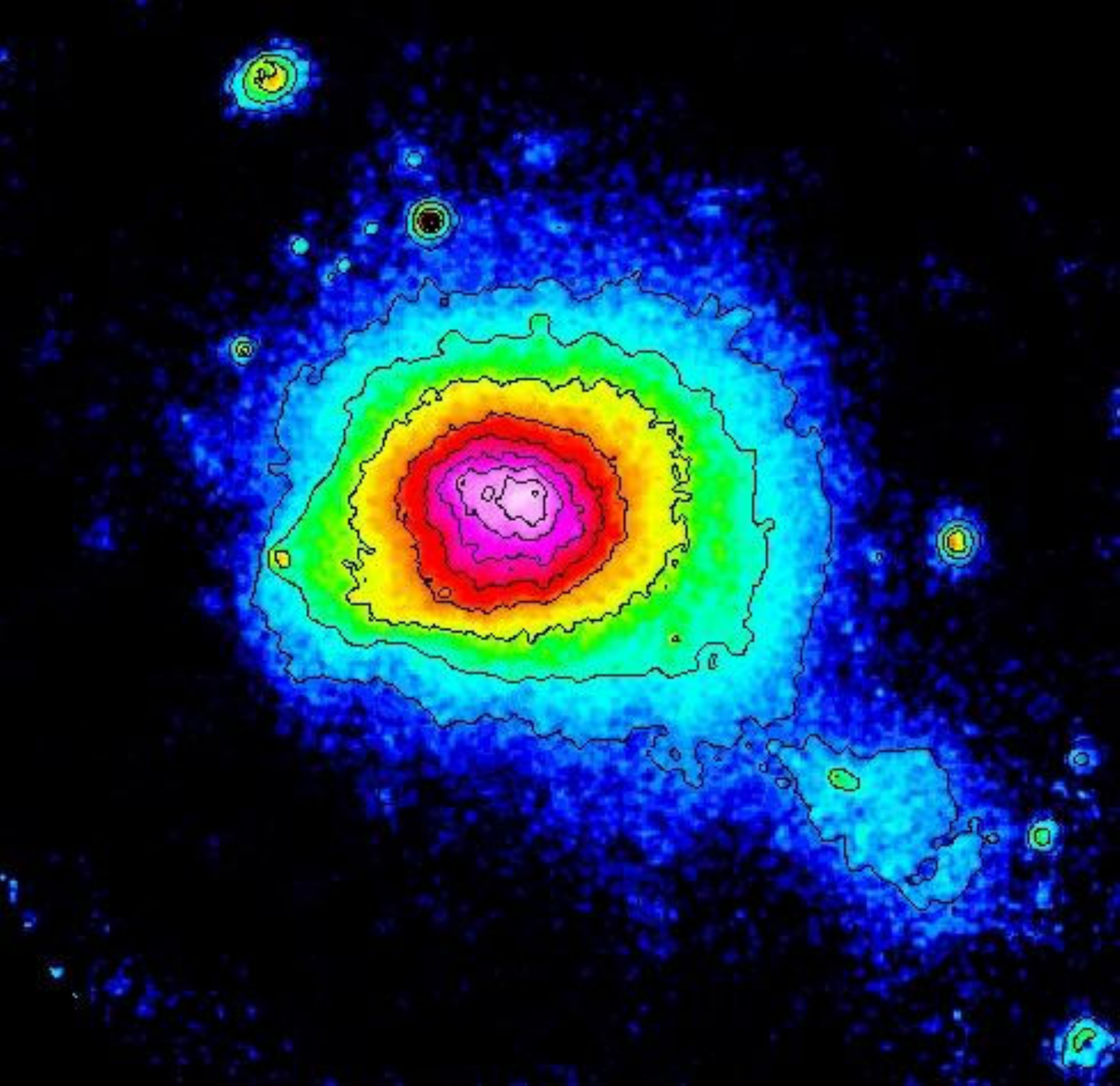}}
\caption{Left panel: M87 in optical and a X-ray map (right figure). Fabricant et al. (1980)\cite{fabricant}. Right panel: the Coma Cluter}
\end{figure}

X-ray emission from hot-gas in elliptical galaxies and clusters is another important evidence of DM existence. 
The first detection was around M87 in Virgo (Byram\cite{byram}; Bradt\cite{bradt}) and in Coma and Perseus (Fritz\cite{fritz}; Gursky\cite{gursky}; Meekins\cite{meekins}). For a spherical system in hidrostatic equilibrium the mass is given by:

\begin{equation}
M(r)=\frac{kT}{\mu m_H} \frac{r}{G} (-\frac{d ln \rho_{gas}}{d ln r}
-\frac{d ln T}{d ln r})
\end{equation}
where k is the Boltzmann constant, T is the temperature of the gas, $\mu$ is the mean molecular
weight of the gas and $m_H$ is the proton mass
A model often used to model the gas density $\rho_{gas}$ is the $\beta$ model 
obtained for an isothermal spherical gas cloud in hydrostatic
equilibrium if the volume density of galaxies is described by a King profile, and is given by:
\begin{equation}
\rho_{gas}=\rho_{0}/[1+(r/R_c)^2]^{3 \beta/2}
\end{equation}
whose parameters $\beta$ and $R_c$ are determined analyzing the X-ray surface brightness profile obtained from X-ray
image analysis. 
After $\rho_{gas}$ and $T(r)$ are obtained through X-ray observations, it is possible to get
the total system mass $M(r)$. 
An interesting example is M87. Fabricant et al. (1980)\cite{fabricant} (see Fig. 3) found a value for the total mass of $1.7-2.4 \times 10^{13} M_{\odot}$, while the gas mass was $ \simeq 10^{12} M_{\odot}$. X-ray observations have shown that the mass of ellipticals is much larger than the visible mass, and $M/L=30 h_0-200 h_0$. Similar behaviour is observed in small galaxies groups (e.g., \cite{david} with  $M/L>100 h_0$), and clusters of galaxies (e.g., 
\cite{sand}).

Other evidences of DM evidence come from gravitational lensing, based on the bending of light-rays passing through a gravitational field. The effect was firstly described by Newton (1704), calculated in Newtonian mechanics by von Soldner (1801)\cite{soldner} and 
in general relativity by Einstein (1915), whose value is twice the amount predicted by von Soldner\cite{soldner}.
Gravitational lensing can be divided in three classes: strong lensing, weak lensing, and micro lensing.

\begin{figure}
\resizebox{8.65cm}{!}{\includegraphics{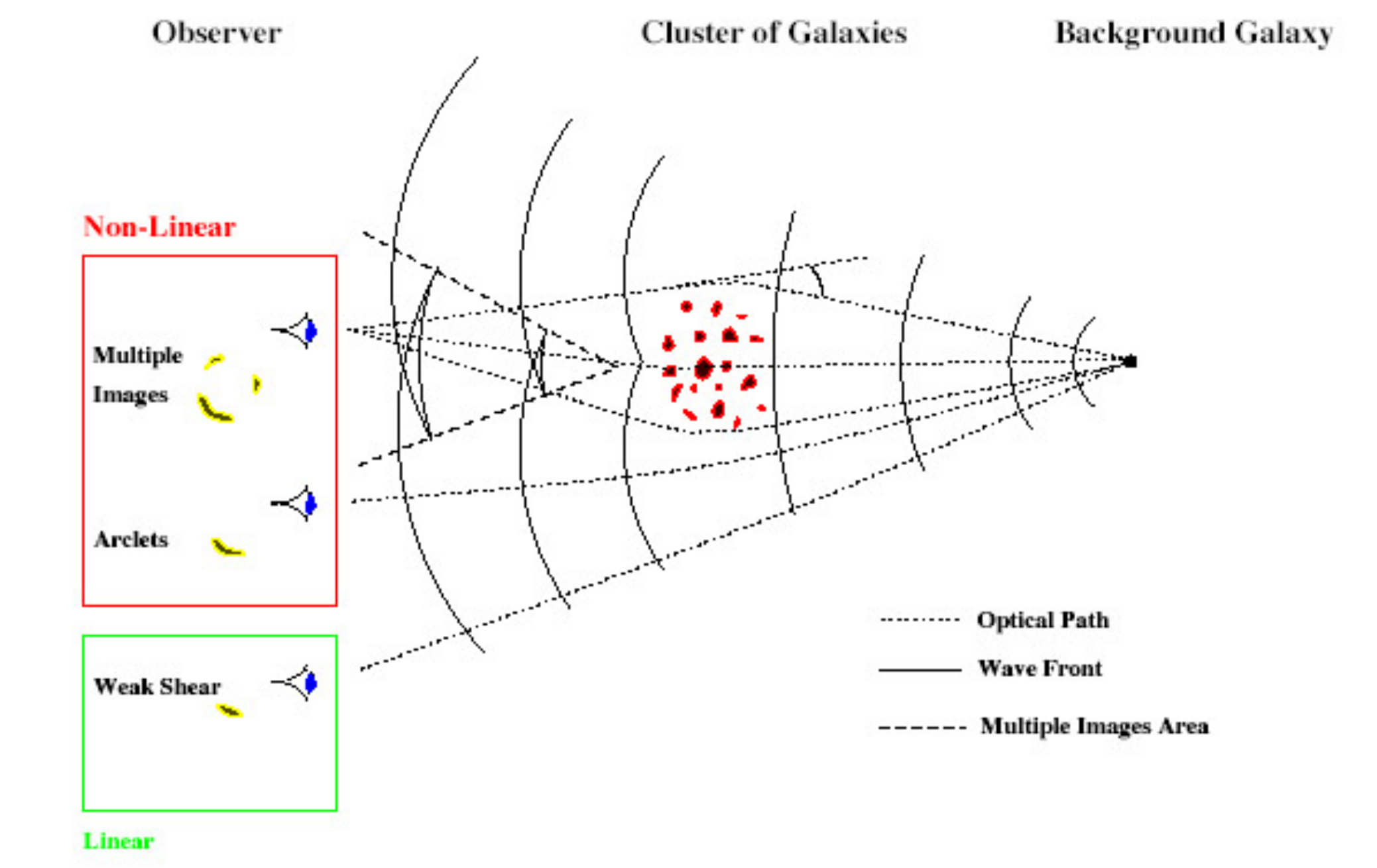}}
\resizebox{5.65cm}{!}{\includegraphics{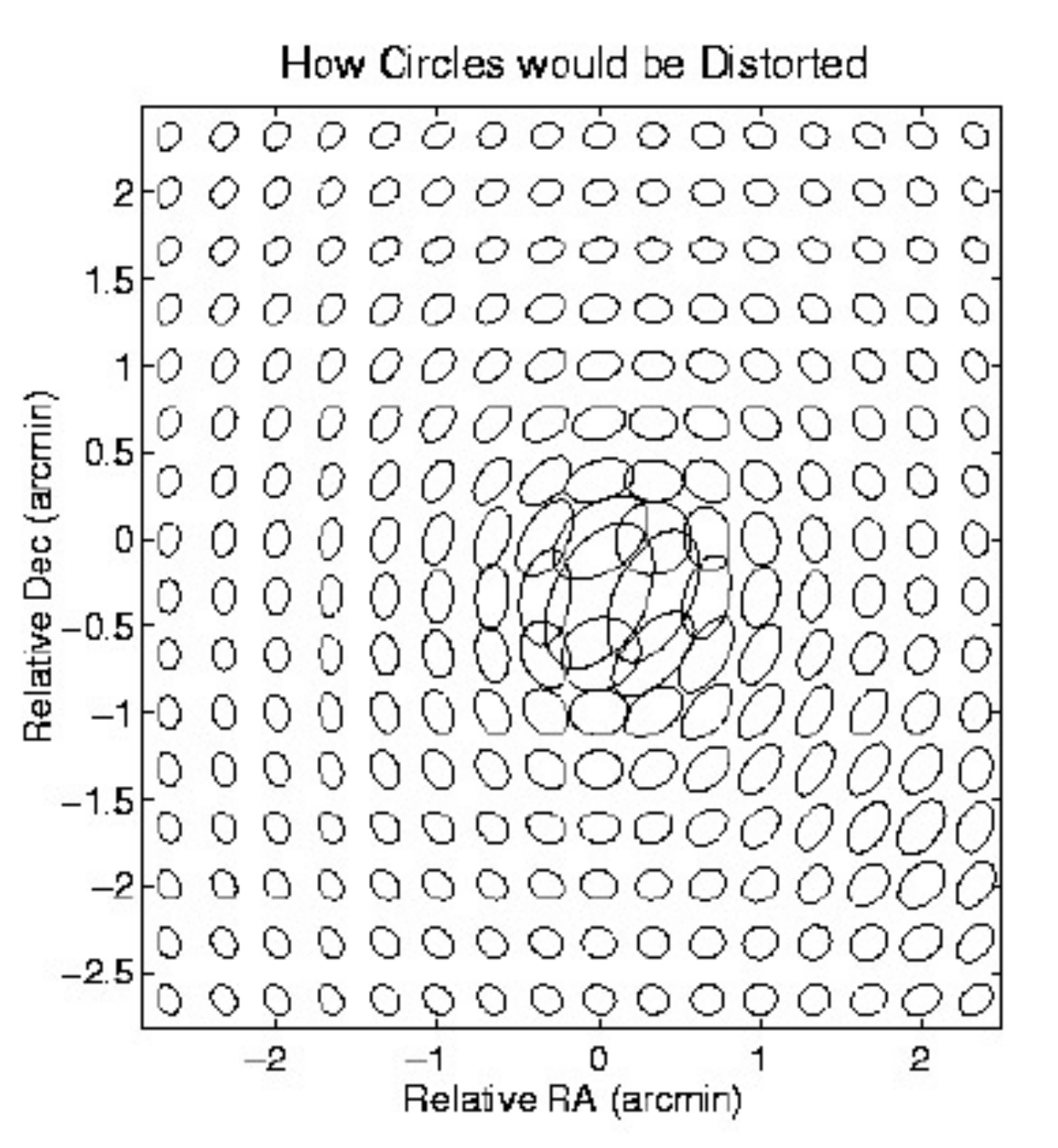}}
\resizebox{6.65cm}{!}{\includegraphics{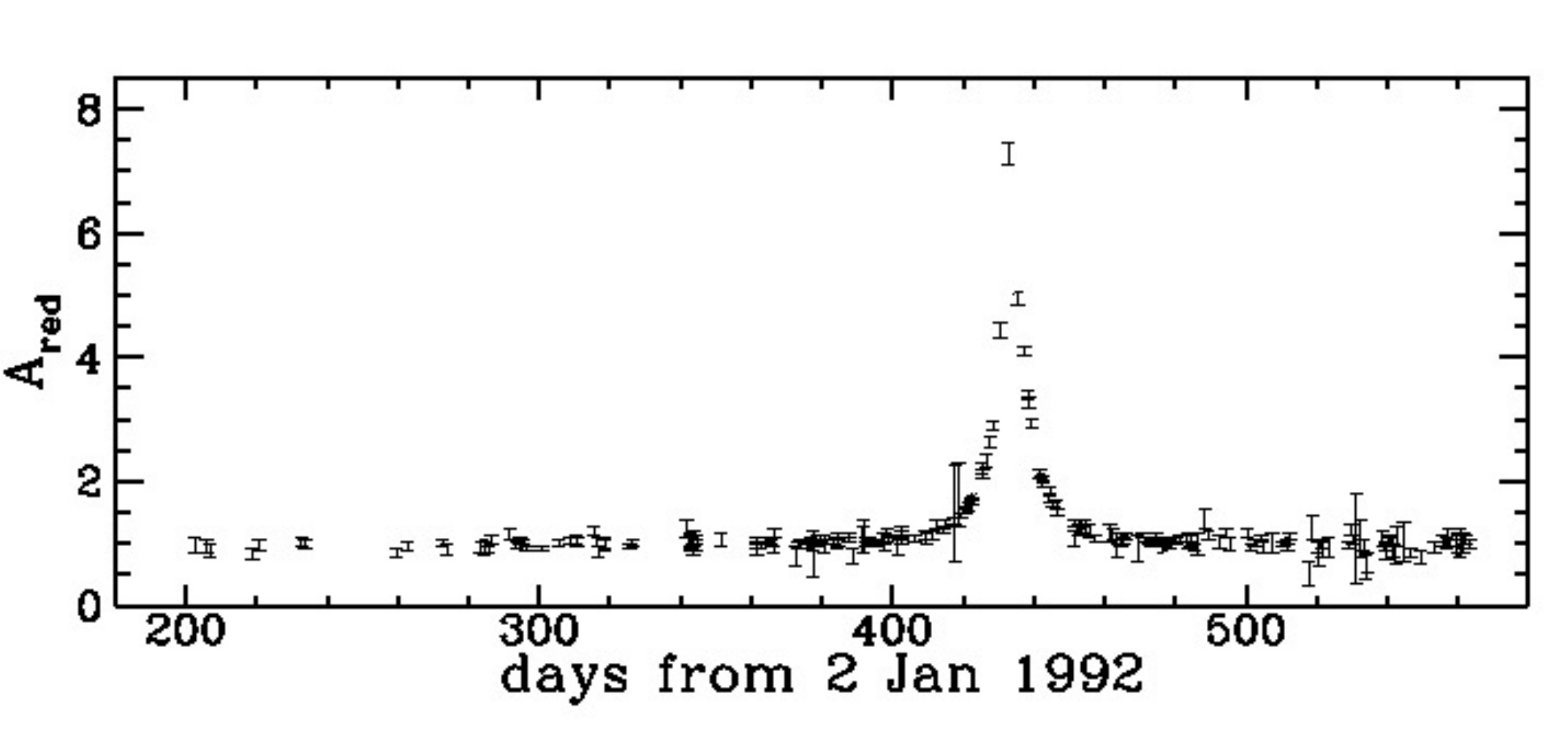}}
\caption{Left panel: gravitational lensing in clusters. Scheme of image formation in strong, intermediate, and weak lensing (from Kneib \& Natarajan 2012\cite{kneib}). Central panel: distorsion by weak lensing. Right panel: microlensing event.}
\end{figure}

In strong lensing photons are subject to a strong gravitational potential causing large deflection angles(see Fig. 4). Original light source may appear as a ring around the massive lensing object (Einstein Ring), or arc segments. Multiple images of an object are visible. The effect was discussed for the first time by Chwolson (1924)\cite{chwolson} even if it is commonly associated with Einstein (1936)\cite{einstein}, and was confirmed in 1979: "Twin QSO" SBS 0957+561, while the first complete Einstein ring, designated B1938+666, was observed in 1998. Strong lensing of distant galaxy by a cluster can be used to determine the cluster mass, while if several backround galaxies are lensed one can constraint $\Omega_{\Lambda}$ and $\Omega_m$\cite{blandford}.

In weak lensing (see Fig. 4), the deflection is through a small angle when the light ray can be treated as a straight line and single images form.
The effect was observed by Tyson et al. (1990)\cite{tyson} (coherent alignment of the ellipticities of the faint blue galaxies behind both Abell 1689 and CL 1409+52). 
Weak lensing has many applications. It can be used, in galaxies by galaxies lensing to study galactic halos properties, while many clusters lenses can be used to get $\Omega_m$ (see the next section).
As already reported in the introduction, combining weak lensing and X-ray observations, one can use clusters collisions to
get informations on the nature of DM. In the case of the Bullet Cluster (see Fig. 5), a small $7 \times 10^{13} M_{\odot}$
sub-cluster collided with a velocity of $4500 \pm 1000$ km/s (velocity indicated from
X-ray shock Mach number) with a $2 \times 10^{15} M_{\odot}$ cluster. The hot gas forming most of the clusters' baryonic mass was shocked and decelerated, whereas the galaxies in the clusters proceeded on ballistic trajectories.
Gravitational lensing shows that most of the total mass also moved ballistically, indicating that DM self-interactions are indeed weak. Non-radial separation of dark matter and the (dominant) baryonic mass (in the gas)
effectively rules out modifications of Newtonian gravity as explanation of dark matter. Separation of dark matter and gas gives direct constraints on the DM-DM cross-section (c.f. terrestrial experiments give only DM-B cross-section), which is 
of the order of $\sigma/m<0.7 cm^2/g$ (Randal et al. (2008)\cite{randall}). 
The weak lensing of large scale structure (LSS), the so called Cosmic Shear (see the following), detected for the first time in 2000 and of great importance as an evidence of DM existence and to put constraints on $\Omega_m$\cite{waerbeke}.

Microlensing (see Fig. 4) happens if the mass of the lensing object is very small (e.g., planets, white or brown dwarf), and one will merely observe a magnification of the brightness of the lensed object. Petrou (1981)\cite{petrou} and Paczynski (1986)\cite{paczynski} discussed the possible detection of DM in the form of compact objects, and the second one founded the in 1992 OGLE microlensing experiment. EROS, MACHO, OGLE, MOA and SuperMACHO collaborations have monitored millions of stars in the Magellanic Clouds (EROS-2 monitored over 6.7 years $33 \times 10^6$ stars) to search for microlensing events caused by such objects.
MACHO experiment (Alcock et al 1996) concluded that 20\% of halo can be due to MACHOs while according to EROS2+EROS1 the quantity is smaller ($< 8\%$). Machos in the mass range $0.6 \times 10^{-7}<M< 15 M_{\odot}$ are ruled out as the primary occupants of the Milky Way Halo (Tisserand et al. 2007\cite{tisserand}). 

\begin{figure}
\resizebox{8.65cm}{!}{\includegraphics{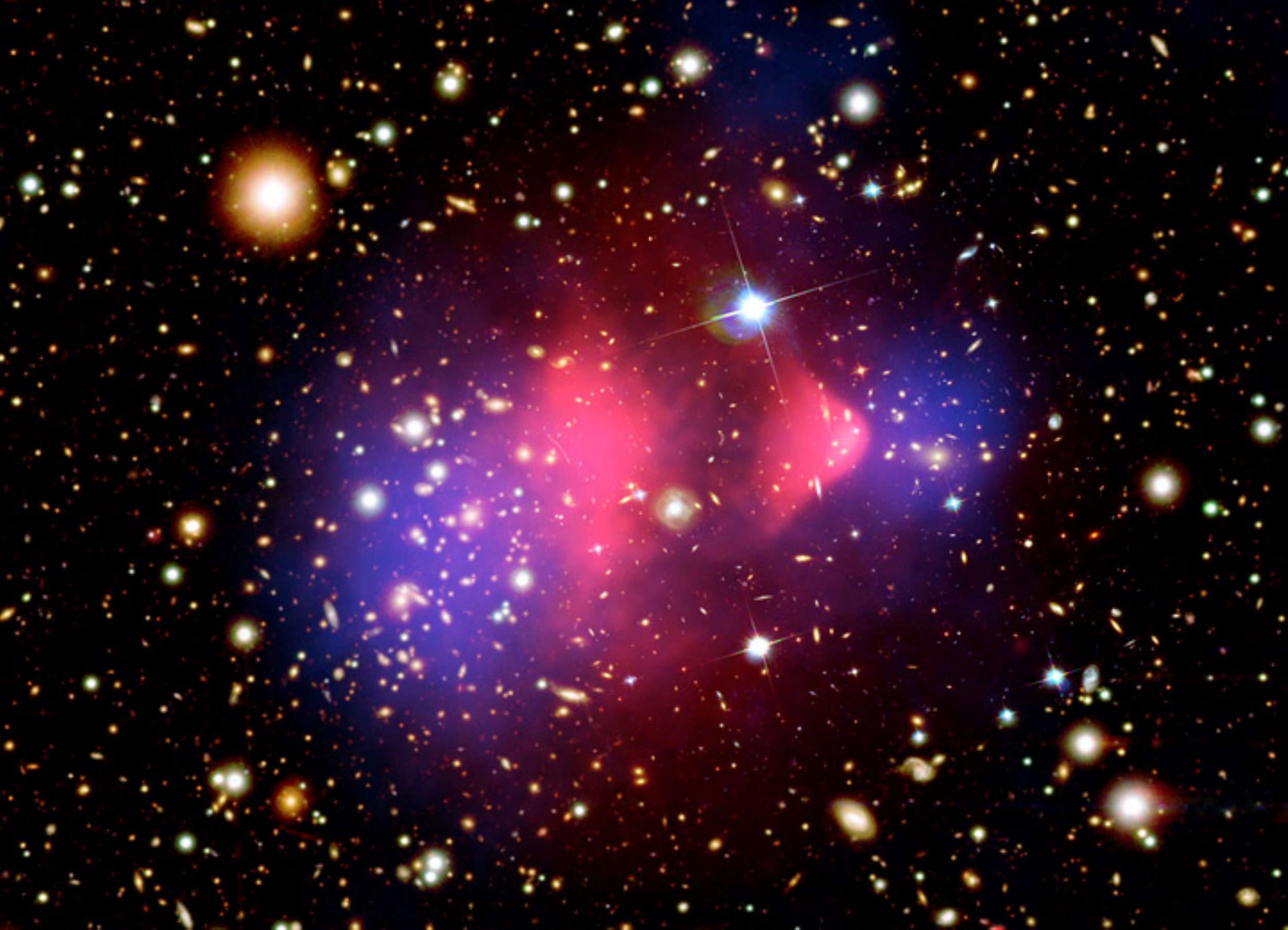}}
\resizebox{8.65cm}{!}{\includegraphics{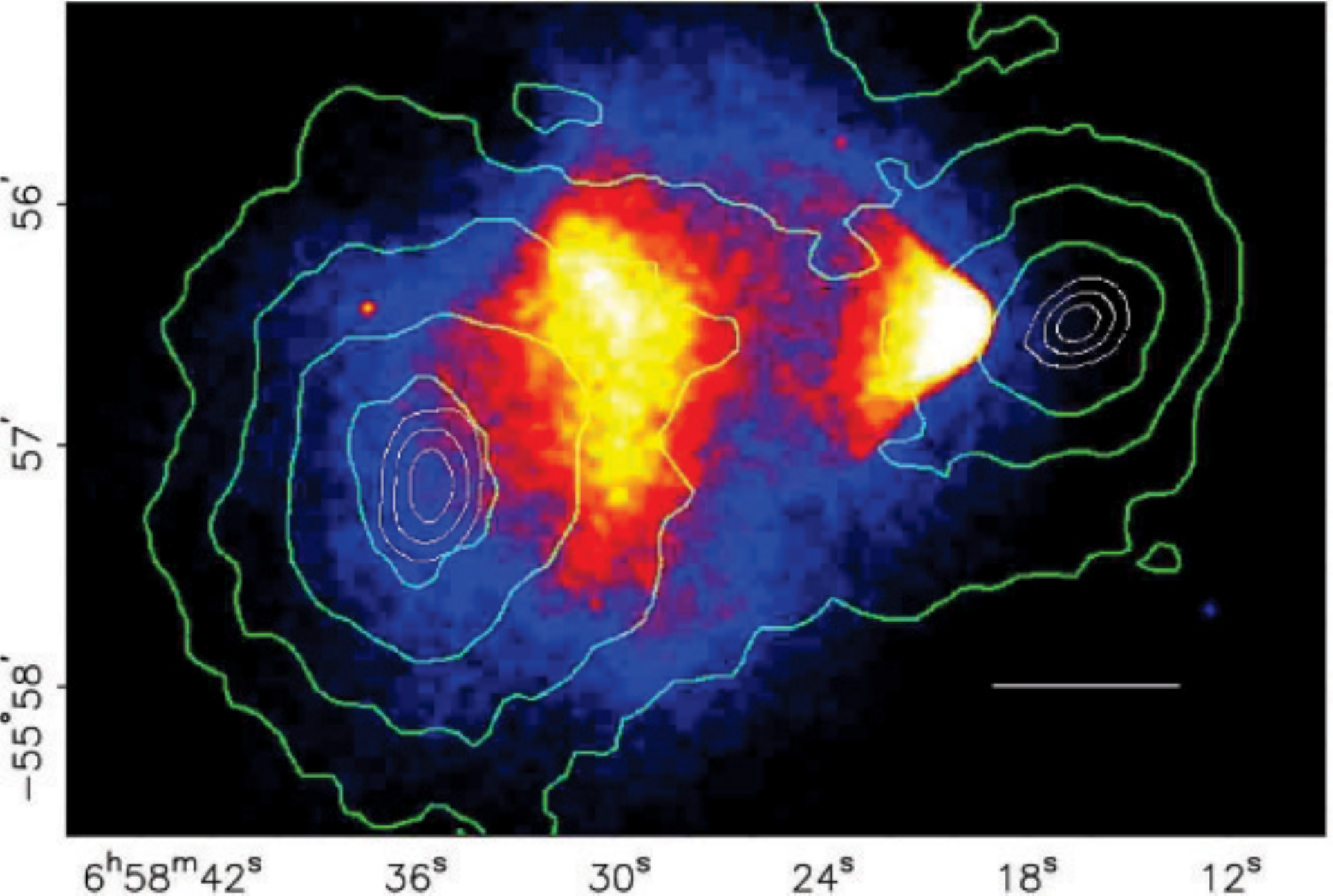}}
\caption{Left panel: the Bullet cluster (1E 0657-56) consists of two colliding cluster of galaxies. 
The smaller cluster that traversed the larger cluster is on the right. X-ray images come from Chandra (red), and the DM (blue) is obtained through lensing.
Studies of the Bullet cluster, announced in August 2006, provide the best evidence to date for the existence of DM. The spatial offset of total mass-baryonic mass peaks cannot be explained using modified gravity, at $8 \sigma$ level.
Right panel: Bullet cluster, mass density contours, in green, obtained through weak lensing, superimposed over photograph got from HST. (From Clowe\cite{clowe})
}
\end{figure}

In the following, before discussing the Cosmic Shear, I summarize the formalism of gravitational lensing.

As well known photons trajectories are bended by gravitational fields. The photon path is a null geodesic, $ds^2=0$. 
Using the Schwarzschild solution
\begin{equation}
ds^2=c^2 (1+2 \Phi/c^2)dt^2-(1-2 \Phi/c^2)dl^2   \hspace{0.5cm} dl^2=dx^2+dy^2+dz^2
\end{equation}
being $c$ the velocity of light and $\Phi$ the gravitational potential of the mass, we get
\begin{equation}
dt=1/c\sqrt{\frac{1-2 \Phi/c^2}{(1+2 \Phi/c^2)}} dl \simeq 1/c (1-2 \Phi/c^2) dl
\end{equation}
 
According to Fermat's principle photons only follow optical paths with extrema propagation time. The paths are those stationary with respect to a small variation $\delta t$. 
Since 
\begin{equation}
ct= \int (1-2 \Phi/c^2) dl
\end{equation}
this equation corresponds to a light beam propagating in a trasperent medium with refraction index
\begin{equation}
n= (1-2 \Phi/c^2) 
\end{equation}
Applying Fermat's principle, we can derive the value of the angle of deflection:
\begin{equation}
\alpha= -2/c^2 \int_S^0 \Delta_\perp \Phi dl
\end{equation}
This is the general expression of a deflection angle for thin lenses in the weak field limit.
In general relativity the deflection angle is given by (see Fig. 6, left panel, for notation)
\begin{equation}
  \hat\alpha = \frac{4GM}{c^2\,\xi}\;.
\label{defless}
\end{equation}
The deflection angle can also be written as
\begin{equation}
\hat{\vec\alpha}(\vec\xi) = \frac{4G}{c^2}
\int d^2 \xi' \Sigma(\vec \xi)
\frac{\vec \xi-\vec \xi'}{|\vec \xi-\vec \xi'|^2}
\label{eq:3.4}
\end{equation}
where 
\begin{equation}
 \Sigma(\vec \xi) \equiv \int dz \,\rho(\xi_1,\xi_2,z)
\label{surfa}
\end{equation}
is the surface mass density, namely the projected density on a plane orthogonal to the light ray. 

The lens equation connecting the true position $\vec\beta$ of the source with the angular positions $\vec\theta$ at which
the observer sees the source, is given by
\begin{equation}
  \vec\beta = \vec\theta-\frac{D_\mathrm{ds}}{D_\mathrm{s}}\,
  \hat{\vec\alpha}(D_\mathrm{d}\vec\theta) \equiv
  \vec\theta-\vec\alpha(\vec\theta)\;,
\label{lens}
\end{equation}
In the case $\beta=0$, using the lens equation Eq. (\ref{lens}), recalling Eq. (\ref{defless}) and
using $\vec\xi=D_\mathrm{d}\vec\theta$ we get the angular radius of the Einstein ring
\begin{equation}
\theta_E= \alpha \frac{D_{ds}}{D_s}
\end{equation}
having mass
\begin{equation}
M_E=\theta^2_E \frac{c^2}{4G} \frac{D_d D_s}{D_s}
\end{equation}

Multiple images are allowed if the equation has more solutions. The possibility of formation of multiple images 
can be quantified through the convergence
\begin{equation}
  \kappa(\vec\theta) = 
  \frac{\Sigma(D_\mathrm{d}\vec\theta)}{\Sigma_\mathrm{cr}}
\end{equation}
where $\Sigma_\mathrm{cr}$ is the critical surface mass density
\begin{equation}
    \Sigma_\mathrm{cr} = \frac{c^2}{4\pi G}\,
  \frac{D_\mathrm{s}}{D_\mathrm{d}\,D_\mathrm{ds}}\;,
\label{eq:3.7}
\end{equation}
In the case $\kappa\ge 1$ (.~$\Sigma\ge\Sigma_\mathrm{cr}$), 
the lens gives rise to multiple images. 
The deflection angle can also be expressed in terms of the convergence term:
\begin{equation}
\vec\alpha(\vec\theta) = \frac{1}{\pi}\,
  \int d^2\theta'\,\kappa(\vec\theta')\,
  \frac{\vec\theta-\vec\theta'}{|\vec\theta-\vec\theta'|^2}\;.
\label{def_con}
\end{equation}
that can be written as the gradient of a two-dimensional potential $\vec\alpha=\nabla\psi$, with
$\psi$ satisfying $\nabla^2\psi(\vec\theta)=2\kappa(\vec\theta)$.

Equation~(\ref{def_con}) implies that the deflection angle can be
written as the gradient of the deflection potential,
\begin{equation}
\psi(\vec\theta) = \frac{1}{\pi} \int d^2\theta'
\kappa(\vec\theta') \ln|\vec\theta-\vec\theta'|
\label{eq:3.9}
\end{equation}

\begin{figure}[htp]
\centering
\begin{tabular}{cc}
\resizebox{4.65cm}{!}{\includegraphics{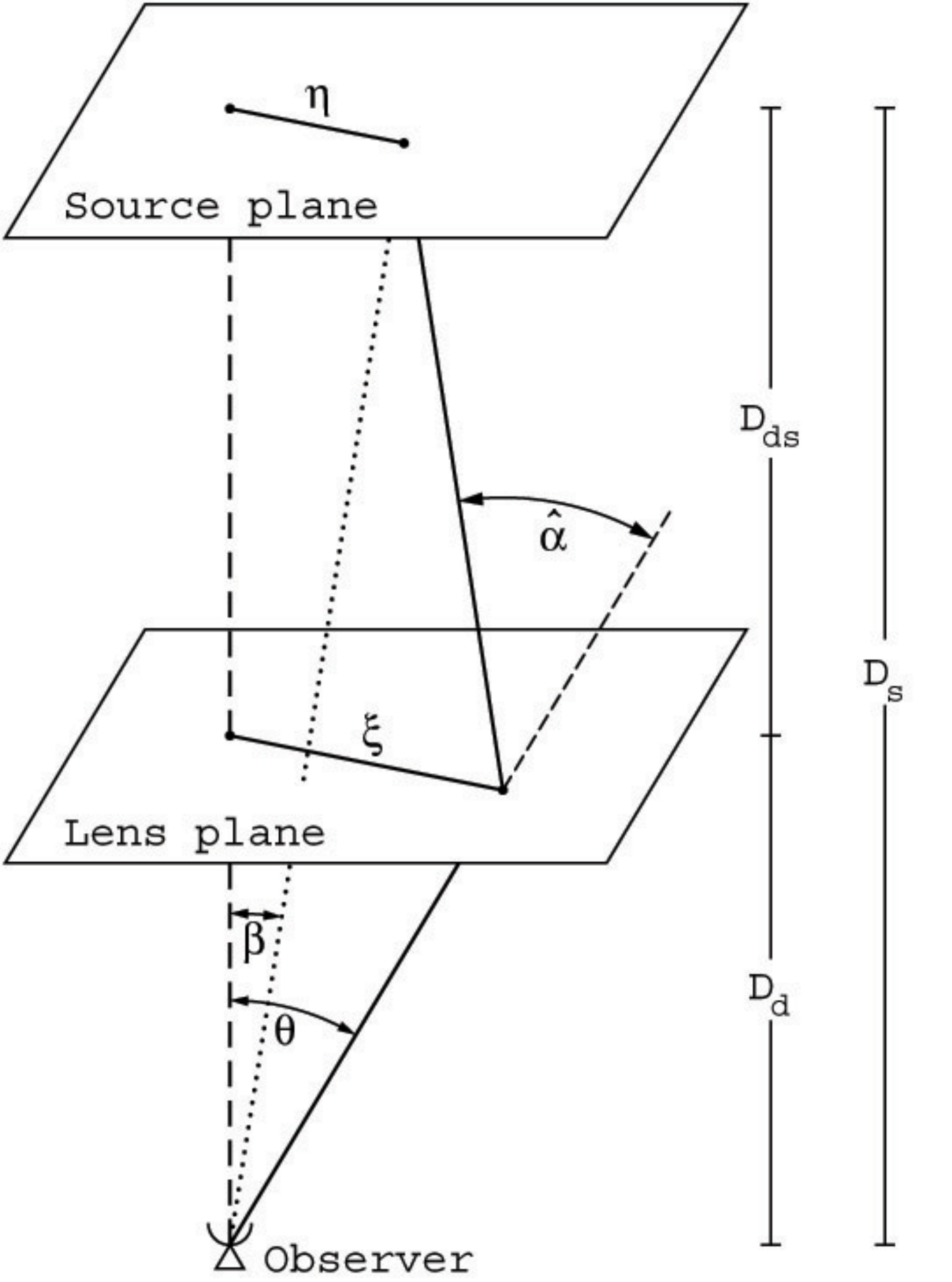}}&
\resizebox{4.65cm}{!}{\includegraphics{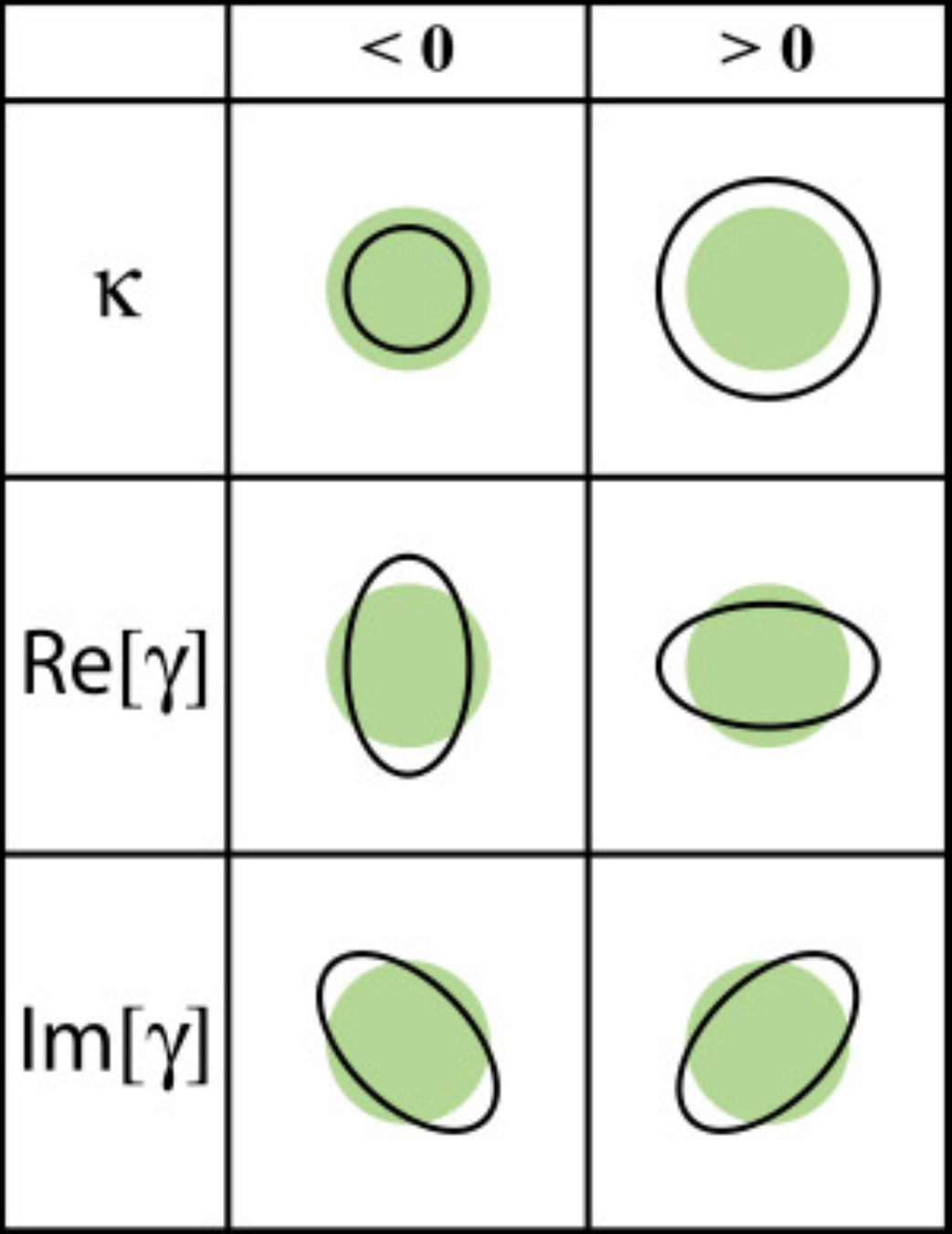}}\\
\resizebox{6.65cm}{!}{\includegraphics{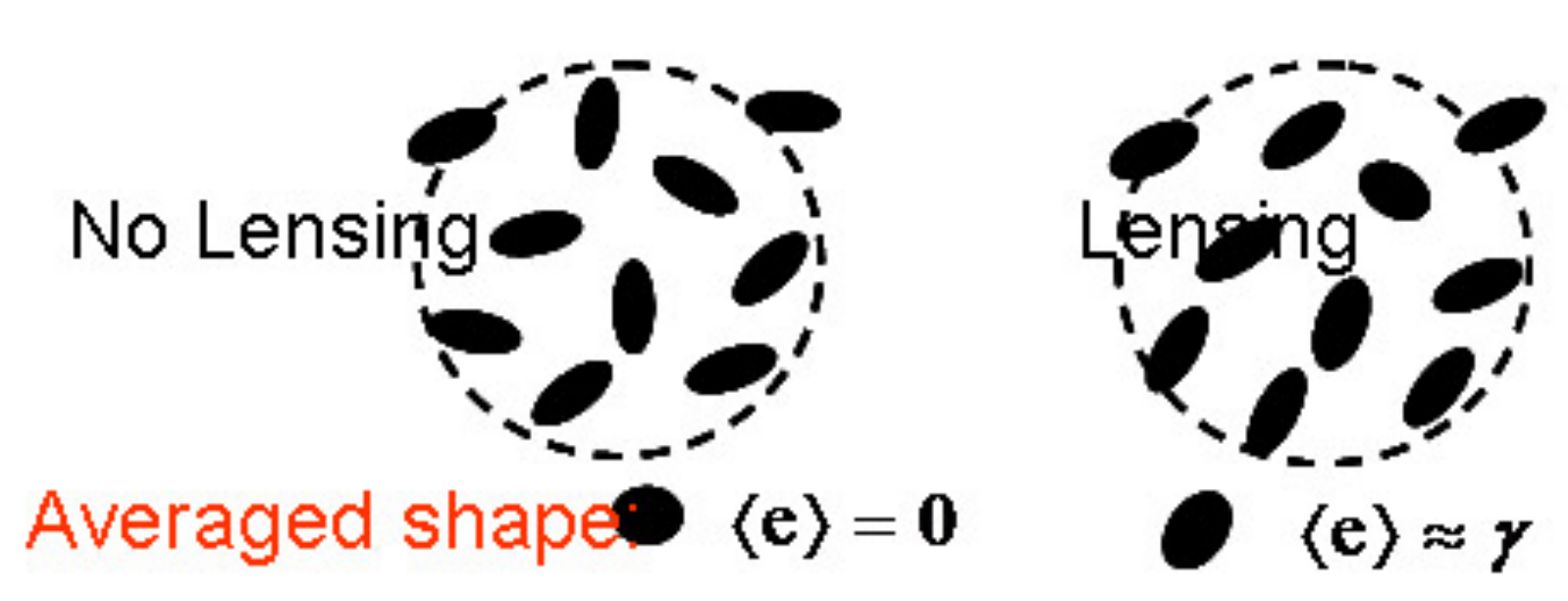}}&
\resizebox{8.65cm}{!}{\includegraphics{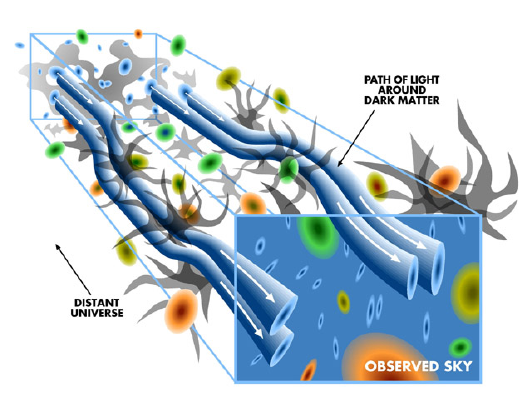}}\\
\caption{Top left panel: notation used in the gravitational lensing formalism. Top right panel: effect of convergence $\kappa$, and shear on a circular source (solid green circle). Bottom left panel: connection among eccentricity and shear. Bottom right panel: artistic representation of the cosmic shear. The picture shows the path of light around the DM of the LSS (large scale structure).
}
\end{tabular}
\end{figure}

\begin{figure}
\resizebox{8.65cm}{!}{\includegraphics{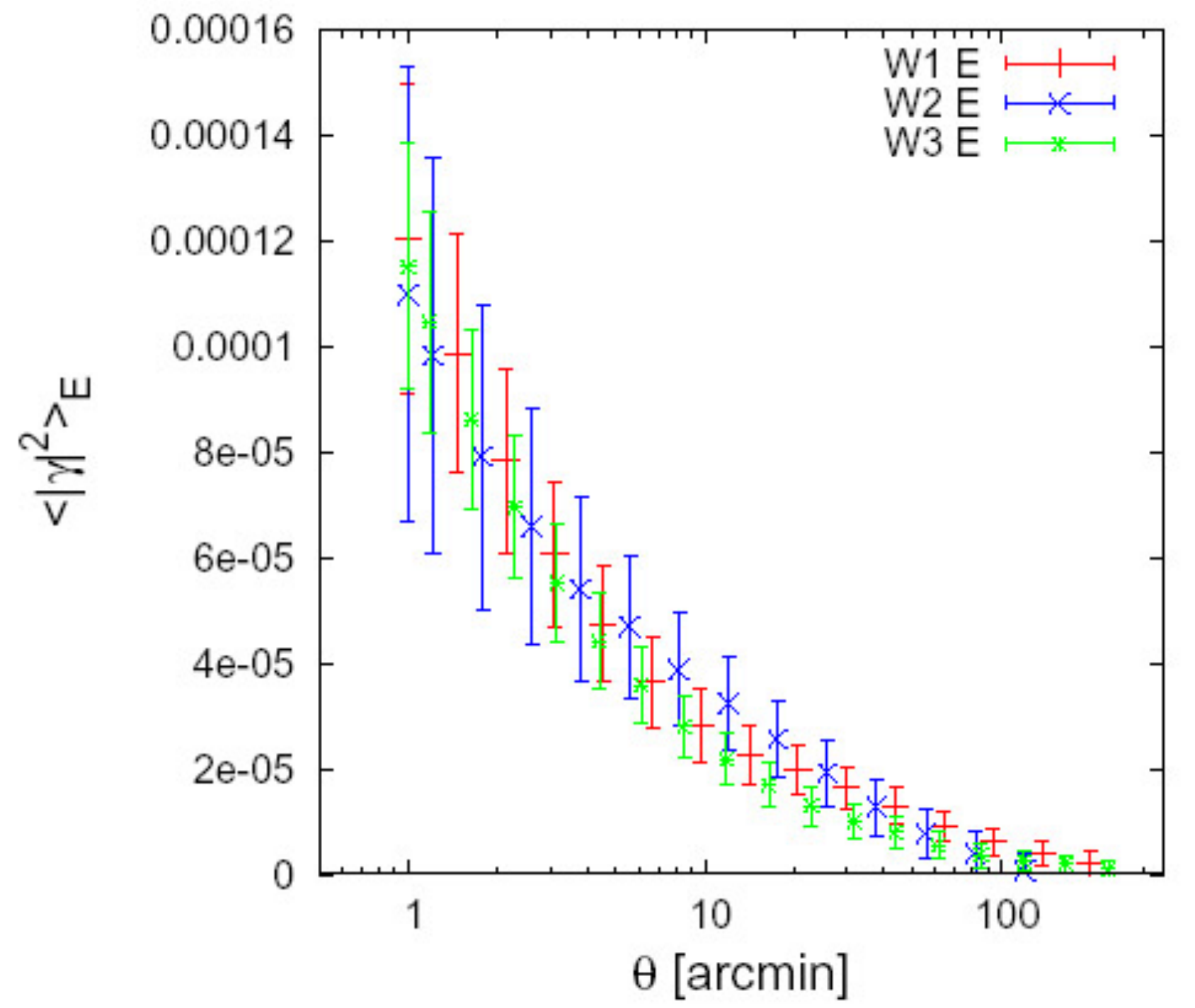}}
\resizebox{7.65cm}{!}{\includegraphics{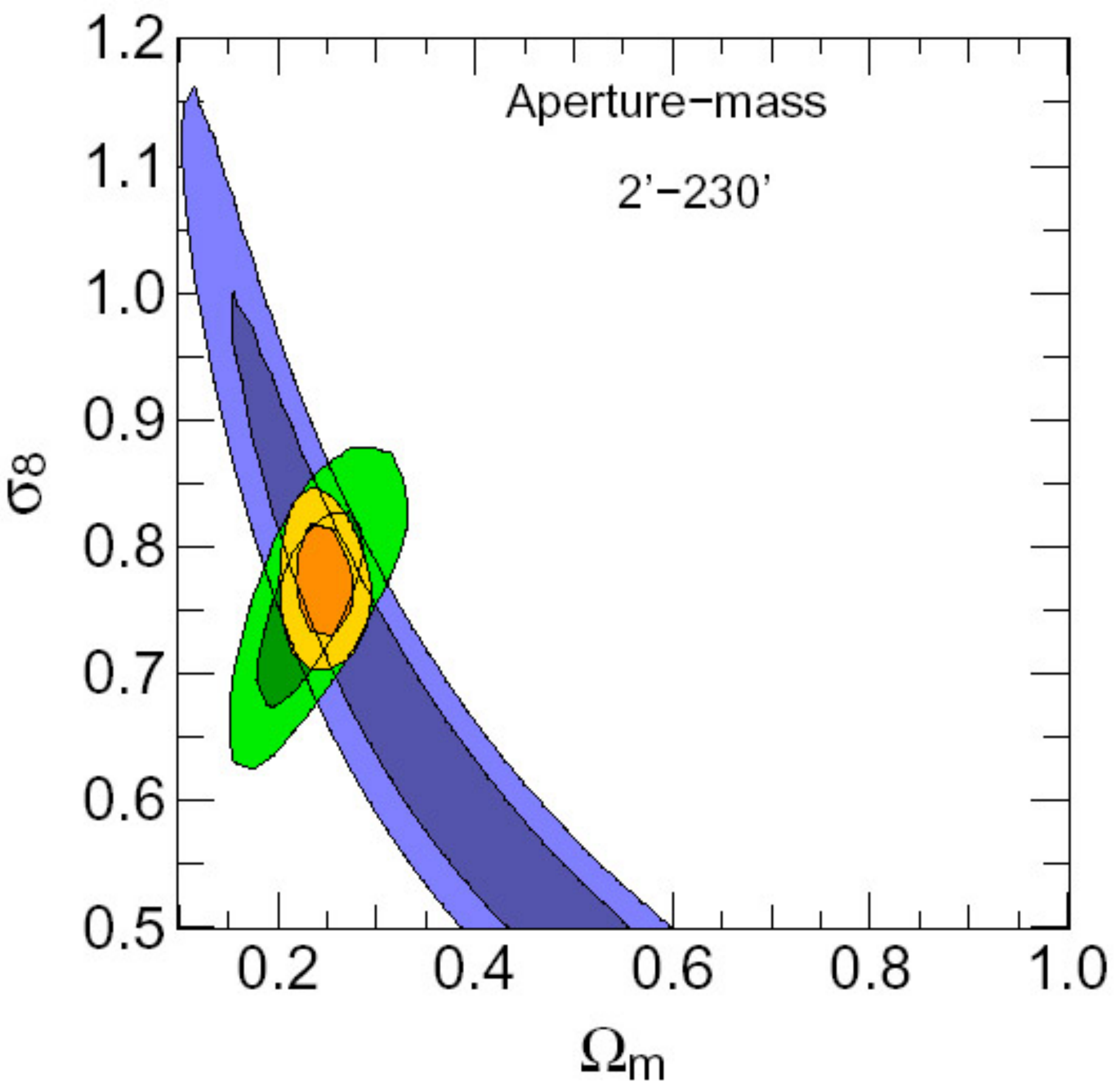}}
\caption{Left panel: Very weak lensing in the CHFTLS: Top-hat E-mode shear signals of the wide fields W1, W2, and W3, up to 200', 120', 230', respectively. Right panel: ($1, 2 \sigma$) comparison of 
CFHTLS result (purple) from Fu et al. (2007)\cite{fu}, with WMAP3 (green contours). In orange the combined WMAP3 and CHFTLS contours. (From \cite{fu}). }
\end{figure}

The effect of lensing is that of producing a distorsion of the images, namely a change of the size, connected to the convergence term, and of the shape, connected to the shear term (see Fig. 6). Surface brightness is conserved\footnote{Surface brightness conservation is due to absence of absorption and emission processes, and Liouville's theorem.}.
The Jacobian matrix describes the image distorsion in the linear regime  
\begin{equation}
  \mathcal{A}(\vec\theta) =
  \frac{\partial\vec\beta}{\partial\vec\theta} =
  \left(\delta_{ij} -
    \frac{\partial^2\psi(\vec\theta)}{\partial\theta_i\partial\theta_j}
  \right) = \left(
    \begin{array}{cc}
      1-\kappa-\gamma_1 & -\gamma_2 \\ 
      -\gamma_2 & 1-\kappa+\gamma_1 \\
    \end{array}
  \right)\;
\label{jacobian}
\end{equation}
where $\gamma\equiv\gamma_1+\mathrm{i}\gamma_2 =
|\gamma|\mathrm{e}^{2\mathrm{i}\varphi}$,
are the components of the shear. 

The, lens transform circumferences in ellipses with semi-major axes:
\begin{equation}
a= 1/(1-\kappa-\gamma); \hspace{0.5cm} b= 1/(1-\kappa+\gamma)
\end{equation}
and moreover produce a magnification of the image represented by
\begin{equation}
A^{-1}=\frac{1}{det A}=\frac{1}{(1-\kappa)^2-|\gamma^2|}
\end{equation}
which depends on both shear and $k$.

Ellipticity is defined as:
\begin{equation}
\epsilon=\frac{1-b/a}{1+b/a} e^{2i\phi} \hspace{0.5cm} \chi=\frac{1-(b/a)^2}{1+(b/a)^2} e^{2i\phi} \hspace{0.5cm} 
\epsilon=\frac{\chi}{1+(1-|\chi^2|)^{1/2)}} e^{2i\phi}
\end{equation}
and the reduced shear
\begin{equation}
|g|=|\epsilon|=\frac{|\gamma|}{1+(1-\kappa)} 
\end{equation}

While, as already seen, the strong lensing regime is defined as $\kappa \leq 1$, the weak lensing regime is characterized 
by $\kappa<<1$, $\gamma <<1$, and $g^2 \simeq 0$. In this regime
\begin{equation}
g \simeq \gamma/(1-\kappa) \simeq \gamma
\end{equation}
and 
\begin{equation}
\chi \simeq \chi^S+2(g-\chi^S \mathcal{R} (g \chi^*))
\end{equation}

\begin{figure}
\resizebox{9.65cm}{!}{\includegraphics{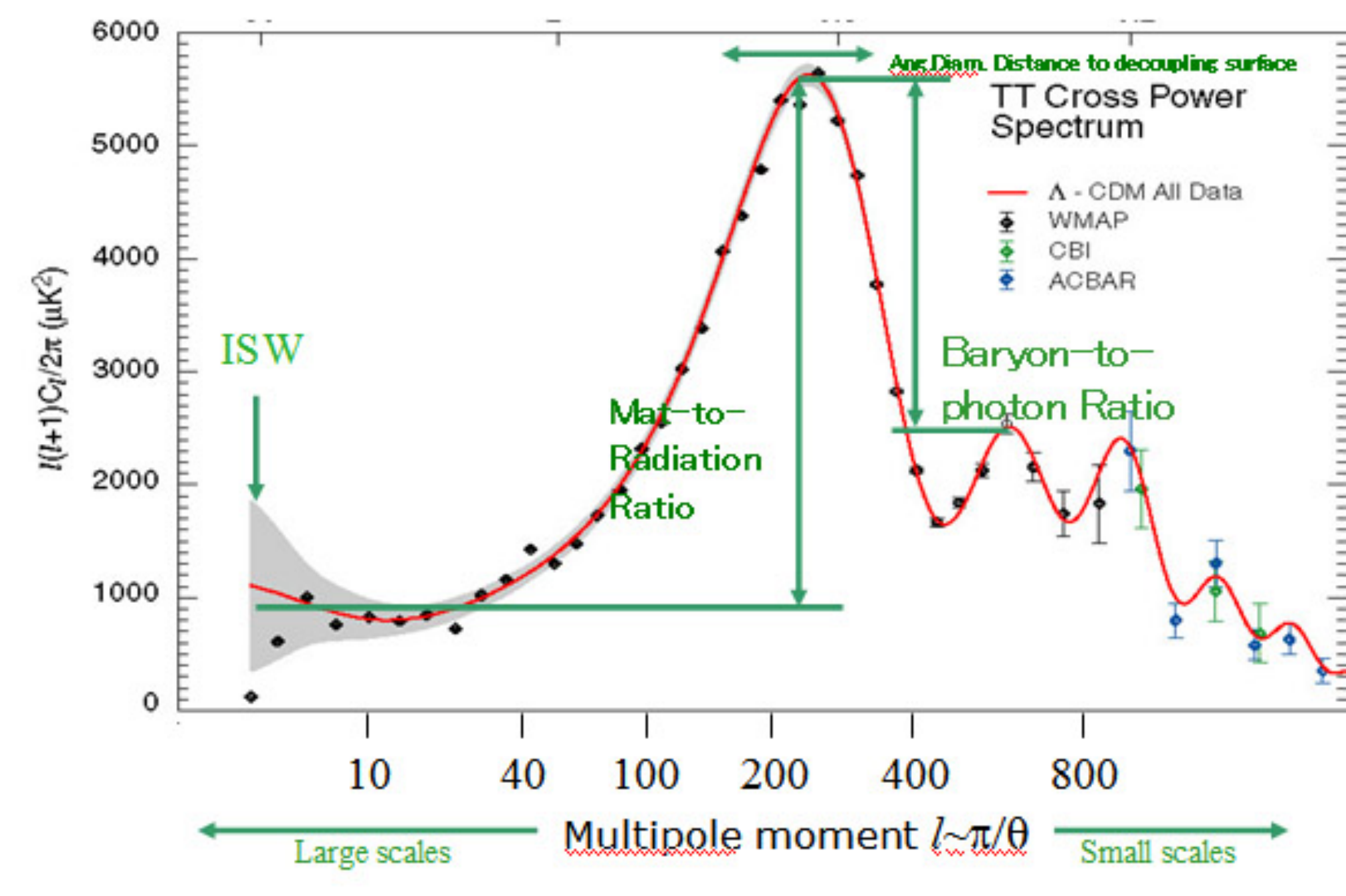}}
\resizebox{12.4cm}{!}{\includegraphics{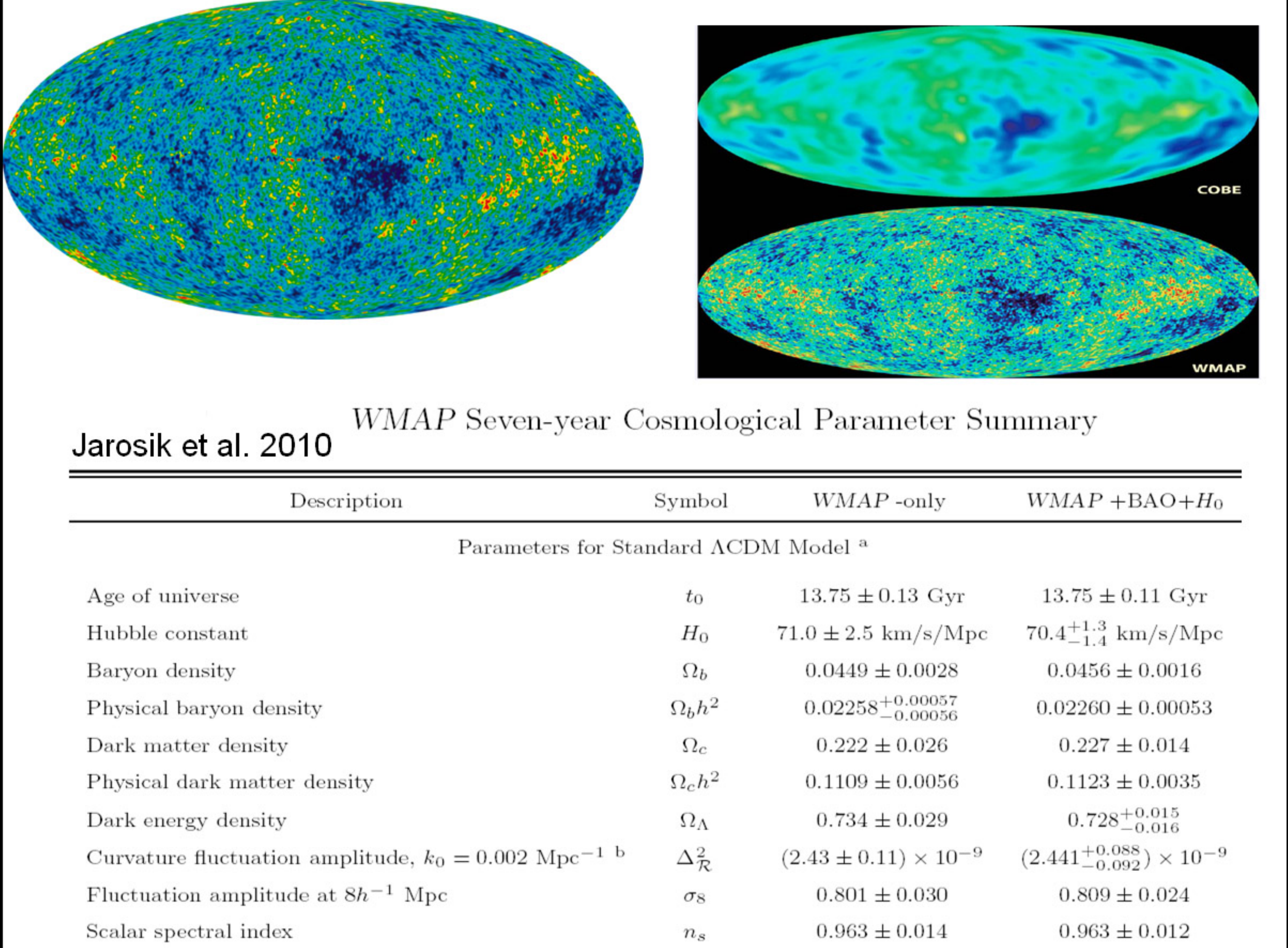}}
\caption{Left panel: WMAP 7-year temperature (TT) power spectrum (adapted from Jarosik et al.\cite{jarosik}). The solid line is $\Lambda$CDM best fit. Right panel: WMAP and COBE sky, snd summary of the WMAP seven-year cosmological parameters (from Jarosik et al.\cite{jarosik}).}
\end{figure}

\begin{figure}
\resizebox{6.65cm}{!}{\includegraphics{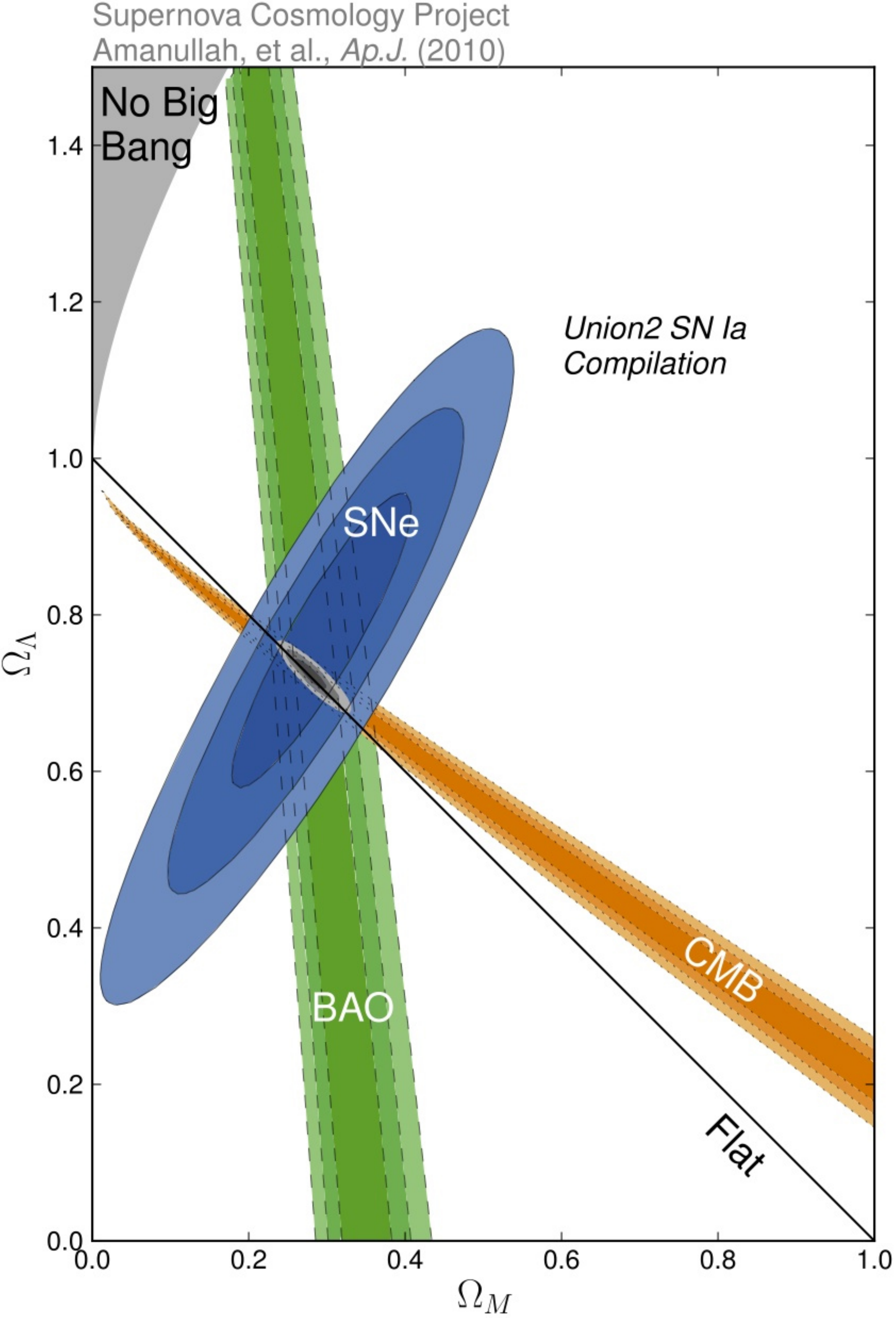}}
\caption{$\Omega_{\Lambda}$-$\Omega_M$ constraints by combining CMB, BAO and SNe (\cite{amanullah})}
\end{figure}

\begin{figure}
\resizebox{8.65cm}{!}{\includegraphics{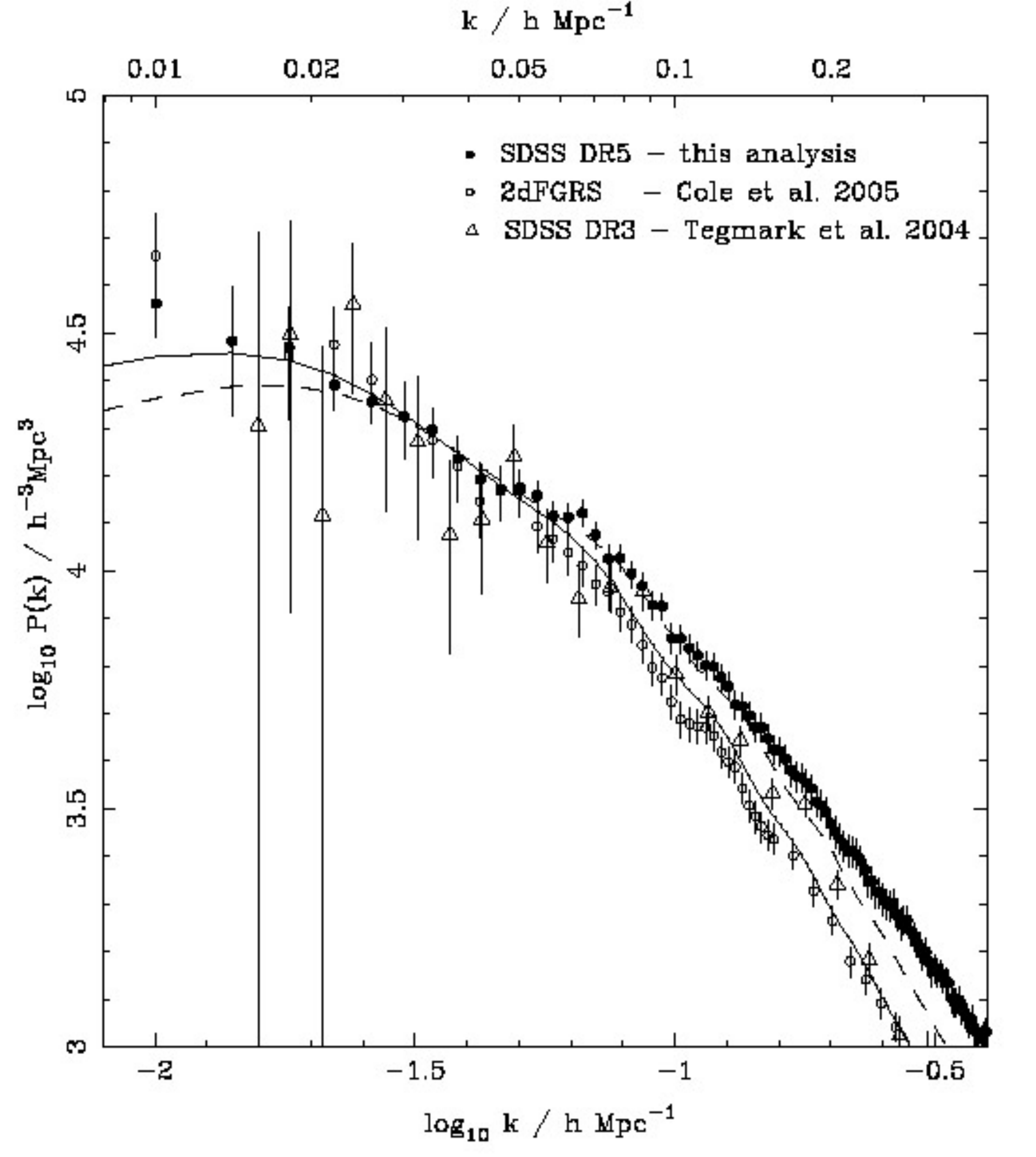}}
\caption{SDSS DR5 power spectrum compared with previous determinations. 
Dashed line: best-fit model (CDM models) over $0.01 < k < 0.06 h/Mpc$. Solid line: best-fit model over $0.01 < k < 0.15 h/Mpc$.
(from Percival et al. (2007)\cite{percivall}).
}
\end{figure}

Since one expects that sources orientation
is isotropically distributed $<\chi^S> \simeq 0$ then $<\chi> \simeq 2 \gamma =2 g$, and the image ellipticity is an unbaiased
estimate of shear. 

Weak lensing can be used to determine the mass or density profile of a structure. Writing the shear as a tangential part, $\gamma_t$, and curl
$\gamma_x$
\begin{equation}
\gamma=\gamma_t+\gamma_x
\end{equation}
and using the fact that for a circularly symmetric lens the curl vanish and the tangential part is
\begin{equation}
<\gamma_t> \equiv \frac{\overline{\Sigma}(R)-\Sigma(R)}{\Sigma_c(R)}
\end{equation}
with the mean projected mass density interior
to the radius R given by
\begin{equation}
\overline{\Sigma}(R)= 2 \int_0^R x \Sigma(x)dx =\frac{M(R)}{4 \pi R^2}
\end{equation}
and $R= \theta D_{d}$. The DM distribution is obtained by fitting the observed shear with a chosen density profile with 2 free parameters.

As seen, gravitational lensing acts as a coordinate transformation that distorts the images of background objects near a
foreground mass. The transformation can be split into two terms, the convergence and shear (tangential
stretching). To measure this tangential alignment, it is necessary to measure the ellipticities of the background
galaxies and construct a statistical estimate of their systematic alignment. 
The fundamental problem is that galaxies are not intrinsically circular, so their measured ellipticity is a combination of their intrinsic ellipticity and the gravitational lensing shear. Typically, the intrinsic ellipticity is much greater than the shear (by a factor of 3-300, depending on the foreground mass). The measurements of many background galaxies must be combined to average down this "shape noise". The orientation of intrinsic ellipticities of galaxies should be almost entirely random, so any systematic alignment between multiple galaxies can generally be assumed to be caused by lensing.
Another major challenge for weak lensing is the correction for the point spread function (PSF) due to instrumental and atmospheric effects, which causes the observed images to be smeared relative to the "true sky". 
This smearing tends to make small objects more round, destroying some 
of the information about their true ellipticity
The PSF typically adds a small level of ellipticity to objects in the image, which is not at all random, and can in fact mimic a true lensing signal. Even
for the most modern telescopes, this effect is usually at least the same order of magnitude as the
gravitational lensing shear, and is often much larger.  

Gravitational lensing by large-scale structure (LSS) is termed Cosmic Shear (see Fig. 6, bottom right). It produces observable pattern of alignments
in background galaxies, but this distortion is only $\simeq 0.1\%-1\%$ - much more subtle than cluster or
galaxy-galaxy lensing. The effect was observed for the firs time in 2000\cite{wittman,bacon,kaiser,vanwa}, and started to be used to put constraints
to $\Omega_M$ (see Fig. 7).

Of utmost importance in determining $\Omega_m$, $\Omega_{\Lambda}$ have been the CMBR experiments
whose value can be better constrained combining CMBR data with data coming for example from SN Ia, which at
high redshifts give information on a combination of $\Omega_m$, $\Omega_{\Lambda}$, and baryonic acoustic oscillations (BAO). 

The temperature anisotropies are expanded in spherical armonics $Y_{\ell m} (\theta, \phi)$: 
\begin{equation}
\frac{\delta T}{T}(\theta, \phi) = \sum_{\ell = 2}^{+\infty}\sum_{m=-\ell}^{+\ell} a_{\ell m}Y_{\ell m}(\theta, \phi) 
\end{equation}
$C_\ell $ is the variance of $a_{l m}$ and can be expressed in the form
\begin{equation}
C_\ell \equiv < |a_{\ell m}|^2 > \equiv \frac{1}{2 \ell+1}
\sum_{m=-\ell}^{\ell}|a_{\ell m}|^2.
\end{equation}

In the hypothesys that fluctuations are Gaussian, the power spectrum, giving $C_\ell$ behavior, contains all needed informations of the CMBR.
Typical information that can be obtained from the CMBR (e.g., $\Omega_B$, $\Omega_M$, $\Omega_{\Lambda}$, etc), are obtained fitting the observations through a cosmological model (e.g., the $\Lambda$CDM model) (see Fig. 8).

Measurements of the power spectrum out to multipole moments corresponding to $l \simeq \pi/\theta$ $\simeq 1000$ (see Fig. 8) 
where $l$ is the angular dimension of fluctuations (in rad), have made possible predictions of fundamental cosmological parameters (see Jarosik et al. 2010\cite{jarosik}.). From WMAP 7-year data alone, was obtained, among the others parameters,  $\Omega_b= 0.0449 \pm 0.0028$, $\Omega_m= 0.222 \pm 0.026$, $\Omega_{\Lambda}=0.734 \pm 0.029$, $H_0= 71.0 \pm 2.5$ km/s (see Fig. 8), furtherly improved when combining the CMBR data to BAO and SNIa (Jarosik\cite{jarosik}). \footnote{Other constraints can be obtained also using experiments on smaller scales than WMAP (e.g., ACBAR\cite{kuo}, and CBI\cite{pearson}, or from astronomical measurements of the Lyman $\alpha$ forest (Croft et al.\cite{croft}
or of the LSS power spectrum  (2dFGRS, Percival\cite{percival}).
}
The results testify in favor of a flat Universe dominated by DM and "dark energy", and a baryon density in agrement with the BBN production of D/H. 

As previously reported, an improvement on $\Omega_{\Lambda}$ and $\Omega_M$ estimate can be obtained using data from SNIa. In 1998 
observations of those kind of supernovae suggested that the universe expansion has been accelerating\cite{riess,perlmutter} since $z \simeq 0.5$\cite{riess1}. The deceleration parameter is connected to the combination of $\Omega_{\Lambda}$ and $\Omega_M$ (at $a=1$) by:
\begin{equation}
-q=\Omega_{\Lambda,0}-\Omega_{m,0}/2
\end{equation}
If $\Omega_M$ is known from other measurements one can obtain $\Omega_{\Lambda}$, and the way round, by using the previous equation. Combining CMB data with SNIa one obtains improvement on the cosmological parameters (see Fig. 9).

Always on large scale, comparison of the matter power spectrum obtained from different surveys (e.g., 2dF galaxy GRS, Sloan Digital Sky Survey (SDSS)) with simulations gives values of $\Omega_m$ in agreement with different measurements. 
Percival et al. (2007)\cite{percivall} found $\Omega = 0.22 \pm 0.04$ over $0.01 < k < 0.15$ h/Mpc (see Fig. 10).

Before concluding this section, I want to recall that on very large scales, it is possible to get an estimate of the density parameter from the distribution of peculiar velocities of galaxies and
clusters. On very large scales, $\lambda$, characterized by linear density perturbation $\delta$ growth, peculiar velocity are given by $v \simeq \delta \lambda H \Omega_m^{0.6}$ (Peebles 1980\cite{peebles}). This tehcnique used in the past by several authors 
gave values of $\Omega_m$ in some case close to 1 (Dressler et al. 1987\cite{dressler}; Dekel et al. 1993\cite{dekel}), and in others more in agreement with recent
results, 0.2-0.5 (Willick \& Strauss 1998\cite{willick}).

\section{Dark matter distribution}

While we don't know what the dark matter (DM) is, we have a reasonable idea of how much of it exists in different structures , how it is distributed, and how fast it is moving. In the case of spiral galaxies this information comes from the rotation curves, as previously reported.

\subsection{Spiral galaxies}

Spiral galaxies consist of several distinct components: a flat, rotating disc of (mainly young) stars and interstellar matter, a central stellar bulge of mainly older stars, which resembles an elliptical galaxy, a near-spherical halo of stars, including many globular clusters, and a supermassive black hole at the very center of the central bulge. 
The stellar disc is exponential with surface brightness $I(r)=I_0 e^{-r/R_D}$, $I_0$ being the central surface brightness and $R_D$ the scale radius ($\simeq 3$ kpc for the MW). For some spirals like NGC 300 (see Fig. 11) the exponential disk goes for at least 10 scalelengths, while in cases like M 33 (see Fig 11) the outer disc is truncated. A fundamntal component, for the purpose of DM studies, is the gas component: HI having a flattish radial distribution and deficient in the centre; CO and $H_2$ roughly exponential distributed and having negligible mass (see Wong \& Blitz 2002\cite{wong}) (see Fig. 12).   A rotation curve (RC) of a galaxy is defined "as the trace of velocities on a position velocity (PV) diagram along the major axis, corrected for the angle between the line-of-sight and the galaxy disk" (Sofue \& Rubin 2001\cite{sofue}) (see Fig. 13). A RC is obtained calculating  the rotational velocity of a tracer (e.g. stars, gas) along the length of a galaxy by measuring the doppler shifts, and then plotting this quantity versus their respective distance away from the centers

\begin{figure}
\resizebox{12.65cm}{!}{\includegraphics{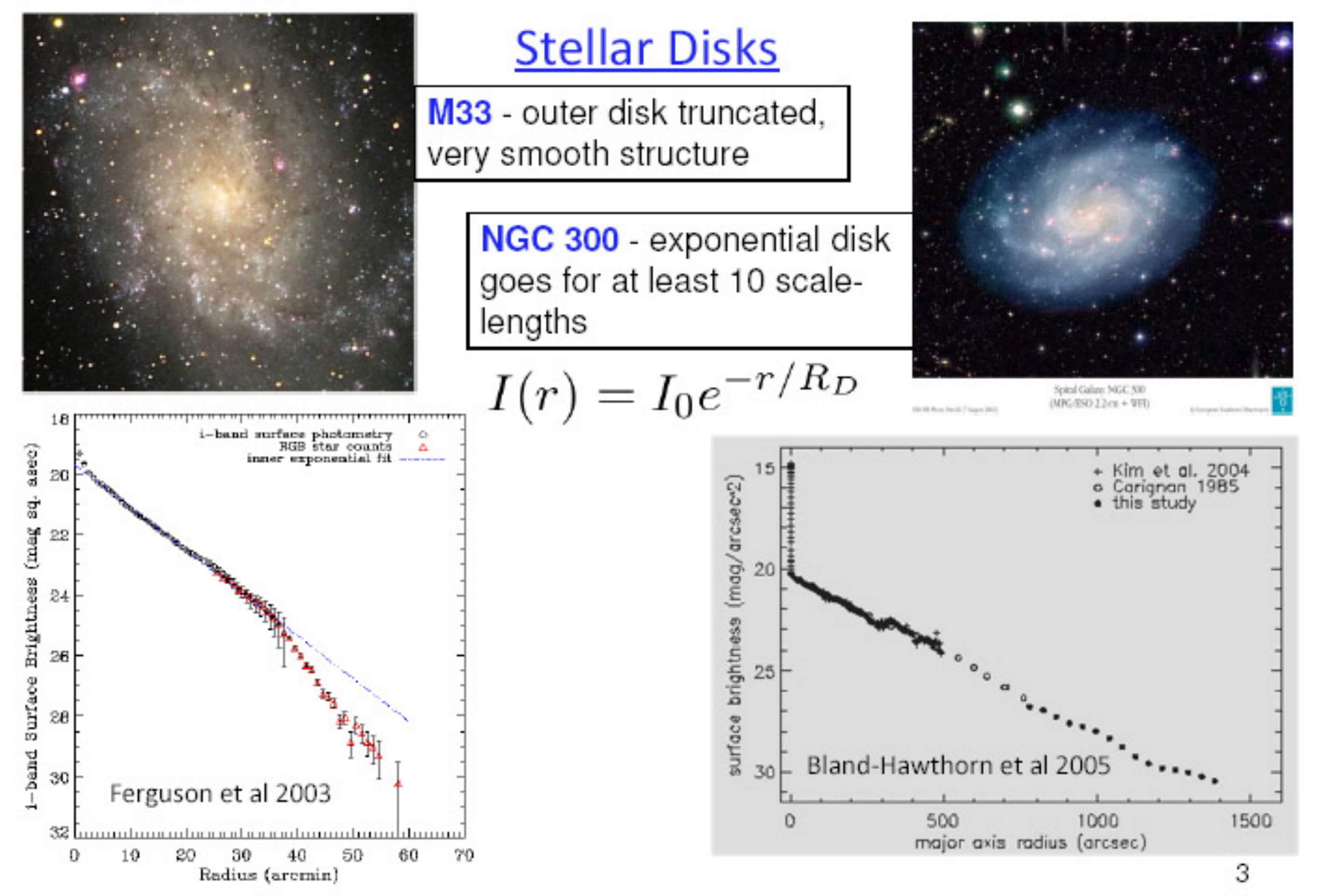}}
\caption{Two examples of spiral galaxies with truncated and non-truncated disks.}
\end{figure}

Tracing the intensity-weighted velocities
\begin{equation}
V_{int}= \int I(v) v dv /\int I(v) dv
\end{equation}
where $I(v)$ is the intensity profile at a given radius as a function of the radial velocity, the rotation velocity is then given by
\begin{equation}
V_{rot}= (V_{int}-V_{sys})/sini
\end{equation}

\begin{figure}
\resizebox{13.65cm}{!}{\includegraphics{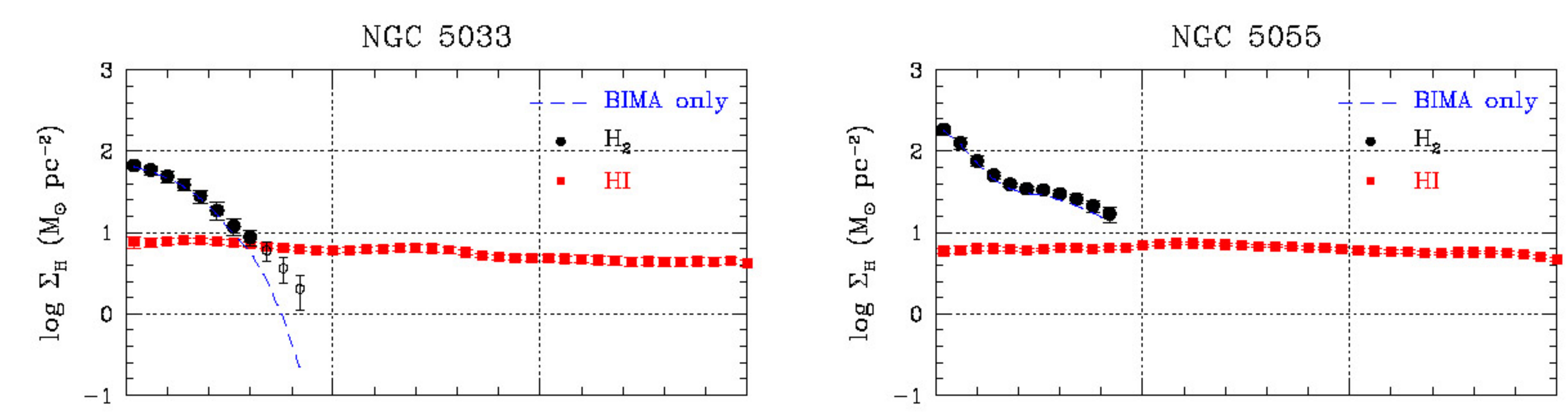}}
\caption{HI and $H_2$ distribution in spirals. From Wong \& Blitz 2002\cite{wong}}
\end{figure}

where i is the inclination angle and $V_{sys}$ the systemic velocity of the galaxy.
Circular velocities can be obtained from spectroscopy from the optical emission lines H$\alpha$ or Na or that of neutral hydrogen (HI), or carbon monoxide (CO). in order to get the distribution of DM in the galaxy one can use several tecniques. The first is the "fit technique".

One writes the rotatio velocity as:
\begin{equation}
V^2_{tot}=V^2_{disk,*}+V^2_{HI}+V^2_{halo}+(V^2_{bulge})
\label{vv}
\end{equation}
where the stellar disc velocity, given by $V^2_{disk,*}=(GM_D/2 R_D)  x^2 B(x/2)$ (where $x=r/R_D$, and $B=I_0K_0-I_1K_1$ ($I_0$, $I_1$, $K_0$, $K_1$ are Bessel's functions) is obtained from I-band photometry, $V^2_{HI}$ is obtained from radio observations of the 21 cm line of HI, $V^2_{halo}$ is modeled starting from a theoretical density profile, and getting $V^2$. The most used profiles are: the Navarro-Frenk-White (NFW) profile, the Einasto profile, the Burkert profile, or the pseudo-isothermal profile (ISO). The quoted profiles are given by (see also Fig. 14)

\begin{figure}
\resizebox{9.65cm}{!}{\includegraphics{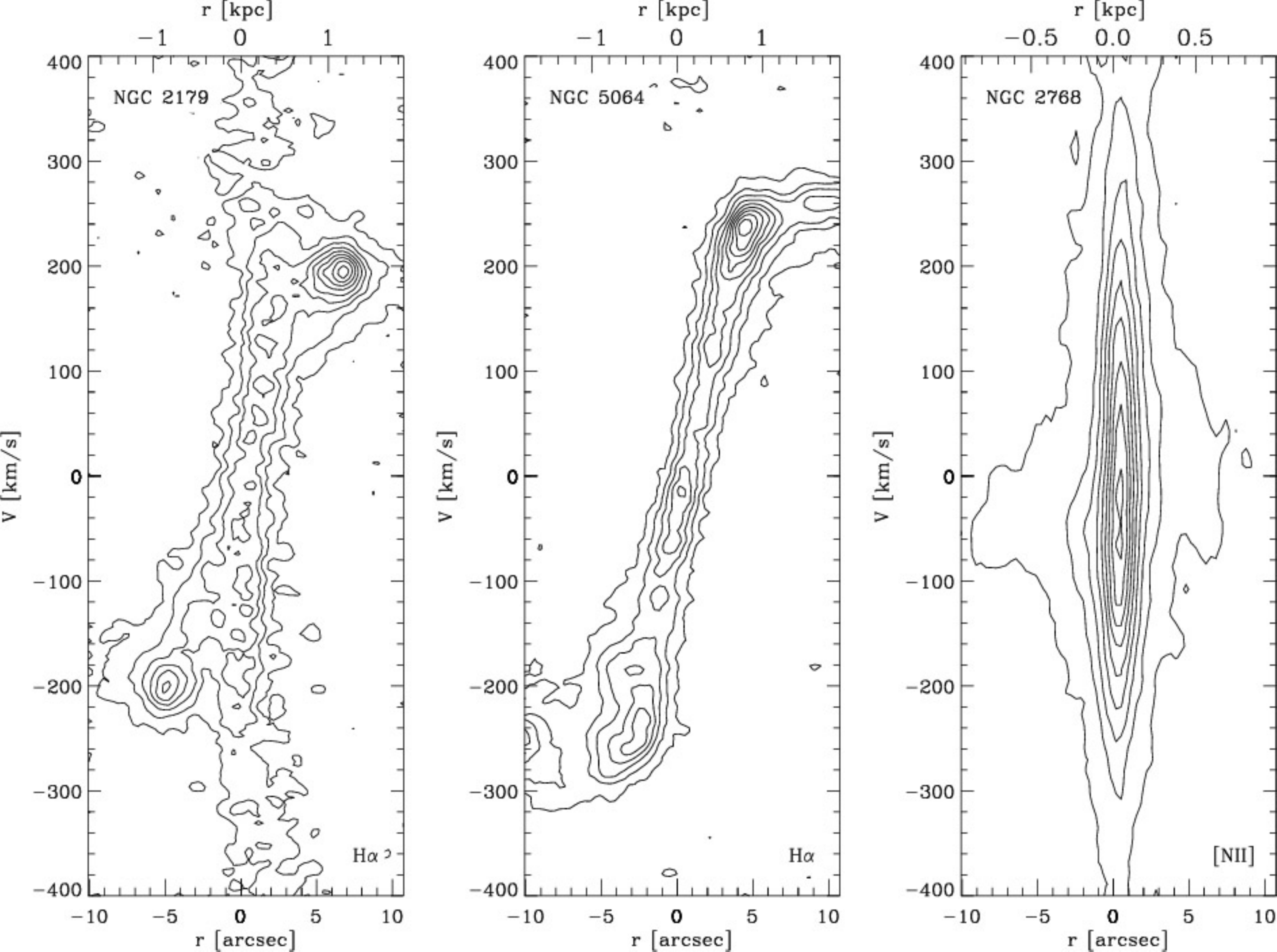}}
\resizebox{6.65cm}{!}{\includegraphics{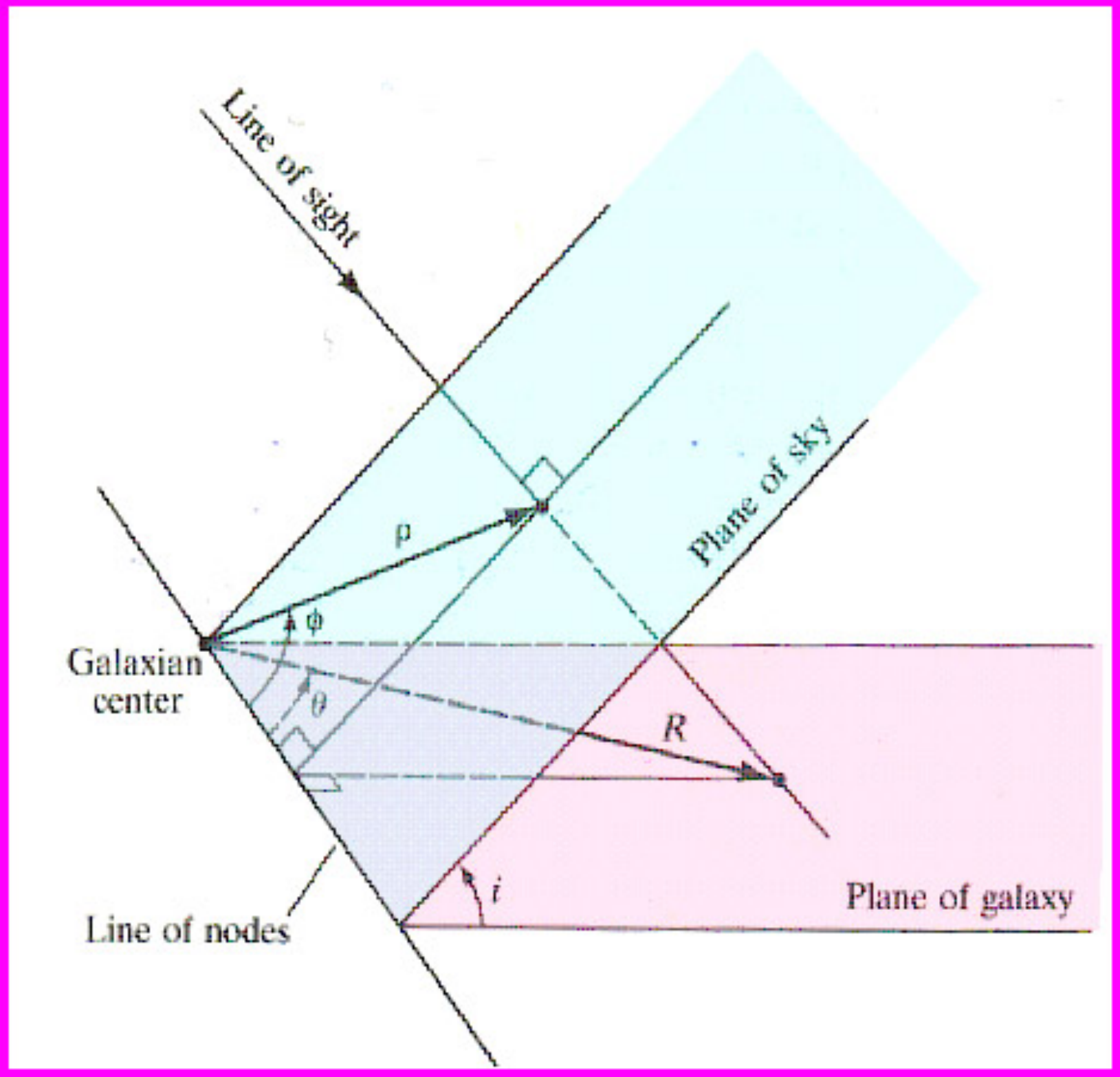}}
\caption{Left panel: example of position velocity diagram. Right panel: correction line-of-sight, plane of the galaxy to get the rotation curve.}
\end{figure}

\begin{equation}
\rho(r)_{NFW}=\frac{\rho_0}{(r/r_s)(1+(r/r_s))^2}
\label{eq:navarr}
\end{equation}
where $\rho_0$ and the "scale radius", $r_s$, are parameters which vary from halo to halo.

\begin{figure}
\resizebox{9.65cm}{!}{\includegraphics{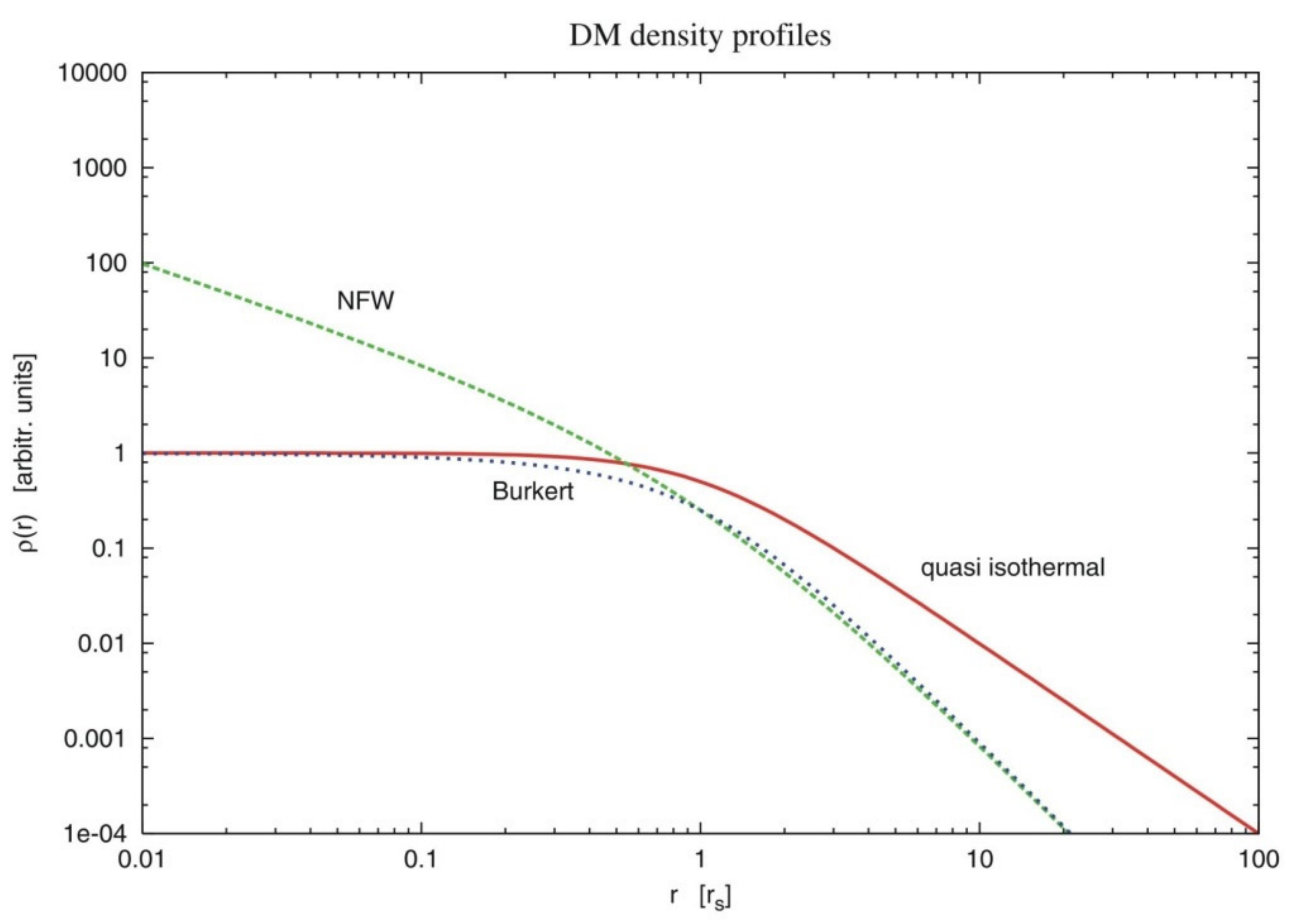}}
\caption{Comparison of different density profiles: NFW, Burkert, ISO.} 
\end{figure}

\begin{figure}
\resizebox{12.65cm}{!}{\includegraphics{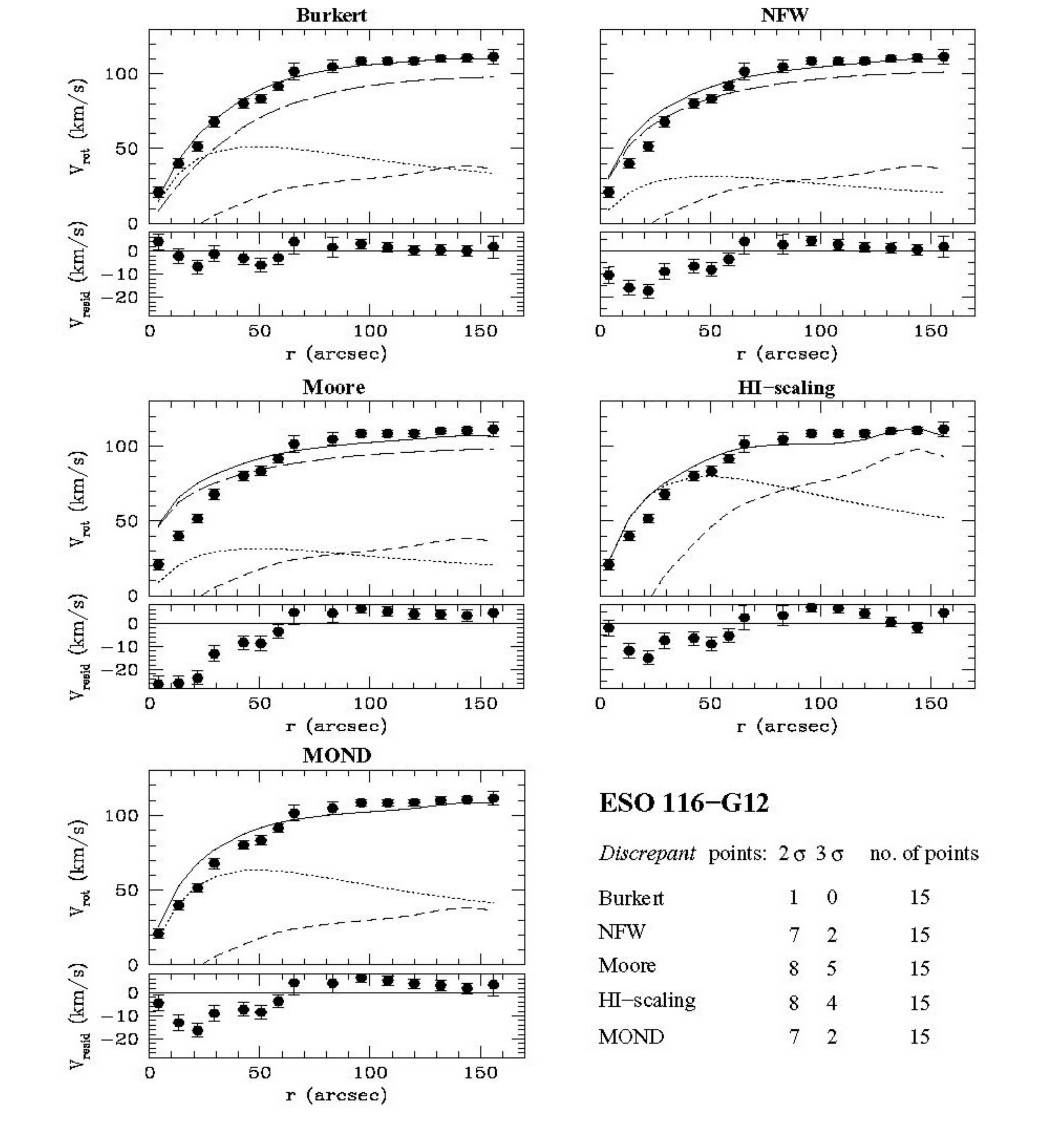}}
\caption{Mass models for the galaxy Eso 116-G12. Solid line: best fitting model;
long-dashed line: DM halo; dotted: stellar disc; dashed: gaseous disc. 
1 kpc corresponds to 13.4 arcsec. Below: residuals: $(V_{obs}-V_{model})$. From Gentile et al. 2004\cite{gentile}} 
\end{figure}

\begin{figure}
\resizebox{8.65cm}{!}{\includegraphics{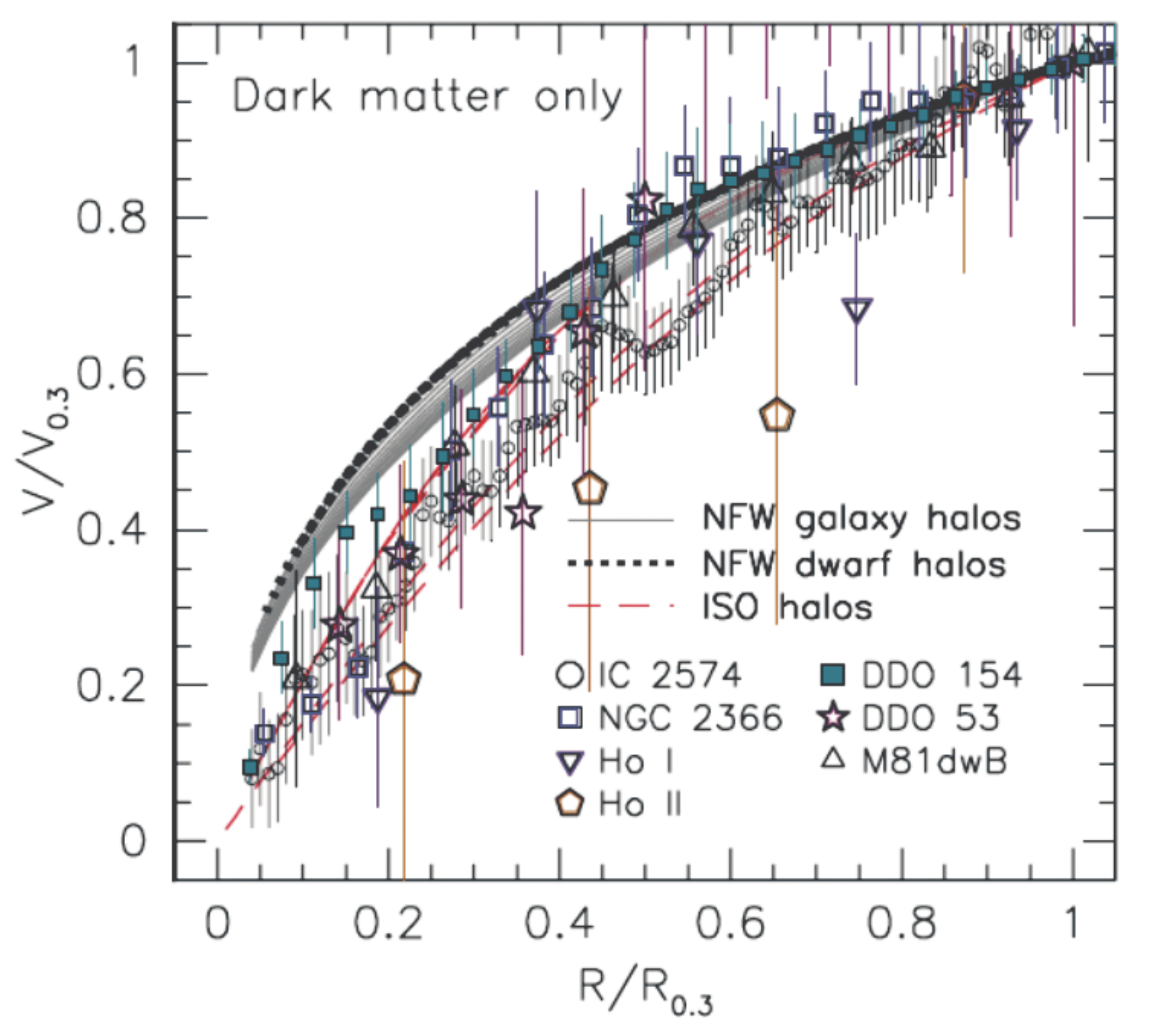}}
\caption{Comparison of seven THINGS dwarfs with NFW model. From Oh et al. 2010\cite{oh}.} 
\end{figure}

\begin{equation}
ln \rho(r)_{Einasto}/\rho_{-2}=\frac{-2/\alpha}{(r/r_{-2})^{\alpha}-1)}
\label{einas}
\end{equation}
where $r_{-2}$ is defined as the radius at which 
\begin{equation}
\frac{d ln \rho}{d ln r}=-2
\end{equation}

\begin{equation}
\rho(r)_{Burkert}=\frac{\rho_0}{(1+r/r_0)(1+(r/r_0)^2)}
\end{equation}
where $\rho_0$, $r_0$, are the central density and radius.

\begin{equation}
\rho(r)_{ISO}=\frac{\rho_0}{(1+(r/r_c)^2)}
\end{equation}

Not taking into account the bulge, the model involving Eq. (\ref{vv}) contains three free parameters: disk mass, halo central density
and core radius (halo length-scale) ($M_D$, $r_0$, $\rho_0$ (if one uses the Burkert model)), 
that can be obtained by best fitting the data to the model.

Other two methods that can be used to get the distribution of DM are the maximum and minimum disk methods (see Battaner \& Florido 2000\cite{battaner}; Simon et al. 2005\cite{simon1}). 
The result of the quoted analysis shows that smaller galaxies are denser and have a higher proportion of dark matter.  
The central density profile is usually well fitted by a Burkert or ISO profile (as shown in Fig. 15, 16) (e.g., Gentile et al. (2004)\cite{gentile}, Oh et al. 2010\cite{oh}) but as shown by de Blok et al. 2008\cite{deblok}, using the THINGS (HI survey of uniform and high quality data) galaxies, for galaxies having $M_B > -19$ the core dominate, and the ISO model fits significantly better than the NFW model, while if $M_B < -19$ the NFW profile or the ISO profile statistically fit equally well.

General results from several samples including THINGS, are that spirals are characterized by small non-circular motions, no DM halo elongation, ISO halos often preferred over NFW with a central core of size $\simeq 2 R_D$. At larger radii the profiles are compatible with NFW. 

For sake of precision, I want to recall that for what concerns the core-like structure of spirals, Simon et al. (2003, 2005)\cite{simon, simon1}
showed that five low-mass spiral galaxies NGC2976, NGC 4605, NGC 5949, NGC 5963, and NGC 6689 have density profiles with inner slope $\alpha$, spanning values from 0 to 1.2.
The large differences in inner slope could be explained in terms of the formation histories and environment role (Del Popolo 2012\cite{delpopolo}). 

\begin{figure}
\resizebox{9.65cm}{!}{\includegraphics{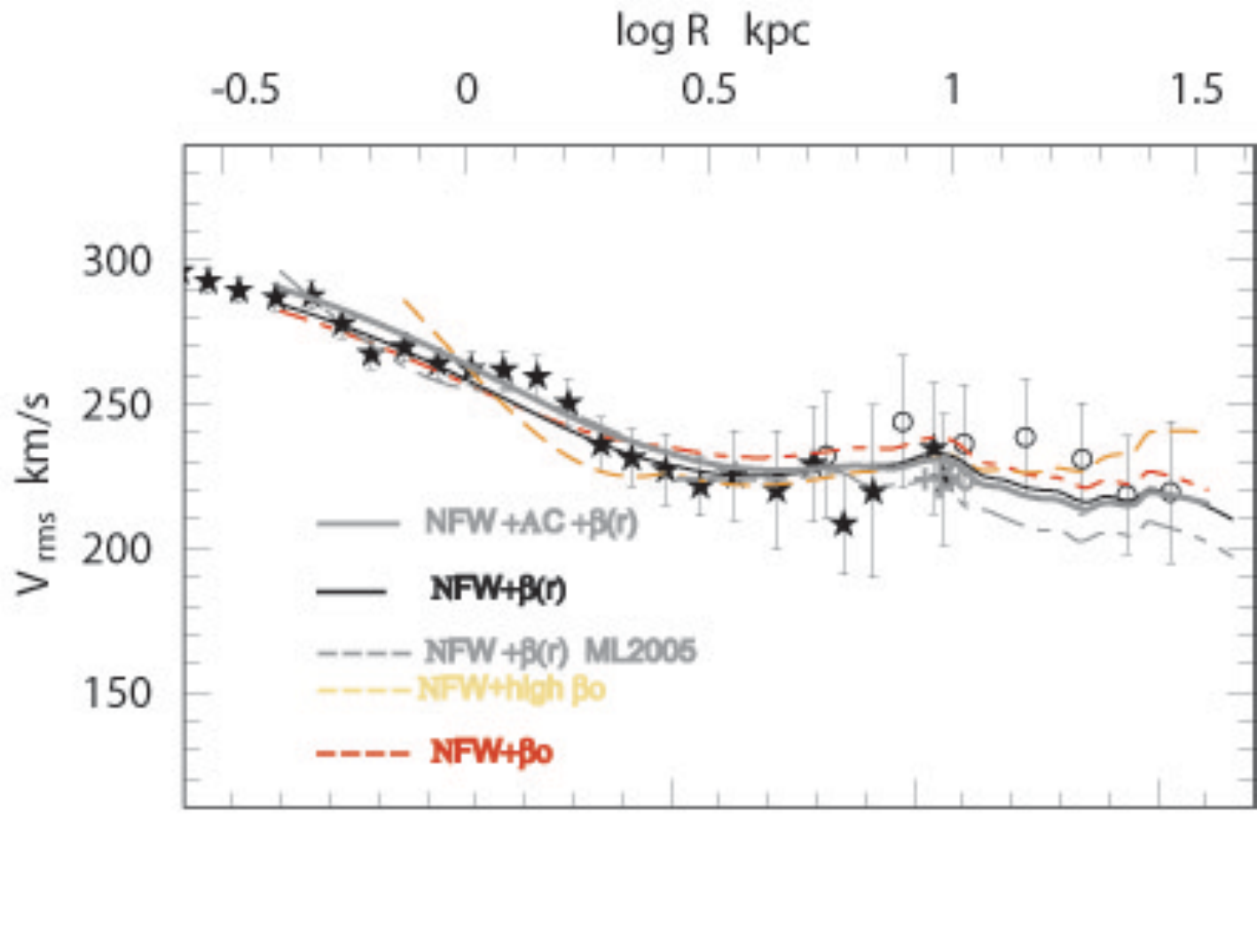}}
\caption{NGC 4374: jeans modelling of PN data with a stellar spheroid plus a NFW halo. NFW stands for (Navarro-Frenk-White), $\beta_0$ is the central value of $\beta(r)$, and is used as a fitting parameter. AC stands for adiabatic contraction, and ML2005 stands for Mamon \& Lokas (2005)\cite{mamon}, 
and means that $\beta(r)$ is chosen from that paper (from Napolitano et al. 2011\cite{napolitano}) 
} 
\end{figure}

\begin{figure}
\resizebox{10.65cm}{!}{\includegraphics{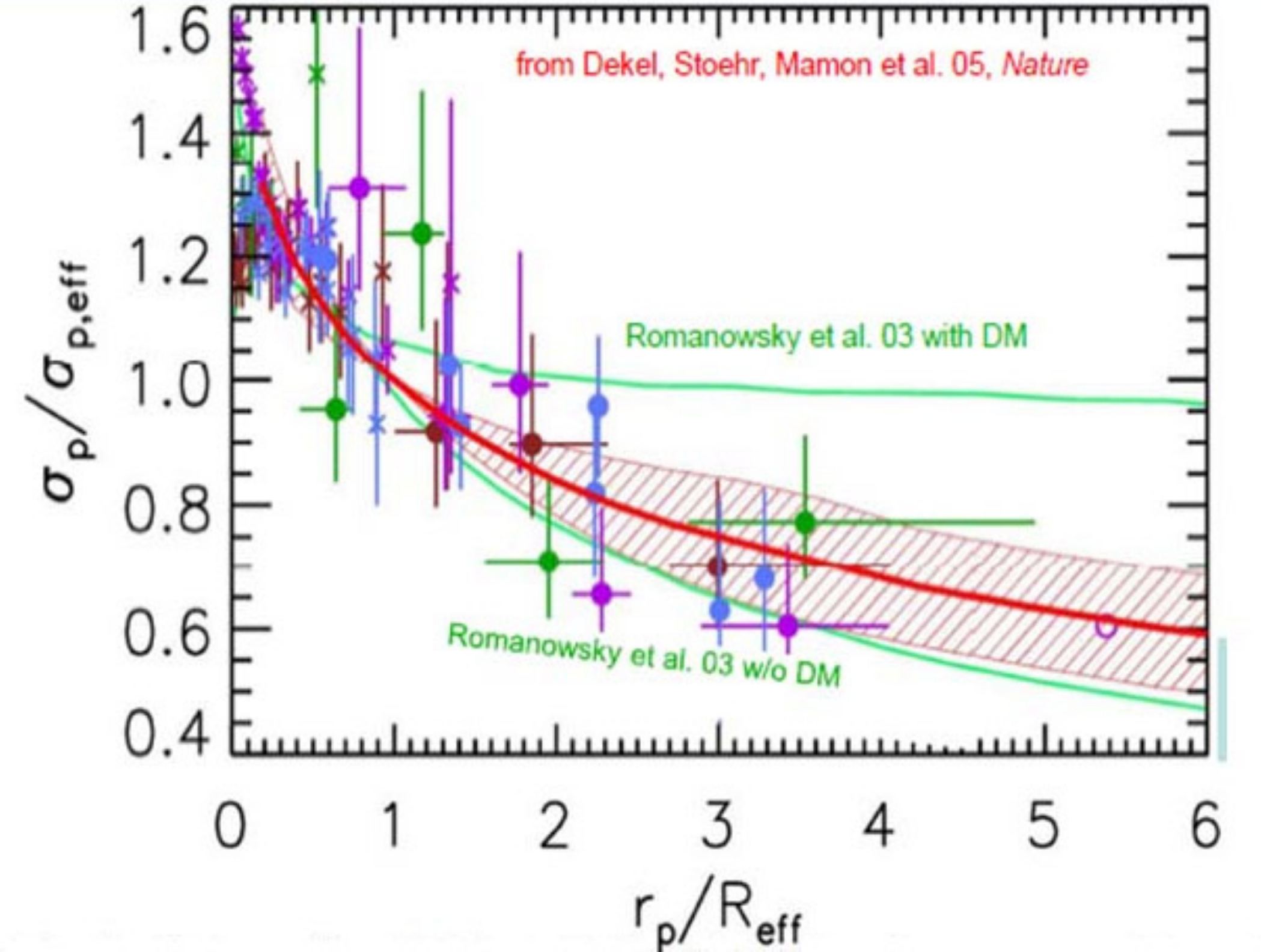}}
\caption{Comparison of observations (symbols) with simulations of spiral galaxies collisions (red curve). 
The green curves were predictions of Romanowsky et al. (2003)\cite{romanowsky}, without and with dark matter. The simulations are binary mergers of spirals made of disk, bulge, gas, and DM halo. The red curve is the mean line-of-sight velocity dispersion profile for the elliptical galaxy remnant, averaged over 10 simulations with different orbital parameters. The hashed region shows the dispersion over the 10 simulations. The galaxies are marked green (821), violet (3379), brown (4494) and blue (4697) with 1s errors; PNs (circles) and stars (crosses).
The plot indicates that one can have strongly decreasing line-of-sight velocity dispersion profiles despite the presence of dark matter (From Dekel et al. 2005\cite{dekell}).} 
\end{figure}

\subsection{Elliptical galaxies}

Elliptical galaxies are characterized by an approximately ellipsoidal shape and a smooth, nearly featureless brightness profile. 
Most elliptical galaxies are composed of older, low-mass stars, with a sparse interstellar medium and minimal star formation activity, and they tend to be surrounded by large numbers of globular clusters. 
The surface brightness of the stellar spheroid follows a Sersic (de Vaucouleurs) law
\begin{equation}
I(R)=I_e e^{-b_n}[(R/R_e)^{1/n}-1]
\end{equation}
that for $n=4$ reduces to
\begin{equation}
ln I(R)=ln I_e +7.669 [1-(r/R_e)^{1/4}]
\end{equation}
where $R_e$ is the effective radius, and n the Sersic index (connected to the light concentration). By deprojecting I(R), we obtain the luminosity density j(r):
\begin{equation}
I(R)= \int^{+\infty}_{-\infty} j(r) dz = 2 \int^{+\infty}_{R}  \frac{j(r) r dr}{\sqrt{r^2-R^2}}
\end{equation}
Assuming radially constant stellar mass to light ratio, $M/L$, we can get the density of the stellar spheroid
\begin{equation}
\rho_{sph}(r)= (M/L)_{\star} j(r)
\end{equation}
The total mass of the galaxy is given by
\begin{equation}
M(r)=M_{sph}+M_h(r)
\end{equation}
being $M_h$ the DM halo component.
In order to determine the total, and DM mass, one has to measure I(r), and the dispersion velocity along the line of sight  (l.o.s.), $\sigma_P$, or more precisely the
aperture velocity dispersion $\sigma_A$\footnote {When observed through an aperture of finite size, the projected
velocity dispersion profile, $\sigma_P$, is weighted on the brightness profile I(R).}, given by:
\begin{equation}
\sigma^2_A(R_A)= \frac{2 \pi}{L(R_A)} \int^{R_A}_{0} \sigma^2_P I(r) R dR 
\end{equation}
where
\begin{equation}
L(R)=2 \pi \int I(r) R dR 
\end{equation}
and
\begin{equation}
\sigma^2_P(R)= \frac{2}{I(R)} \frac{\int^{+\infty}_{R} \rho_{sph}(r) \sigma_r^2 r}{\sqrt{r^2-R^2}} dr
\end{equation}
and 
\begin{equation}
\sigma^2_r(r)= \frac{G}{\rho_{sph}(r)} \frac{\int^{+\infty}_{R} \rho_{sph}(r') M(r')}{r'^2} dr'
\end{equation}

In order to model ellipticals to get the total and DM mass there are several methods, and in general the analysis is more complex than spirals because ellipticals lack neutral gas (except for some ellipticals which has neutral gas in the outermost regions (Battaner \& Florido 2000\cite{battaner})), and DM evidence in elliptical galaxies is less than in spiral galaxies. 
The mass of the stellar spheroidal can be obtained as previously described, while the DM can be derived from:  
1) virial theorem; 2) dispersion velocities of kinematical tracers (e.g., stars, Planetary Nebulas); 3) X-ray properties of the emitting hot gas; 4) combining weak and strong lensing data. 

The virial theorem has been applied by Zwicky to Coma cluster, as already discussed. Defining the virial of Clausius
\begin{equation}
G=\sum_i p_i r_i
\end{equation}
that is, the sum over all the particles of the dot product of each particle's momentum with its position.
Taking the derivative one gets:
\begin{equation}
\frac{dG}{dT}=\sum_i F_i r_i+ 2 T
\end{equation}
where $F_i$ is the total force exerted on the i-th particle. Computing the time average (for a steady state)
\begin{equation}
<T>= -<V>/2
\end{equation}
where <T> is the time average of the total kinetic energy, and <V> is the time average of the total potential energy.
For a spherical, steady-state, elliptical galaxy, the virial theorem reduces to
\begin{equation}
M \simeq 2 R/\sigma^2 
\end{equation}
that allows to calculate the approximate mass of the galaxy once the velocity dispersion is known. 

Concerning the second method, we know that in ellipticals gravity is balanced by pressure gradients, and one can use Jeans Equations to try to determine 
the DM content
\begin{equation}
\frac{1}{\rho} \frac{d (\rho \sigma_r^2)}{dr} +2 \frac{\beta \sigma_r^2}{r}=-\frac{d \phi}{dr}
\label{jeans_eq}
\end{equation}
where $\rho(r)$ is the stellar density, $\beta=1- \frac{\sigma_{\theta}^2}{\sigma_r^2}$, the velocity dispersion anisotropy, and $\phi$ the gravitational potential.

Using the quoted method and Eq. (\ref{jeans_eq}) to infer DM content in the galaxy one encounters couple of complications.
Firstly, the so called mass/anisotropy degeneracy. For a given $\rho(r)$, and $\sigma_r(r)$, two unknown remains in Eq. (\ref{jeans_eq}): M(r), and $\beta(r)$ and one
cannot solve Jeans equation for both, unless one assumes no rotation and makes use of the 4-th order moment (kurtosis) of the velocity distribution (Lokas \& Mamon 2003\cite{lokas}).

The degeneracy can be broken in different ways. One can adopt a $\sigma_r$ depending on some parameters:
\begin{equation}
\sigma_r(r) = \sigma_0 \left[1+\left(\frac{r}{r+r_0}\right)^\eta \right]^{-1},
\label{vdisp}
\end{equation}
where ${\sigma_0,r_0,\eta}$ are the quoted parameters (Napolitano et al. 2011\cite{napolitano}), and also assume a given $\beta(r)$. After projecting  
$\sigma_r$ and $\sigma_\theta$ along the l.o.s., one gets a best fit of the three parameters in Eq. (\ref{vdisp}) to the observed dispersion profile.
Then Eq. (\ref{vdisp}) can be inserted in the Jeans equation 

\begin{equation}
M(r) = -\frac{\sigma_r^2~r}{G}\left( \frac{d\ln \rho_*}{d\ln r}+\frac{d\ln\sigma_r^2}{d\ln r} +2\beta \right) ,
\end{equation}
where $\rho_*(r)$ is the spatial density of the tracers (stars, planetary nebulas (PNe).

Another way is that of choosing a density profile (e.g., NFW) and then exploring a multi-dimensional parameter space, constituted by 
$\beta$ and the parameters of the density profile (see Napolitano et al. 2011\cite{napolitano}).  
In Fig. 17, are displayed the results of such aan analysis. 

A second difficulty in inferring the presence of dark matter halos in ellipticals is connected to the fact that the velocity dispersions of the usual kinematical tracer, stars, can only be measured out to $2R_e$. One can solve this problem using other tracers (e.g., planetary nebulas (PNe)). 

The quoted Jeans analysis has nowdays shown that there exist big DM halos around Ellipticals, however, when it was used initially by 
Romanowsky et al. (2003)\cite{romanowsky}, they concluded that there was a dearth of DM in ellipticals, since the strong decrease in 
the velocity dispersion was interpreted by them as sign of lack of DM. 
Dekel et al. (2005)\cite{dekell} (see Fig. 18) considered binary mergers of spirals and obtained again a decreasing velocity dispersion, even if the merging systems contained DM. 
The inconsistency between kinematical and dynamical models is produced by projection effects and stellar orbits very radial.

Mamon \& Lokas (2005)\cite{mamon} considered a superposition of Sersic models for the stellar mass component of elliptical galaxies with hot gas (from X
rays) and a central black hole (from the Magorrian relation), plus Dark Matter models: NFW, Jing \& Suto, and Einasto. No adiabatic contraction was considered. Their result indicates that while dark matter dominates outside of a few $R_e$, the stellar component dominates inside
$R_e$. Therefore, it is difficult to measure the amount of dark matter in the inner regions of ellipticals.

We previously mentioned two other techniques to study DM in ellipticals: X-ray emission from hot gas, and lensing.
Concerning the first technique we discussed how it works in Sec. 1 (Dark matter evidences). Nagino \& Matsushita (2009)\cite{nagino} (see Fig. 19) derived the mass profiles of 22 ellipticals that were observed through Chandra and XMM-Newton. They concluded that DM is a common feature of ellipticals and that in the inner parts of the galaxies ($r \leq 3 R_e$) DM mass content is smaller to the stellar mass, somehow in agreement with Mamon \& Lokas (2005)\cite{mamon} results. 

\begin{figure}
\resizebox{8.65cm}{!}{\includegraphics{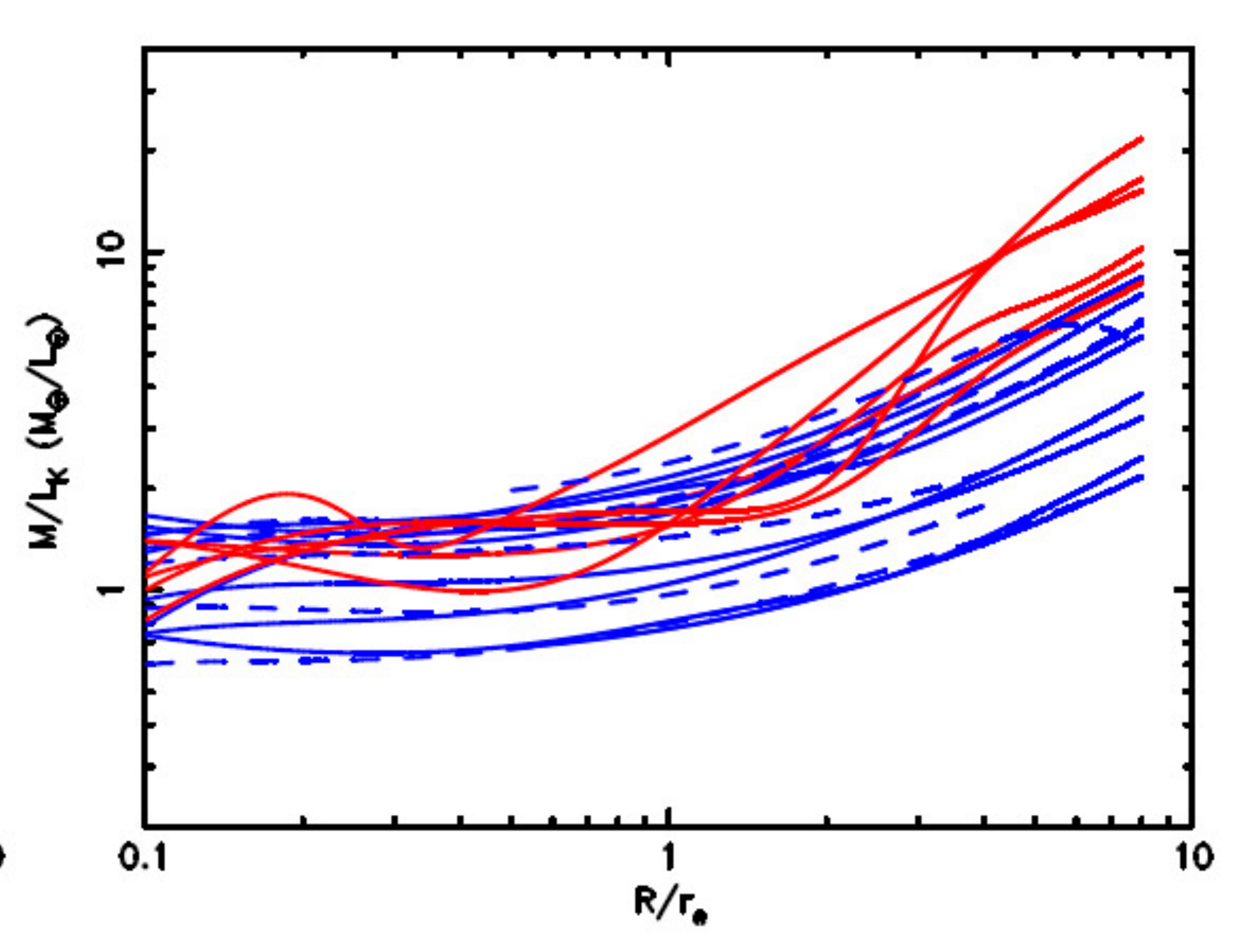}}
\caption{Summary of M/L ratio of 19 of the 22 galaxies (only two groups) studied by Nagino \& Matsushita (2009)\cite{nagino} with
XMM-Newton and Chandra observations. The figure shows that DM is important only at $r>r_e$} 
\end{figure}

Gavazzi et al. (2007)\cite{gavazzi} (see Fig. 20) combined weak and strong lensing to study 22 massive SLACS galaxies modeled as a
sum of stellar component (de Vaucoulers), and a DM halo (NFW). They found that the total
density profile is close to isothermal over $\simeq  2$ decades in radius. The uncertainties in the density profile are larger at $r> 10 $ kpc and much smaller at $r< 10 $ kpc because of strong lensing data in the inner part of the profile.
The transition between star and DM-dominated mass profile occurs close to the mean effective radius.

Concerning lensing, Mandelbaum et al. (2009)\cite{mandelbaum} (see Fig. 21) considered $3  \times 10^7$ sources (SDSS galaxies) and 170 000 lenses (isolated galaxies), and measured the shear around galaxies of different luminosities out to 500 - 1000 kpc
reaching out the virial radius. Both NFW and Burkert halo profiles agree with data. However, using the Jeans analysis or even lensing, usually a NFW model fits well the density profile of ellipticals.


\begin{figure}
\resizebox{8.65cm}{!}{\includegraphics{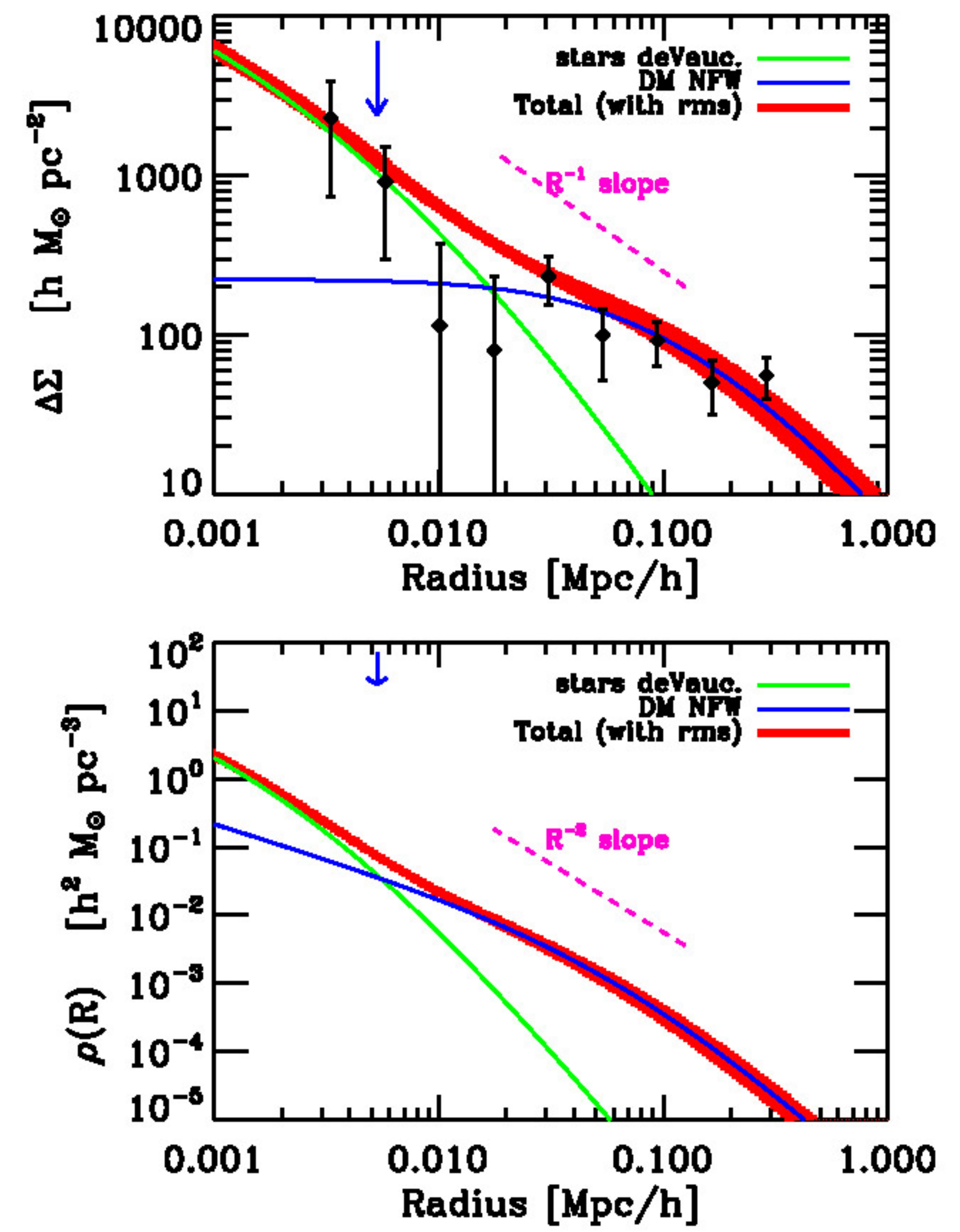}}
\caption{Shear and three-dimensional density profile of 22 massive SLACS galaxies modeled as a 
sum  of stellar component (de Vaucoulers), and a DM 
halo (NFW). From Gavazzi et al. 2007\cite{gavazzi}.} 
\end{figure}

\begin{figure}
\resizebox{8.65cm}{!}{\includegraphics{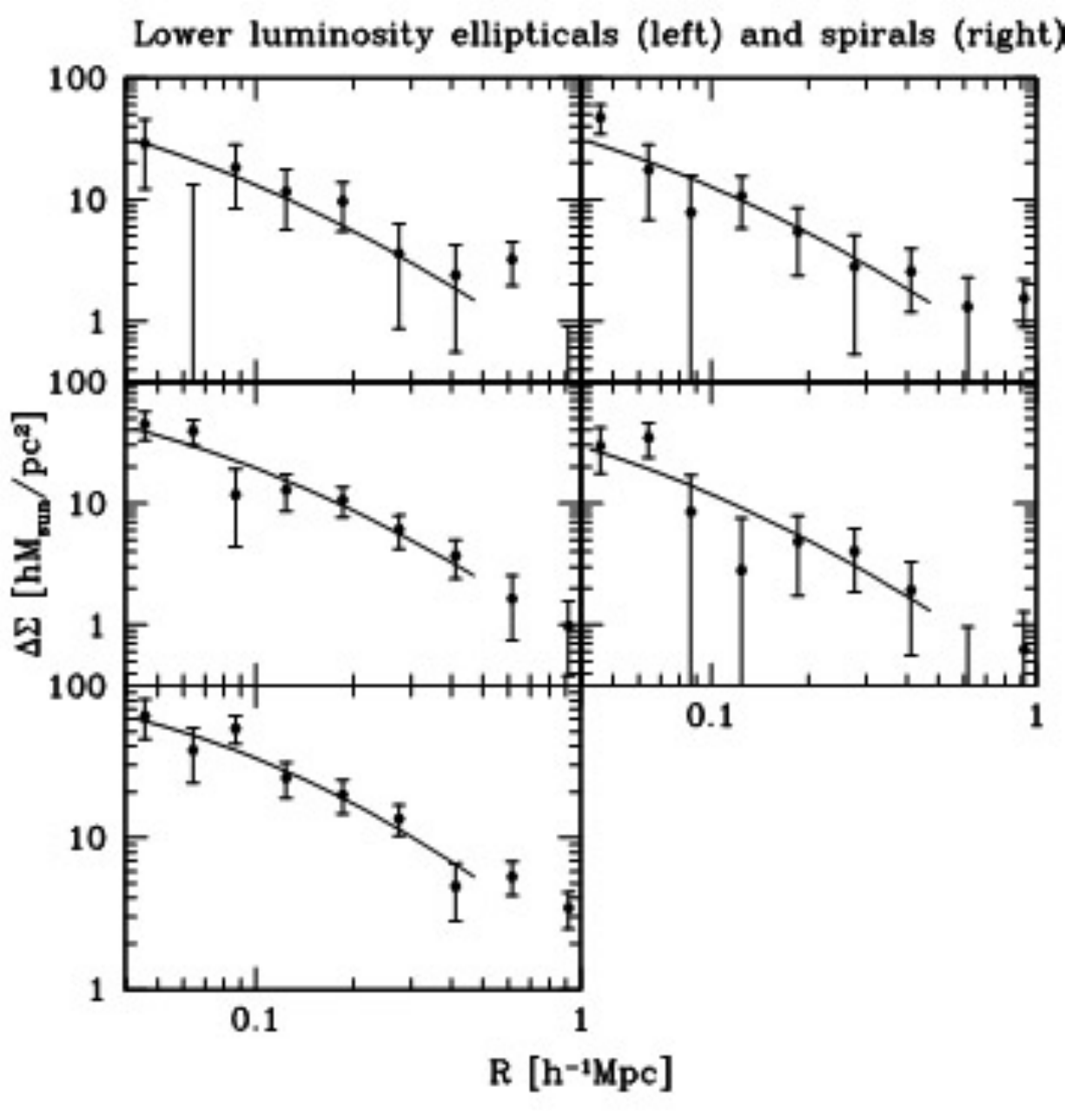}}
\caption{Mandelbaum et al. (2009)  measured the shear around galaxies of different luminosities out to 500 - 1000 kpc 
reaching out the virial radius, although with a not negligible observational uncertainty. The result, in the limits of the uncertainties,
is that both NFW and Burkert halo profiles agree with data.} 
\end{figure}

Koopmans et al. (2006)\cite{koopmans} (see Fig. 22) combined stellar-dynamical analysis and gravitational lensing of a subsample of 15 massive field early-type galaxies from SLACS Survey. The inner mass density profiles of the galaxies are homogeneous and 
the total density profile is isothermal, $\rho 
\propto r^{-2}$. The total mass inside $R_e$ is mostly accounted by the stellar spheroid 

In summary, elliptical galaxies are characterized by a small amount of DM inside $R_e$, by a mass profile compatible with the NFW and in rare cases with Burkert profile. DM is traced out to $R_{vir}$.

\begin{figure}
\resizebox{8.65cm}{!}{\includegraphics{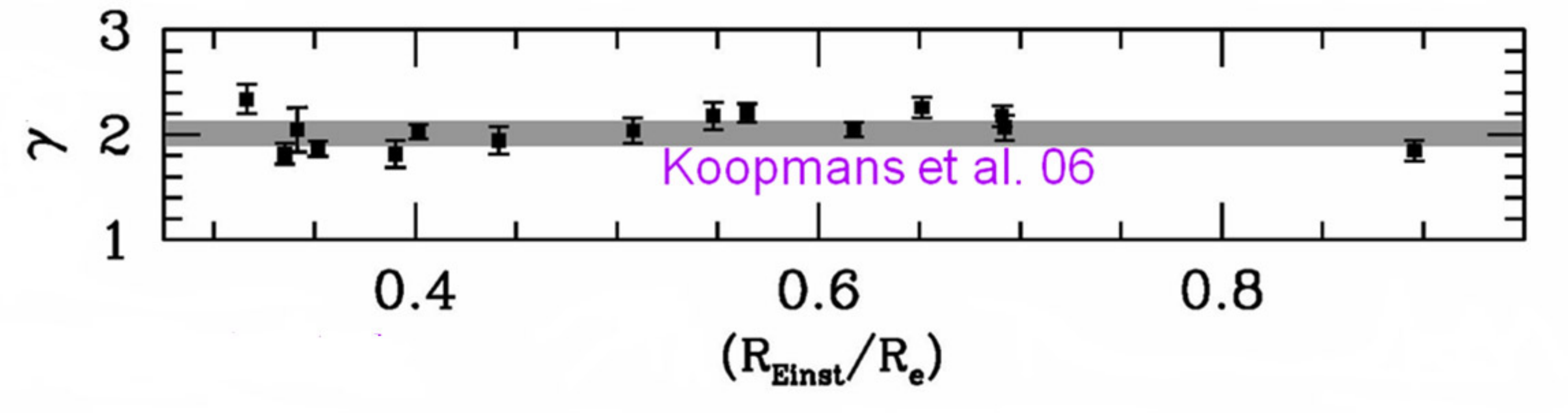}}
\caption{Joint gravitational  lensing and stellar-dynamical analysis of a subsample of 15 massive field early-type galaxies from SLACS Survey. The logarithmic slope $\gamma$ of the profiles is compatible with -2. From Koopmans et al. 2006\cite{koopmans}} 
\end{figure}

\subsection{Dwarf spheroidals}

Dwarf spheroidals (dSphs) are small galaxies, with $M_{tot} \simeq 10^7 M_{\odot}$, low luminosity, $L=2 \times 10^3-10^7 L_{\odot}$, high central dispersion velocity $\sigma_0 \simeq 7-12 $ km/s, and $r_0 \simeq 130-500$ pc. They are gas deficient with often old stellar population, and apparently in equilibrium. 
dSphs have an high $M/L$, ranging from 10 to several hundreds, meaning that if they are really in equilibrium, they are DM dominated (but see Kroupa\cite{kroupa, klessen}). 
Mateo (1998)\cite{mateo1} found an empirical relation connecting the mass-to-light ratio and the total luminosity
\begin{equation}
M/L =2.5+[10^7/(L/L{\odot})]
\end{equation}
implying for the Bo\"otes dSph even $M/L \simeq 610$ (see also M\~unoz et al. (2006)\cite{munozz}.
After the discovery of ultra-faint MW satellites\cite{willman,belokurov,zucker,grillmair,sakamoto,irwin} the range of dSph structural parameters has been extended of 3 orders of magnitudes in luminosity and 1 in radius. 

\begin{figure}
\resizebox{10.65cm}{!}{\includegraphics{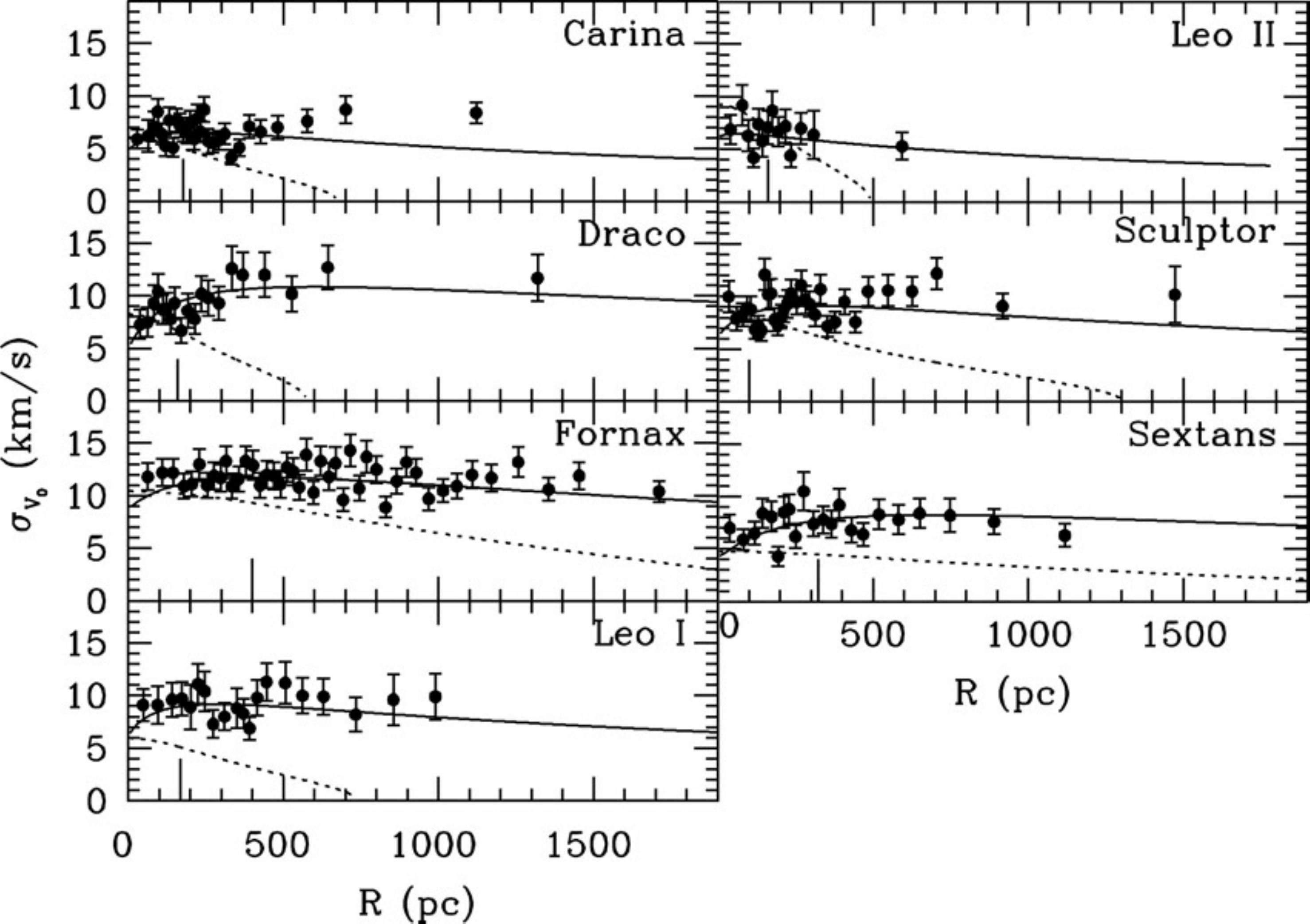}}
\resizebox{10.65cm}{!}{\includegraphics{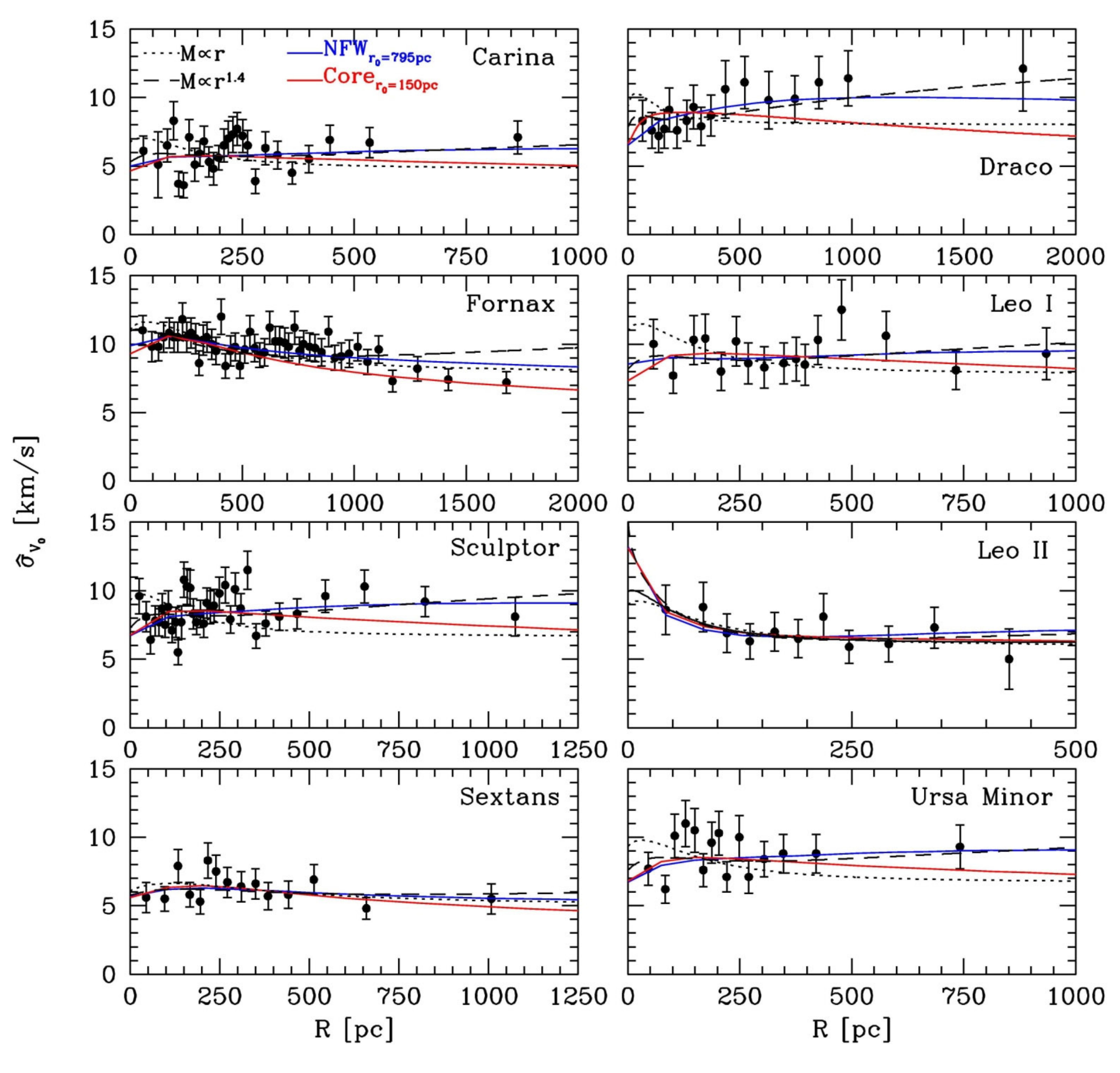}}
\caption{Left panel: dSphs density profiles. The plot shows that the profiles remain flat to large radii (from Walker et al. 2007
\cite{walker}). Right panel: Disperion velocities of dSphs fitted with different mass models, as shown inside the figure. Cored and cusped halos with orbit anisotropy fit dispersion profiles equally well (from Walker et al. 2009\cite{walker1}). } 
\end{figure}

\begin{figure}
\resizebox{9.65cm}{!}{\includegraphics{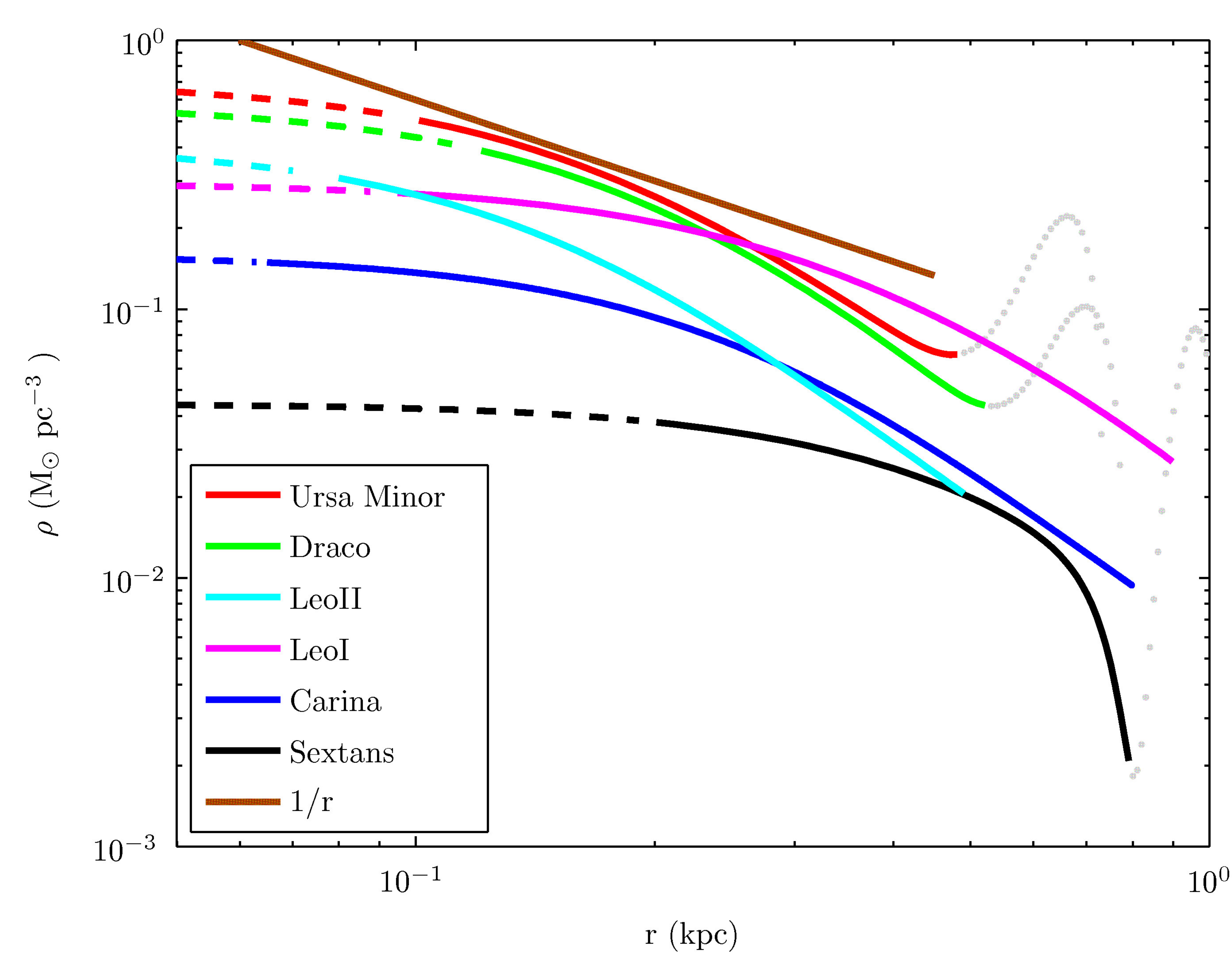}}
\caption{Density profile of the indicated dSPhs obtained Jeans analysis (from Gilmore et al. 2007)\cite{gilmore}.} 
\end{figure}

The high content of DM is deduced from their high velocity dispersion. Aaronson (1983)\cite{aaronson} measured the velocity dispersion of Draco based on observations of 3 carbon stars, and obtained a value of $M/L \simeq 30$. Mateo (1997)\cite{mateo}
obtained the first dispersion profile of Fornax, Walker et al. (2007)\cite{walker} determined the velocity dispersion profiles of Carina, Draco, Fornax, Leo I, Leo II, Sculptor, and Sextans (see Fig. 23), and Walker et al. (2009)\cite{walker1} fitted the previous velocity dispersion (and that of Ursa Minor, too) with different profiles (see Fig. 23). Dispersion velocity profiles remain generally flat to large radius. Cored and cusped halos with orbit anisotropy fit dispersion profiles equally well. 
Using a Jeans analysis, Gilmore et al. (2007)\cite{gilmore} found a cored DM profile for Ursa Minor, Draco, Leo I, Leo II, Carina, and Sextans (see Fig. 24). At a similar result arrived Kleyna et al. (2003)\cite{kleyna} in the case of Ursa Minor. This galaxy would survive for less
than 1 Gyr if the DM core was cusped, and similarly Magorrian (2003)\cite{magorrian} found a inner slope $\alpha=0.55^{+0.37}_{-0.33}$ for
the Draco dSph.  

\begin{figure}
\resizebox{15.65cm}{!}{\includegraphics{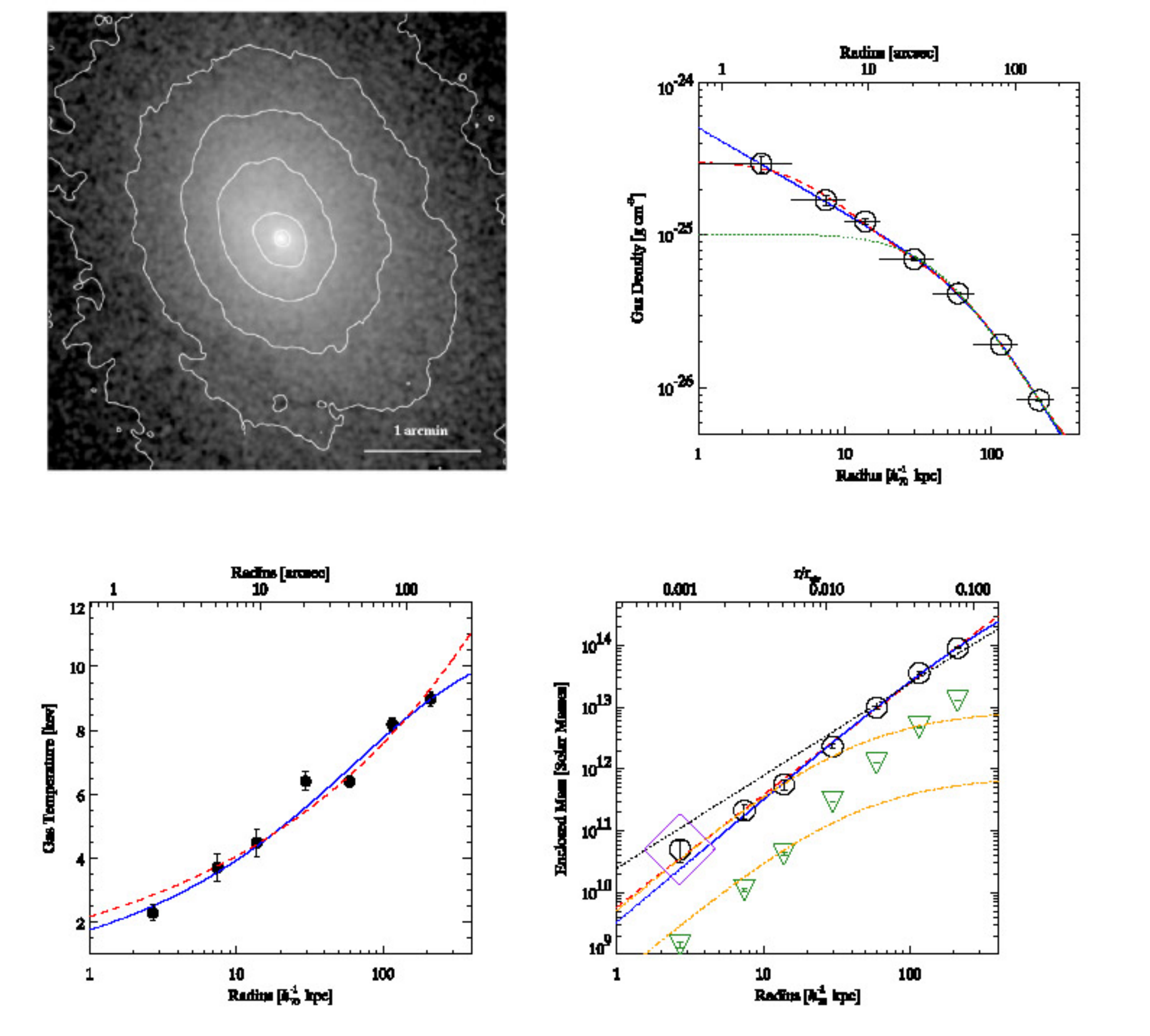}}
\caption{Top Left: Chandra ACIS  image of Abell 2029. Top Right: Radial gas density profile of Abell 2029 (large circles) fit to several standard parameterizations. Bottom Left: The radial temperature profile of Abell 2029, again fit to a standard paramaterization to facilitate the hydrostatic equilibrium analysis. Bottom Right: Total enclosed cluster mass profile. The open circles are the data points and the lines are fits to the data, with the NFW profile being a very good fit. The upside down triangles show the contribution from the cluster gas mass. The bright yellow band shows the possible 
contribution from the cluster BCG,  (From Lewis et al. (2002, 2003)\cite{lewis1,lewis2}).} 
\end{figure}

\begin{figure}
\resizebox{19.65cm}{!}{\includegraphics{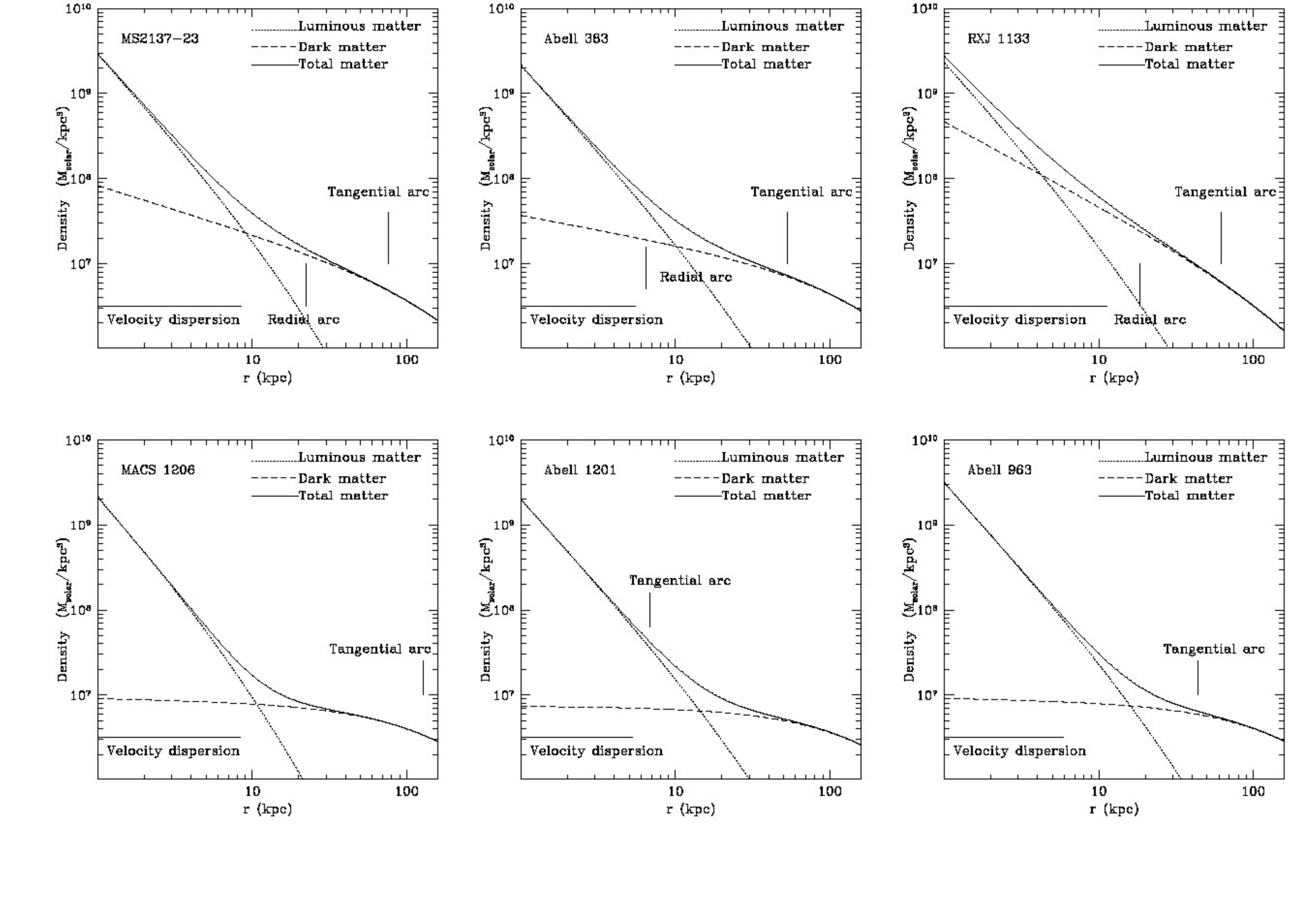}}
\caption{Density profiles of obtained combining lensing and velocity dispersion. Luminous and DM matter are disentangled, and the total mass is also shown (from Sand et al. 2004)\cite{sand1}).
} 
\end{figure}

\begin{figure}
\resizebox{9.65cm}{!}{\includegraphics{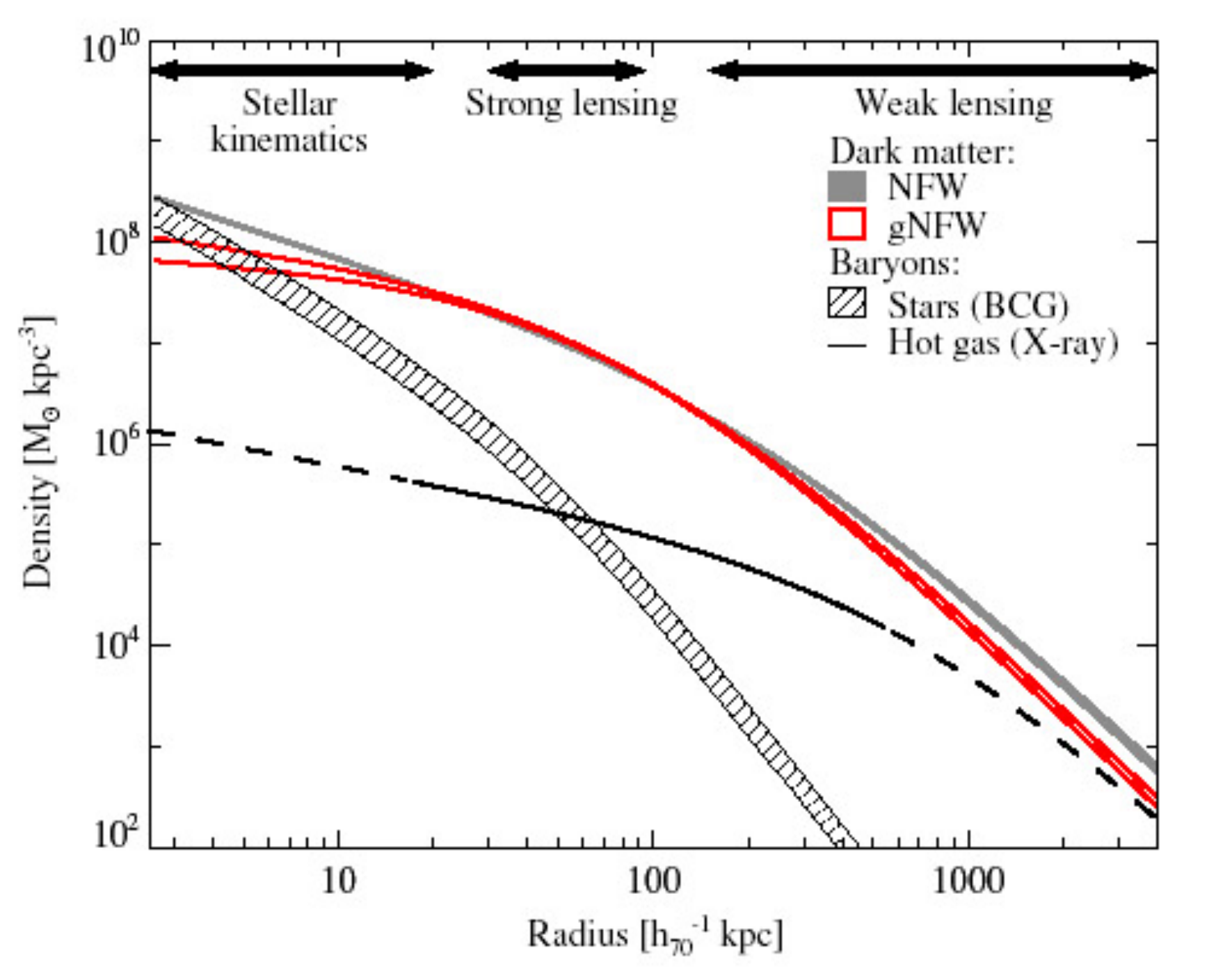}}
\caption{Density profiles of A611 obtained combining strong, weak lensing and velocity dispersion. Baryonic components (stars, gars) are disentangled from DM. The DM profile is fitted with a NFW model (grey) and generalized NFW (gNFW) model (from Newman et al. 2009\cite{newman}).} 
\end{figure}

\begin{figure}
\resizebox{14.65cm}{!}{\includegraphics{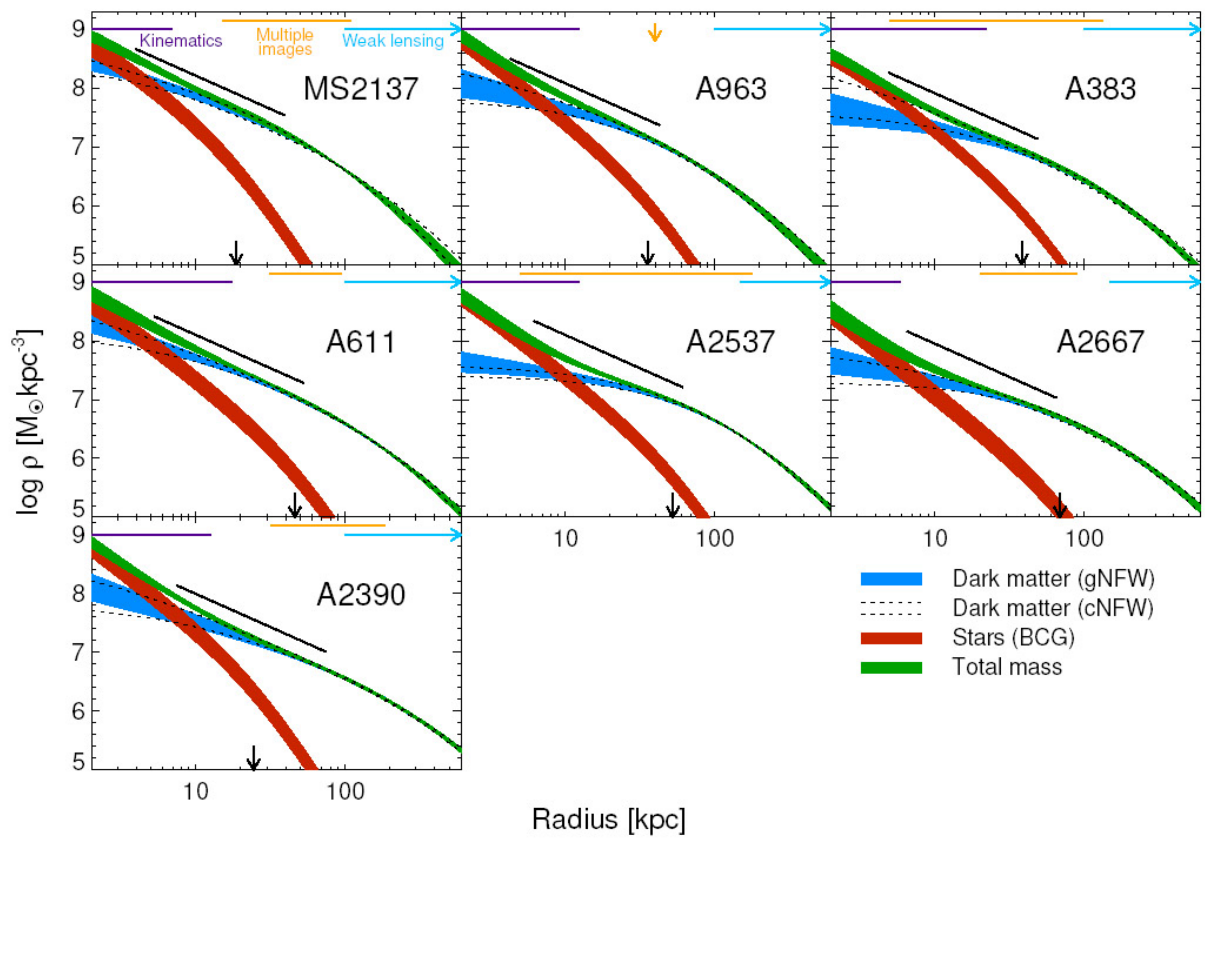}}
\resizebox{6.65cm}{!}{\includegraphics{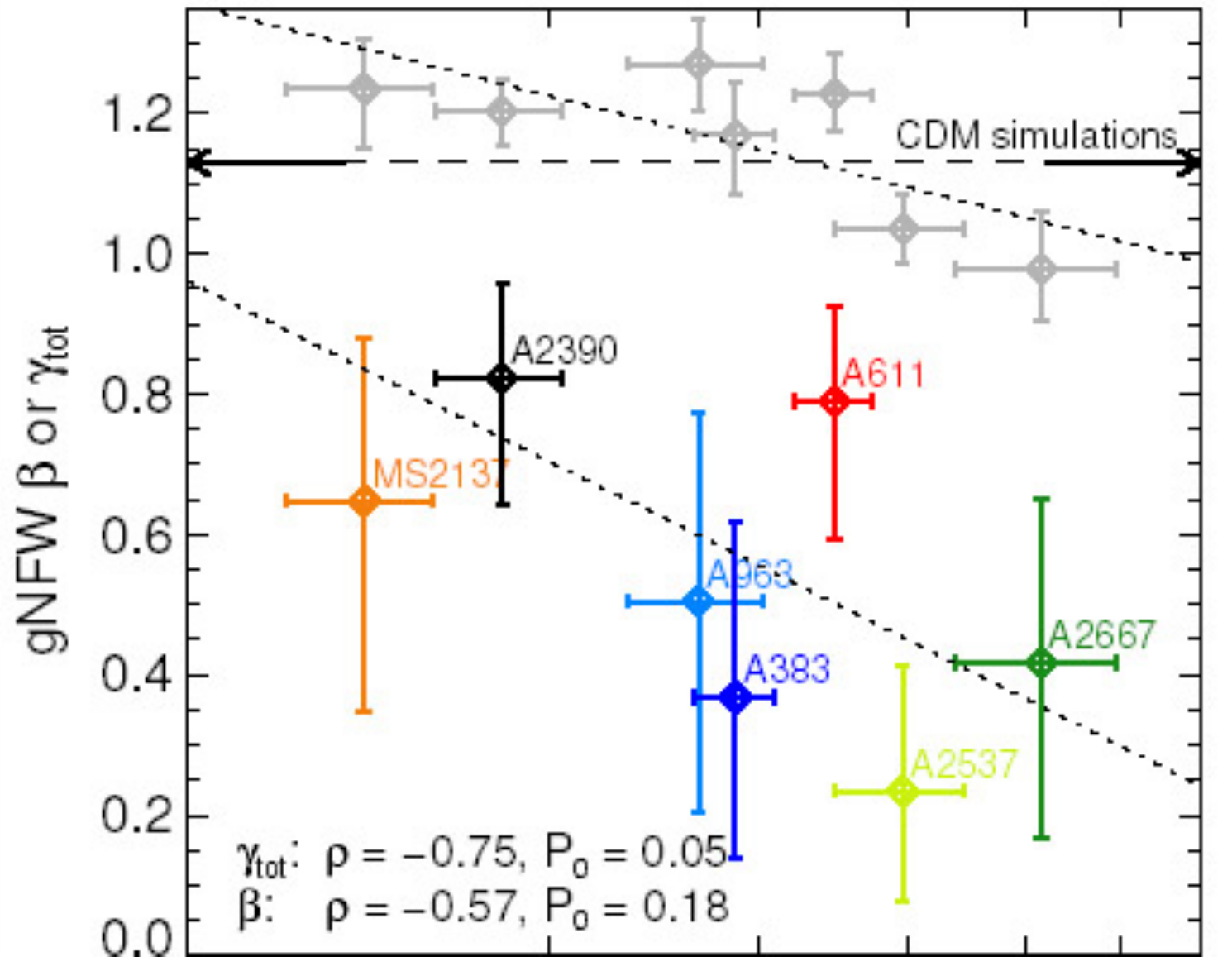}}
\caption{Left Panel: Density profiles of different clusters obtained combining strong, weak lensing and velocity dispersion. Baryonic components are disentangled from DM. The DM profile is fitted with two different NFW models. Right panel: comparison of the slopes of the inner profiles obtained through simulations (grey) and from observations 
(from Newman et al. 2012)\cite{newman2}).} 
\end{figure}

Summarizing, dSphs are DM dominated at any radius, and have a density profile consistent with Burkert's profile.

\subsection{Clusters of galaxies}

Galaxy groups and clusters are the largest known gravitationally bound objects in the universe. 
They are typically  a few Mpcs across, may contain 100-1000 galaxies according to their richness
class, with masses reaching $10^{15} M_{\odot}$, and the distribution of galaxies falls as $r^{1/4}$ (like surface brightness of elliptical galaxies). X-ray studies have revealed the presence of large amounts of intergalactic (intracluster medium). This gas is very hot, between $10^7$ K and $10^8$ K, and hence emits X-rays in the form of bremsstrahlung and atomic line emission. The total mass of the gas is greater than that of the galaxies by roughly a factor of two. However, as already reported this is still not enough mass to keep the galaxies in the cluster. 
In a typical cluster perhaps only 5\% of the total mass is in the form of galaxies, maybe 10\% in the form of hot X-ray emitting gas and the remainder is dark matter. 

The methods used to study the DM in clusters are similar to those used for elliptical galaxies. 
The Jeans analysis is based on Eq. (\ref{jeans_eq}) and assuming 
\begin{equation}
\beta(r)=r^2/(r^2+r_a^2)
\end{equation}
and using the Opsikov-Merritt (Osipkov 1979\cite{osipkov}; Merritt 1985a,b\cite{merrit}) parameterization of anisotropy one gets the projected velocity dispersion

\begin{equation}
\sigma^2_P(R)= \frac{2}{(M/L) I(R)} \frac{\int^{+\infty}_{R} [1-R^2/(r_a^2+r'^2)]\rho(r') \sigma_r^2(r') r'}{\sqrt{r'^2-R^2}} dr'
\end{equation}
which furnish the aperture velocity dispersion $\sigma_A$, as previously described, which can be compared to observations.

The other method is based on the X-Ray emission of ICM, and again is similar to the analysis discussed for ellipticals (see Fig. 25). This technique has several limits.
Using X-ray data alone it is complicated to constraint the mass distribution, especially in the central part of clusters, where cooling flows can disturbe X-ray emission, and the same hydrostatic equilibrium condition is no longer verified (Arabadjis, Bautz \& Arabadjis 2004\cite{arabadjis}). The presence of the BCG (brightest cluster galaxy) is difficult to take into account, and at $r< 50$ kpc substructure or instrumental resolution put limits to the temperature determination (Schmidt \& Allen 2007\cite{schmidt}). Finally, the technique measures the total mass, and to disentangle DM from luminous matter one needs another mass tracer. 
This last problem of X-ray observations is common to another technique used to study cluster mass distribution, namely gravitational lensing. Weak lensing are more suitable to reconstruct the mass distribution in the outer parts of clusters (till 100 kpc) while strong lensing in the inner parts, with limits at 10-20 kpc (Gavazzi 2005\cite{gavazzi1}; Limousin et al. 2008\cite{limousin}).
A pro of lensing is that mass measurement is insensitive to the dynamical state of the cluster.

Both gravitational lensing and X-ray analyses agree in the conclusion that clusters of galaxies contain a large 
quantity of DM, but give conflicting results concerning especially the inner part of the density profile. 
For example, some clusters, studied through lensing, have cuspy inner profiles in agreement with N-body simulations (Dahle et al 2003\cite{dahle}; Gavazzi et al. 2003\cite{gavazzi2}; 
Donnaruma et al. 2011\cite{donnaruma}), but others have much shallower slopes (Sand et al. 2002\cite{sand2}; Sand et al. 2004\cite{sand1}; 
Newman et al. 2009, 2011, 2012\cite{newman,newman1,newman2}) (see Figs. 26, 27, 28) several times in agreement with core-like profiles. 
The conflicting results are due to different reasons (see Del Popolob 2012b\cite{delpopolo1}) but for what concernes the clusters studied in Sand's papers and in those of Newman (see Figs. 26, 27, 28), the main reason for the difference is due to the fact that the authors use a combination of different techniques (weak lensing in the outskirts of the clusters, strong lensing in inner regions, till 20 kpc, and stellar kinematics in the inner region where the BCG is localized). The authors show that using only lensing the inner profile would be different. Another interesting feature shown in Sand's and Newman's papers is the dominance of baryons over DM in the inner part of the clusters. This could explain the discrepancy with N-body simulations always getting cuspy profiles. N-body simulations are not taking into account baryons, and expecting that
a profile obtained using only DM (simulations) agrees with real clusters even in the part where baryons are dominant is at least weird (bizzare).

Also X-ray analyses have led to wide ranging of the value of the inner slope, from -0.6 (Ettori et al. 2002\cite{ettori}) to -1.2 (Lewis et al. 2003) till -1.9 (Arabadjis et al. 2002\cite{arabadjis1}). However, Schmidt \& Allen (2007)\cite{schmidt}, using 34 Chandra clusters concludes that they are in agreement with the NFW profile.

We may conclude this section telling that DM is present from dwarf galaxies scales to large
scales. dSphs are DM dominated with $M/L \simeq 100$. Normal spirals have $M/L$ an order of magnitude smaller
than dSph, and elliptical galaxies have variable content of DM: some as M87 have a very high value of $M/L$, but some could even not contain DM (Battaner \& Florido 2000\cite{battaner}). Inside $R_e$ baryons are dominating. Clusters have high values of 
$M/L >100$ but in the central 10 kpc are dominated by baryons. 


\section{The nature of dark matter}

\begin{figure}
\resizebox{7.65cm}{!}{\includegraphics{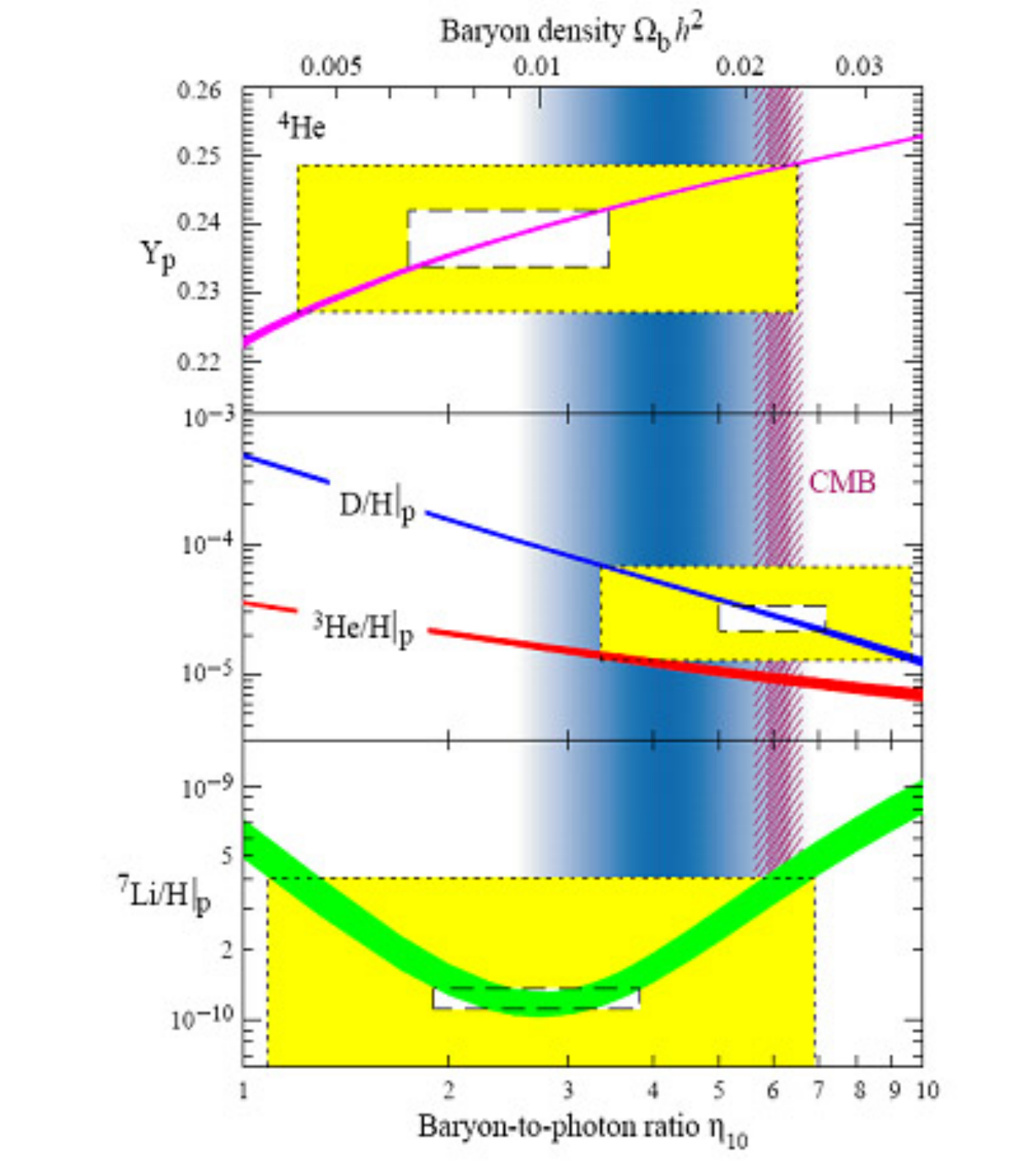}}
\caption{Elements abundance predicted by the standard big bang nucleosynthesis (BBN). The narrow vertical line represents the CMB measure of the cosmic baryon density. Boxes represents the observed abundances of ligh elements.}
\end{figure}

\begin{figure}
\resizebox{10.65cm}{!}{\includegraphics{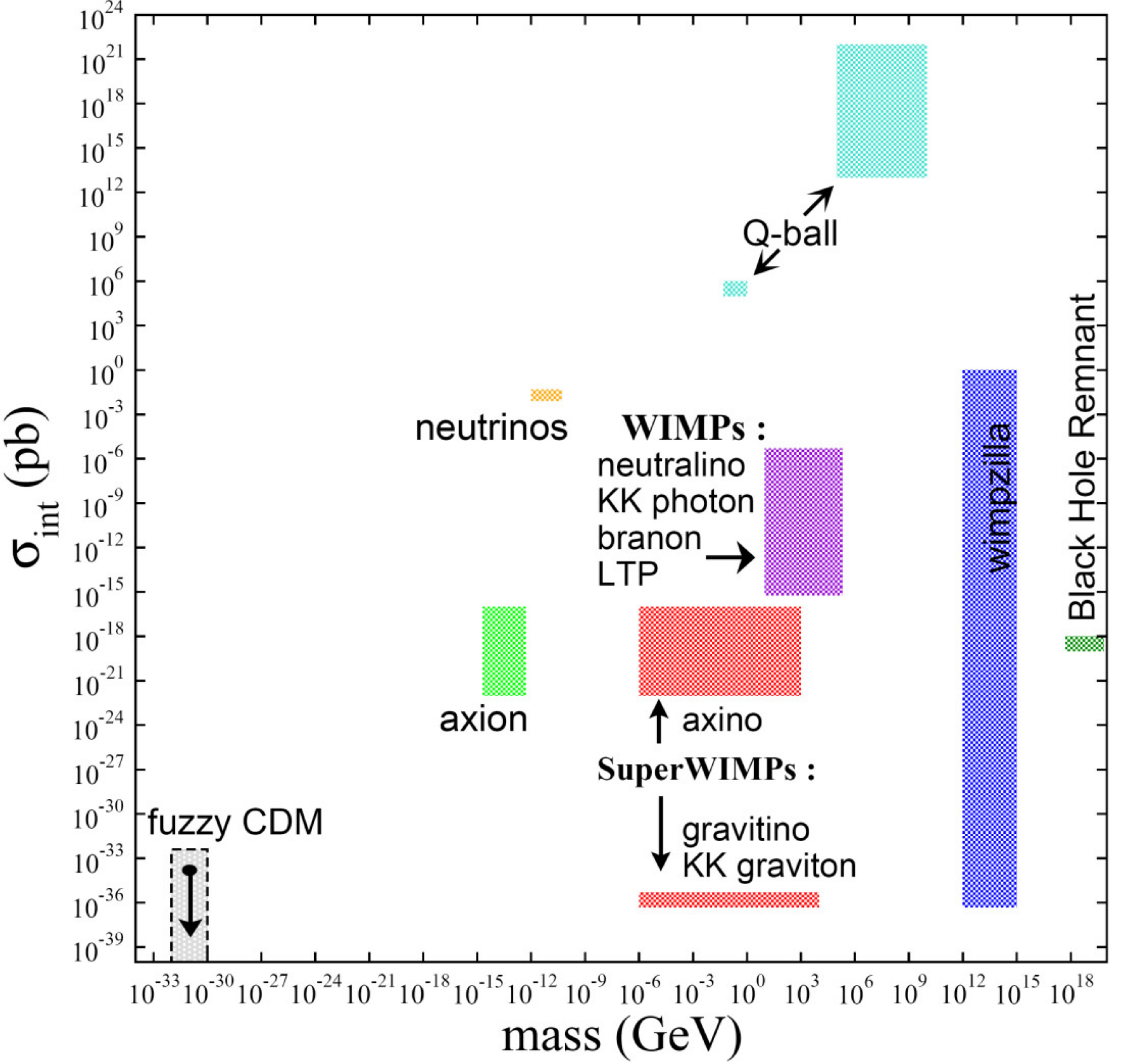}}
\caption{DM candidates.} 
\end{figure}

Till now we discussed the historical evidences for DM, and its distribution in structures. The main question is: What is DM?
The attempts to answer this question have a long history. We are not certain on the nature of DM, but most probably it is made of particles. At the same time we can estimate the total baryonic matter of the universe by studying Big Bang nucleosynthesis.  
This is done by connecting the observed He/H ratio of the Universe today to the amount of baryonic matter present during the early hot phase when most of the helium was produced.  
Big Bang Nucleosynthesis (BBN) (deuterium abundance) (see Fig. 29) and cosmic microwave background (e.g., WMAP) together with BAO (Baryonic Acoustic Oscillations) and supernovae measure a baryon contribution  $\Omega_B=0.0456 \pm 0.0016$, and a DM contribution of $\Omega_M =0.227 \pm 0.014$. The previous result shows that baryons are too few to explain all DM. Moreover baryons are unable to drive galaxy formation since decouple too late from photons, and 
gravitational instabilities have not enough time to grow. The value of the baryonic density contrast, $\delta \rho$  is given by
\begin{equation}
\frac{\delta \rho}{\rho} \leq A_{\lambda} 2 \times 10^{-3}
\end{equation}
where $A_{\lambda}=1-10$ is a scale dependent growth factor.
At the same time contribution to the density parameter coming from visible matter is $\Omega_{lum} \simeq 3 \times 10^{-3}$ (Persic \& Salucci 1992\cite{persic}) or $0.02$ (Fukugita, Hogan \& Peebles (1998)\cite{fukugita}) (including plasmas in groups and clusters). 
Since $\Omega_{B}>\Omega_{lum}$ baryonic DM, even if it is not the principal component of DM it has to exist (we already discussed the MACHO, EROS1, EROS2 limits). If baryonic DM is not the fundamental component of DM the question is what is the main constituent of it? 
One of the first authors who had some ideas on DM nature was Rees (1977)\cite{rees}. According to him DM could be of a "more
exotic character" (e.g., small rest mass neutrinos). In other words, DM could be made of particles. 

In the last case, the main problem is to understand which kind of particles are suitable to make DM. A good DM candidate must have the following characteristics:\\
1) it must be non-baryonic. \\
The reasons for this come from the BBN, CMBR limits already discussed, and from structure formation. As already told, baryonic dark matter is unable, alone, to give rise to structures since they are coupled to baryons till the decoupling epoch (380000 yrs after big bang) and from that time on there was not enough time for baryons to form structures.\\ 
2) Stable (protected by a conserved quantum number). \\
For obvious reasons.\\
3) No charge, and no colour (namely weakly interacting). \\
If DM is non electrically neutral it could scatter light and then it would not be dark. If it had colour it would be endowed with strong interactions.
In other terms DM must be weakly interacting.\\
4) The relic abundance must be compatible to observations, and cannot be hot (relativistic at decoupling). \\
If DM were relativistic at decoupling, Silk damping would erase the small scale power in the spectrum, while galaxy two pint correlation function is indicating a large amount of power on small scale. As a consequence, at matter-radiation equality epoch, particles must be non-relativistic.

\begin{figure}
\resizebox{8.65cm}{!}{\includegraphics{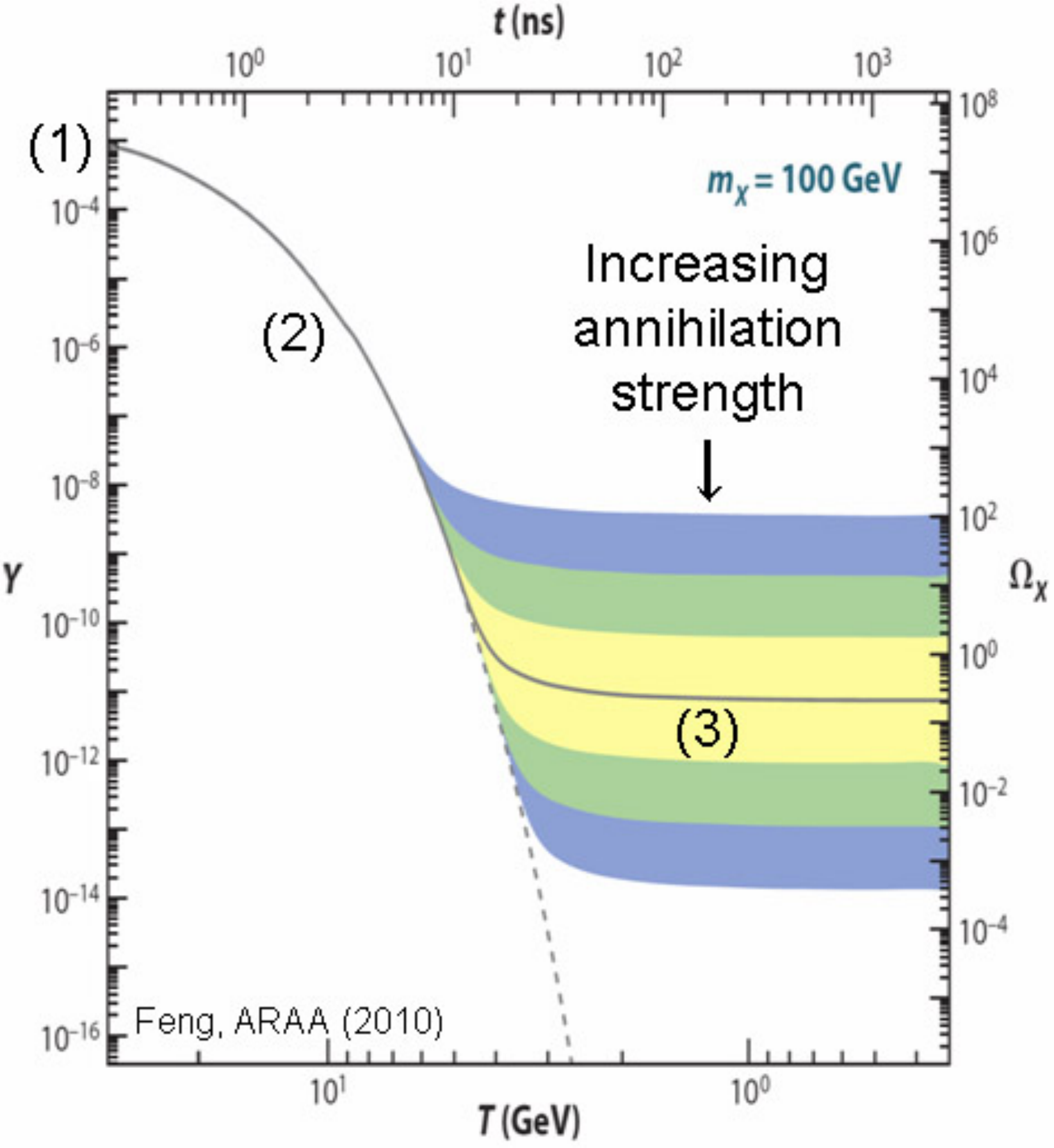}}
\caption{WIMPS density evolution with temperature, and final freez-out (from Feng 2010\cite{feng})} 
\end{figure}

 Using the previous arguments and having a look in the standard model (SM) particles list, one concludes that we need new particles.  
The list, or zoo of DM candidates (see Fig. 30), as it is often called, includes a large number of particle species ranging from $10^{-6}$ eV (axions) to $10^{16}$ GeV 
(WIMPzillas): SM neutrinos (but see the following), sterile neutrinos, axions, supersymmetric candidates (neutralinos, sneutrinos, gravitinos,axinos), candidates from the Little Higgs model, light scalar DM, Kaluza Klein states, Wimpzillas (or superheavy DM), Qballs etc. 
In the following discussion, I will concentrate the attention mainly on supersymmetric candidates, and Kaluza Klein states, and I will just hint on the other candidates. However, before going on I would like to stress that we cannot be certain that DM is made of a single particle species. We know that SM neutrinos constitute a part of DM, and this means that at least we need two candidates particles for DM, and moreover in N=2 supersymmetry the coesistence of two stable DM particles is allowed (C. Boehm, P. Fayet and J. Silk\cite{boehm}).

\subsection{Dark matter formation}

As already stressed, we do not have certainties on how DM formed, however we know that relics of weakly interacting massive particles
(WIMPs) should be a byproduct of a hot phase in early universe\cite{srednicki,gondolo,kolb}.
Let it $X$ be a stable particle interacting with a SM particle $Y$ according the process $X\bar{X} \leftrightarrow Y \bar{Y}$. 
The number density, $n_X$, of a species $X$ is described by the Boltzmann equation:
\begin{equation}
\frac{dn_X}{dt} + 3 H n_X = -<\sigma_{X\bar{X}} |v|> (n^2_X - n^2_{X,\,{\rm eq}}),
\label{boltzman}
\end{equation}
being $<\sigma_{X\bar{X}} |v|>$ the product of $\sigma_{X\bar{X}}$ (thermal averaged cross section), and the relative velocity, and $H$ is the expansion rate of the Universe.
When the temperature, $T$, is very large, $T \gg m_X$, the WIMPS density is given by the equilibrium density
\begin{equation}
n_{X, \, {\rm eq}} = g_X \bigg(\frac{m_X T}{2 \pi}\bigg)^{3/2} e^{-m_X/T},
\end{equation}
where $g_X$ is the number of internal degrees of freedom of $X$. When $T \ll m_X$ the equilibrium density is small, and decreases further because of    
the terms $<\sigma_{X\bar{X}}|v|> n_X^2$ and $3Hn_X$. When $n_X$ is small the dilution due to Hubble expansion becomes much more important than the annihilation term and the comoving number density of particles remains fixed (freeze-out) (see Fig. 31).

In other terms, at temperatures higher than $m_X$, creation and annihilation processes had the same efficience, and $X$ density was large. With time the Universe expands and its temperature decreases to values smaller than $m_X$. The annihilation process continues but the creation one is (exponentially) suppressed. In order the particles are not destroyed, it is necessary that some effect stops the annihilation. This is furnished by the Universe expansion which produces a dilution of the species density and as a consequence particles stops to interact with each other surviving up to the present.   

Numerically solving the boltzman equation is possible to obtain the relics density 
\begin{equation}
\Omega_X h^2 \approx 0.1 \, \bigg(\frac{x_{\rm FO}}{20}\bigg) \bigg(\frac{g_{\star}}{80}\bigg)^{-1/2} \bigg(\frac{a+3b/x_{\rm FO}}{3\times 10^{-26} {\rm cm}^3/{\rm s}}\bigg)^{-1}.
\end{equation}
where $g_{\star}$ is the number of external degrees of freedom, $M_{P}$ is the Planck mass, 
$a$ and $b$ are two terms in the (non-relativistic) expansion, $<\sigma_{X\bar{X}} |v|> = a + b <v^2> + \mathcal{O}(v^4)$, and   
the freeze-out temperature is given by
\begin{equation}
x_{\rm FO} \equiv \frac{m_X}{T_{\rm FO}} \approx \ln\bigg[ c(c+2) \sqrt{\frac{45}{8}} \frac{g_X}{2\pi^3} \frac{m_X M_{\rm Pl} (a+6b/x_{\rm FO})}{g^{1/2}_{\star} x^{1/2}_{\rm FO}} \bigg]. 
\label{xfo}
\end{equation}
where $T \equiv m_x/x$, and $c \simeq 0.5$.

The previous calculation is no longer correct if are present one or more particles sharing a quantum number with the relic particle 
and also having a mass similar to the relic particle\cite{griest} (see \cite{bertone} for a treatment of this issue).

An important issue, is that for a particle with a GeV-TeV mass, with an annihilation cross section of $<\sigma v> \simeq$ pb
one obtain a thermal abundance equal to the observed dark matter density. At the same time, we know that generic weak interaction yields:
$<\sigma v> \simeq \alpha^2 (100 GeV)^{-2} \simeq 3 \times 10^{-26} cm^3/s \simeq$ pb

It is not clear if this occurrence dubbed the "WIMP miracle" is a numerical coincidence or an indication that dark matter originates from electro-weak (EW) physics.
However, it is often used to conclude that DM probably consists of particles with interactions and masses in the weak-scale.

\subsection{Dark matter candidates}

\subsubsection{Neutrinos}

Neutrinos has been considered for a long time candidates for DM\cite{schramm}.
Light neutrinos with masses $m < 30$ eV are an example of Hot Dark Matter (HDM). They are relativistic at decoupling, erase density
perturbation through free-streaming. Light neutrinos are ruled out as main component of DM because they give rise to very large structures that later in a top-down process of fragmentation gives rise to smaller scale structures like galaxies. In this scenario, galaxies form late at $z \leq 1$ in contraddiction with observations. 

In SM no right-handed states are possible and this brings to the consequence that neutrinos cannot have mass, unless one adds the quoted right-hand state, resulting in the generation of a Dirac mass for neutrinos. Adding a further term (see Olive 2003\cite{olive}) a Majorana mass can also be generated. 

The calculation of the neutrinos relic abundance is different for neutrinos with mass, $m_\nu$, larger than 1 MeV and smaller that the quoted value (which is the temperature at which decoupling happens) (see Fig. 32). Neutrino with mass $> 1$ MeV annihilate before decoupling, while this does not happen for neutrinos with masses $< 1$ MeV. For neutrinos with $m_\nu< 1$ MeV the condition that the universe have an age $>12$ Gyr leads to the constraint 
$\Omega h^2=\Sigma m_\nu/91.5 eV \leq 0.3$ which in terms of total mass reads $m_{tot} \leq 28$ eV for Majorana neutrinos and $m_{tot} \leq 14$ eV for Dirac neutrinos. HST key project data SIa, BBN\cite{lewis_bridle} brings to $m_{tot} \leq 0.3$ eV. Concerning neutrinos with $m_\nu > 1$ Mev, we have that $\Omega_\nu h^2 \propto 1/(<\sigma_{ann} v) \simeq \/m_\nu^2$, that combined with the constraint $\Omega h^2 \leq 0.3$ leads to a lower limit to $m_\nu $, namely 
$m_\nu \geq 3-7$ GeV, according to the the Dirac or Majorana nature of the neutrino. LEP constraints on neutrinos flavours $N_\nu=2.9841 \pm 0.0083$ excludes $m_\nu \leq 45$ GeV implying $\Omega h^2 \leq 0.001$ for masses 45 GeV$<m_\nu< 100$ TeV. Dirac neutrinos with 10 GeV $<m_\nu < 4.7$ TeV are excluded by Lab constraints\cite{ahlen,caldwell,beck}. Majorana neutrinos have $\Omega h^2 \leq 0.001$ for $m_\nu> 45$ GeV, and then they are not interesting from the cosmological point of view. 
Neutrino mixing, LEP limit on neutrino species light weakly interacting neutrinos leads to 0.05 eV $< m_{tot}< 8.4$ eV. Experimental and observational data leads to conclude that weakly interacting neutrinos have $0.0005 < \Omega_\nu h^2 < 0.09$.

\begin{figure}
\resizebox{8.65cm}{!}{\includegraphics{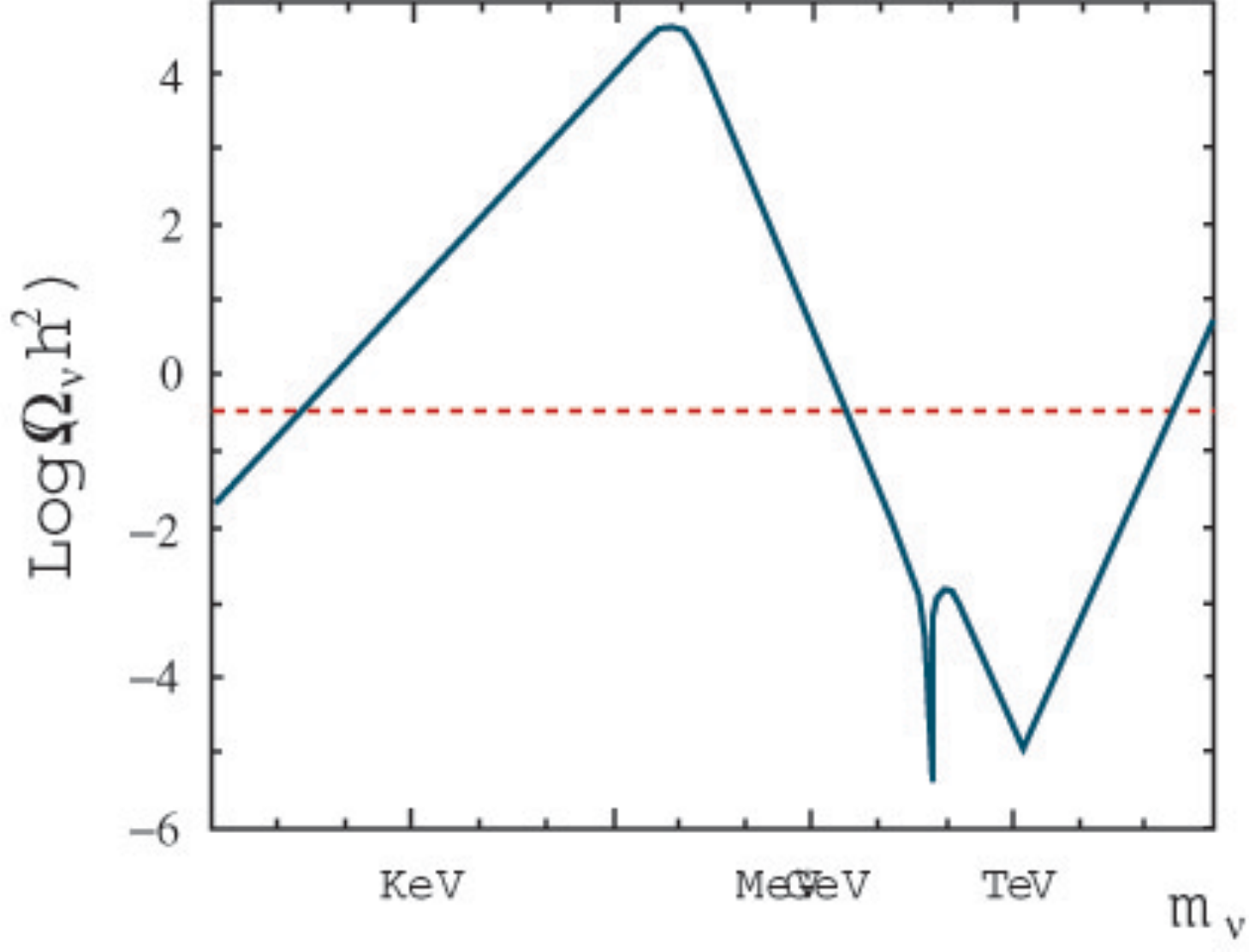}}
\caption{Relic density of Dirac neutrinos 
(From K. Kainulainen and K. A. Olive\cite{kainulainen})} 
\end{figure}

\subsubsection{Axions}

Experiments show no CP violation in the QCD sector but natural terms in QCD Lagrangian 
\begin{equation}
L_{QCD}=L_{PERT}+ \overline{\Theta} g^2/(32 \pi^2) G^{a \mu \nu} \tilde{G}_{a \mu \nu}
\end{equation}
(where $L_{PERT}$ is the remaining part of the Lagrangian, $G^{a \mu \nu}$ is the gluon field strength tensor, and $g$ the coupling constant)
are able to break the CP simmetry (e.g., a nonzero choice of the $\Theta$ angle). 
This induces a large electric dipole moment for the neutron, $d_{n_{QCD}} \simeq 10^{-15}$ e cm, while observations give $d_{n_{data}} \simeq 10^{-26}$ e cm. This question constitutes what is known as the strong CP problem. 
Promoting $\Theta$ to a dynamical variable, and postulating a spontaneously broken global $U(1)_{PQ}$ symmetry (at a scale $f_A$), 
the theta-term is effectively driven to zero, solving the problem\cite{peccei,peccei1}.
The associated pseudo-Nambu Goldstone boson is the axion, having a mass $M_A \simeq 0.62$ eV $10^7 GeV/f_A$. The value of the relic density depends on the assumption made on the way it is produced. Axions are characterized by extremely weak interactions, and are non-thermal relics. Stellar cooling, SN 1987A data, microwave cavity experiments, polarization experiments, constraints axion mass as $m_A< 0.01$ eV. An anomalous signal was reported by the Italian collaboration PVLAS (Polarizzazione del Vuoto con LASer), and was interpreted as a possible hint of its existence (Zavattini\cite{zavattini,rabaden}).
However, Robilliardet al. (2007)\cite{robilliard} showed with a 94\% confidence level that the results of\cite{zavattini} were incorrect.
Zavattini et al. (2006)\cite{zavattini1} concluded after performing an experiment with higher accuracy, that the axion interpretation of the anomalous signal was ruled out.

\subsubsection{Sterile neutrinos}

This kind of hypothetical particle was somehow discussed in the section dealing with neutrinos, and proposed by Dodelson and 
Widrow (1993)\cite{dodelson}. It is a light right-handed neutrino that do not interact via any of the fundamental interactions of the SM, apart from mixing, and except gravity. Limits on the particle mass come from WMAP result relative to reionization. They imply that DM structures were already formed at $z>20$. This result is contraddicted if the mass of the particle is smaller than $\simeq 10$ keV\cite{yoshida, shi}.

\subsubsection{Dark matter from Little Higgs models}

As we will see in the next sections, one of the problems of the SM, namely the "hierarchy problem" can be solved 
introducing supersimmetry. In the "little Higgs" models\cite{arkani, arkani1,arkani2,arkani3} the Higgs boson of the SM is a pseudo-Goldstone boson, 
arising from some global symmetry breaking at a $\simeq 10 $ TeV energy scale.
The "hierarchy problem" is solved in these models by a spontaneous breaking of such approximate global symmetries to stabilize the mass of the Higgs boson(s), responsible for electroweak symmetry breaking. In some of the little Higgs models a stable scalar particle exist with the correct observed DM density\cite{birkedal}.

\subsubsection{Light scalar dark matter}

Lee \& Weinberg (1977)\cite{lee} showed that in the case of DM with fermi interactions cannot
have a mass smaller than a few GeV. This limit can be avoided if DM has other characteristics. Boehm, Ennsslin, \& Silk (2002)\cite{boehm_es}
, and Boehm \& Fayet (2002)\cite {boehm_fayet}
proposed scalar candidates in the mass range 1-100 MeV. This candidate could explain the 511 keV line observed by INTEGRAL from the galactic bulge generated through WIMPS annihilation into positrons, which would produce by a subsequent annihilation the line observed in gamma-rays\cite{boehm_etc} (see the section on detection of DM).

\subsubsection{Superheavy dark matter}

Superheavy dark matter candidates, generically called "Wimpzillas"\cite{kolb_etc, chang}, are candidates with mass $>10^{10}$ GeV. 
These candidates during freez-out were not in thermal-equlibrium, and their relic abundance depends from their production cross section,  
not by their annihilation cross section. These particle could be gravitationally produced at the end of inflation\cite{kolb_etc1}. A motivation to wimpzillas is connected to cosmic rays observed at energy above the GZK (Greisen-Zatsepin-Kuzmin) cut-off ($\simeq 5 \times 10^{19}$ eV). The GZK limit is a theoretical upper limit on the energy of cosmic rays (CR) coming from "distant" sources. The limit is set by slowing-interactions of cosmic ray protons with the microwave background radiation over long distances ($\simeq 50$ Mpcs). The quoted Ultra-GZK Cosmic Rays could be produced through 
annihilation or decay of superheavy dark matter particles (e.g., P. Blasi, R. Dick and E. W. Kolb\cite{blasi}).



\subsubsection{Supersimmetric dark matter}

There are severeal reasons to consider supersymmetry one of the most insteresting extensions of the SM. 
Weakscale supersymmetry furnish a solution to the hierarchy problem, linked to the huge difference between 
electroweak and Planck energy scales\cite{maiani,thooft,witten,amaldi}, enables gauge-coupling unification\cite{dimo_etc}
and also furnish a stable dark matter candidate\cite{pagels}.

Concerning the hierarchy problem, SM predict very precisely the results of experiments, and this high precision requires calculations of higher orders. 
The mass of each particle is constituted by their "bare" mass and radiative corrections. Fermion mass corrections increase logarithmically, while scalars 
mass increase quadratically (with corrections at 1-loop). The mass of the Higgs boson similarly to other scalars mass can be written as:
\begin{equation}
M^2_H=M^2_{H,bare}+\delta M_H^2=M^2_{H,bare}+\lambda^2_f/(8 \pi^2) \Lambda^2
\end{equation}
being $\Lambda$ a high-energy cut-off, which is the scale at which effects of the "new physics" have a role. If for example 
$\Lambda \simeq M_{P} \simeq 10^{19}$ GeV, the Higgs mass would be huge, while it is expected to be of the order of the electroweak scale $M_W \simeq 100$ GeV. This will ruin the electroweak scale stability. The problem can be solved introducing a "supersymmetry"
\begin{equation}
Q|boson>=|fermion>; Q|fermion>=|boson>
\end{equation}
changing fermions into bosons and vice versa.
In simple words, supersymmetry is an extension that creates 'superpartners' for all SM particles: squarks, gluinos, 
sleptons, higgsinos, etc,
are superpartners of quarks, gluons, leptons, higgs boson, etc. 

Assuming that new particles with spin different by one half and similar mass exist, the correction to the Higgs mass is  
\begin{equation}
\delta M^2 \simeq \frac{\lambda_f^2}{8 \pi^2} (\Lambda^2+m_B^2)-\frac{\lambda_f^2}{8 \pi^2} (\Lambda^2+m_F^2) \simeq \frac{\lambda_f^2}{8 \pi^2} (m_B^2-m_F^2)
\end{equation}
since $ |m_B^2-m_F^2| < $ 1 TeV, the Higgs mass divergence is cancelled. 

Before going on, I would like to enter a bit more in detail on the "hierarchy poblem" and the quadratic divergences. A new point of view\cite{aoki}
is that there are two different hierarchy problems: 1) Why is the EW scale so small compared to the cut-off scale? This is connected to
quadratic divergence. 2) Why is the EW scale so small compared to the GUT scale (if exists). This is connected to logarithmic divergence. From the Wilsonian RG point of view, one can give arguments that quadratic divergence is not generated by radiative corrections and it is not the real issue of hierarchy problem, so that we need to take care of only logarithmic divergences. This broadens the possibilities of model constructions beyond SM\cite{aoki}.

As already told, the second reason for which SUSY is interesting, is that with its introduction at TeV scale the three gauge interactions of the SM, which define the electromagnetic, weak, and strong interactions, are merged at a scale $M_U \simeq  2 \times 10^{16}$ GeV, into one single interaction characterized by one larger gauge symmetry and thus one unified coupling constant.

The third important reason that makes SUSY interesting is that it furnish a stable dark matter candidate. 
Incorporating supersymmetry into the SM requires doubling the number of particles 
and with the addition of new particles, there are many possible new interactions. The simplest possible supersymmetric model consistent with the Standard Model is the Minimal Supersymmetric Standard Model (MSSM) which can include the necessary additional new particles that are able to be superpartners of those in the SM.
The absence of interactions producing rapid proton decay led to assume the conservation of R-parity in the MSSM:
\begin{equation}
R = (-1)^{2s+3B+L}
\end{equation}
where s is the spin, B the baryon number, and L the lepton number. $R = +1$ for all SM particles and $R = -1$ for all superpartners.
The conservation of the quoted R-parity requires creation and destruction of superpartners in pairs, and as a consequence exists at least one SUSY particle, the lightest SUSY particle (LSP), that is stable\cite{goldberg}.
The identity of the LSP depends from the details SUSY is broken. 

MSSM, despite its name, contains over a hundred (more precisely 124) new parameters. In order to simplyfy the model is assumed that all gauginos (superpartner of a gauge field) masses receive the same mass, $m_{1/2}$. Another assumption is that scalar masses have a common value $m_0$ at the GUT scale. In order to obtain the gauginos mass and scalar masses at the EW scale one uses a set of Renormalization Group Equations (RGEs). 
For what concernes the Higgs sector, in the MSSM we have $\mu$ (Higgs mixing mass parameter), two Higgs doublets, which are two vevs (vacuum expaction values), namely the ratio of the two vevs, $tan \beta=v_1/v_2$ and another related to the mass of the Z boson.

\begin{figure}
\resizebox{8.65cm}{!}{\includegraphics{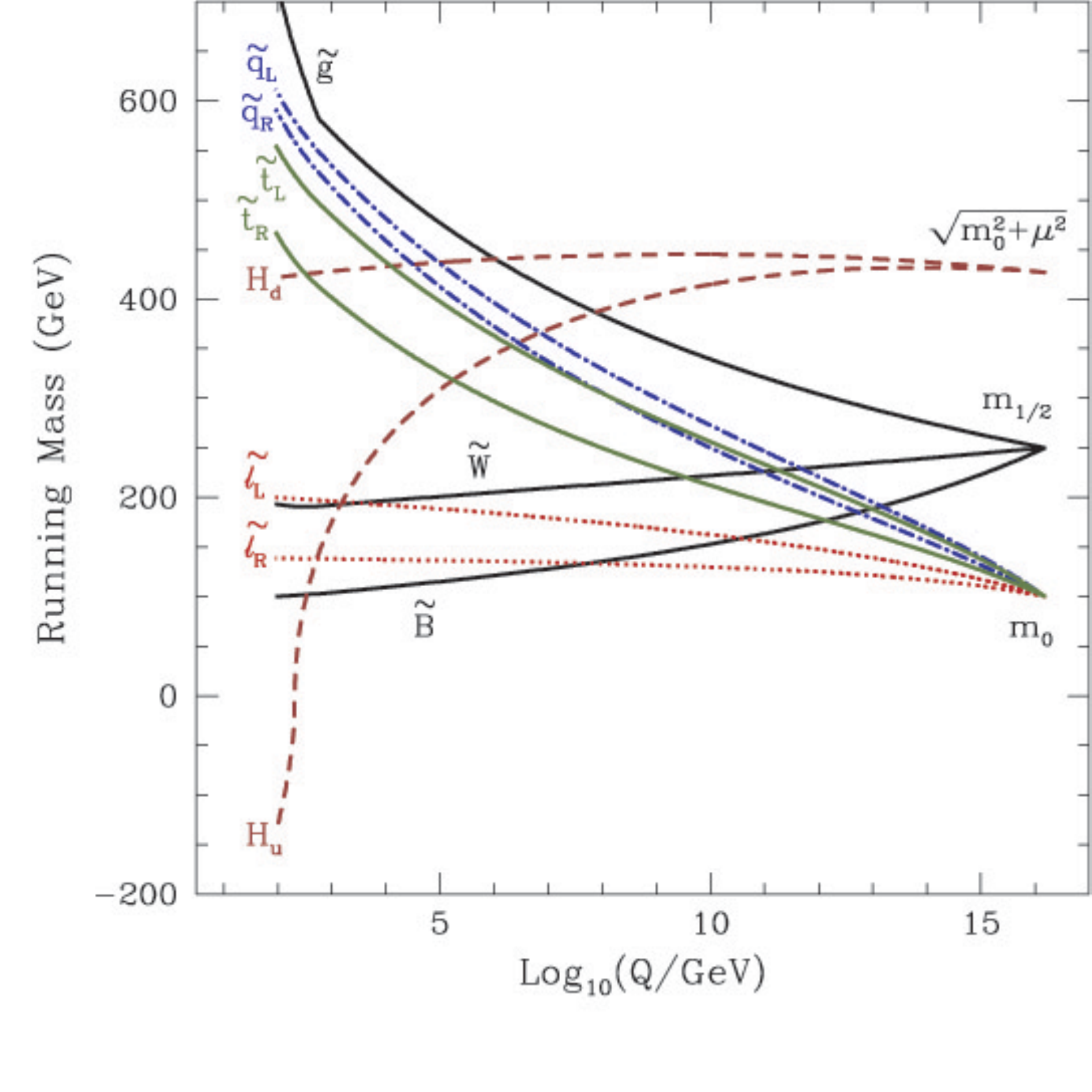}}
\caption{RG evolution of the mass parameters in the CMSSM (from Olive 2003)\cite{olive}).} 
\end{figure}

Then, a further simplification can be obtained if the supersymmetry breaking Higgs soft masses are unified at GUT scale, and have mass $m_0$. With these assumptions, we obtain the so called CMSSM (constrained MSSM) model characterized by the parameters: $m_{1/2}$ (gauginos mass); $m_0$ (scalars mass); $A_0$ (soft breaking trilinear coupling constant); $tan \beta=v_1/v_2$; sign(${\mu}$) (signum of the Higgsino mass parameter).
Setting the values of these limited number of parameters, all of the masses of the supersymmetric particles at the weak scale can
be determined through RG evolution of the mass parameters (see Fig. 33). 
Another simplified model, focusing on of supersymmetric phenomenology is the so called phenomenological MSSM (pMSSM (see \cite{bertone}).

The only electrically neutral and colorless superparnters in MSSM are the four neutralinos, 
three sneutrinos, and the gravitino. Each one of the four neutralinos is a linear combination of the neutral fermions\cite{ellis,goldberg1}, namely the superpartner of the 3-rd component of the $SU(2)_L$ gauge boson (named wino, $\tilde{W}^3$), the superpartner of of the $U(1)_Y$ gauge field corresponding to weak hypercharge (named the bino, $\tilde{B}$), and the neutral Higgsinos, $\tilde{H_1}$, and $\tilde{H_2}$. 
The lightest of the four states is referred to as neutralino:
\begin{equation}
\chi_0=N_{11} \tilde{B}+ N_{12} \tilde{W}^3 +N_{13} \tilde{H_1} +N_{14} \tilde{H_2}
\end{equation}
$|N_{11}|^2+|N_{12}|^2$ is the gaugino fraction and  $|N_{13}|^2+|N_{14}|^2$ the higgsino fraction of the lightest neutralino. 
The $N_{ij}$ are the elements of the unitary matrix, $N$, that diagonalizes the neutralino mass matrix, $M_{\chi^0}$, that allows to get the masses and mixings of these four states: 
\begin{equation}
M_{\chi^0}= \,\,\,\,\,\,\,\,\,\,\,\,\,\,\, \,\,\,\,\,\,\,\,\,\,\,\,\,\,\, \,\,\,\,\,\,\,\,\,\,\,\,\,\,\, 
\,\,\,\,\,\,\,\,\,\,\,\,\,\,\, \,\,\,\,\,\,\,\,\,\,\,\,\,\,\, \,\,\,\,\,\,\,\,\,\,\,\,\,\,\, \,\,\,\,\,\,\,\,\,\,\,\,\,\,\, \,\,\,\,\,\,\,\,\,\,\,\,\,\,\, \,\,\,\,\,\,\,\,\,\,\,\,\,\,\, \,\,\,\,\,\,\,\,\,\,\,\,\,\,\,
\end{equation}
\begin{eqnarray}
\arraycolsep=0.01in
\left( \begin{array}{cccc}
M_1 & 0 & -m_Z\cos \beta \sin \theta_W^{} & m_Z\sin \beta \sin \theta_W^{}
\\
0 & M_2 & m_Z\cos \beta \cos \theta_W^{} & -m_Z\sin \beta \cos \theta_W^{}
\\
-m_Z\cos \beta \sin \theta_W^{} & m_Z\cos \beta \cos \theta_W^{} & 0 & -\mu
\\
m_Z\sin \beta \sin \theta_W^{} & -m_Z\sin \beta \cos \theta_W^{} & -\mu & 0
+\end{array} \right)\nonumber \;, 
\end{eqnarray}
in the $\widetilde{B}$-$\widetilde{W}^3$-$\widetilde{H}_1$-$\widetilde{H}_2$ basis.
In the matrix, $\theta_W$ is the Weinberg angle, and $M_1$, and $M_2$ the masses of the bino and wino. 
The other parameters in the matrix have been described previously.

The neutralino is a thermal relic, a Majorana fermion, neutral and colorless with weak-type interactions, stable if R-parity is conserved.
As all Majorana fermions (or particles), it is a fermion that is its own antiparticle\footnote{Majorana particles were hypothesised by Ettore Majorana in 1937\cite{majorana}. No elementary fermions are known to be their own antiparticle, though the nature of the neutrino is not settled and it might be a Majorana fermion. In condensed matter physics, Majorana fermions exist as quasiparticle excitations in superconductors and can be used to form Majorana bound states. 
}.

Since they are Majorana fermions, neutralinos, should annihilate each other, when they get very close to other neutralinos. 

After neutralinos general annihilation cross-section is known\footnote{Annihilation proceeds mainly through sfermion exchange, and the bino is the LSP in much of the parameter space of interest}, the relic abundance is obtained solving the Boltzmann equation similarly to the case of massive neutrinos and given by\cite{ellis, goldberg1}:
\begin{equation}
\Omega_{\chi} h^2 \simeq 1.9 \times 10^{-11} (T_{\chi}/T_{\gamma})^3 N_f^{1/2} \frac{GeV}{a x_f +1/2 b x_f^2}
\end{equation}
where $f$ stands for freeze-out and $a$ and $b$ comes from the expansion $\sigma v = a+bx+. . ..$, and $T_{\gamma}$ the photon temperature.

Since the relic abundance is larger than what expected from observations, over most of the supersymmetric parameter space, it is necessary to consider only peculiar regions in which neutralino annihilation efficiency is high (see Fig. 34).

\begin{figure}
\resizebox{12.65cm}{!}{\includegraphics{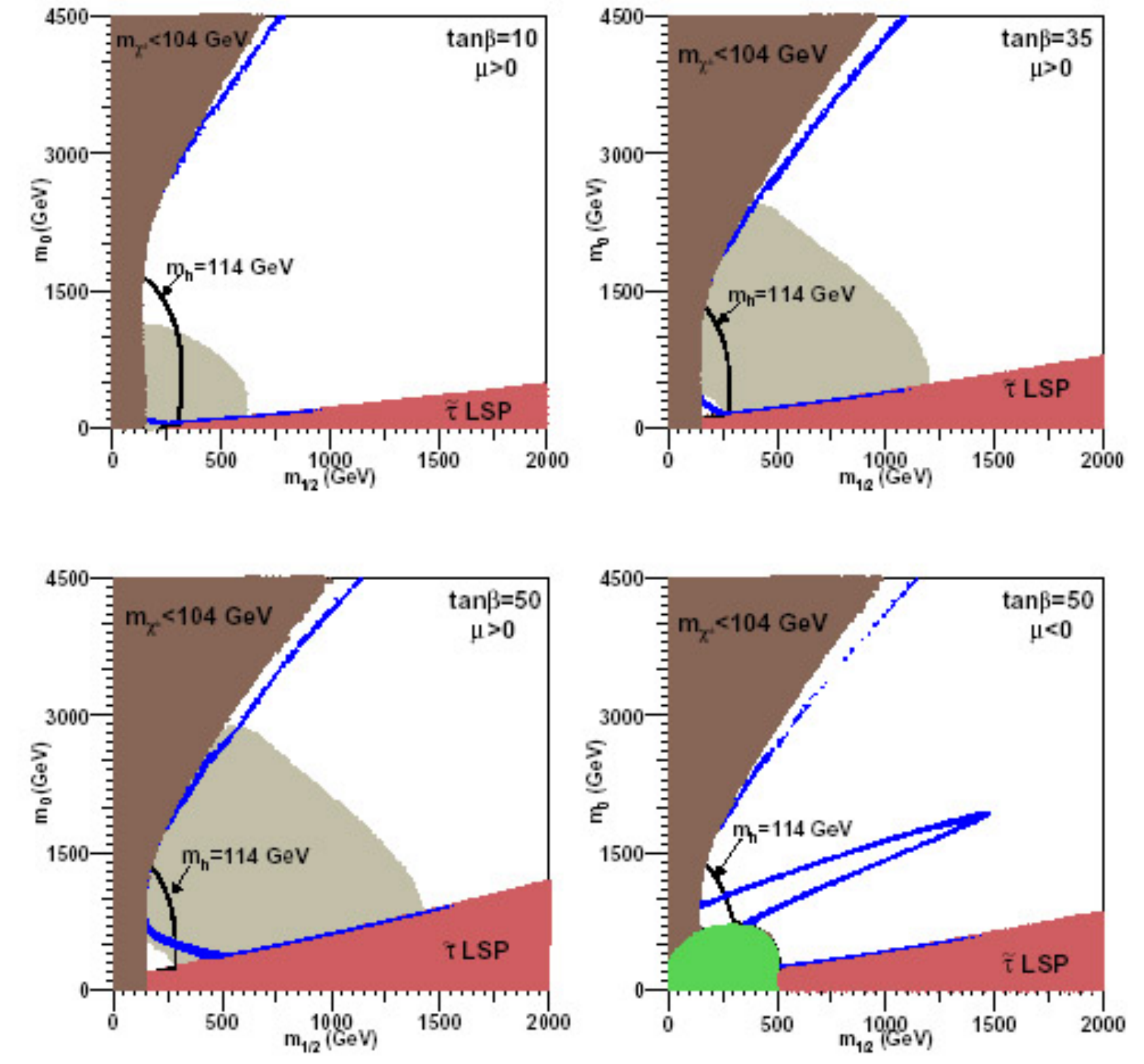}}
\caption{CMSSM space: representative region. Blue regions: neutralino density in agreement with DM abundance. Lower right region: contains a $\tilde{\tau}$, and then disfavored. However, just outside of this region the slightly larger mass of $\tilde{\tau}$ with respect to the LSP (neutralino) gives rise to neutralino-$\tilde{\tau}$ coannihilation.
Upper left shaded region (focus point):disfavored by the LEP bound on chargino (particle which refers to the mass eigenstates of a charged superpartner (i.e., any electrically charged fermion predicted by SUSY)).
The solid line with $m_h=114$ Gev is the LEP bound on light Higgs (nowadays known to be $\simeq 125$ GeV). Light shaded region: region favored by muon's magnetic moment. Lower right frame: A-funnel region, viable region along the CP-odd Higgs resonance 
($m_{\chi_0} \simeq m_A/2$). The CSSM parameters $A_0$ is set $=0$ and $\mu>0$ (from Hooper 2009\cite{hooper}).} 
\end{figure}

Before concluding this section on SUSY DM, I want to add some comments on the results of LEP results of 2012. Direct constraints from the LHC's CMS and larger ATLAS detectors have not seen any sign of sparticles.They saw only particles inside the SM. 
Mixed signals have also come from studies of the Higgs boson, which behaves just as SM predicts with no sign of particles outside the SM influencing it. As I discussed before, SUSY predicts other Higgs particles $h$, $H$, $H^{\pm}$, and $A$ (the MSSM Higgs bosons), not yet discovered. Another problem adds to the previous ones. Physicists from the LHCb experiment, clocked  a rare decaying process, namely the one from $B_S$ meson to a muon-antimuon pair, which has a probability of 1 over $ \simeq 3 \times 10^8$ to happen\cite{chalmers}. This probability would increase to 1 over $ \simeq 3 \times 10^6$ with "a little help" of sparticles. However, the $B_S$ behaves as no sparticle exists, and this could be a further damaging result for SUSY. In conclusion, SUSY has much less chance than before the experiments to be correct, in its actual form or better in its CMSSM incarnation. The previous experiments on Higgs boson or the $B_S$ meson behavior, gives indirect answer to SUSY existence, and contrarily to what one can think, these results (paradoxically) are more significative than directly looking for superparticles without finding them. If sparticles existed, independently from their mass, their effects should be visible (similar story of the unseen planets and stars discussed in the Introduction, but whose effects made clear their existence). 
Another issue to add, is that one reason to introduce SUSY was that of stabilizing the EW scale, but even in this 
there is disagreement about whether superheavy sparticles can offer enough stabilization.

Scientists working in the area, are waiting for the experiments that will be performed, probably, in the early 2015, when LHC will reopen and will operate at 13 TeV instead of the actual 8 TeV. The "upgraded" LHC would have some more chances to "directly" detect sparticles.
Unfortunately, a troubling possibility is that sparticles have masses some TeVs larger than the energies the "upgraded" LHC can 
catch, or even too heavy to be measured in "terrestrial" colliders. However, as I stressed before, in order to be sure of a particle existence, one does not necessarily need to chase it directly, it should give signs of its existence indirectly, at least, before one tried to "see" it.    

All the previous discussion, affects somehow "DM theory". Too heavy sparticles could also be not convincing candidates for DM. We recall, however, that DM is not necessarily made of SUSY sparticles.

\subsubsection{UED and KK dark matter}

Even if our universe is characterized by three spatial dimensions and one time dimension, at higher scales other dimensions could exist. 
In extra dimensions models, the usual (3+1) dimensions, called brane, are embedded in a spacetime having $3+\delta+1$ spacetime, the so called bulk. 
SM fields are confined in the brane while gravity propagates in the extra dimensions. In this kind of models, the hierarchy problem is solved in different ways. 
For example in the Arkani-Hamed, Dimopoulos and Dvali\cite{arkani4} (ADD) scenario, the extra dimensions are compactified on different topologies (e.g., circle) of scale $R$, with the effect of lowering the Planck scale energy close to the EW. Another way to reach the same goal is to introduce extra dimensions with large curvature, like in the Randall and Sundrum\cite{randall_su} (RS) scenario. 
Compactification of extra dimensions gives rise to a quantization of momentum, $p^2 \simeq 1/R^2$, of the fields propagating in the bulk, and the apparition of a set of Fourier expanded modes (Kaluza-Klein (KK) states, for each bulk field. 
Particles moving in extra dimensions appear as heavy particles with masses $m_n=n/R$ (a set of copies of normal particles).
The new states have the same quantum numbers (e.g., charge, color, etc). 

A peculiar case of extra dimensions, are the Universal Extra Dimensions (UED). In this case, all the SM fields propagates in flat extra dimensions. 
The universal extra dimensions are assumed to be compactified with radii much larger than the traditional Planck length, but smaller than in the ADD model, $ \simeq 10^{-18}$ m. Extra dimensions are characterized by $R \simeq $1/TeV.

If extra dimensions are compactified around a circle or torus, the extradimensional momentum conservation implies conservation of the KK number
n, and the lightest 1-st level KK state is stable\cite{kolb_sla}.

However, in order to obtain chiral fermions at zeroth KK level, the extra dimension is compactified on an $S^1/Z_2$ orbifold (orbit-fold)\footnote{
A compactification on an orbifold, $S^1/Z_2$, a circle $S^1$ with the extra identification of y with -y, corresponds to the segment $y \in  [0, \pi R]$, a manifold with boundaries at $y=0$ and $y = \pi R$ }.
The orbifold leads to the violation of KK number conservation but can leave a "remnant" of the quoted simmetry, usually named  KK parity. 
The conservation of the KK-parity implies that interactions require an even number of odd KK modes, leading to the lightest Kaluza-Klein particle (LKP)
stability.

Theories with compact extra dimensions can be written as theories in ordinary four dimensions by performing a Kaluza Klein (KK) reduction.
Consider as an example a field theory of a complex scalar in flat five-dimensional (5D) spacetime. The action will be given by
\begin{equation}
S=\int dx^4 \int_{-\pi R}^{\pi R}  dy L_{SM} (x,y)
\end{equation}
As an example, we may suppose that the fifth dimension is compact with the topology of a circle $S^1$ of radius R, which
corresponds to the identification of y with $y + 2\pi R$.
In such a case, the 5D complex scalar field can be expanded in a Fourier series:
\begin{equation}
\phi(x^{\mu},y)=\sum_{n=-\infty}^\infty \frac{1}{\sqrt{2 \pi R}} \phi^{(n)} (x^\mu) e^{iny/R}
\end{equation}
Zero modes are identified with SM fields and one gets
\begin{equation}
S=\int dx^4 \int_{-\pi R}^{\pi R}  dy L_{SM} (x,y)=\int dx^4 [L_{SM}^{(0)}(x)+.... ]
\end{equation}
which is a 4D theory. The parameters in UED models are completely specified
in terms of the SM parameters. There are only three free parameters in minimal UED model:
$R$ (size of extra dimension), $\Lambda$ (cut-off scale), $m_h$ (Higgs boson mass).

\begin{figure}
\resizebox{6.65cm}{!}{\includegraphics{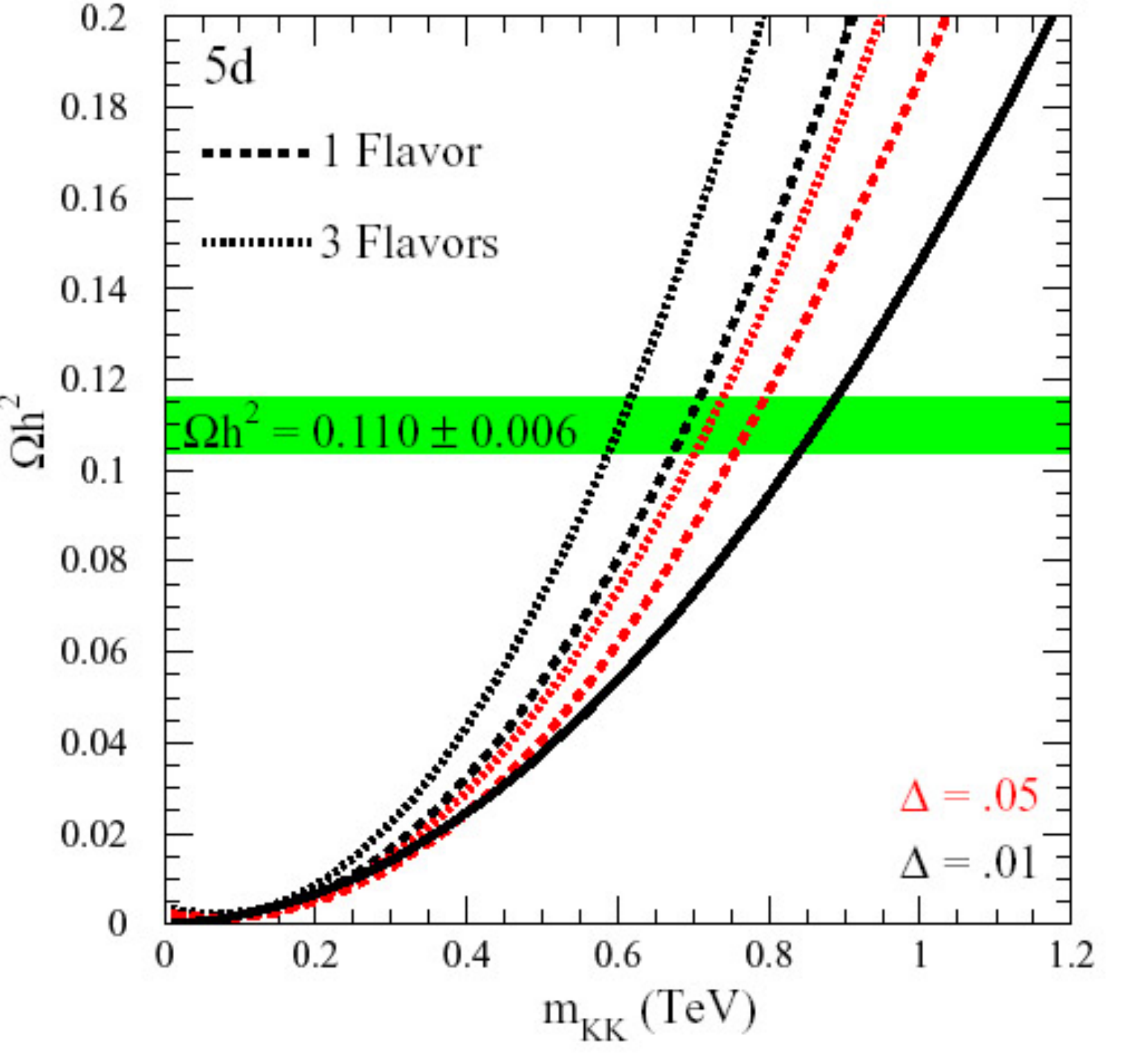}}
\caption{KK relic abundance. Solid line $B^{(1)}$ alone. Dashed line (dotted line): $B^{(1)}$ plus the effect of KK leptons 5\% (1\%) heavier that the LKP.
Horizontal band: measured DM abundance (from Hooper 2009)\cite{hooper}). }
\end{figure}

The LKP's candidates are the first KK excitation of the hypercharge gauge boson, the $B^{(1)}$ state (more commonly indicated with: first KK excitation of photon); KK excitation of: Z, neutrinos, Higgs boson, graviton.
The first level KK excitation of Higgs is not a likely candidate because of its large zero-mass mode, while direct detection excludes KK sneutrinos (as sneutrinos and fourth generation Dirac neutrinos).
Consequently the LKP is most likely associated with the $B^{(1)}$ state\cite{cheng,schmaltz}, or a mixture of $B^{(1)}$ and KK Z 
(Olive 2003\cite{olive}).

Since KK leptons freeze-out quasi independently from the LKP, they have a larger relic abundance with respect other states (see Fig. 35).

\subsection{Dark Matter detection}

There are different ways to try to verify the existence of the zoo of particles about which we discussed. Here we discuss the direct and indirect detection methods. 

The idea at the base of direct detection is that since DM (WIMPs) fill in our galaxy, and one can try to detect these particles when they pass through Earth, studying the DM scattering with nuclei\cite{drukier, wasserman}.

In the case of elastic scattering, a WIMP interact with a nucleus causing it to recoil. In inelastic scattering the WIMP interact with the target's orbital electrons exciting them or giving rise to ionization of the target.  
Scattering can be also spin dependent or independent. In spin dependent scattering, interactions are originated from couplings to the nucleon spin, and the cross section is proportional to $J(J+1)$. In spin-independent scattering the cross-section is proportional to $A^2$, where $A$ is the atomic mass of the nuclei of the target. 
For a WIMP of $\simeq 100 $ GeV, and velocity $10^{-3}$ c, the recoil energy is $\simeq 1-100$ kev.
Experiments which attempt to detect dark matter particles through their elastic scattering with nuclei (normal
matter recoiling from DM collisions), include CDMS, XENON , ZEPLIN, EDELWEISS, CRESST, CoGeNT, DAMA/LIBRA, COUPP, WARP, and KIMS .
Considering a spin-independent experiment the rate observed in a detector is $R= \sigma_A I_A$, where
\begin{equation}
\sigma_A=\frac{4 \mu_A^2}{\pi} [Z f_p+(A-Z)f_n]^2 
\end{equation}
where $f_p$ and $f_n$ are the WIMP's couplings to protons and neutrons $\mu_A= M_\chi M_A/(M_\chi+M_A)$, $A$, and $Z$ are the mass and atomic number, respectively, and 
\begin{equation}
I_a =N_T n_\chi \int dE_R \int_{v_{min}}^{v_{esc}} d^3 v f(v) \frac{m_A}{2 v \mu_a^2} F_A^2(E_R) 
\end{equation}
where $N_T$ is the number of target nuclei, $n_\chi$ is the local DM number density, $E_R$ is the recoil energy,
$f(v)$ is the DM velocity distribution, and $F_A$ is the nuclear physics form factor. 
Results are typically reported assuming $f_p=f_n$, so $\sigma_A \simeq A^2$ , and scaled to
a single nucleon. Elastic scatter of neutralinos with quarks happens through s-channel squark exchange or 
t-channel CP-even Higgs exchange. In scattering dominated by heavy Higgs exchange with mass (e.g.) 200 GeV, the calculated cross section of a $\simeq 100$ GeV neutralino with a nucleon is $10^{-5}-10^{-7}$ pb, for $|\mu| \simeq 200 $ GeV, and 
$10^{-7}-10^{-9}$ pb for $|\mu| \simeq 1$ TeV.  In the case of a target of Germanium (e.g., CDMS or Edelweiss experiments) a WIMP with a cross section $10^{-6}$ pb would produce an elastic scattering per kilogram-day
of exposure. In the case of Kaluza-Klein DM, the KK-Nucleon cross sections is very small (see Hooper 2009\cite{hooper})
). Detection of KK DM would require ton-scale detectors. 

The previous values of the cross section WIMP-Nucleon, are theoretically calculated. Several experiments has put 
constraints on the same quantity. For example Aprile et al. (2011)\cite{aprile} (see Fig. 36) showed the limits from different experiments, (e.g,  XENON100 (2010), EDELWEISS (2011), CDMS (2009), CDMS (2011) and XENON10 (2011), Cogent, DAMA) and compared to predictions of the CMSSM, assuming that WIMPs are isothermally distributed in a halo with $v_0= 220$ km/s, and assuming a galactic escape velocity of $v_{esc}= 544^{+64}_{-46}$ km/s, and a density 0.3 GeV/$cm^3$ (0.008 $M_{\odot}/pc^3$).
The limits on cross section obtained are $\sigma= 1-10$ zb. 
Collision rate should change as Earth's velocity adds constructively/destructively with the Sun's, this should produce an annual modulation.
(Drukier, Freese, Spergel 1986\cite{drukier}). 
Researchers of DAMA experiment claimed to have seen a modulation
with $T \simeq 1$ yr, reaching its maximum on June 2 (8.9 $\sigma$ signal) (see Fig. 37). However, the claimed modulation has not
been seen in CDMS or Edelweiss experiments, which have higher sensitivity. 
Several studies have attempted to reconcile the DAMA modulation signal with the null results of other direct-detection experiments~\cite{dama_diversi}. 
For example a WIMP with mass of several GeV scattering elastically can satisfy DAMA result being consistent with CoGENT, CRESST, CDMS, and XENON results. 

\begin{figure}
\resizebox{9.65cm}{!}{\includegraphics{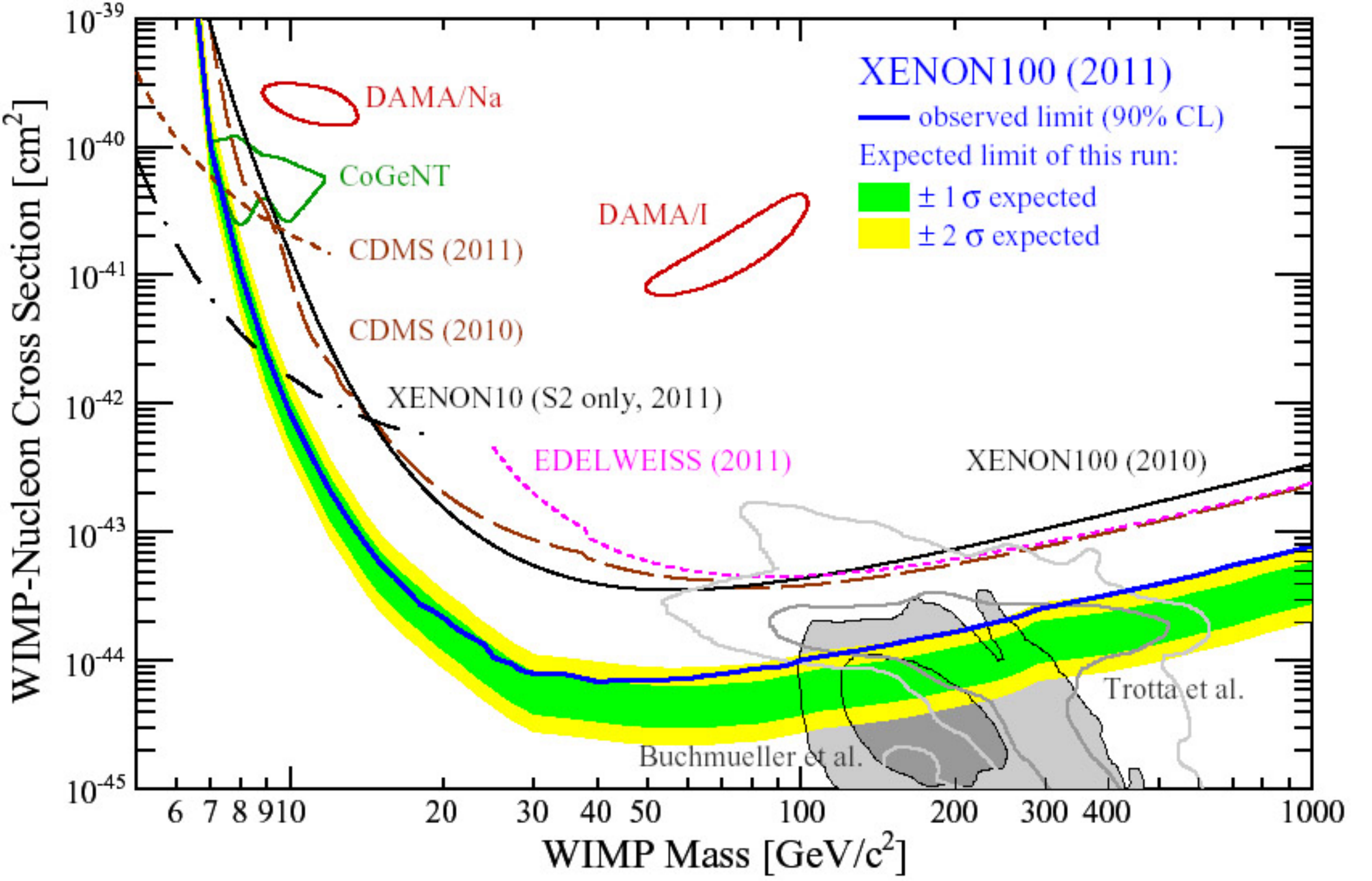}}
\caption{Spin-independent elastic WIMP-nucleon cross-section as function of WIMP mass (from Aprile et al. 2011)\cite{aprile}).}
\end{figure}

\begin{figure}
\resizebox{6.65cm}{!}{\includegraphics{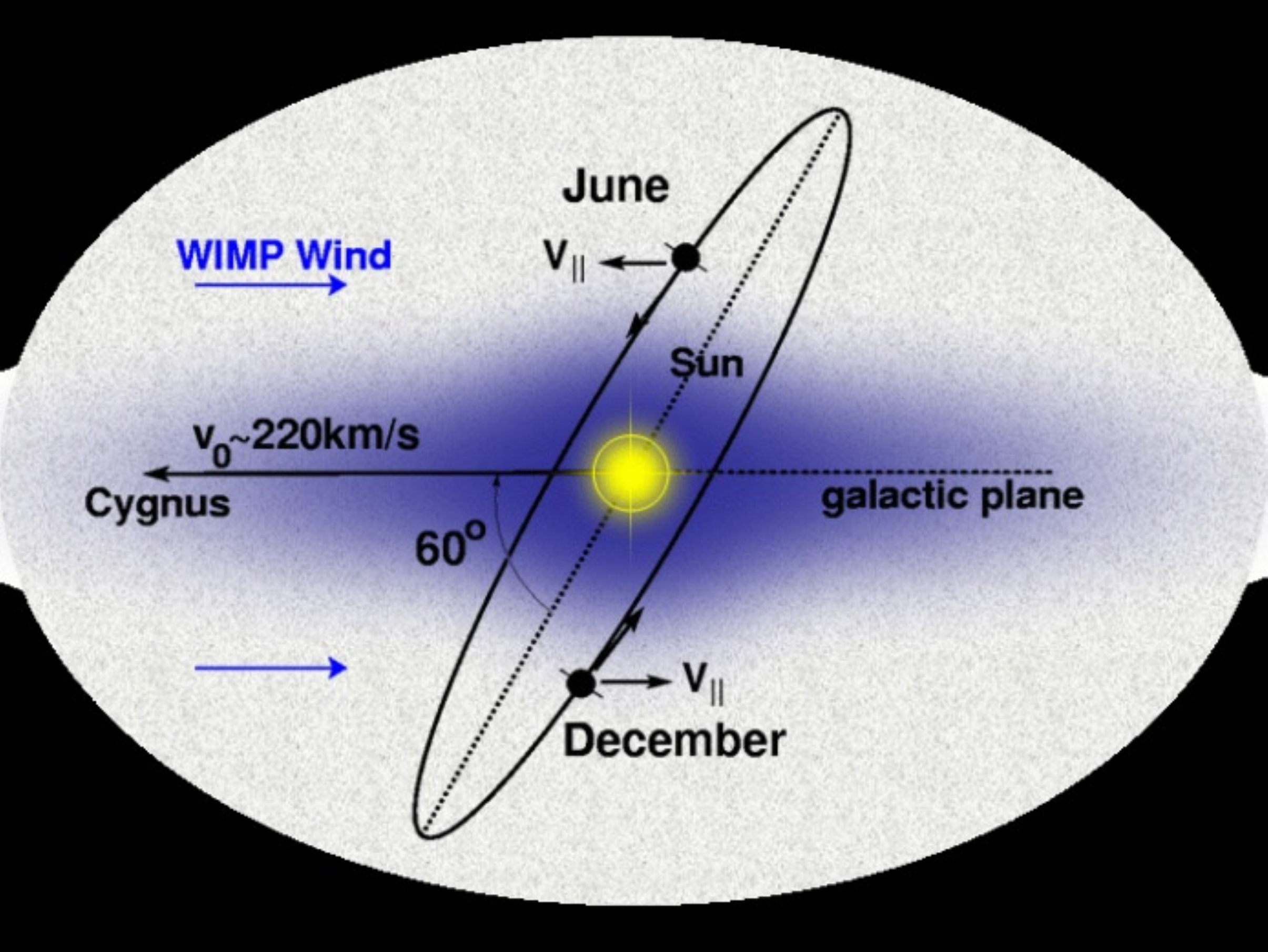}}
\resizebox{14.65cm}{!}{\includegraphics{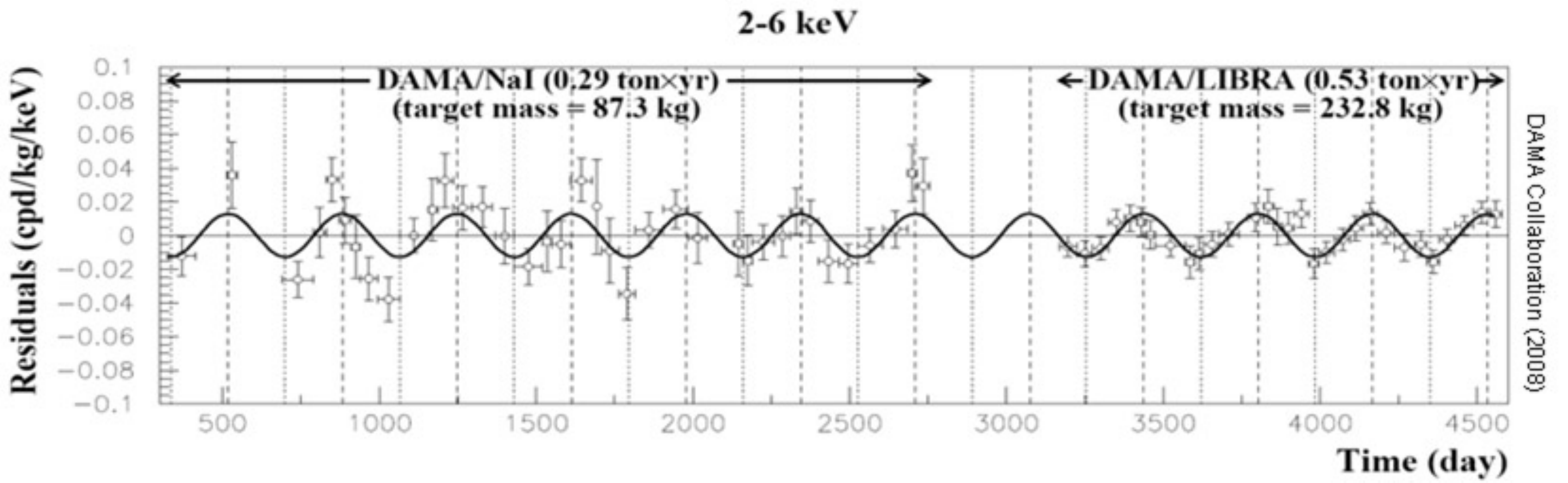}}
\caption{DAMA experiment (from Bernabei et al. (2008)\cite{bernabei}}
\end{figure}

\subsubsection{Indirect detection}

The second method used to detect DM is indirect detection, which tries to detect WIMPs annihilation products like electrons, positrons, antiprotons, neutrinos, and gamma rays. Gamma rays can be also produced through direct decay of WIMPs
to gamma rays: $\chi \chi -> \gamma \gamma$, $\gamma$Z, or $\gamma$h, giving rise to a gamma-ray line, which if observed would be a very strong evidence in favor of DM annihilation. 
Typical final states include heavy fermions: $b \overline{b}$, $t \overline{t}$, $\tau^{+}
\tau^{-}$, gauge or Higgs bosons: $W^{+} W^{-}$, $ZZ$, $hA$, $HA$, $ZA$, $Zh$, $ZH$, $W^{\pm} H^{\pm}$\footnote{$h$, $H$,
$H^{\pm}$, and $A$ are the MSSM Higgs bosons.}

Synchrotron and Inverse Compton
Relativistic electrons up-scatter
starlight to MeV-GeV energies, and
emit synchrotron photons via
interactions with magnetic fields

In the case of KK dark matter annihilation channels are $\tau^+ \tau^-$, $\mu^+ \mu^-$, and $e^+ e^-$\cite{servant,cheng}.

\subsubsection{Gamma Rays produced by annihilation of WIMPs}

The use of gamma rays as indirect search of WIMPs has some advantages on other indirect techniques: a) they are not influenced by magnetic fields and as a consequence they can be used to localize their sources; b) they are not attenuated and spectral information is not lost.  

Fig. (38) plots the gamma ray spectrum corresponding to different modes of annihilation

The places were is more probable to detect indirectly DM are high density regions like: the Galactic centre\cite{buckley,berezi}, dwarf satellite galaxies, black holes neighborhood, nearby galaxies. 

The prospects for this depend, however, on a
number of factors including the nature of the WIMP, the distribution of dark
matter in the region around the Galactic Center, and our ability to understand the astrophysical backgrounds present.

In order to determine the gamma ray flux coming from DM annihilation one needs the product of WIMP's annihilation cross section with the relative velocity, $<\sigma_{XX} |v|>$, the DM density in terms of the distance from the Galactic center, $\rho(r)$, and the gamma ray spectrum 
$dN_{\gamma}/dE_{\gamma}$ generated by WIMP annihilation. 
The flux can be written as:
\begin{equation}
\Phi_{\gamma}(E_{\gamma},\psi) = \frac{1}{2}<\sigma_{XX} |v|> \frac{dN_{\gamma}}{dE_{\gamma}} \frac{1}{4\pi m^2_X} \int_{\rm{los}} \rho^2(r) dl(\psi) d\psi.
\label{flusso}
\end{equation}
where los stands for line of sight, and $\psi$ is the is the angle measured from the Galactic centre direction. 
Averaging over $\psi$ Eq. (\ref{flusso}), one gets:
\begin{equation}
\Phi_{\gamma}(E_{\gamma}) \approx 2.8 \times 10^{-12} \, \rm{cm}^{-2} \, \rm{s}^{-1} \, \frac{dN_{\gamma}}{dE_{\gamma}} \bigg(\frac{<\sigma_{XX} |v|>}{3 \times 10^{-26} \,\rm{cm}^3/\rm{s}}\bigg)  \bigg(\frac{1 \, \rm{TeV}}{m_{\rm{X}}}\bigg)^2 J(\Delta \Omega, \psi) \Delta \Omega,
\label{flux2}
\end{equation}
where $J(\Delta \Omega, \psi)$ is the average of $J(\psi)$
\begin{equation}
J(\psi) = \frac{1}{8.5 \, \rm{kpc}} \bigg(\frac{1}{0.3 \, \rm{GeV}/\rm{cm}^3}\bigg)^2 \, \int_{\rm{los}} \rho^2(r(l,\psi)) dl.
\label{jpsi}
\end{equation}
on the solid angle $\Delta \Omega$.  

\begin{figure}
\resizebox{5.65cm}{!}{\includegraphics{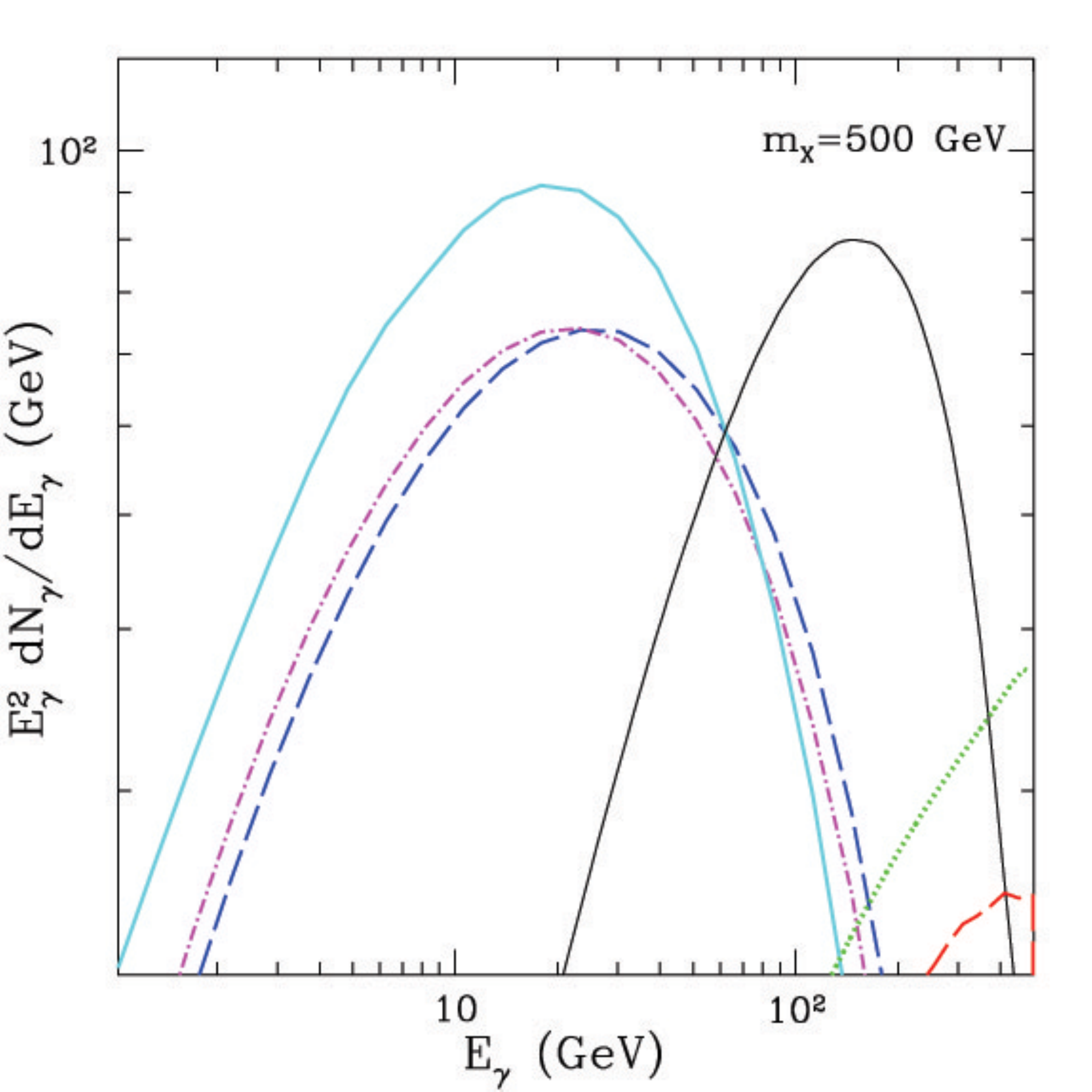}}
\resizebox{5.65cm}{!}{\includegraphics{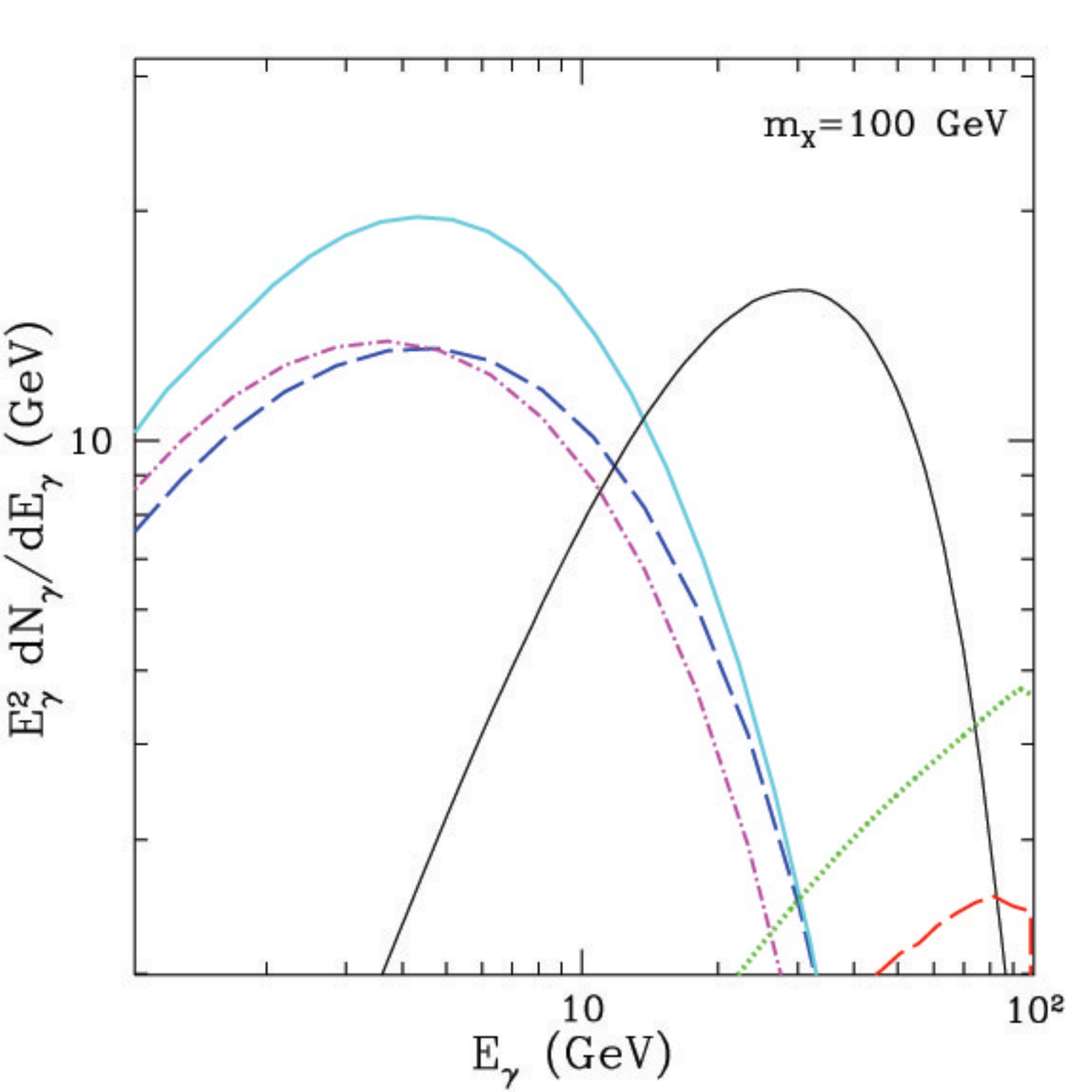}}
\caption{WIMP annihilation gamma ray spectrum: 500 GeV (left) 100 GeV (right), from Hooper (2009)\cite{hooper}
. Annihilation modes: $\mu^+ \mu^-$ (red dashed); 
$\tau^+ \tau^-$ (black solid); $ZZ$ (magenta dot-dashed), $e^+ e^-$ (green dotted), $W^+ W^-$ (blue dashed), and $b \bar{b}$ (solid cyan).}
\label{spectra}
\end{figure}

As we already reported, we need $\rho(r)$ to obtain the flux. The density profile used may be one of those described in section 2 (Dark matter distribution), but often a generalized NFW profile (Zhao et al. (1996)\cite{zhao}, Kravtsov et al. (1998)\cite{kravtsov})) is used
\begin{equation}
\rho(r) = \frac{\rho_0}{(r/R)^{\gamma} [1 + (r/R)^{\alpha}]^{(\beta - \gamma)/\alpha}} \,,
\label{profile}
\end{equation}
where $\rho_0 \sim\,$0.3 GeV/cm$^3$, and $R \sim 20$ kpc. 
The NFW profile has $\gamma=\alpha=1$, $\beta=3$, and inner slope ($\rho \propto r^{-1}$) while the Moore profile, characterized by a steeper inner profile ($\rho \propto r^{-1.5}$) than the NFW model has  $\gamma=\alpha=1.5$, $\beta=3$ (see Fig. 39 for more density profiles). 
For different density profiles the flux changes, it is larger for cuspy profiles and smaller for cored ones. So, one expects a larger value of the flux for the Moore profile than a NFW profile. The expected value of $J$, for the two profiles is: $J(\Delta \Omega=10^{-5} \, {\rm sr}, \psi=0) \sim 10^5$ for the NFW profile, and $\sim 10^8$ for Moore profile. From the above discussion, it is clear that the distribution of DM in the central part of 
the Galaxy is of fundamental importance, and at the same time that it is difficult to predict the DM distribution in the inner part of the Galaxy.

\begin{figure}
\resizebox{8.65cm}{!}{\includegraphics{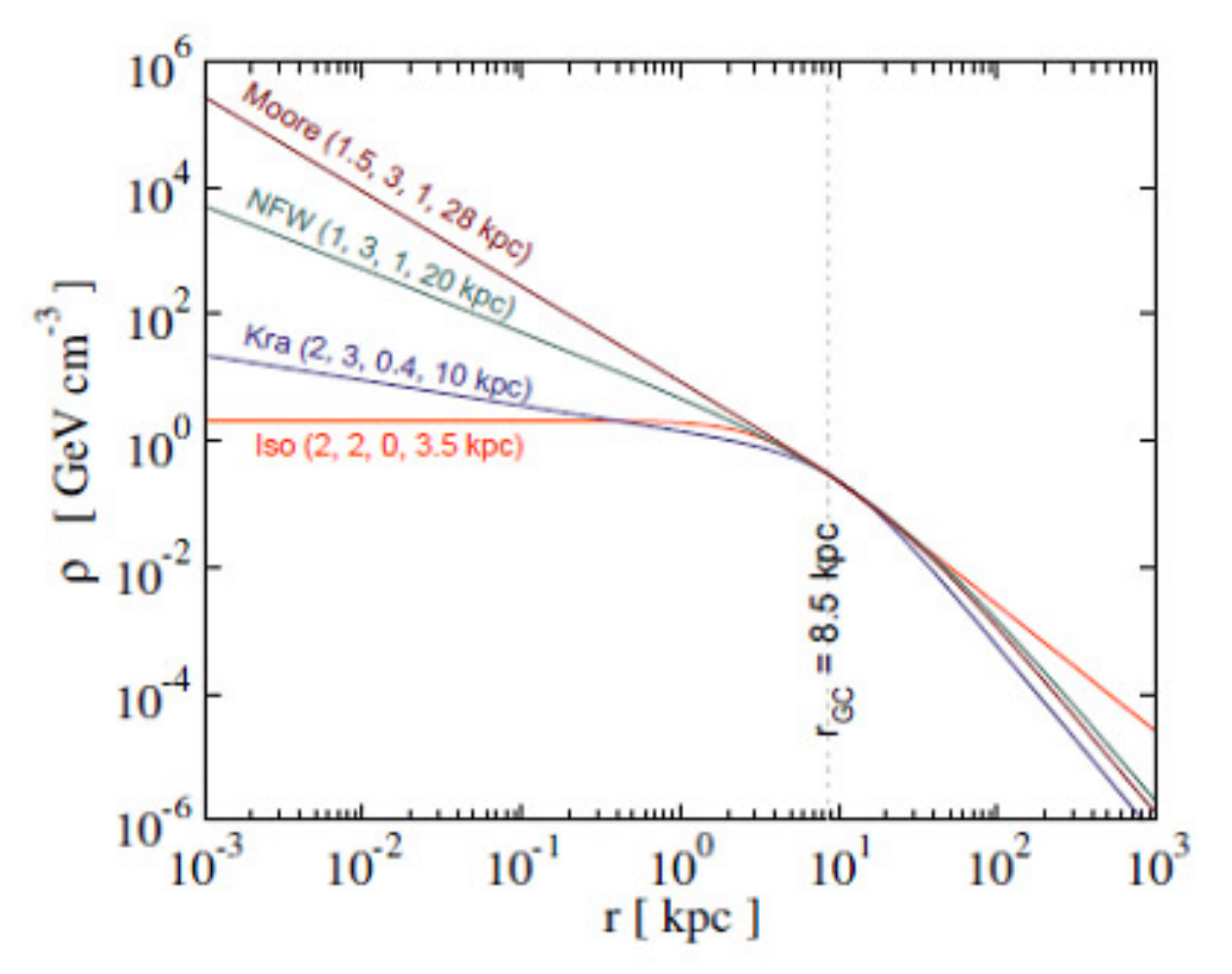}}
\caption{Comparison of different density profiles used in the calculation of X-ray emission from DM annihilation. Moore profile has $\alpha=\gamma=1.5$, and $\beta=3$. NFW profile has $\alpha=\gamma=1$, and $\beta=3$. Kra stands for Kravtsov with $\alpha=2$, $\beta=3$, and $\gamma=0.4$. ISO has $\alpha=\beta=2$, and $\gamma=0$.}
\label{spectra}
\end{figure}

N-body simulations can predict DM distribution to $\simeq 100$ pc (Stadel et al. 2009\cite{stadel}), and moreover in the inner part of the Galaxy the role of baryons 
is important, and not taken into account by simulations.  
Another problem is to disentangle gamma ray emission coming from astrophysical sources and those caused by DM. 
For example, an emission of very energetic gamma rays, observed by HESS, coincident with the center of the Galaxy, characterized by a power-law 
spectrum in the range 160 GeV-20 TeV, is incompatible with a DM interpretation.  

Gamma ray detection coming from DM annihilation can be detected through ground based telescopes (Atmospheric Cerenkov Telescopes) (e.g., MAGIC, HESS, and VERITAS), or space telescopes (e.g., Fermi). Ground based telescopes study small angular regions with greater exposure of space telescopes which study a larger fraction of the sky. 
The Fermi-LAT instrument\cite{atwood}, explores the gamma-ray sky in the 20-300 GeV range with a sensitivity ten times larger than EGRET\cite{link} Fermi's "forefather". Targets of telescope are the Galactic center, the MW dsPhs, the MW halo, 
the isotropic gamma-ray background radiation (IGRB) generated by unresolved haloes. 

Mazziotta et al. (2012) used Fermi-LAT data of the MW and the dSphs of the MW, and an approach model-indipendent to set upper limits to the DM annihilation cross sections for four channels, $b \overline{b}$, $\mu^{+} \mu^{-}$ $\tau^{+} \tau^{-}$, 
and $W^{+} W^{-}$. In the case of the dSphs the upper limits 
on $<\sigma v>$, in the range $\simeq 5- \simeq 25$ Gev, are below the prediction of the canonical thermal relic scenario ($3 \times 10^{-26} cm^3/s$) (see Fig. 40).

\begin{figure}
\resizebox{8.65cm}{!}{\includegraphics{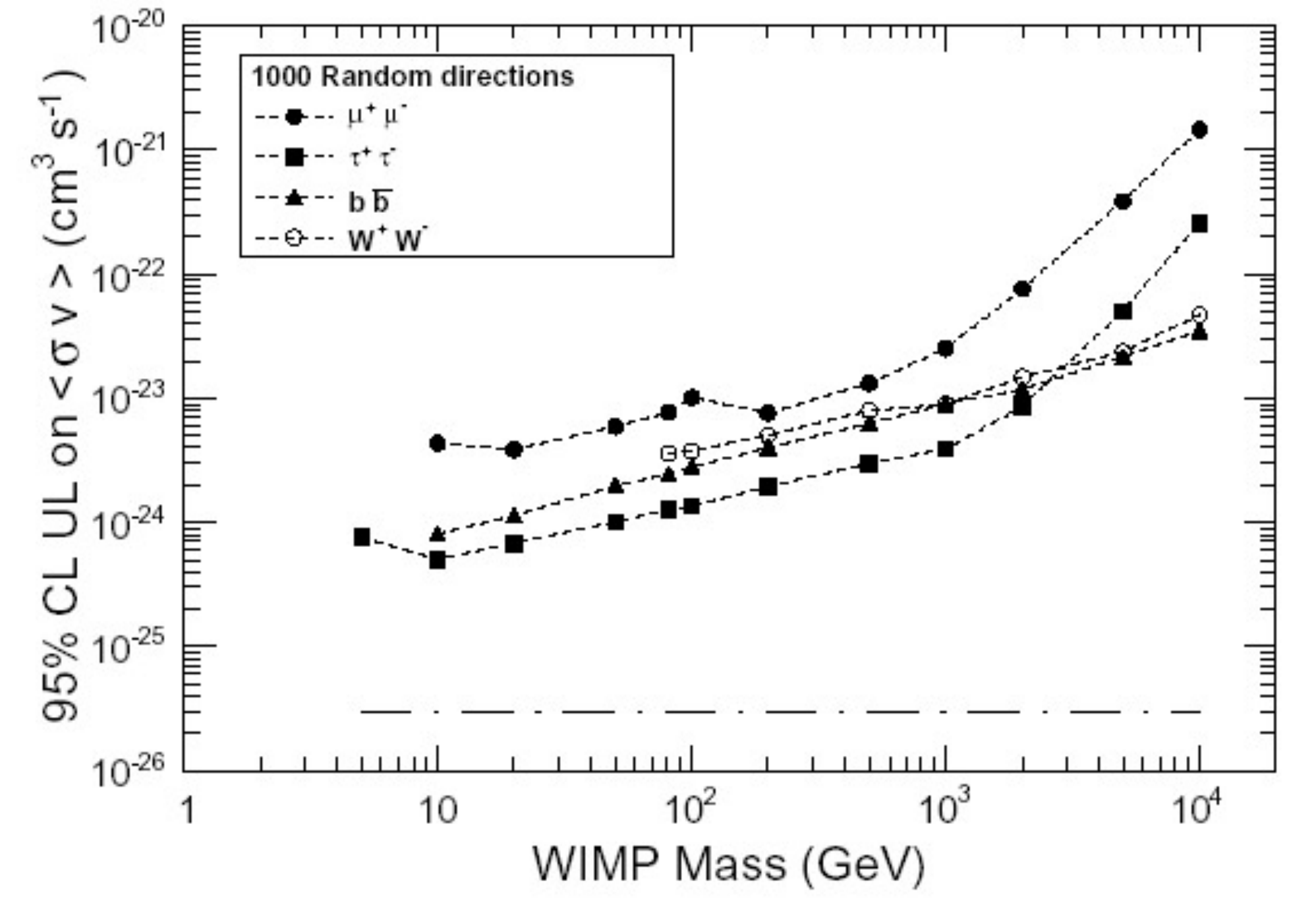}}
\resizebox{8.65cm}{!}{\includegraphics{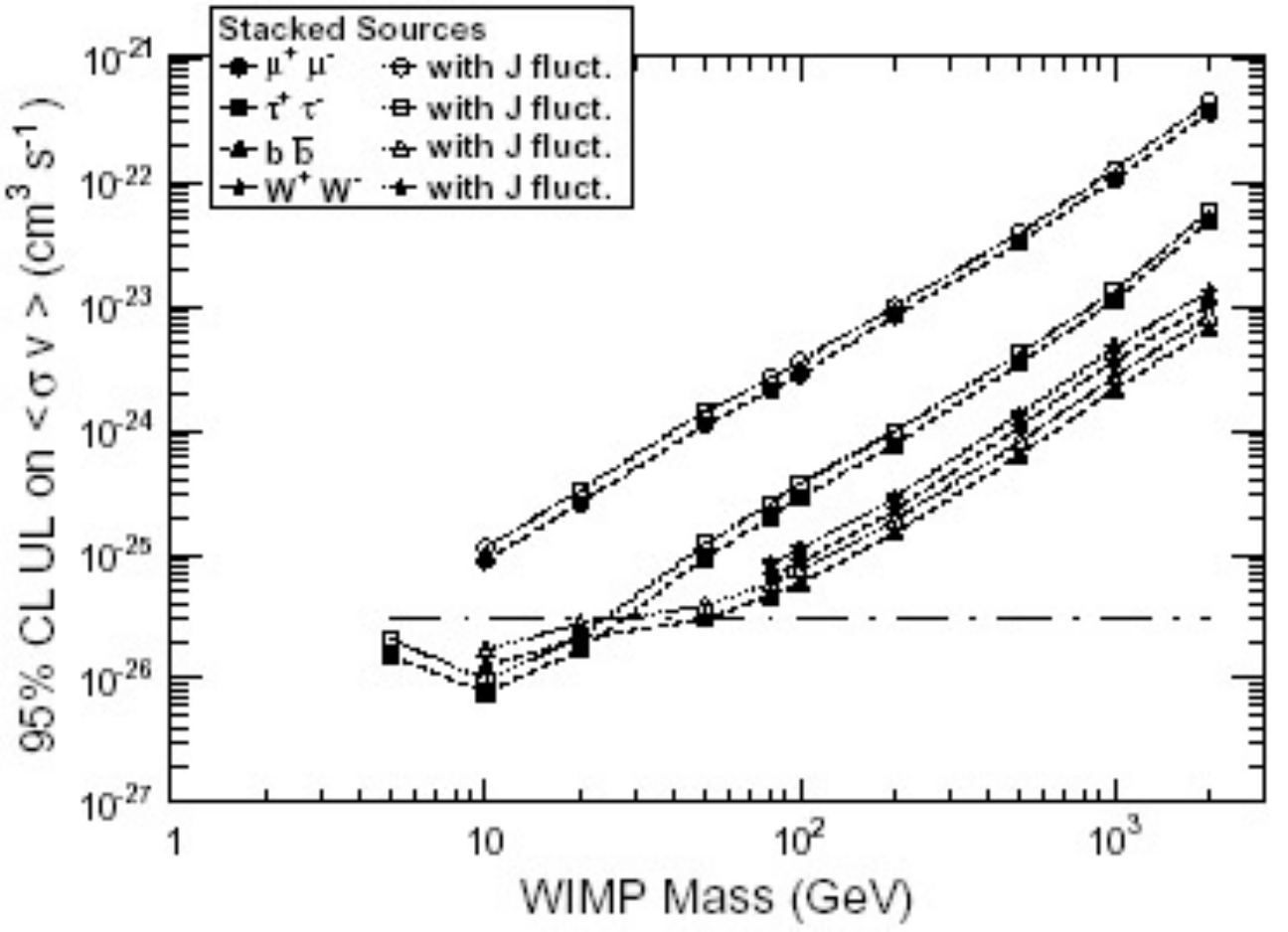}}
\caption{Upper limits to $<\sigma v>$ for different annihilation channels, obtained through Fermi-LAT,in the case of MW (left panel) and MW dSphs (right panel). The dot dashed line correspond to annihilation cross section of in the canonical thermal relic WIMP scenario ($3 \times 10^{-26} cm^3/s$ (Maziotta et al. (2012)\cite{mazziotta}}).
\label{spectra}
\end{figure}

Abazajian \& Kaplinghat (20012)\cite{kaplin}
 used the second year diffuse Galactic map, and the point source catalog of Fermi-LAT of the Galactic center ($7^0 \times 7^0$) (see Fig. 41). The gamma ray-spectrum is compatible with annihilation of DM particles of masses 10 GeV-1 TeV to $b \overline{b}$ quarks, and particles of 10-30 GeV annihilation to $\tau \overline{\tau}$ leptons. However, spectrum and intensity of emission are compatible with emission from millisecond pulsars, and moreover a part of the emission region is in conflict with constraints coming from MW dSphs obtained by means of Fermi.

\begin{figure}
\resizebox{8.65cm}{!}{\includegraphics{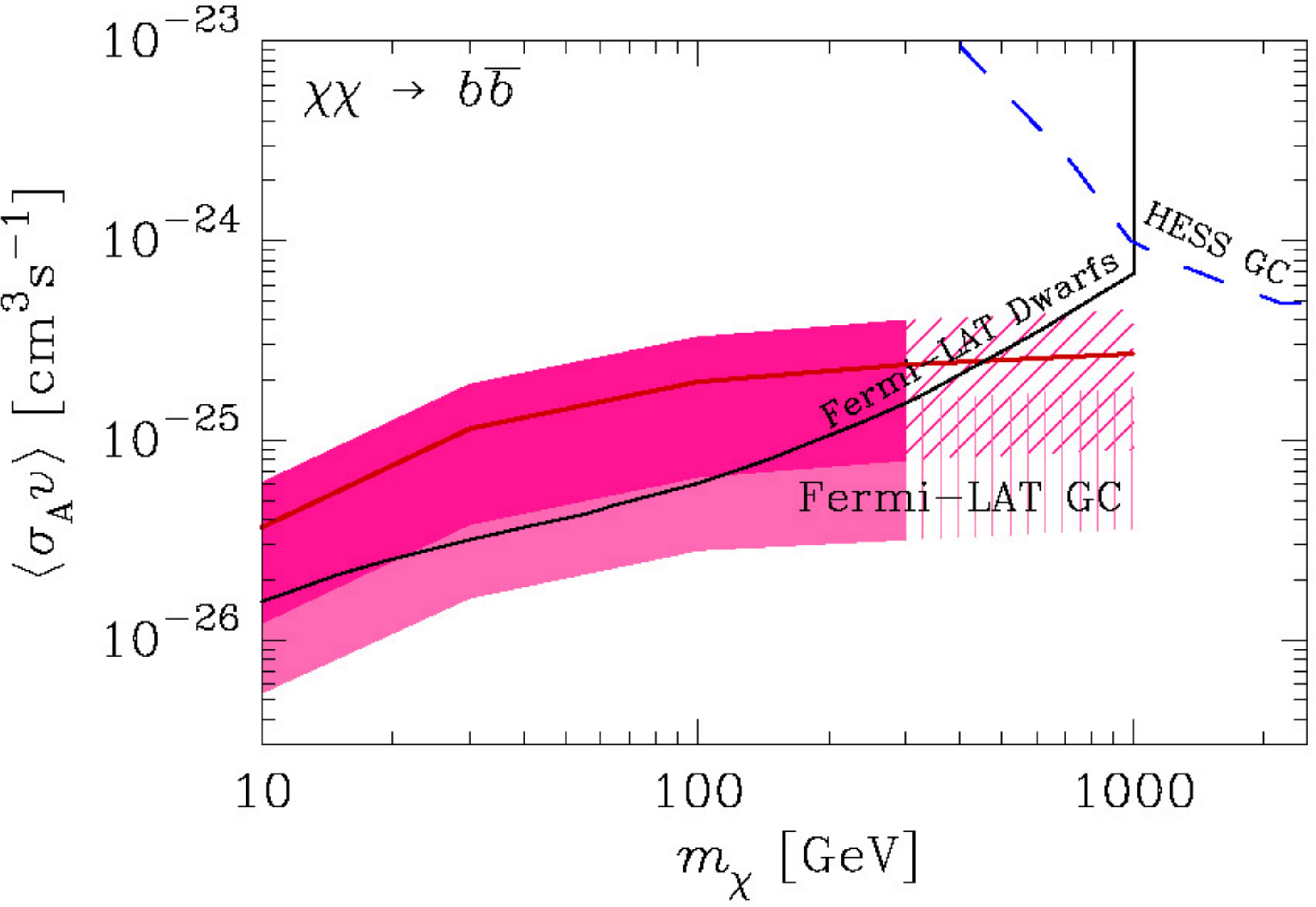}}
\resizebox{8.65cm}{!}{\includegraphics{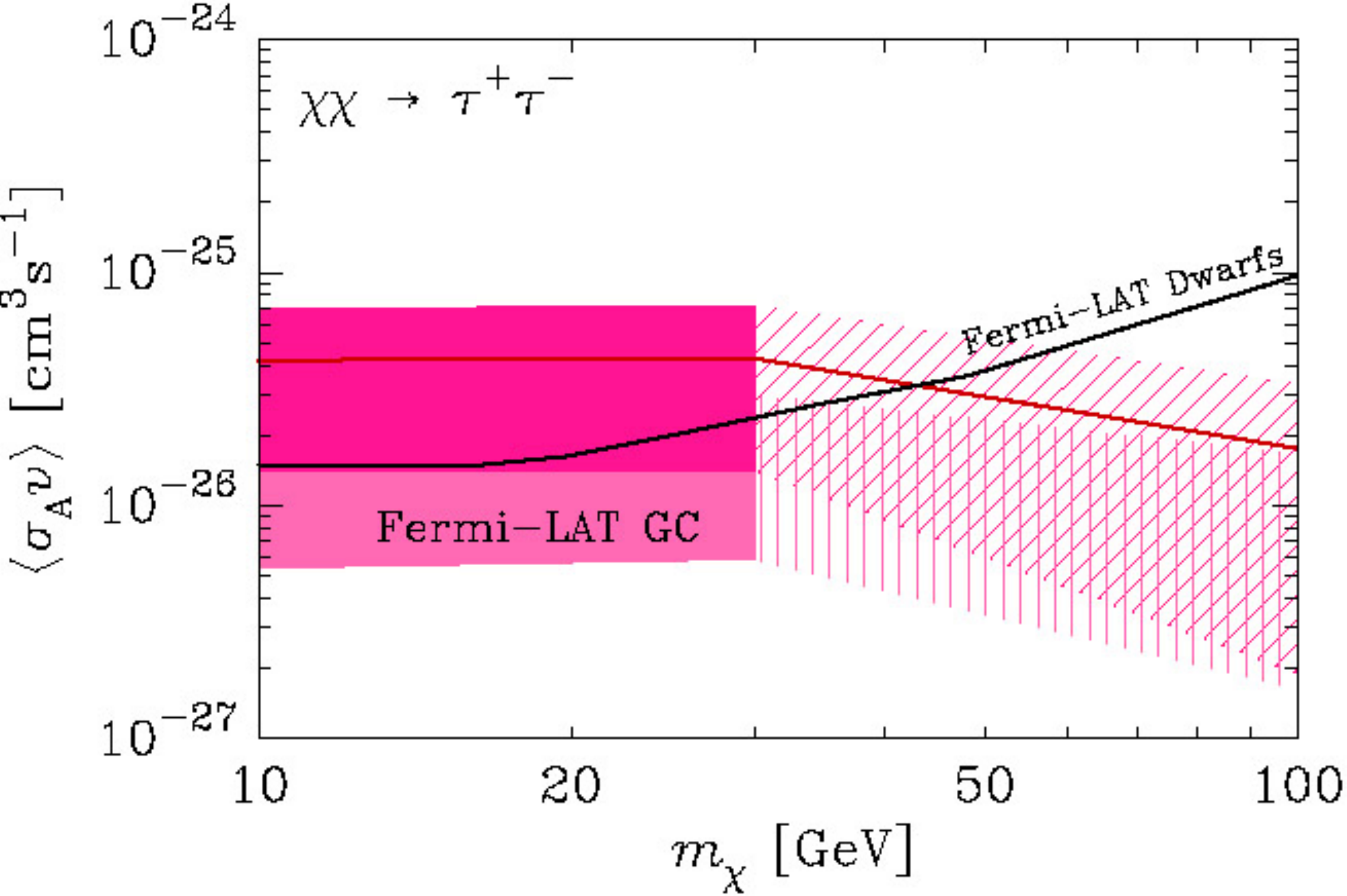}}
\caption{
Limits to $<\sigma v>$ for two different annihilation channels, obtained through Fermi-LAT, from GC (from 
Abazajian \& Kaplinghat 2012.)\cite{kaplin}}
\label{spectra}
\end{figure}

Moving to larger scales, Fermi-LAT and EGRET measured the Isotropic Gamma Ray Background (IGRB). 
The flux from DM induced extragalactic photons can be expressed as\cite{ullio}
\begin{equation}
\frac{d\phi_\gamma}{dE_0} = \frac{\langle\sigma v \rangle}{8 \pi} \frac{c}{H_0} \frac{\bar{\rho}_0^2}{m_{DM}^2} \int{dz (1+z)^3} \frac{\Delta^2(z)}{h(z)} \frac{dN_\gamma(E_0(1+z))}{dE} e^{-\tau(z,E_0)}, \label{eq:1}
\end{equation}
where $\tau(z,E_0)$ is the optical depth, $c$ is the speed of light, $m_{DM}$ the DM mass, $h(z)=\sqrt{\Omega_M(1+z)^3 + \Omega_\Lambda}$, $dN_\gamma/dE$ the gamma-ray spectrum, while  $\Delta^2(z)$ (see \cite{ullio}), is the enhancement factor of annihilation due to DM clustering. 
This last quantity is obtained from N-body simulations (with the two limits before indicated). 
The spectrum shape depends on the particle physics, the attenuation effects, and the way DM structures evolve
with $z$ (N-body simulation). 

Fermi-LAT and EGRET measurements of the IGRB have been compared to DM induced spectra 
from $\mu^{+} \mu^{-}$, $b \overline{b}$, and $\gamma \gamma$ annihilation channels\cite{abdo}, to set upper limits on the annihilation cross section for the three quoted annihilation channels (see Fig. 42).  

\begin{figure}
\resizebox{10.65cm}{!}{\includegraphics{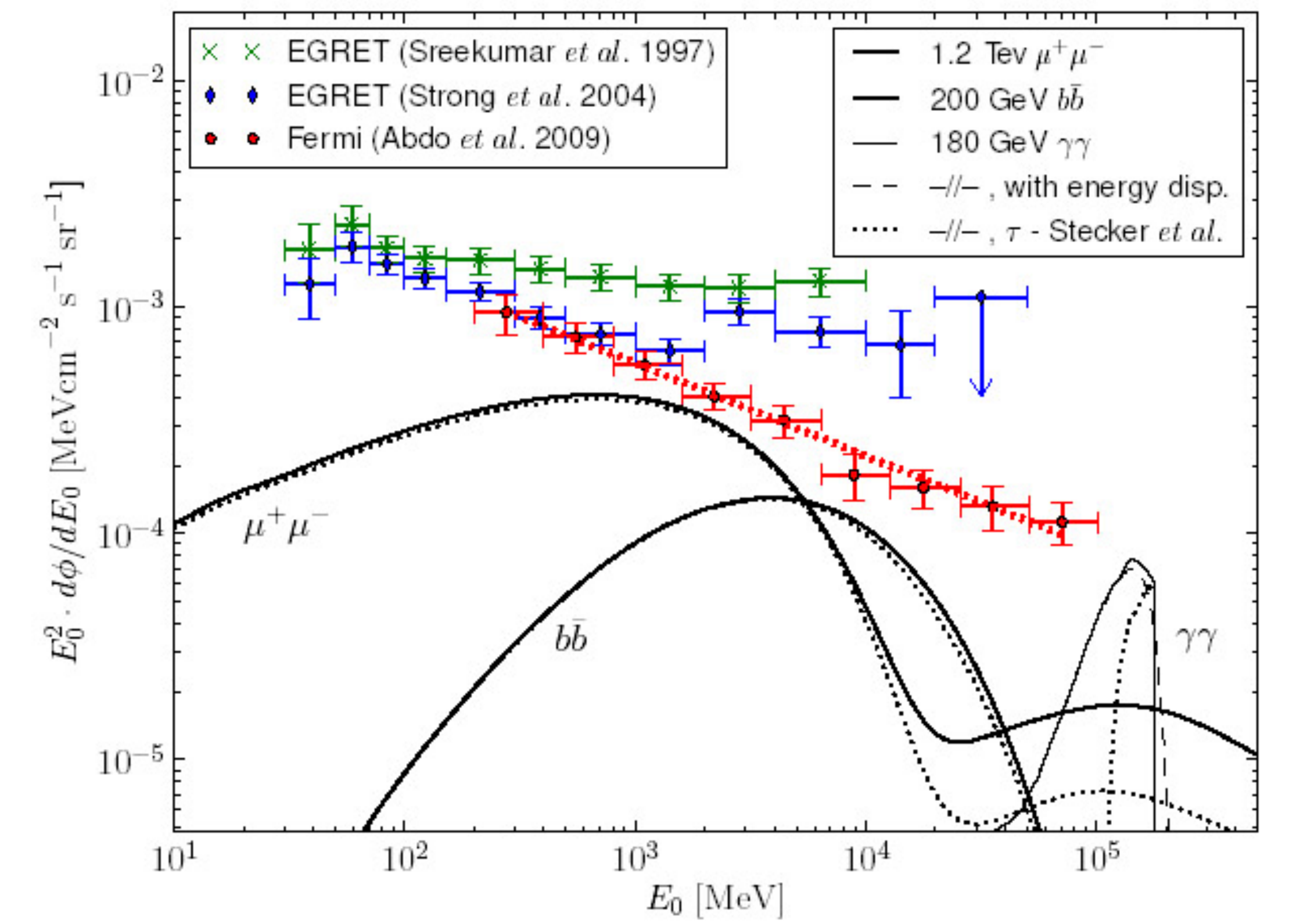}}
\caption{IGRB measurements by Fermi-LAT and EGRET, the lines represent three types of gamma-ray spectra by DM annihilation (from Abdo et al. 2010)\cite{abdo}).}
\label{spectra}
\end{figure}

Concerning the astrophysical contribution to the isotropic diffuse signal, AGNs have been the favored candidates, but according to Fermi measurement of blazar luminosity function they can make up maximally 30\% of the extragalctic signal\cite{fermi_lat}. Concerning star forming galaxies (SFG), Fields et al\cite{fields_etc}, based partly on Fermi measurement of the Galaxy diffuse emission conclude that SFG could make up most of the extra-galactic signal at lower energies. 

Han et al. (2012)\cite{han}, using 3-yr Fermi-LAT data, observed evidence of extended gamma-ray excess emission within three degrees of the center of Coma, Virgo, and Fornax (see Fig. 43). Interpreting the emission as annihilation emission from DM, the models preferred by data are a) a 20-60 GeV particle annihilating into
the $b \overline{b}$ final state, or b) a 2-10 TeV particle annihilating into $\mu^{+} \mu^{-}$ final state.  

\begin{figure}
\resizebox{9.65cm}{!}{\includegraphics{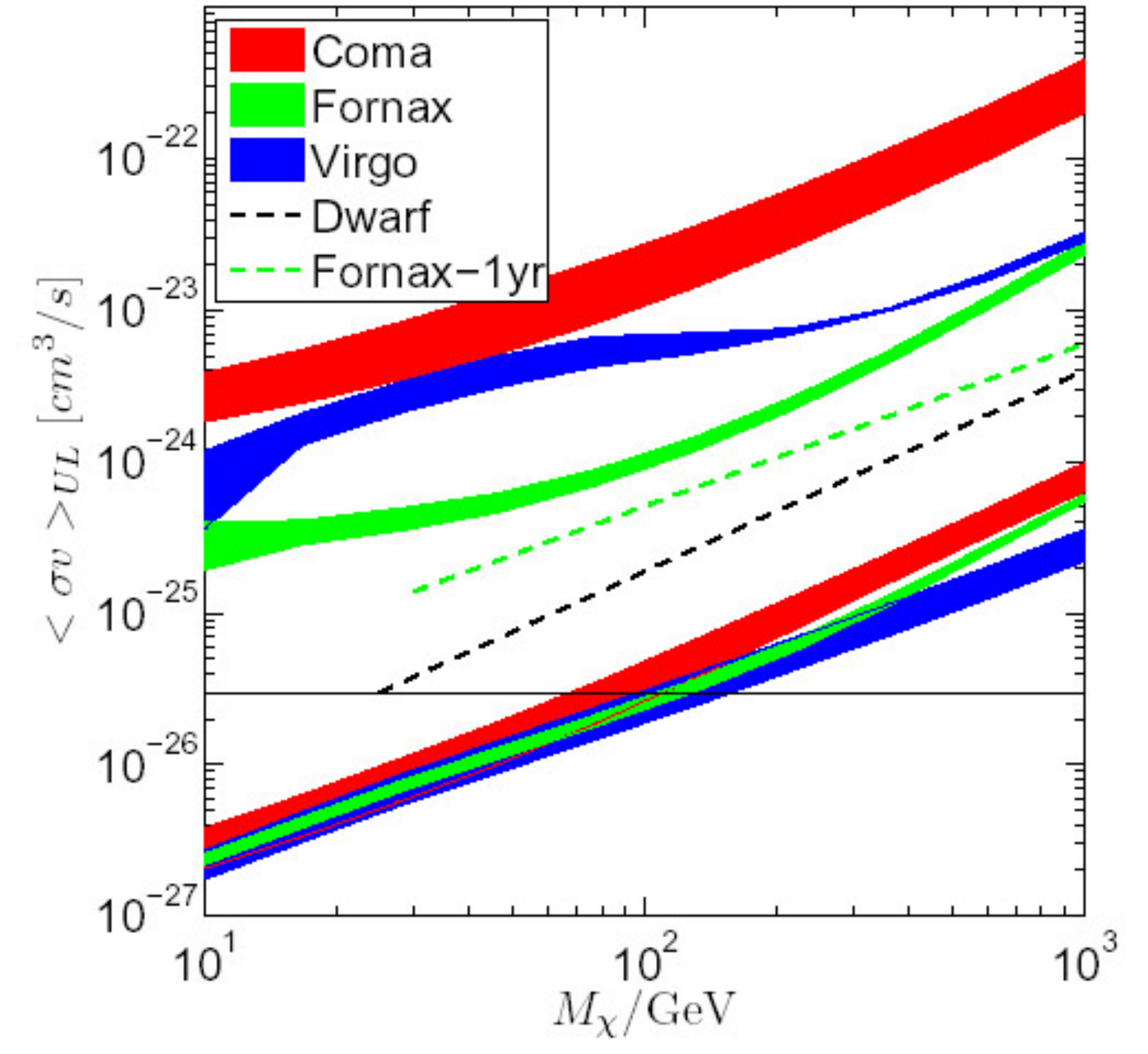}}
\caption{Upper limit to $b \overline{b}$ DM annihilation cross-section from clusters of galaxies (from Han et al. 2012\cite{han}).
}
\end{figure}

\subsubsection{WIMPs annihilation and Charged Cosmic Rays}

WIMP annihilations throughout the galactic halo can produce charged cosmic rays, including protons, antiprotons, electrons, and positrons. 
PAMELA experiment observed a rise in the cosmic ray positron fraction (the ratio $e^+/(e^++e^-)$ starting at 10 GeV and measured till $\simeq 100$ GeV\cite{adriani}, in agreement with previous observations from AMS-01 and HEAT, which was confirmed and extended to $\simeq 200$ GeV by Fermi\cite{ackerman1} (see Fig. 44).
Interpreting the observations in terms of DM annihilation, the excess can be generated by a 3 TeV DM particle annihiliating into 
$\tau^{+} \tau^{-}$\cite{cirelli}. A feature in the CR electron spectrum starting at 300 GeV, peaking at 600 GeV and ending at 800 GeV was observed by ATIC\cite{chang} and also observed by Fermi\cite{abdo1,bergstrom} (see fig. 45)). It has been interpreted as annihilation of a KK dark matter particle with mass $\simeq 600$ GeV.  

Even if the energy spectrum shape of the previous observations is consistent with WIMP dark matter candidates, in order to produce the PAMELA and ATIC signals WIMPs it is necessary that annihilation happens mostly in the charged leptons channel. In this way there will not be overproduction of CR antiprotons, and the resulting spectrum is hard enough. However, the annihilation rate needed to explain the observations should be from two to three orders of magnitude larger than expected from a thermal relic. A solution to the problem could be a boosting of the annihilation rate by local inhomogeneities, or one could suppose that the production mechanism is non-thermal (see Hooper 2009\cite{hooper} for other solutions). 

\begin{figure}
\resizebox{7.65cm}{!}{\includegraphics{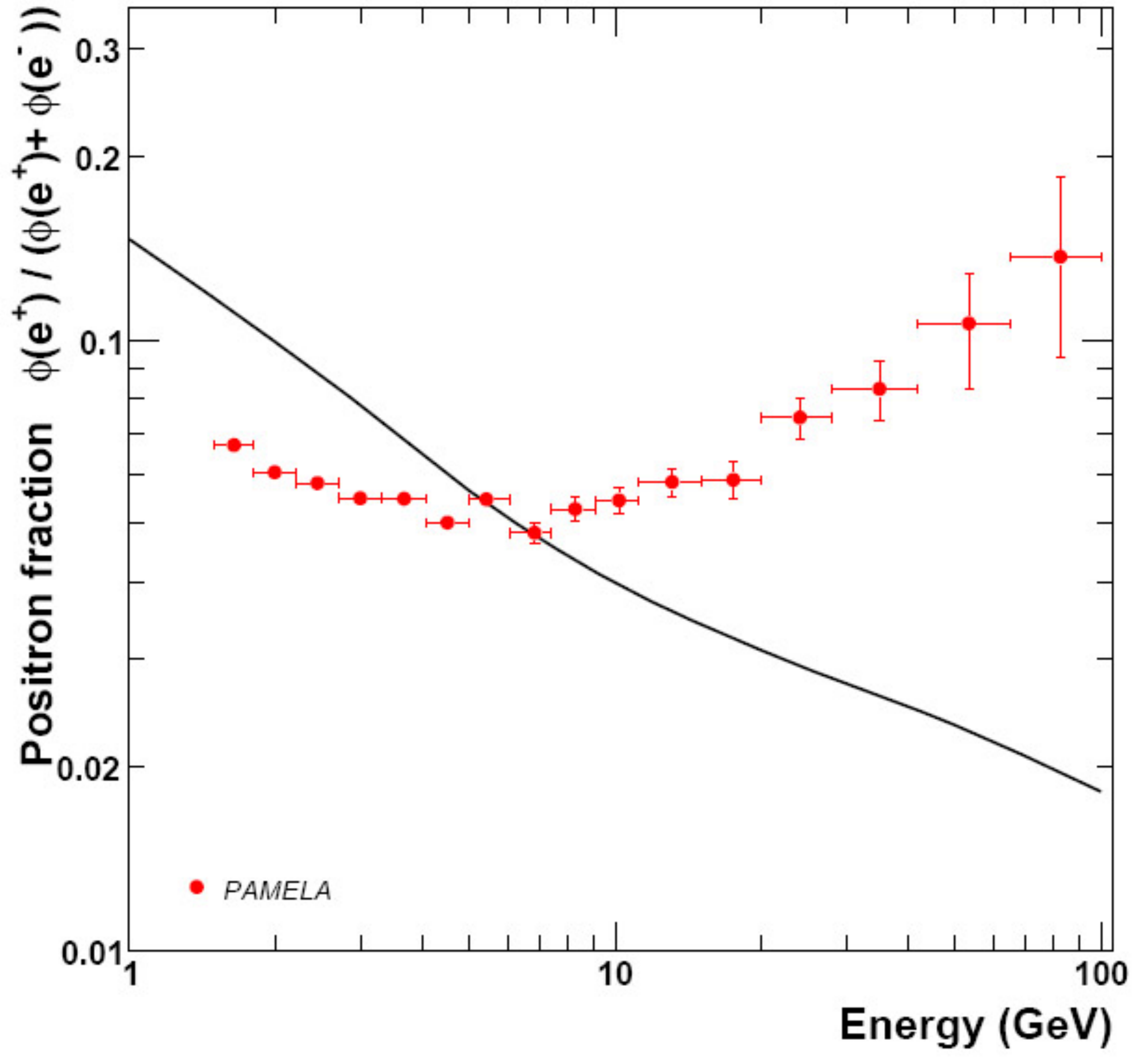}}
\resizebox{8.65cm}{!}{\includegraphics{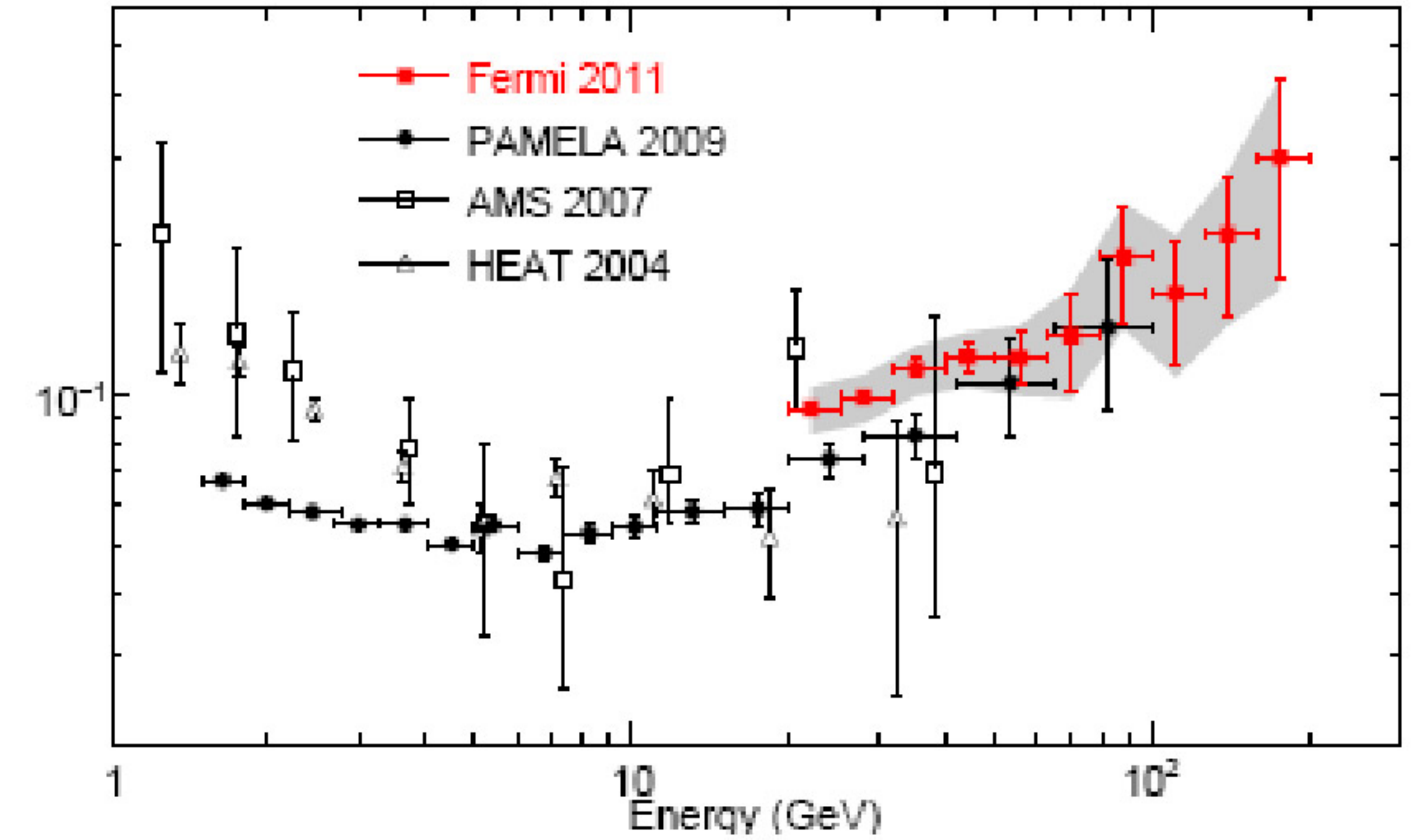}}
\caption{Left Panel: positron fraction from Pamela\cite{adriani}. The solid line is a spectrum from GALPROP (Moskalenko \& Strong) for pure secondary production of positrons during CR propagation in MW. Right panel: data from Fermi 
2011\cite{ackerman1}, Pamela, AMS, and Heat.}
\end{figure}

\begin{figure}
\resizebox{9.65cm}{!}{\includegraphics{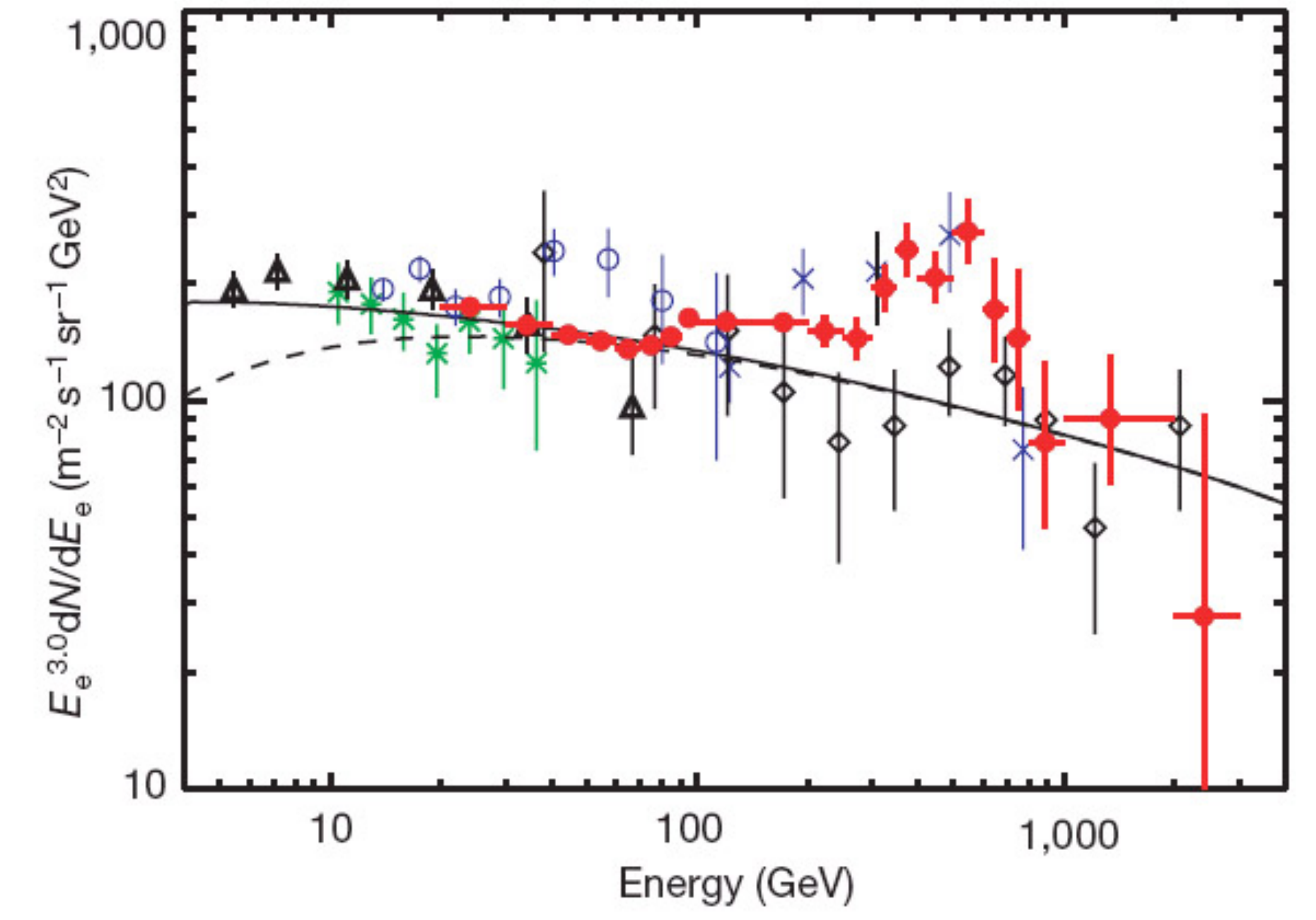}}
\resizebox{9.65cm}{!}{\includegraphics{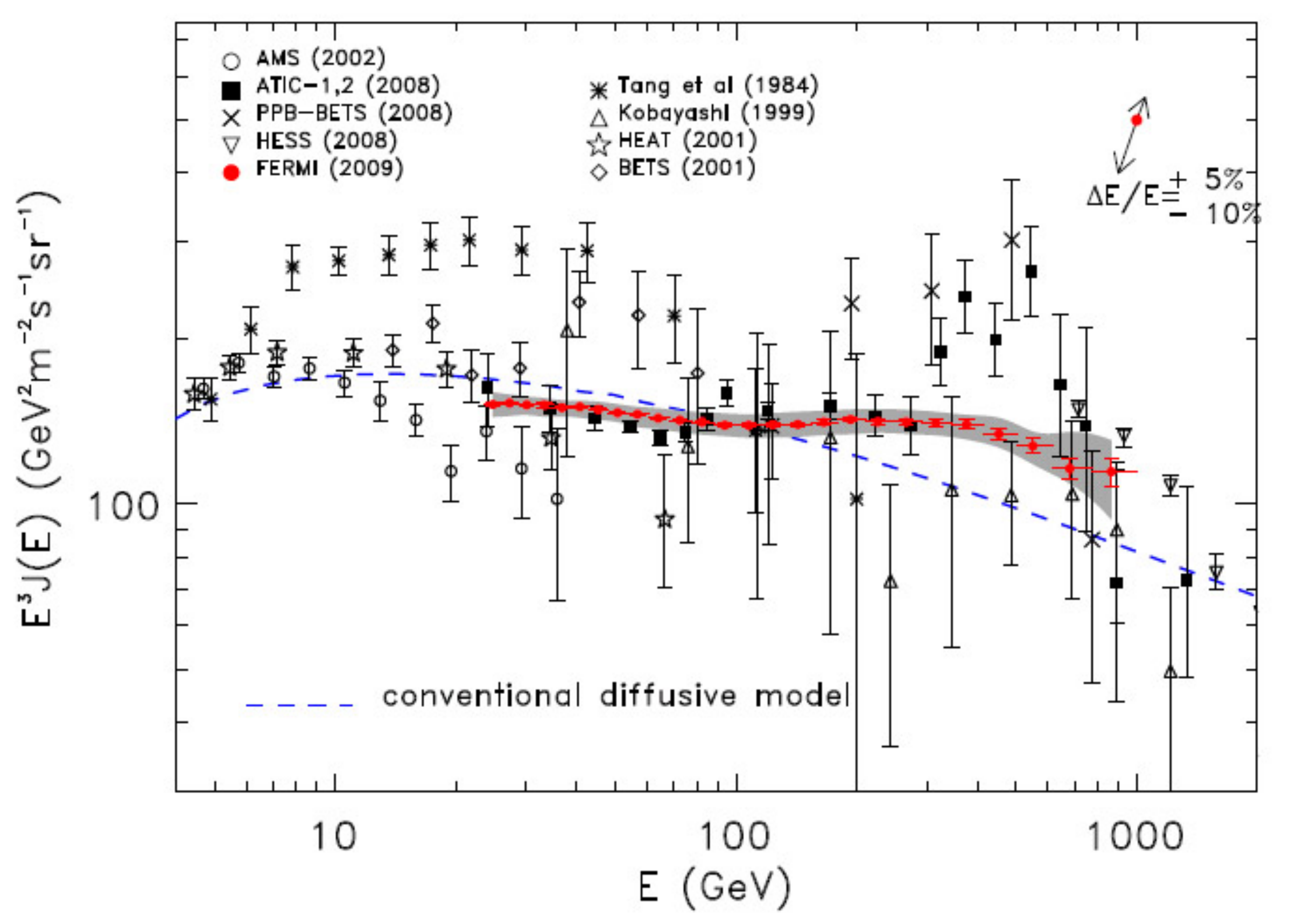}}
\caption{Total electron+positron spectrum. Left panel: data from ATIC. It is clearly visible the bump at $\simeq 600$ GeV. The lines are theoretical models results\cite{chang}. Right panel: Fermi excess compared with other previous data( from \cite{abdo1})}
\end{figure}

Conversely, the quoted emission excess could be produced by Pulsars\cite{hps, profumo1, aharonian, zhang, buesching} (see Fig. 46).
However, the edge in ATIC excess is easier to explain with DM particle particles annihilating directly to $e^+ e^-$ producing an edge in the CR spectrum of $e^-$, that drops abruptly at energy $E_e = m_X$. Pulsars, instead, produce more regular spectra. 

\begin{figure}
\resizebox{9.65cm}{!}{\includegraphics{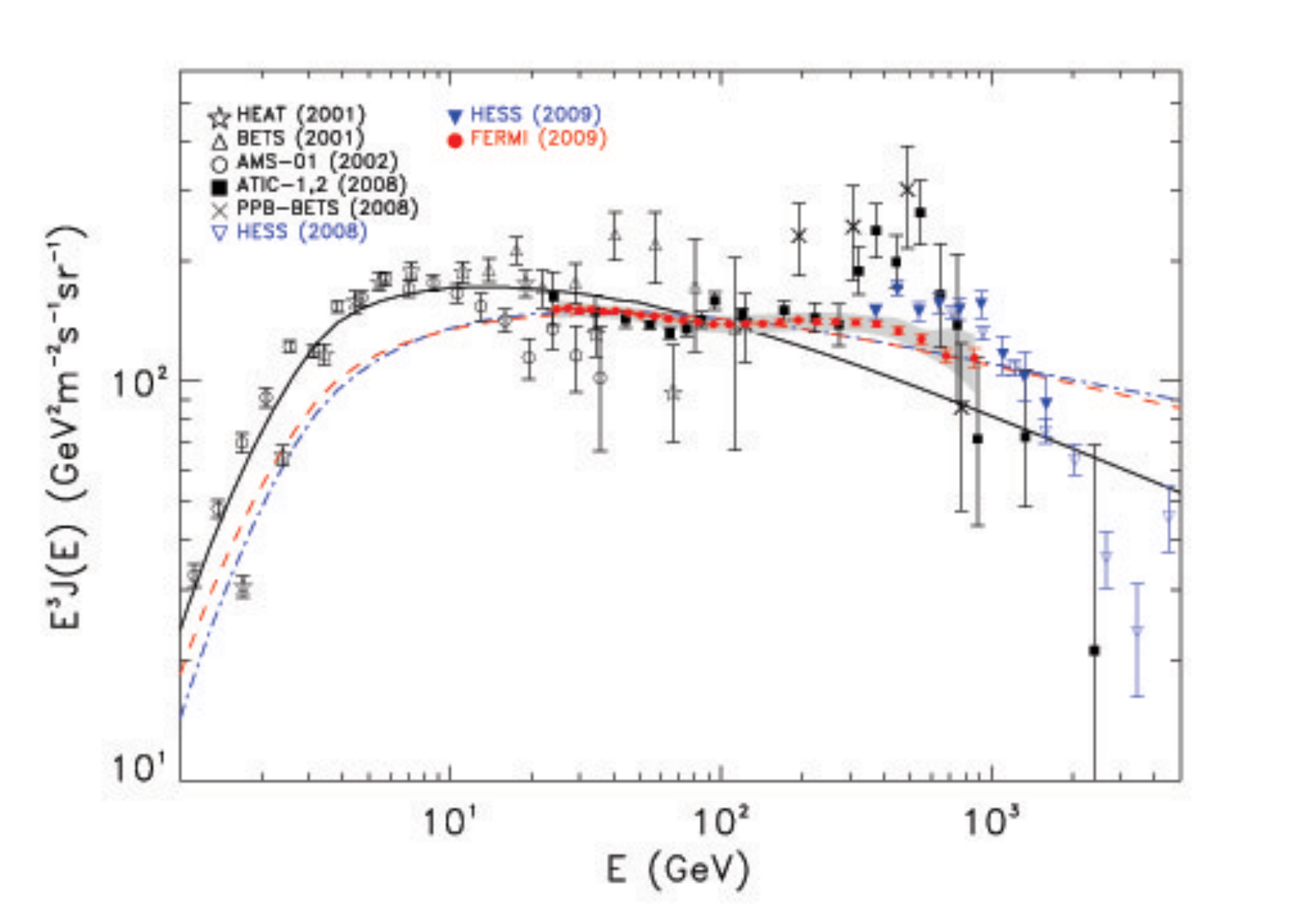}}
\resizebox{9.65cm}{!}{\includegraphics{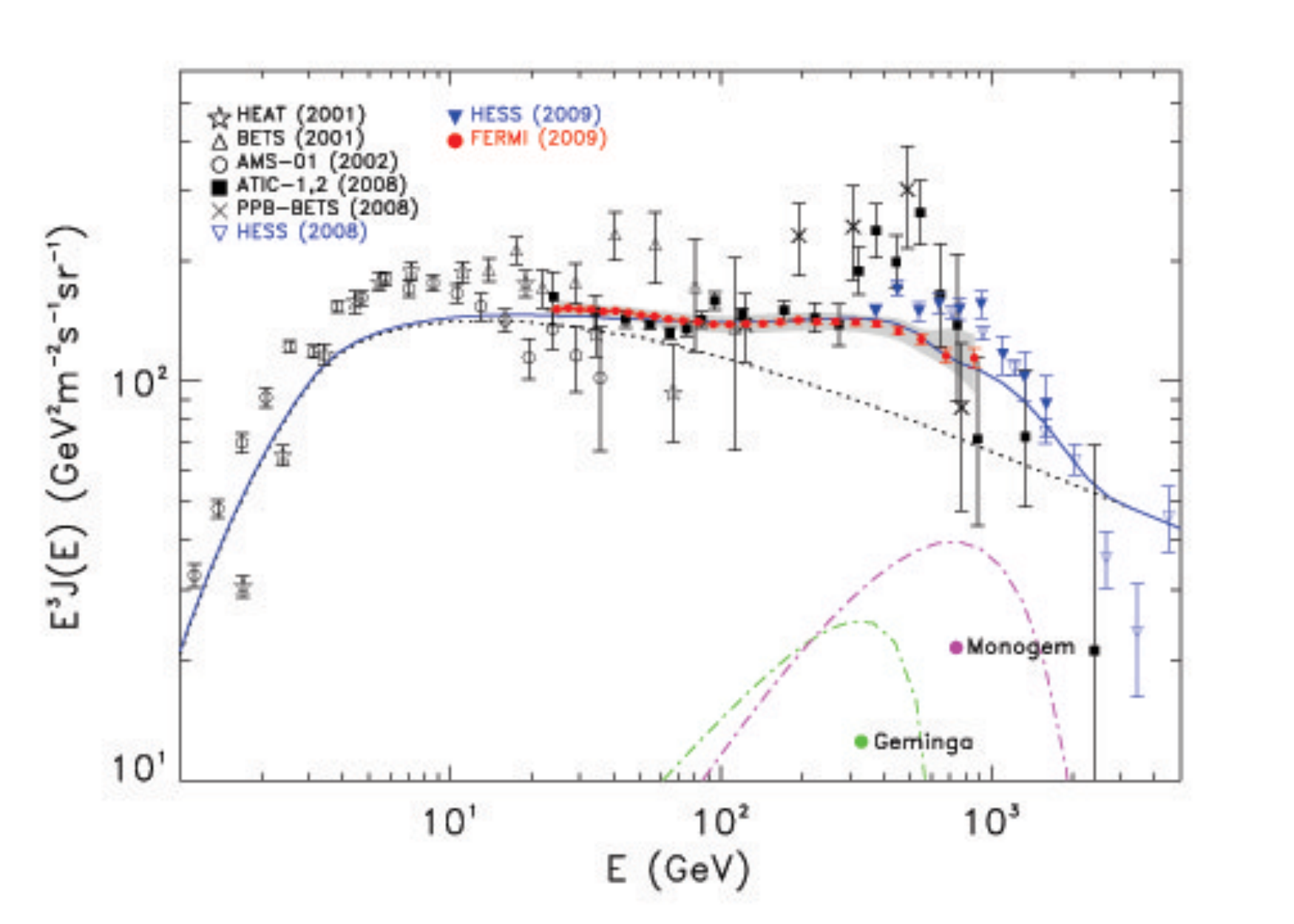}}
\caption{Total electron+positron spectrum. Left panel: Fermi excess compared with other previous data( from \cite{grasso}). 
Black continuous line: conventional model\cite{strongg}) to fit pre-Fermi data model.
Gray band: systematic errors on the cosmic ray electrons (CRE) spectrum measured by Fermi-LAT.
Red dashed, and blue dot-dashed lines, models having modified injection to fit Fermi-LAT CRE data.  
Right panel: electron-plus-positron spectrum (blue continuous line) from pulsars from 
the ATNF catalogue\cite{manchester}, summed to the large-scale Galactic component (GCRE). 
Colored dot-dashed lines: Monogem and Geminga pulsars. Black-dotted line: the GCRE, computed with GALPROP.
Gray band: systematic errors on the CRE Fermi-LATdata. 
}
\end{figure}

A microwave excess emission from the Galactic center was observed by WMAP\cite{dobler,fink, hfd} and confirmed by 
Plank\cite{ade} (see Fig. 47). The "WMAP haze", as it has been dubbed, has been interpreted as syncroton emission from electrons/positrons. 
Initially it was interpreted as thermal bremsstrahlung from gas with  temperature $10^4-10^6$ K. This interpretation is ruled out by the lack of a corresponding H$\alpha$ recombination (X-ray) line. The emission appears to be hard synchrotron emission from a new
population of energetic electrons/positrons in the inner galaxy, too hard to be supernovae shocks, and too extended to be a singular event (GRB, etc).
In summary it is very difficult to explain astrophysically.
In 2004, Doug Finkbeiner\cite{fink} proposed the explanation that WMAP Haze could be synchtrotron from electrons/positrons produced in dark matter annihilations in the inner galaxy. If one assumes a NFW profile, a WIMP mass of 100 GeV and an annihilation cross section of
$3 \times 10^{-26} cm^3/s$, the total power in dark matter annihilations in the inner 3 kpc of the Milky Way is $\simeq 1.2 \times 10^{39} GeV/s$, and the total power of the WMAP Haze is between $0.7 \times 10^{39}-3 \times 10^{39} GeV/s$, in agreement with DM annihilation.
When the effects of diffusion are accounted for, a cusped halo profile, with $r \propto R^{-1.2^{+0.1}_{-0.1}}$, is consistent with the Haze morphology.
For a typical 100-1000 GeV WIMP, the annihilation cross section needed is within a factor of 2-3 of the value needed to generate the
density of dark matter thermally ($3 \times 10^{-26} cm^3/s$). So, no boost factors are required.

\begin{figure}
\centering
\resizebox{14.65cm}{!}{\includegraphics{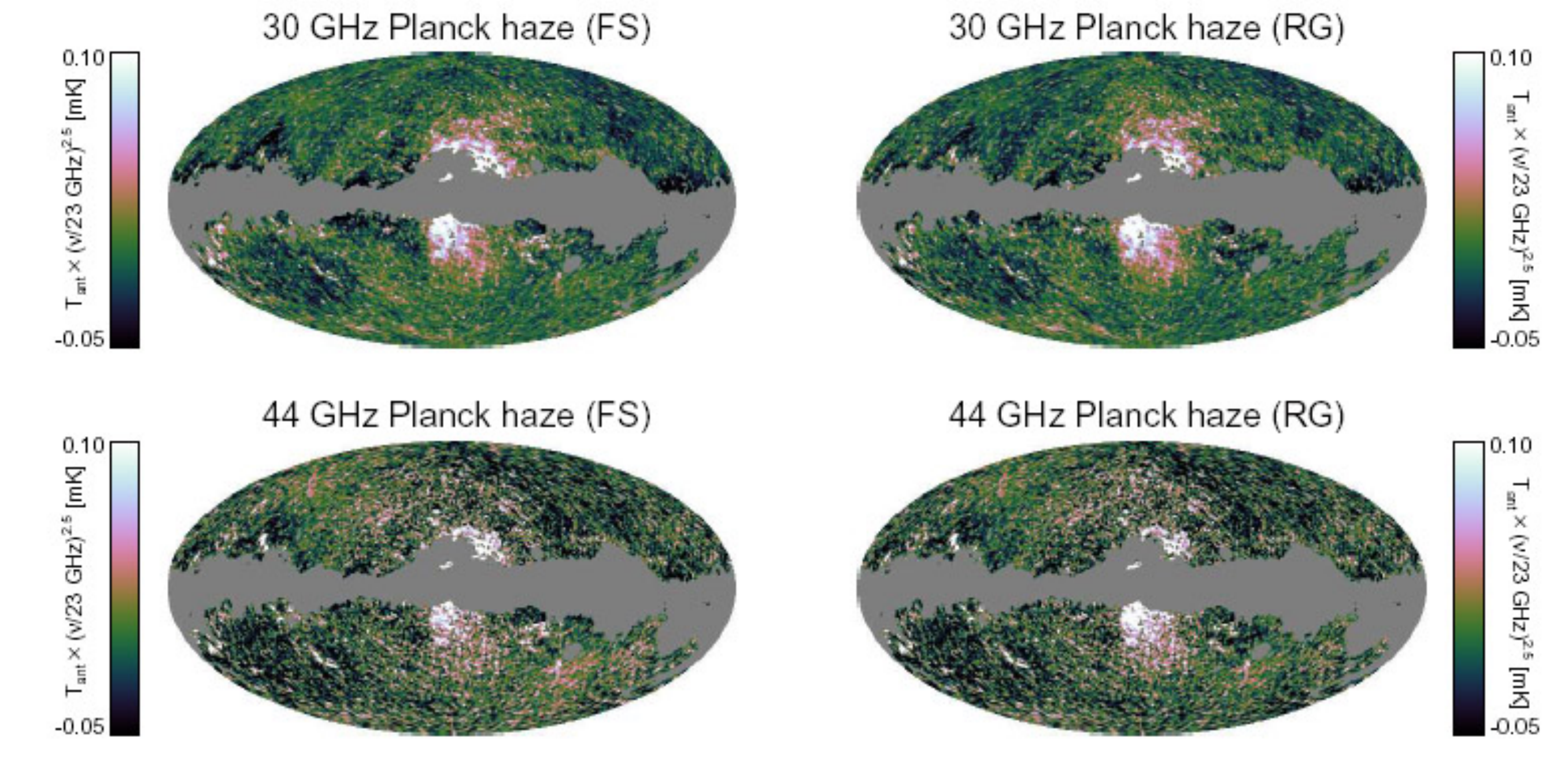}}
\caption{WMAP haze seen by Planck (from Ade et al. 2012\cite{ade})}
\end{figure}

In the following will recall some older possible evidences for DM detection. 
%
%

In 2003, the INTEGRAL/SPI spectrometer observed a bright 511 keV emission from the bulge of the
Milky Way ($1.3 \times 10^{43}$ positrons injected persecond). The morphology is Gaussian, spherically symmetric
It is challenging to explain 511 signal with non-exotic astrophysics. Type Ia supernovae are unable to generate the
observed injection rate (too few escape). Hypernovae (type Ic SNe) or gamma ray bursts could potentially generate enough
positrons if high estimates for rates are considered. However, even if the injection rate is sufficient, a
mechanism is required to transport it from disk to bulge, and this appears to be somewhat difficult.
A DM particle with mass 1-10 MeV annihilating to $e^+ e^-$\cite{bo} could generate both the 511 keV emission and the measured relic abundance. In any case, is difficult to construct a viable particle physics model with a MeV WIMP. As already discussed\cite{boehm_es,boehm_fayet}, were proposed scalar candidates in the mass range 1-100 MeV, that can generate the quoted integral line
(see Hooper 2009\cite{hooper} for other solutions to this problem).  

%


\section{Small scale problems of the $\Lambda$CDM}

Despite the successes of $\Lambda$CDM on large and intermediate scales, serious issues remain on smaller, galactic and sub-galactic, scales. In particular:
1) The missing satellite problem, namely the too high predicted number of small haloes from simulations of the $\Lambda$CDM model in comparison with observations. 2) The Too Big To Fail problem: dissipationless $\Lambda$CDM simulations predict central densities of MW dSPhs too large with respect to subhaloes obtained in simulations. 3) The angular momentum catastrophe, namely the low angular momentum of baryons obtained in simulations with the consequence to get galaxies with a small radius of discs. 4) The Cusp/Core Problem, namely while simulations predict cuspy dark matter haloes,  
dwarf galaxies, and Irregulars, DM dominated, show cored profiles.

\subsection{The missing satellite problem}

Concerning the missing satellite problem, Moore et al (1999)\cite{moore1} noticed that the number of subhaloes within galactic and cluster mass halos was much larger than the number predicted by N-body simulations. In MW the number of satellites with circular velocities larger than Draco and Ursa-Minor (i.e. bound masses $>10^8 M_{\odot}$ and tidally limited sizes $>$ kpc) were around 500, and we know that the MW dSphs are much less (see Fig. 48, top left panel). All subsequent simulations showed this problem. Basically all cosmological simulations predict that there are at least one order of magnitude more small subhalos (dwarf galaxies) around Milky Way like galaxies than what is observed (e.g. Via Lactea simulation\cite{diemand}
The problem was slightly mitigated after the discovery of ultra-faint MW 
satellites\cite{willman,belokurov,zucker,grillmair,sakamoto,irwin}.
The idea on which the solutions to the problem are based is that only some satellites are visible, and the proposed solutions to the problem can be classified as: a) satellites having the largest masses before accretion (LBA) by MW, and which could resist tides\cite{diemand1}. 
b) The ones which acquired gas before re-ionization and formed stars: Earliest Forming (EF)\cite{bullock,moore1} (see Fig. 48, 
top right, and bottom left). These models are based on reionization suppression\cite{bullock,moore2}. 

Other possible solutions studied are stellar and supernova feedback (e.g. \cite{dekel1,mori}), and
gas stripping by ram pressure (e.g., \cite {mayer}) as favorite suppression mechanisms. 
An interesting paper is that of\cite{simon_geha}, in which they corrected for the  Ultra-faint dwarfs with $M/L \simeq 1000$ discovered (from SDSS data). In the case reionization happened at $z= 9-14$, protostructures could not attract enough baryonic matter to create a visible dSPh. In this case the 
problem is almost solved (see Fig. 49, bottom right panel).

\subsection{The Too Big To Fail problem}

The second problem, Too Big To Fail, is connected to the previous one. The problem arose from analyses of the Aquarius and Via Lactea simulations. 
Each simulated halo had  $\simeq 10$ sub-halos that were too massive and
dense with respect to MW dSphs that they would appear to be too big to fail to form lots of stars. The TBTF problem is
that none of the observed satellites of the Milky Way or Andromeda have stars moving as fast
as would be expected in these densest sub-halos\cite{boylan, boylan1} (see fig. 50).

\begin{figure}[htp]
\centering
\begin{tabular}{cc}
\resizebox{8.6cm}{!}{\includegraphics{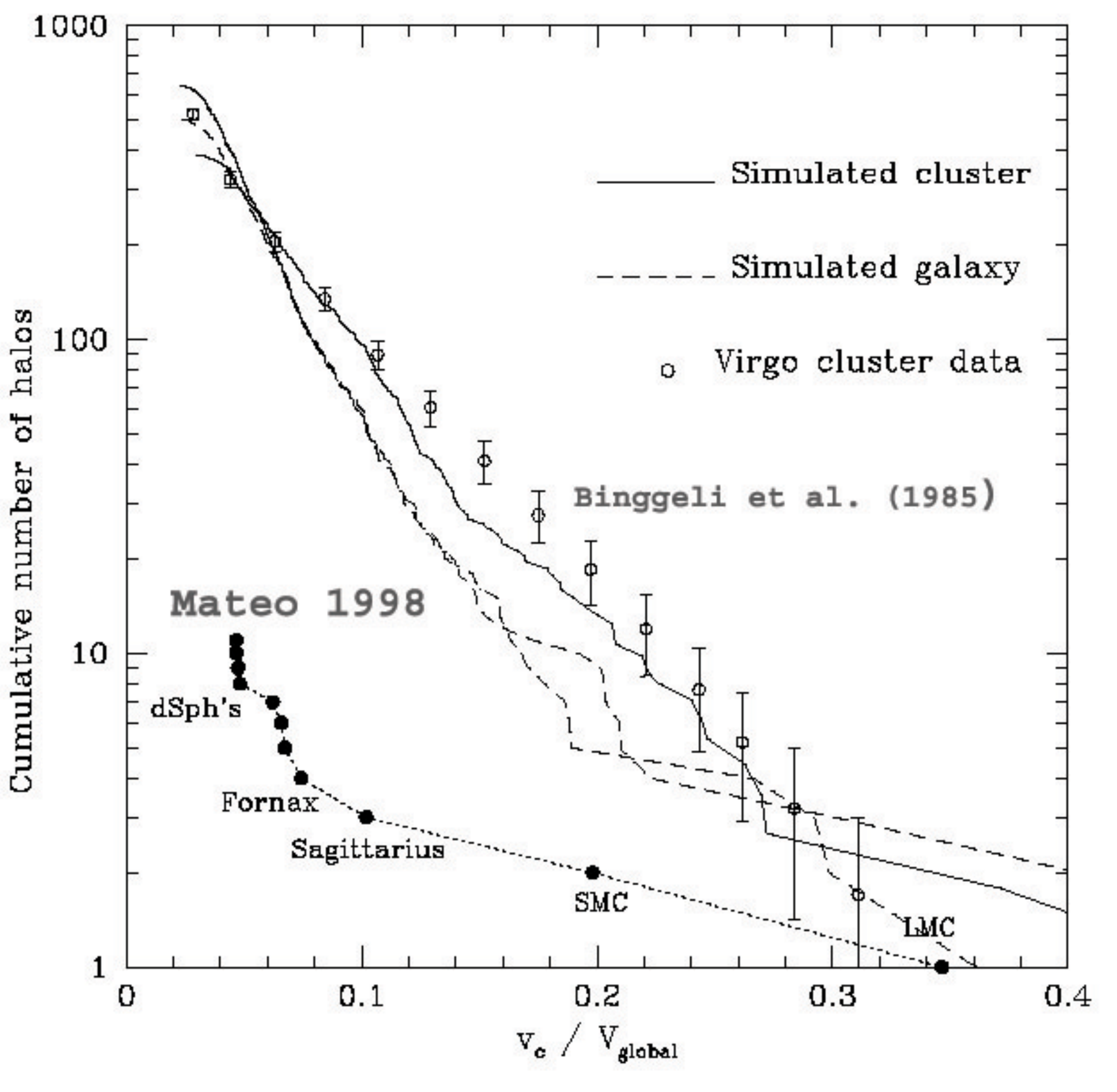}}&
\resizebox{8.6cm}{!}{\includegraphics{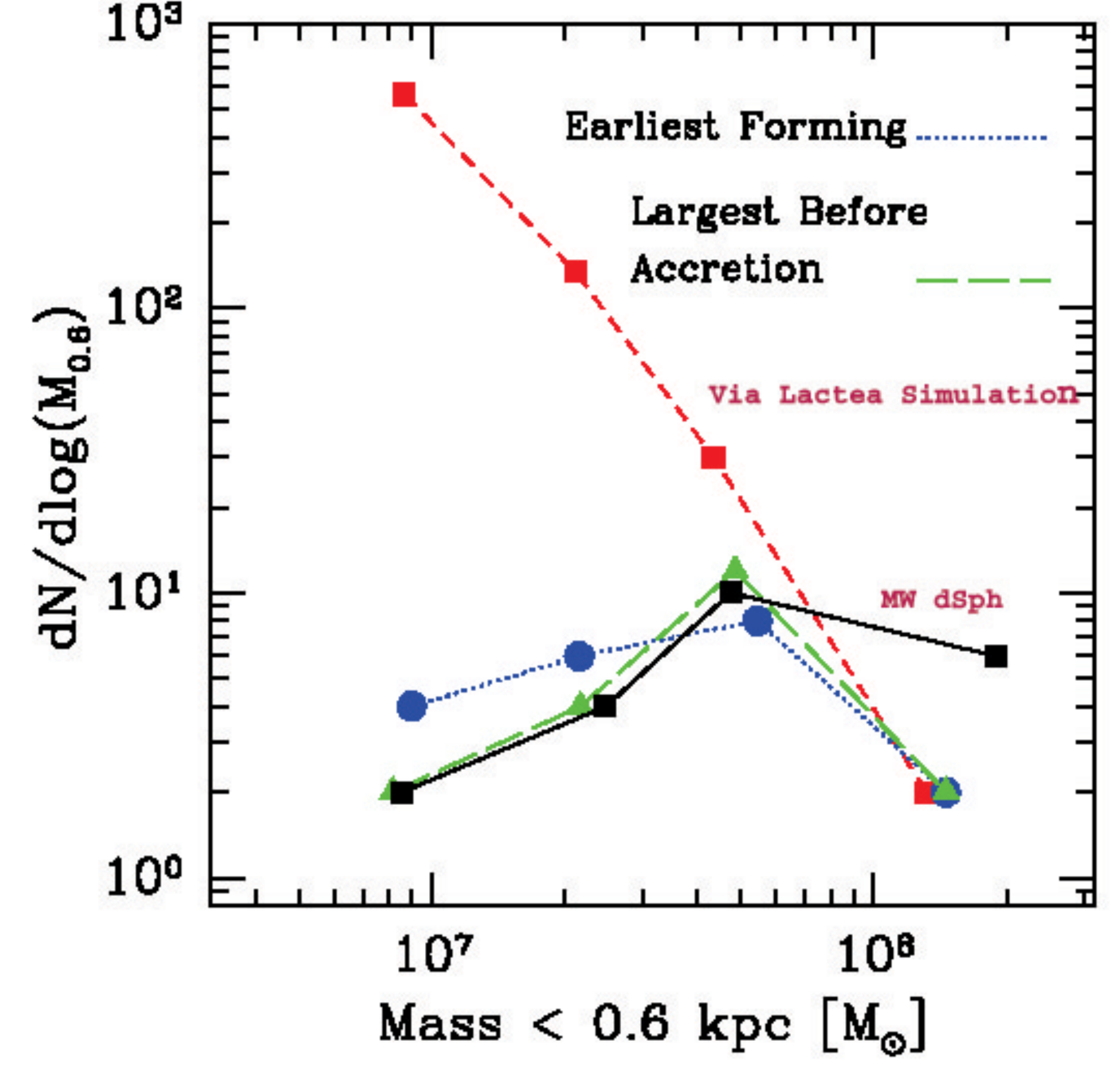}}\\
\resizebox{8.6cm}{!}{\includegraphics{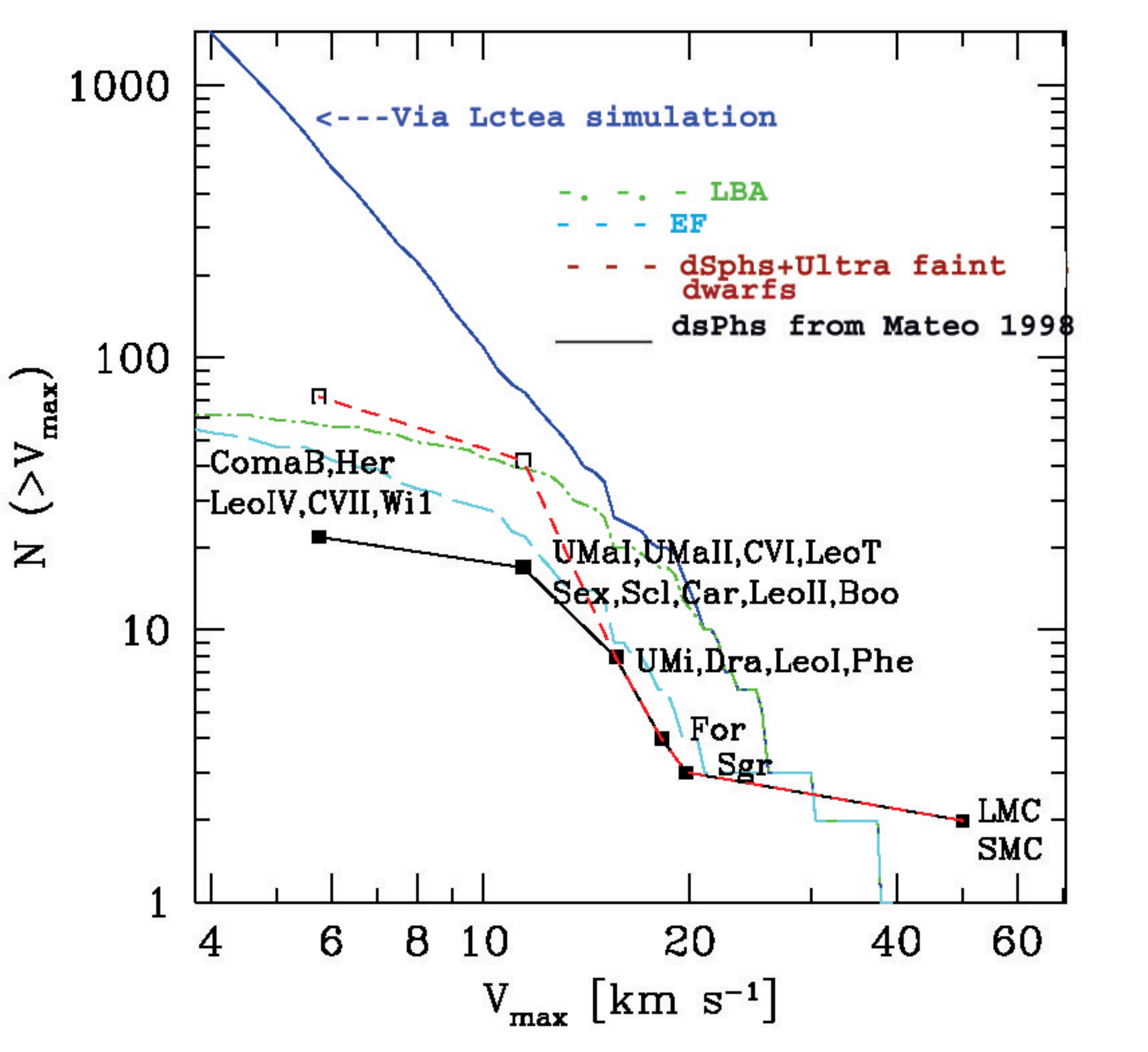}}&
\resizebox{8.6cm}{!}{\includegraphics{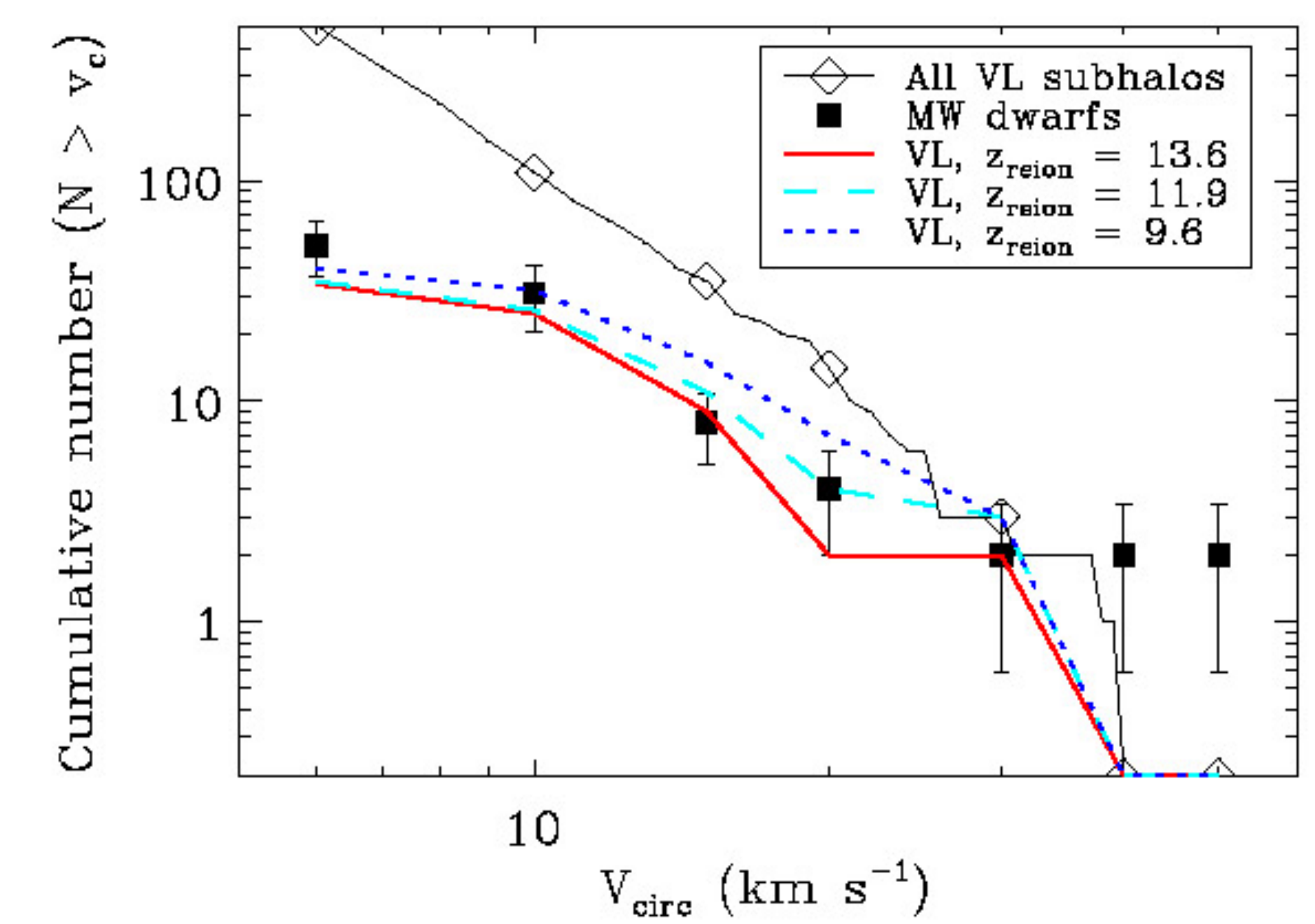}}\\
\caption{Top left panel: Comparison of simulation of Moore et al. (2009)\cite{moore1} simulations for dSphs and clusters with observations from Mateo 1998\cite{mateo1}, and \cite{binggeli} (from Moore et al. 2009\cite{moore1}). Top right panel: Comparison of Via Lactea Simulation with the MW dSphs, the EF, and LBA scenario (from Strigari et al. 2007\cite{strigari}. Bottom left panel: Comparison of Via Lactea Simulation with the MW dSphs, MW dSphs+ Ultra-faint dwarfs, the EF , and LBA scenario (from Madau et al. 2008\cite{madau}). Bottom right panel: comparison of Via Lactea Simulation with the MW dSphs+ Ultra-faint dwarfs, and Via Lactea subhaloes at redshifts 9.6, 11.9, 13.6 (from Simon \& Geha 2007\cite{simon_geha}).
}
\end{tabular}
\end{figure}


Inclusion of baryonic physics can create shallower slopes of the dark matter densities in the centers of low-mass galaxies reducing or
solving the discrepancy between cuspy profile predicted in N-body simulations and flat ones seen in observation. One possible solution is that showed by Brooks et al. (2012)\cite{brooks} using a suggestion of Zolotov et al. (2012)\cite{zolotov}. The last author proposed a correction to the velocity in 1 kpc 
\begin{equation}
\Delta (v_{1kpc})= 0.2 v_{infall}-0.26 km/s; \hspace{0.5cm} 20 km/s<v_{infall}< 50 km/s
\end{equation}
that must be applied to the central parts of low-mass galaxies, in order to  take into account tidal stripping enhancement due to baryons, and the effect of supernova feedback flattening the cusp in the haloes (see fig. 51).

\begin{figure}
\resizebox{14.65cm}{!}{\includegraphics{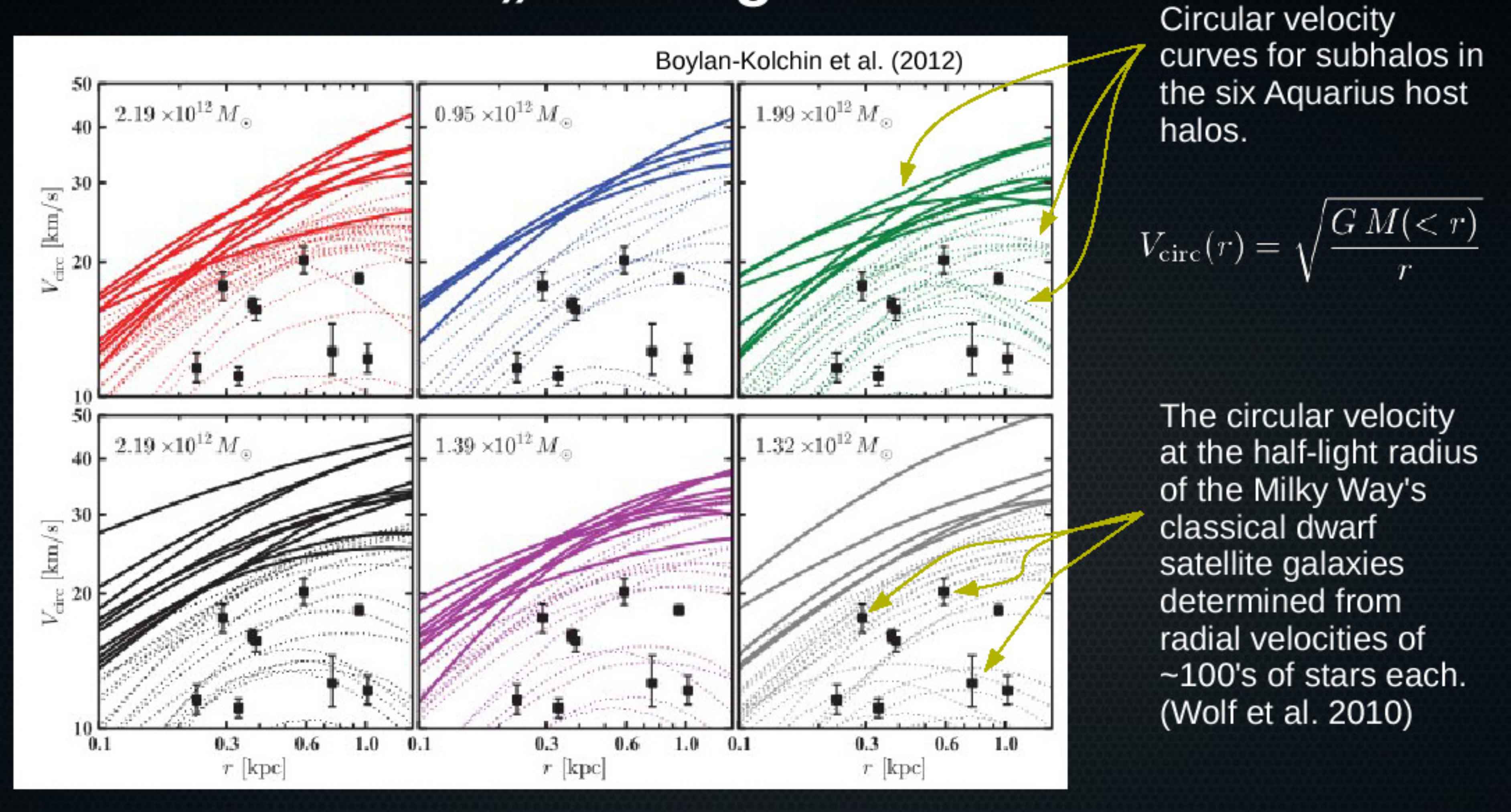}}
\caption{the Too Big To Fail problem (from Boylan-Kolchin et al. (2012)\cite{boylan1}).}
\label{spectra}
\end{figure}

\begin{figure}
\resizebox{12.65cm}{!}{\includegraphics{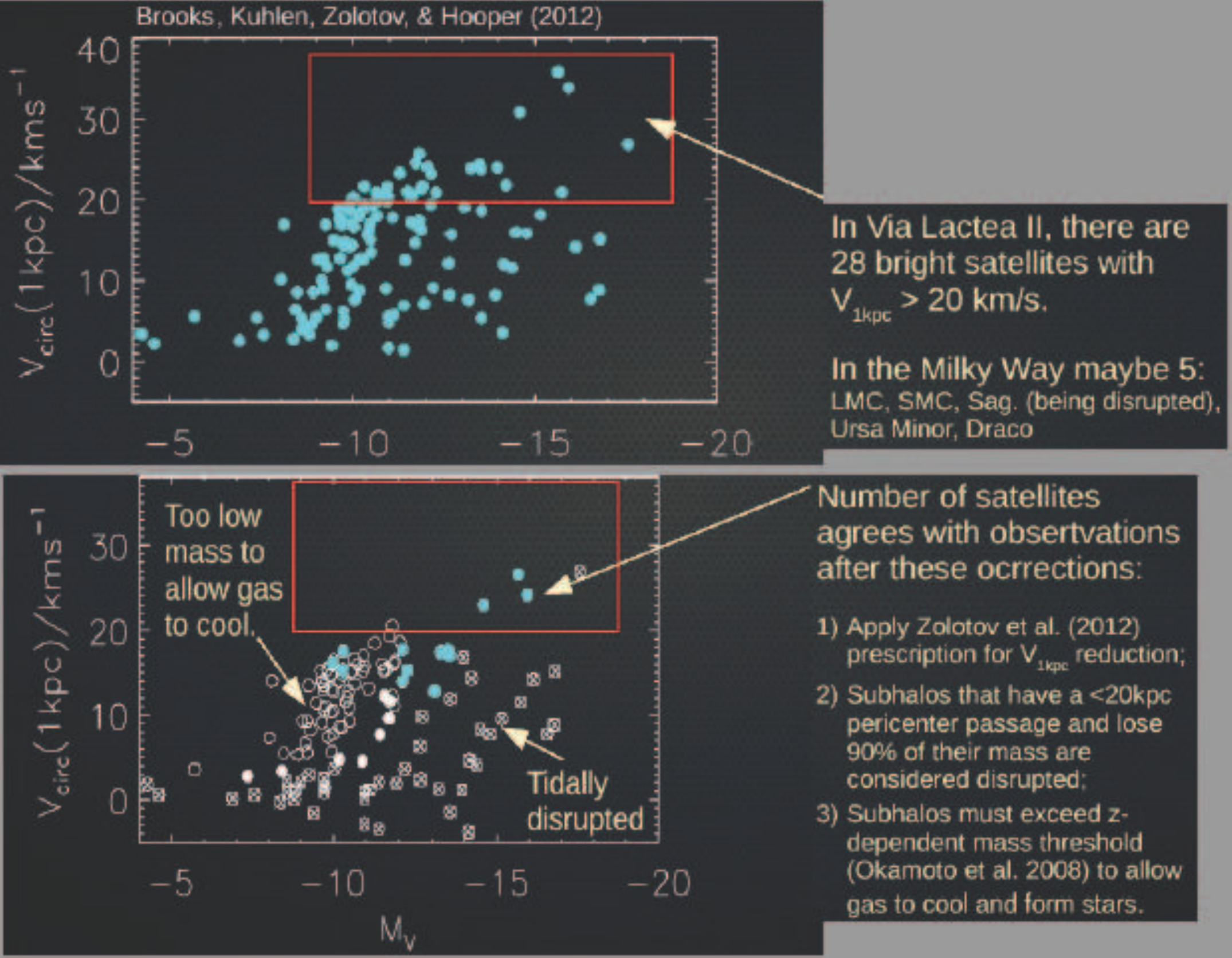}}
\caption{A solution to the Too Big To fail problem through baryon physics (from Brooks et al.\cite{brooks}).}
\label{spectra}
\end{figure}

\subsection{The angular momentum catastrophe}

The third problem is the "angular momentum catastrophe". In the standard model of disk formation one assumes detailed Conservation of Angular Momentum (Mestel 1963), that baryons initially trace dark matter\cite{fall}, adiabatic contraction\cite{blumenthal} is taken into account, a realistic Halo Profile\cite{mo} is needed, bulge is assumed to form from disk instabilities\cite{mo,bosch} and finally the receipt is closed by adding supernova feedback\cite{bosch1}. Unfortunately this Standard Model has problems. Hydrodynamical simulations show that the angular
momentum of the baryons is not conserved during collapse, baryons have $\simeq 10\%$ of the angular momentum of observed 
disks\cite{navarro_benz,steinmez_navarro,sommer_larson_etc},
and also the distribution of specific angular momentum in N-body simulations does not agree with observations (j-profile mismatch)\cite{bullock,bosch1}.
Other problems add to the previous: the spread in disk sizes seems to be narrower
then the spread in the spin parameter, $\lambda$, values\cite{lacey}. Major mergers should lead to spheroids, but they also
have the highest $\lambda$ values\cite{gardner}.

\begin{figure}
\resizebox{8.65cm}{!}{\includegraphics{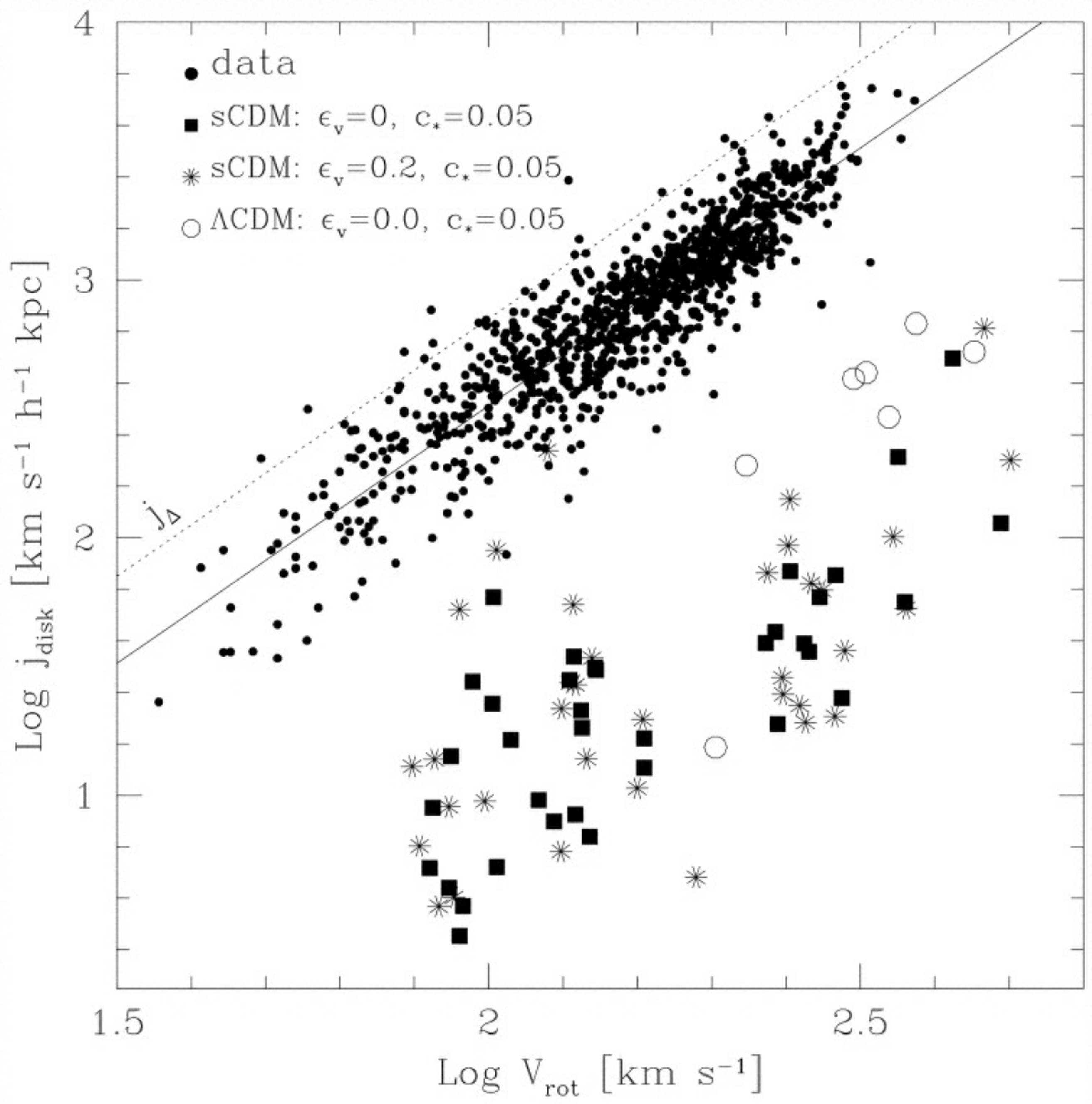}}
\resizebox{8.65cm}{!}{\includegraphics{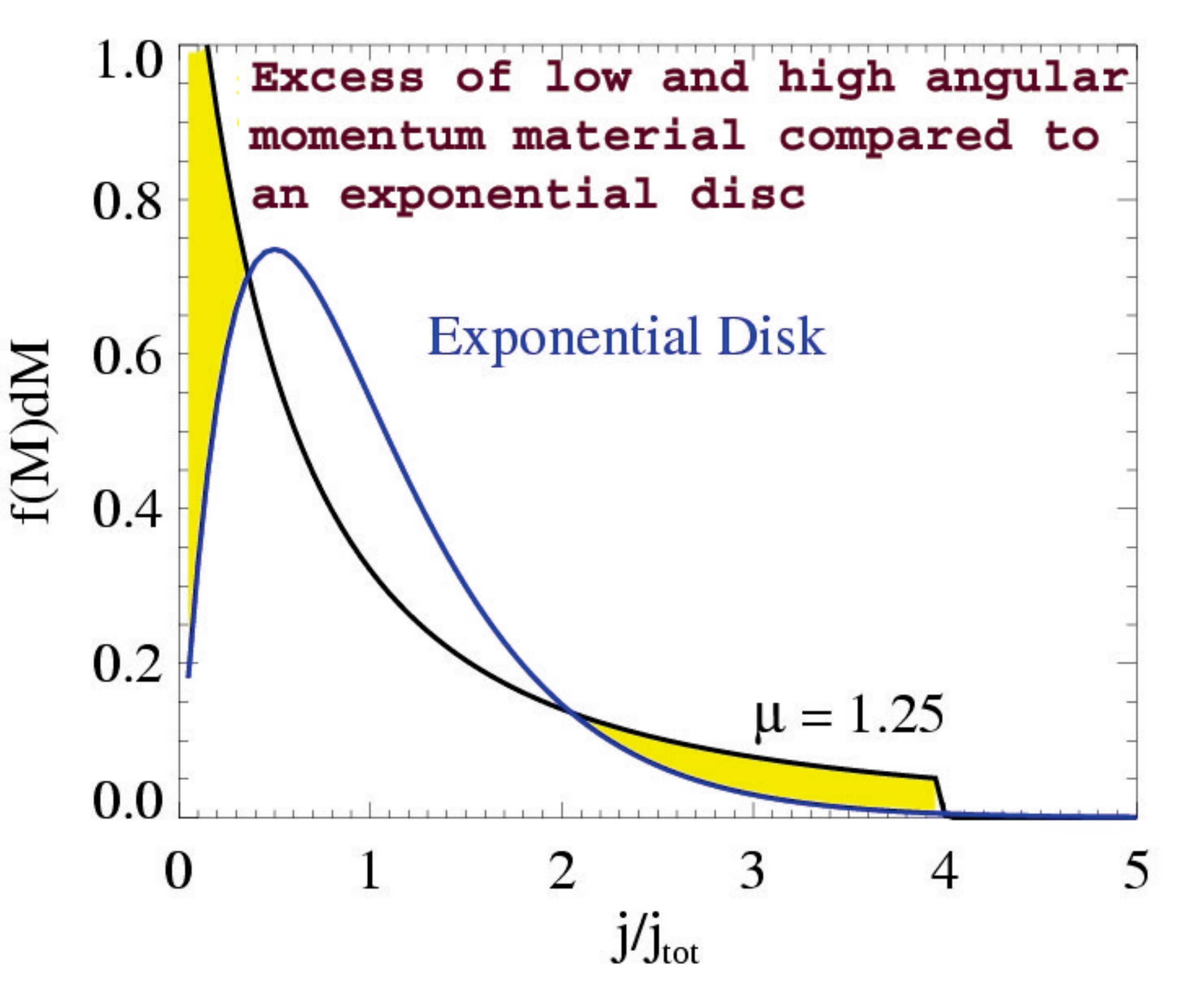}}
\caption{Left panel: Comparison of the observed and simulated specific angular momentum (from Navarro \& Steinmetz 2000. Right panel: Comparison of specific angular momentum distribution into an exponential disc and the excess of low and high angular momentum material in simulations}
\label{spectra}
\end{figure}

The loss of angular momentum, as been dubbed "angular momentum catastrophe" (see Fig. 51). Angular momentum is possibly lost during repeated collisions through dynamical friction or other mechanisms\cite{bosch2,navarro_steinmez}.

This problem has been also associated with the problem of "over-cooling" also seen in hydrodynamical simulations. If the baryons cool rapidly and sink to
the centers of dark halos, then they will loose their angular momentum. Following the illustration in Fig. (52, left and central panel) from Maller \& Dekel (2002)\cite{maller}, we see that if a satellite overcools, it becomes resistant to tidal stripping. It will then infall in the center of the halo and will transfer its angular momentum to the halo through dynamical friction. The satellite is dominated by DM in its outer parts, and this DM is stripped in the external parts of the halo.
As a consequence it will keep a part of its angular momentum. 
\begin{figure}
\resizebox{15.65cm}{!}{\includegraphics{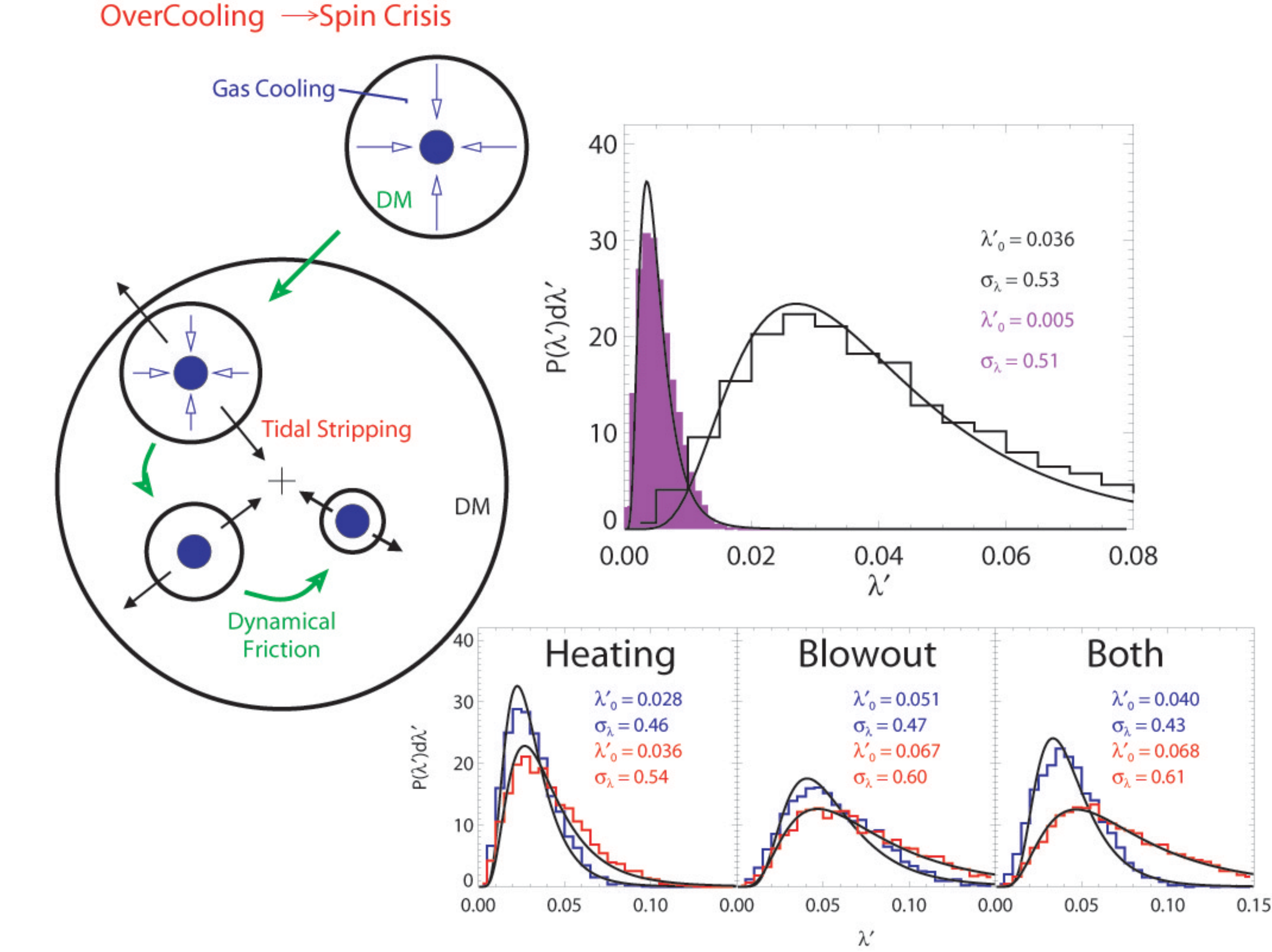}}
\caption{Left panel: From Over cooling to angular momentum catastrophe. Central panel: Effect of over-cooling on dark matter and baryons (purple) spin distribution.  Right panel: effect of heating, blowout and heating+blowout on bright galaxies (blue histograms) and dwarf galaxies (red histograms) (from Maller \& Dekel 2002).}
\label{spectra}
\end{figure}

In order over-cooling is prevented, it is necessary that in the system exists some form of heating. Adding supernova feedback solve the problem\cite{maller}. Supernova transfer energy to the interstellar medium, and as a result gas is removed from small halos (which by merging giving rise to the low specific angular momentum of the halo) with the result that baryons with low specific angular momentum are eliminated. Loss of angular momentum of baryons due to
dynamical friction is reduced by the combination of tidal stripping and the heating with consequent gas puffing up in larger haloes (see Fig. 53, bottom).  

\subsection{The Cusp/core problem}

The fourth problem, we mentioned is the Cusp/Core problem. Flores \& Primack\cite{flores} and Moore (1994)\cite{moore94} found that DDO galaxies have rotation curves ruling out cuspy profiles and that their density profiles are well approximated by isothermal profiles. 
The problem is that dissipationless simulations of the CDM produce cuspy profiles. Navarro, Frenk, \& White (1996, 1997)\cite{nfw,nfw1}, showed that DM profiles 
are cuspy, with inner density $\rho \propto r^{-1}$ (see Eq. \ref{eq:navarr}), that they are universal, namely independent from the cosmology, and from the scale. 
Higher resolution simulations by\cite{moore98,fukushige} found a different inner slope, $\rho \propto r^{-1.5}$, while\cite{jing,ricotti1,ricotti2,ricotti3, delpopolo_n} found evidences for the non-universality of the NFW profile\footnote{Jing \& Suto 2000\cite{jing}
 found different -1.5 for galaxies, -1.3 for groups, and -1.1 for clusters.}. In Del Popolo\cite{delpopolo_n,delpopolo_n1}, it was shown that the non-universality is connected to the baryons presence in the inner parts of the structures. 
More recent simulations\cite{stadel,navarro10} showed that the density profile is better approximated by an Einasto profile (Eq. \ref{einas}), characterized by a slope that becomes shallower toward the center of the clusters. In the case of Stadel et al. (2008)\cite{stadel} simulations the slope at 120 pc is -0.8. However, from the observational point of view, the inner profile of LSBs, dwarf Irr, dSphs, which are DM dominated are characterized by a cored profile
\cite{burkert1,deblok_bosma,swaters,deblok,delpopolo2,oh,denaray} (see fig. 53).

\begin{figure}
\resizebox{7.65cm}{!}{\includegraphics{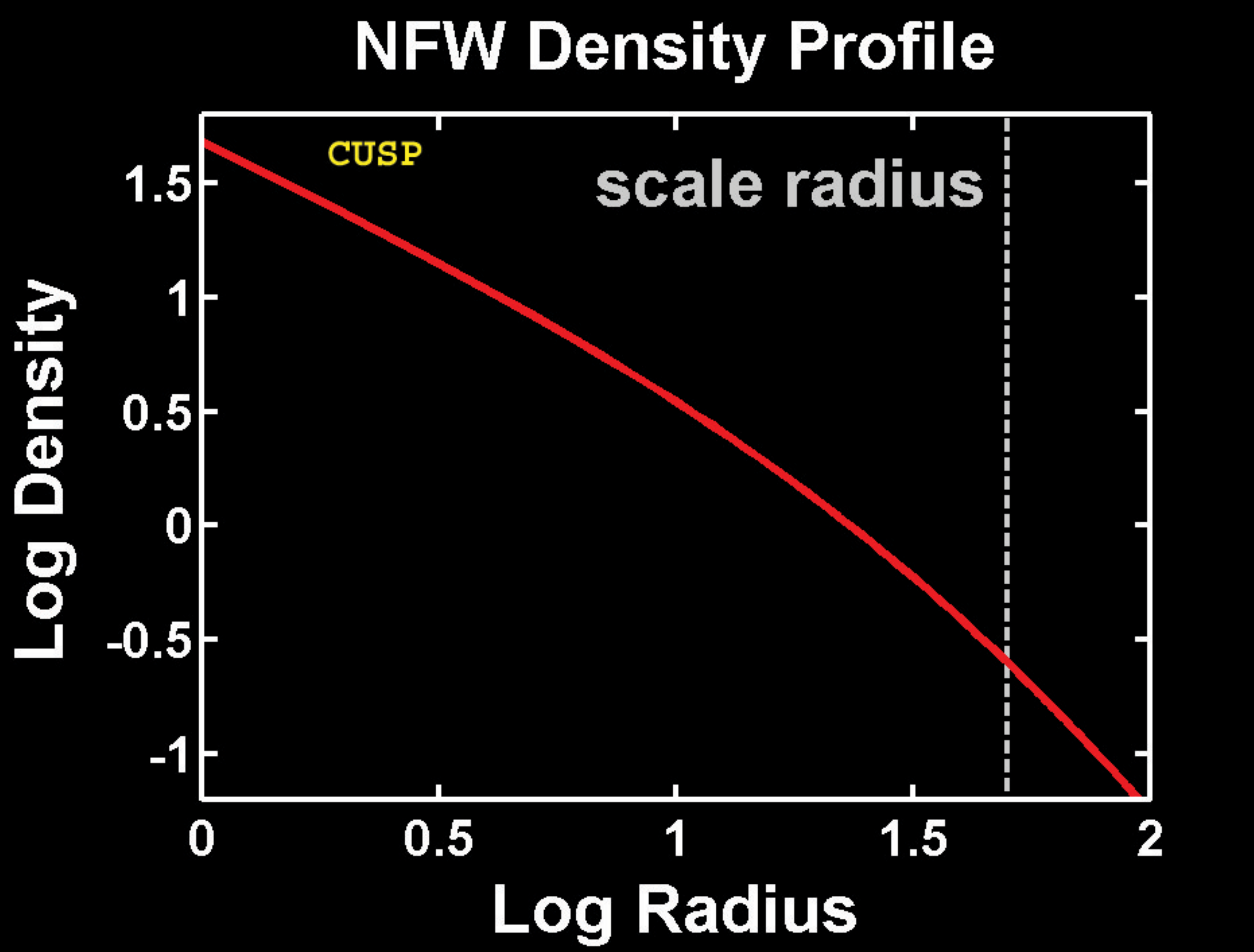}}
\resizebox{7.65cm}{!}{\includegraphics{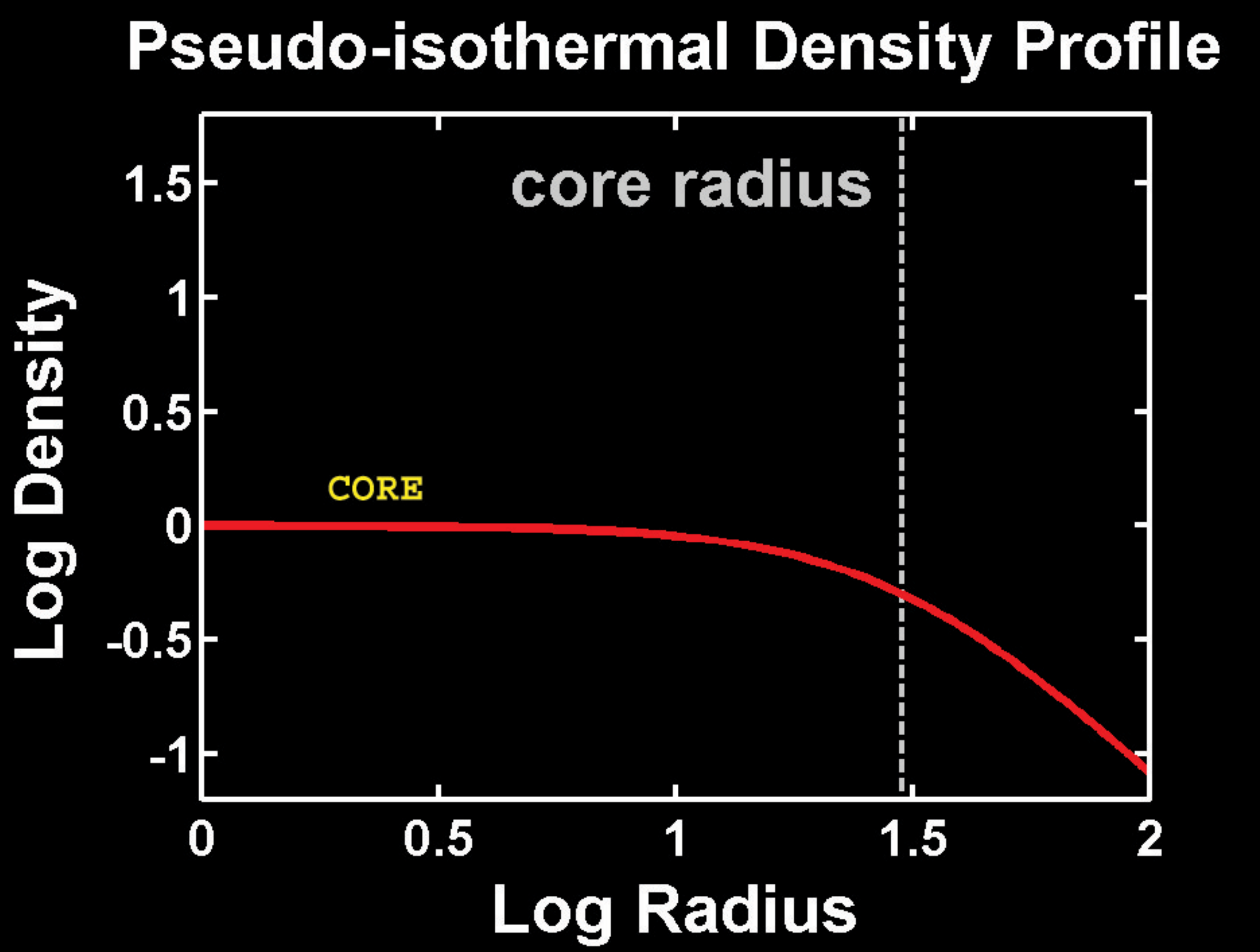}}
\caption{The Cusp/Core problem}
\label{spectra}
\end{figure}

\begin{figure}
\resizebox{10.4cm}{!}{\includegraphics{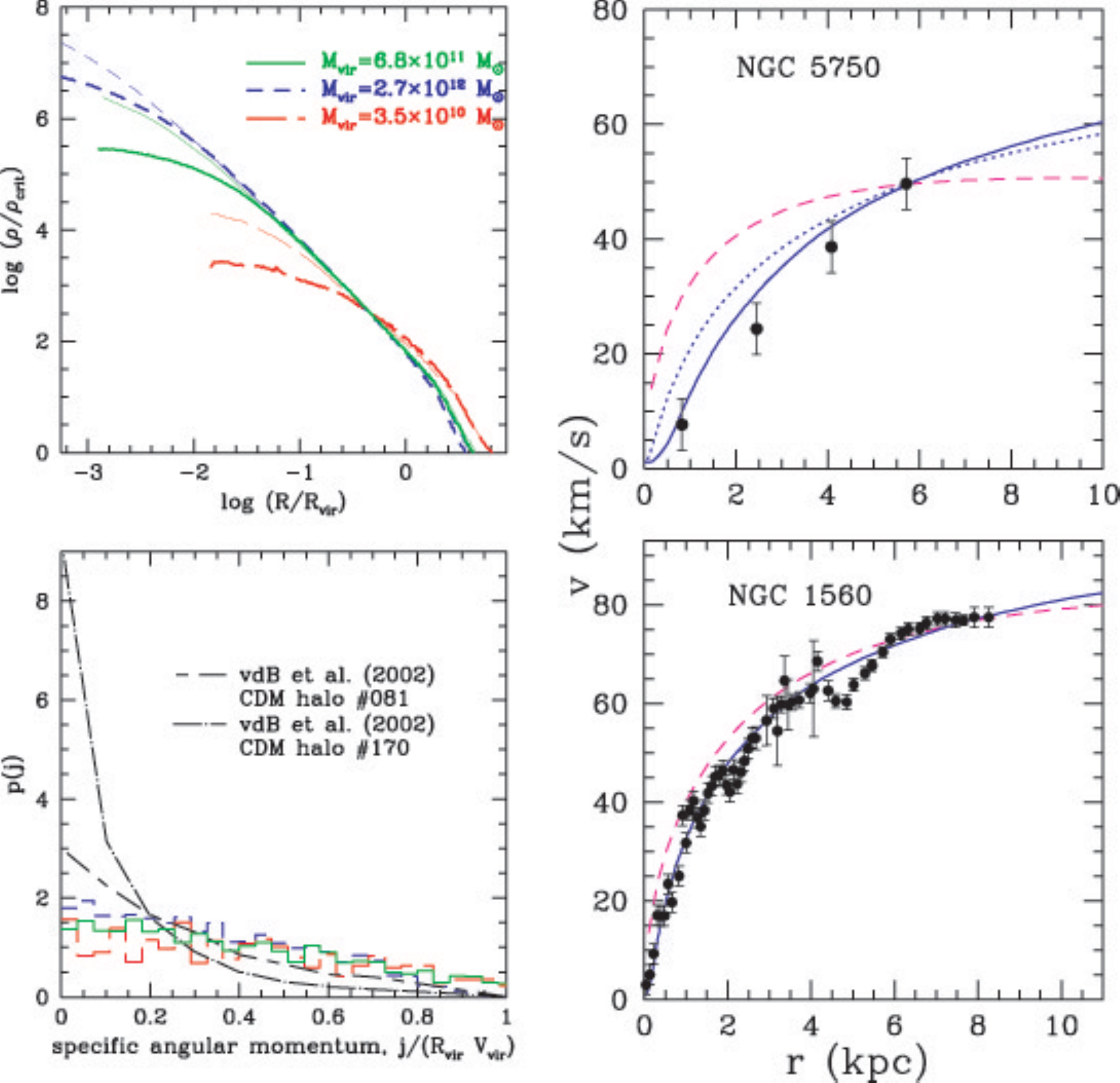}} 
\caption{Top left panel: density profiles for different masses. Bottom left panel: comparison of the specific angular momentum in simulations from
\cite{vdb} (lines) and semianalytical model result (histograms). Top and bottom right panel: comparison of the semyanalitical model (solid line), the NFW model (dashed lines) with two rotation curves (from \cite{williams}.}
\label{spectra}
\end{figure}

\begin{figure}
\resizebox{11.4cm}{!}{\includegraphics{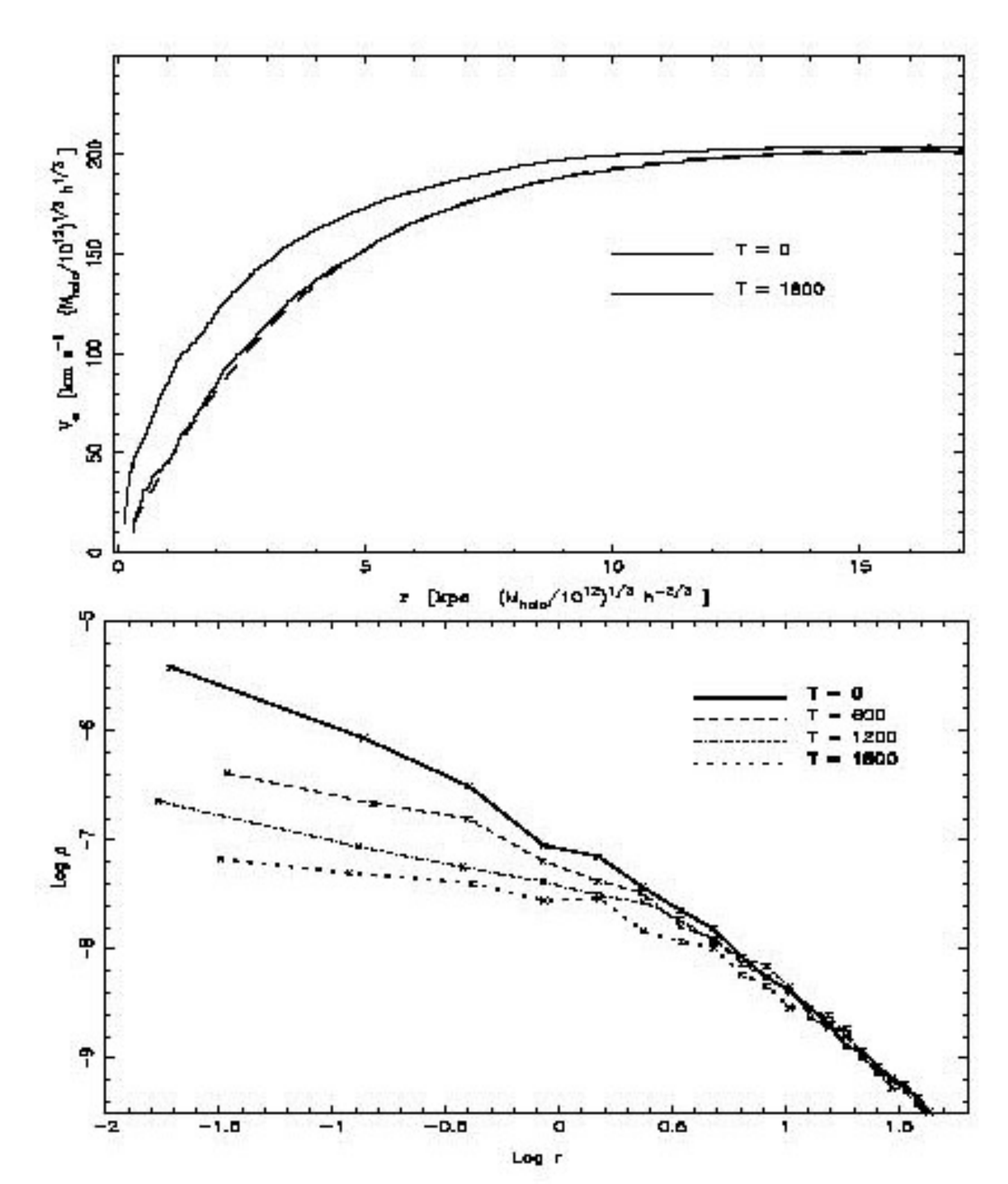}} 
\caption{Top panels: evolution of the rotation curves with time and for different configuration of the system. Bottom panels: same as the top pannels but for the density profile (from \cite{el1}).}
\label{spectra}
\end{figure}

Gentile et al.\cite{gentile,gentile1,gentile2} decomposed the total rotional curves of some spiral galaxies in stellar, gaseous, and dark matter components (see Fig. 15). Fitting the density with various models they found that constant density core models are preferred over cuspy profiles. Similar result was obtained by Oh et al. (2010)\cite{oh} using 7 dwarf galaxies from the THINGS (The HI Nearby Galaxy Survey) galaxies (see Fig. 16). However, as already reported, de Blok et al. (2008)\cite{deblok} using THINGS galaxies showed that high mass spiral galaxies with $M_B < -19$ have profiles equally well fitted by the NFW profile or a Pseudo Isothermal (ISO) profile, while
low mass spirals with $M_B > -19$ prefer a ISO model. Simon et al. (2005)\cite{simon1} studied the low mass spirals NGC 2976, NGC 6689, NGC 5949, NGC 4605, and NGC 5963 finding a large scatter of the inner slope $\alpha$, compatible with a cored profile, $\alpha \simeq 0.01$, for NGC 2976, and a cuspy one $\alpha \simeq 1.28$ for NGC5963. The other three galaxies had
$\alpha \simeq 0.80$ (NGC 6689), $\alpha \simeq 0.88$ (NGC 5949), and $\alpha \simeq 0.88$ (NGC 4605). 
In other terms, if a large part of dwarfs are well described by cored profiles, some others are not.

As previously described, a similar problem is also presenting galaxy clusters. Sand et al. (2004)\cite{sand1} combining weak lensing, strong lensing, and velocity dispersion studies of the stars of the BCG (Brightest Central Galaxy) found that of the clusters MACS 1206, MS 2137-23, RX J1133, A383, A1201, A963, only RX J1133 had a profile compatible with the NFW model, and similar studies of Newmann et al. (2009)\cite{newman}
(A611), Newman et al. (2011)\cite{newman1}(A383), and Newman et al. (2012)\cite{newman2}
 (MS2137, A963, A383, A611, A2537, A2667, A2390). For sake of precision, I want to recall that Donnaruma et al. (2011)\cite{donnaruma} found a cuspy profile for A611 combining strong lensing and X-ray observations, and this is not the only discrepancy in the study of the same cluster. 
In general gravitational lensing yields conflicting estimates sometime in agreement with numerical simulations (Dahle et al 2003\cite{dahle}; Gavazzi et al. 2003\cite{gavazzi2}; Donnaruma et al. 2011\cite{donnaruma}) or finding much shallower slopes (-0.5) (Sand et al. 2002\cite{sand2,sand1}; Newman et al. 2009, 2011, 2012)\cite{newman,newman1,newman2}. X-ray analyses have led to wide ranging of value of the slope from: -0.6 (Ettori et al. 2002\cite{ettori}) to -1.2 (Lewis et al. 2003) till -1.9 (Arabadjis et al. 2002\cite{arabadjis1}), or in agreement with the NFW profile (Schmidt \& Allen 2007\cite{schmidt}; 34 Chandra X-ray observatory Clusters) Newman et al. 2012\cite{newman2}

\subsubsection{Proposed solutions to the Cusp/Core problem.}

Many solutions have been proposed to solve the Cusp/Core problem and in general the small scale problems of the $\Lambda$CDM. While a decade ago some authors (e.g., \cite{vandenbosch,vdb_swa}) pointed the finger against observations, claiming that the inconsistence could be due to poor resolution or to a not proper way of taking account of systematic effects (non-circular motions, beam smearing, off-centering) which tend to systematically lower slopes. Moreover, according to \cite{hayashi} error bars were large enough so that the cores are favored but cusps can usually not be ruled out. 
Several authors (\cite{power,navarro04,hayashi}) suggested ways
to reconcile simulations with the observations. According to them, the new simulations, performed in 2004 
(\cite{power,navarro04,hayashi}), are in better agreement with observations, since they become progressively shallow from the virial radius inwards.
DM haloes are non-spherical and 3D. Comparing rotation speeds of gaseous disks to spherically averaged
circular velocity of DM haloes, one should expect differences. In other words, the observational disagreement would be with the fitting formulae, rather than with simulated haloes (\cite{hayashi}). Nowadays it is well known, that even with very high resolutions, the minimum inner slopes is $-0.8 $ (\cite{stadel}), and at the same time it is clear that high resolution observations 
can distinguish cored and cuspy haloes by deriving their asymptotic inner slopes from rotation curve data\cite{denaray}.

Another possibility is a failure of the CDM model or problems with simulations\cite{deblok1,deblok3,borriello}. However, modern simulations doe not suffer of problems like lack of resolution, relaxation, and overmerging, like in the past. Convergence tests performed by\cite{diemand}, showed that N-body simulations are finding correctly the CDM density profiles. However, N-body simulations are dissipationless, baryons are not taken into account, and we know that in the inner regions of galaxies (inner kpc) baryons are not negligible, and that in the central 10 kpc of clusters baryons are dominating over DM (Sand et al. 2004\cite{sand1}; Newmann 2009, 2011, 2012\cite{newman,newman1,newman2}). So, we need SPH simulations in order to study the inner parts of galaxies and clusters, supposing that we know enough of baryon physics. The other possibility, as mentioned, is that the CDM model is wrong, and has been speculated that other forms of DM (warm\cite{som_dol}, fuzzy\cite{hu}, 
repulsive\cite{goodman}, fluid\cite{peebles2000}, annihilating\cite{kap}, decaying\cite{cen1}, or self-interacting\cite{sperg_ste}) could solve the small scale problems of the CDM model. 
Another solution is to modify gravity (e.g., $f(R)$, $f(T)$ theories, MOND).
$f(R)$, and $f(T)$ theories are a types of modified gravity theories, generalization of Einstein's General Relativity. $f(R)$ are theories 
defined by a different function of the Ricci scalar. $f(R)$ theory was first proposed in 1970 by Buchdahl\cite{buchdahl}. Starobinsky\cite{starobinsky}
turned the field into an active research field. The $f(T)$ has been introduced to explain Universe acceleration without using dark energy (see 
\cite{bengochea}). 
Finally, the Modified Newtonian Dynamics, introduced in 1983 by Milgrom\cite{milgrom} as a way to model rotation curves of galaxies.

However, before throwing away a model that explains many of the observations and features of our Universe, it would be wise
to verify if the small scale problems could be connected to some piece of local physics that we do not understand correctly or that we are not introducing in our calculations. 

Possible solutions of the Cusp/Core problem saving the $\Lambda$CDM model are based on the idea that some mechanism, that could "heat" the DM, would produce a inner flatter density profile. Example of these mechanisms are the effect of a rotating bar, transferring angular momentum from baryons to DM through dynamical friction\cite{el1,el2,delpopolo2}, AGN, gas bulk motions generated by supernova explosions\cite{mashchenko,governato}

Since N-body simulations does not take into consideration baryon physics, some semi-analytical models have been used to study the Cusp/Core problem. 

Several authors\cite{nusser,hiotelis,delliou,ascasibar,williams} used secondary infall models to discuss the role of angular momentum in structure formation, arriving to the conclusion that 
the larger is the angular momentum of a proto-structure the flatter is its inner density profile, finding agreement with the rotation curves of dwarfs\cite{williams}, and showing that the specific angular momentum acquired in the semyanalitical model is larger than that of simulations (see
 Fig. 54). El-Zant et al. (2001, 2004)\cite{el1,el2}, showed that clumps of baryons loose energy that is trasferred,  through dynamical friction, and deposited in the DM component of the system producing a flattening or erasing of the cusp, both in dwarf galaxies and in clusters of galaxies (see Fig. 55). 
Other authors studied the baryons effects only through adiabatic contraction of DM haloes\cite{blumenthal,gnedin} that produce a steepening of the density profiles. In one paper, Del Popolo (2009)\cite{delpopolo2} took all the quoted effects into account, in a secondary infall model, namely: ordered angular momentum acquired by the proto-structure through tidal torques, random angular momentum, energy and angular momentum exchange between baryons and DM through dynamical friction, and adiabatic contraction. The paper showed that the discrepancy between N-body simulations and observed density profiles is due to the fact that in the inner part of the proto-structure the role of baryons is not negligible, and comparing dissipationless simulations to real structures containing baryons is not correct (see Fig. 56, 57). In Del Popolo (2012)\cite{delpopolo}, the model was applied to dwarfs of galaxies showing that the formation history, the content of baryons, and the environment influence their density profiles. While on average dwarfs are well fitted by Burkert's profiles, in some case the NFW profile is a good fit, as shown in de Blok et al. (2008)\cite{deblok}, and Simon et al. (2005)\cite{simon1}. In Del Popolo (2012)\cite{delpopolo1}, were studied the density profiles of clusters and compared with Sand et al. (2004)\cite{sand1}, and Newmann et al. (2009, 2011, 2012)\cite{newman,newman1,newman2} observations, finding good agreement (see Fig. 58).

\begin{figure}
\resizebox{8.4cm}{!}{\includegraphics{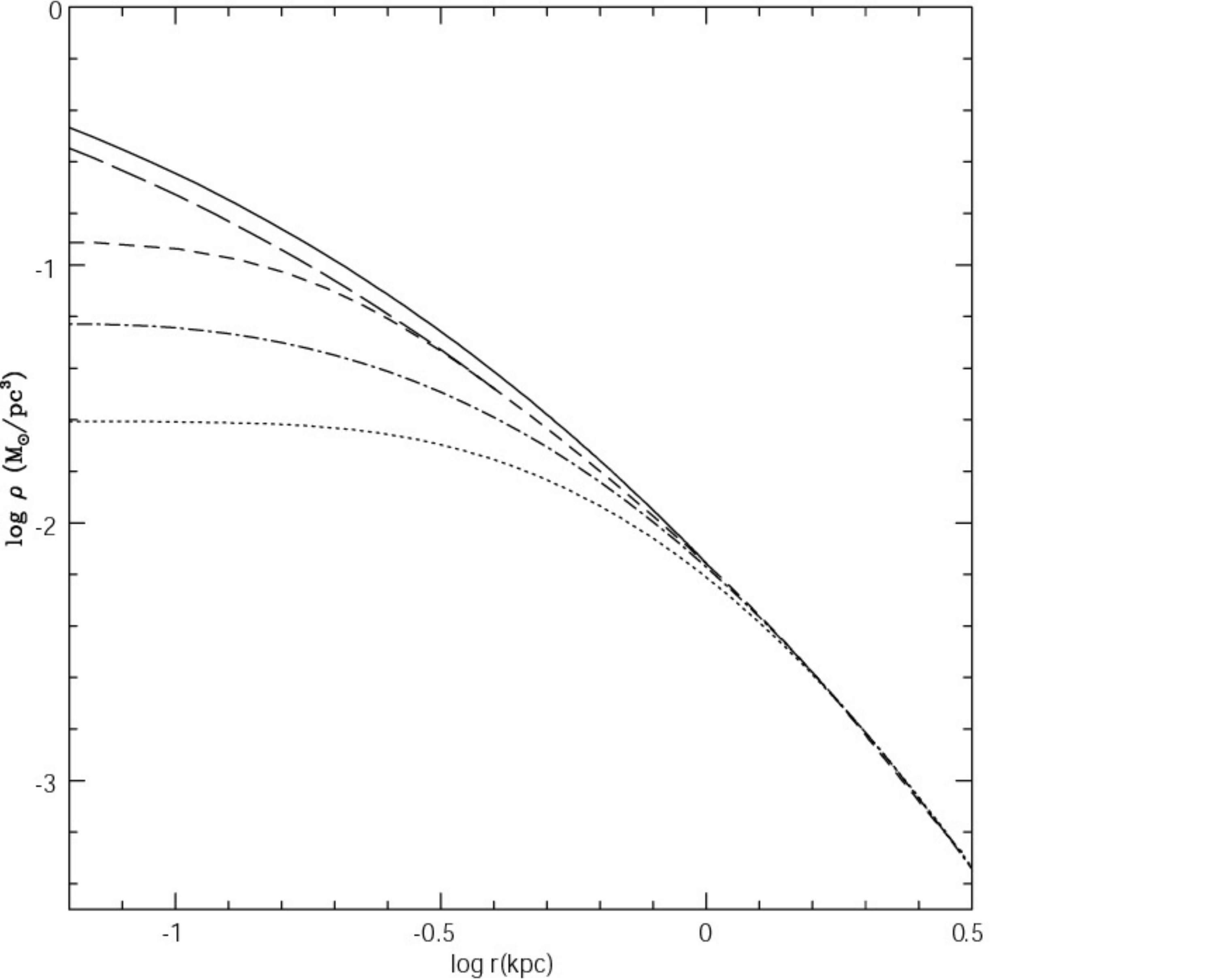}} 
\resizebox{7.1cm}{!}{\includegraphics{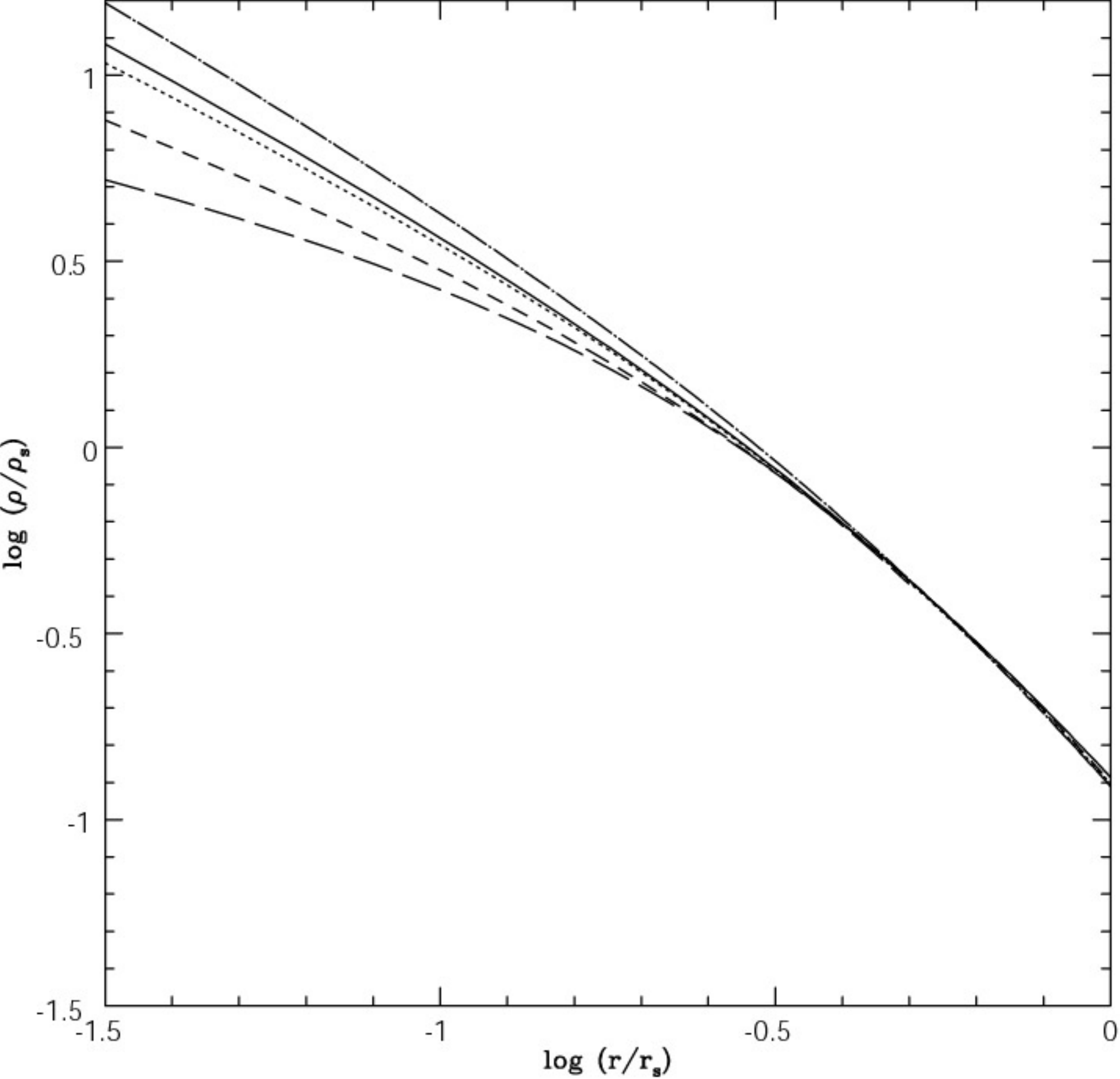}} 
\caption{Left panel: evolution of a density profile of a $10^8 M_{\odot}$ galaxy from $z=10$ to $z=0$. Right panel: evolution of a density profile of a $10^{14} M_{\odot}$ cluster from $z=3$ to $z=0$ (from Del Popolo (2009)\cite{delpopolo2}.}
\label{spectra}
\end{figure}

\begin{figure}
\resizebox{14.4cm}{!}{\includegraphics{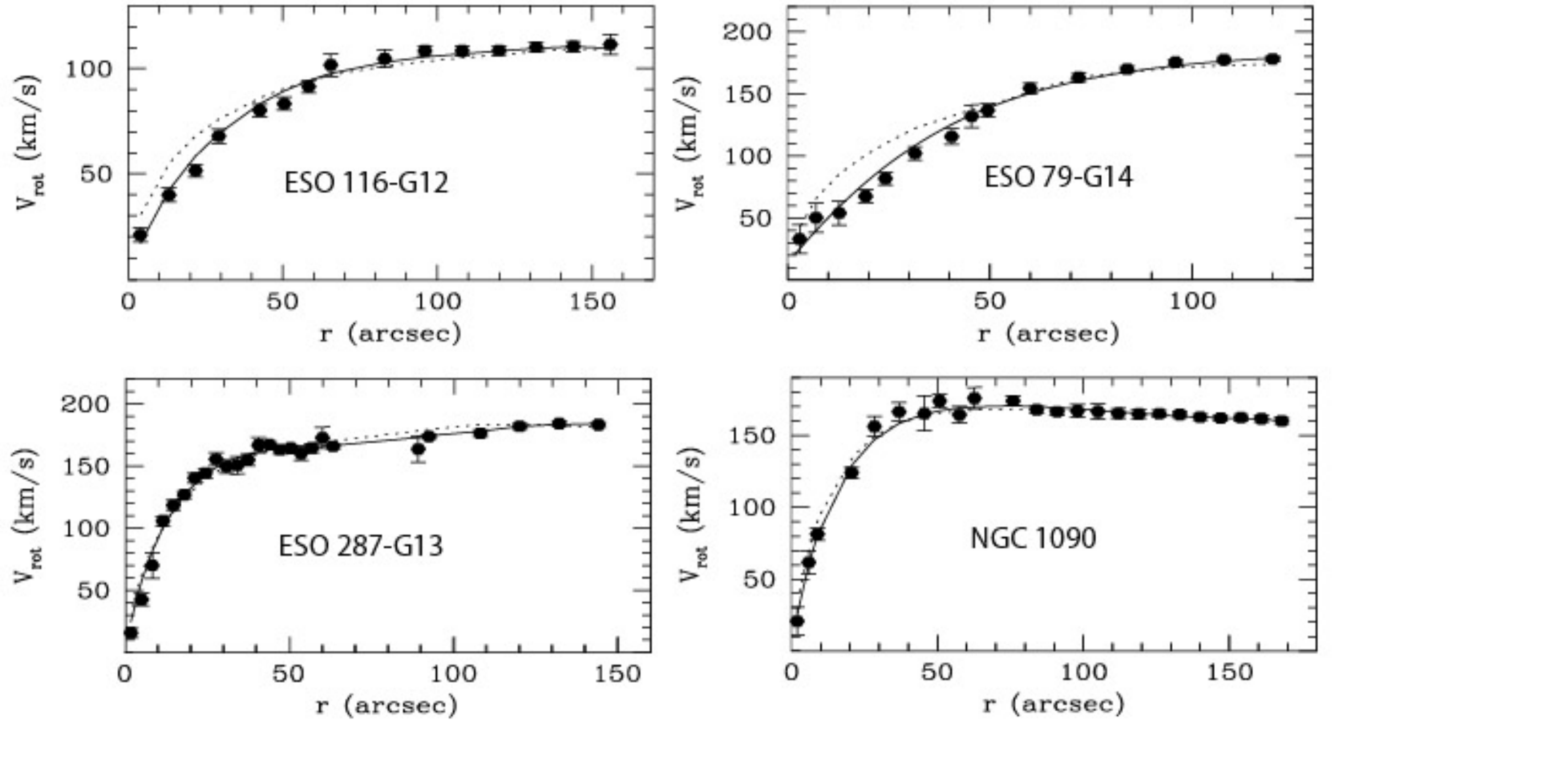}} 
\caption{Comparison of rotation curves obtained semianalitically (solid lines) with NFW model (dashed lines) (from del Popolo (2009)\cite{delpopolo2}).}
\label{spectra}
\end{figure}

\begin{figure}
\resizebox{14.4cm}{!}{\includegraphics{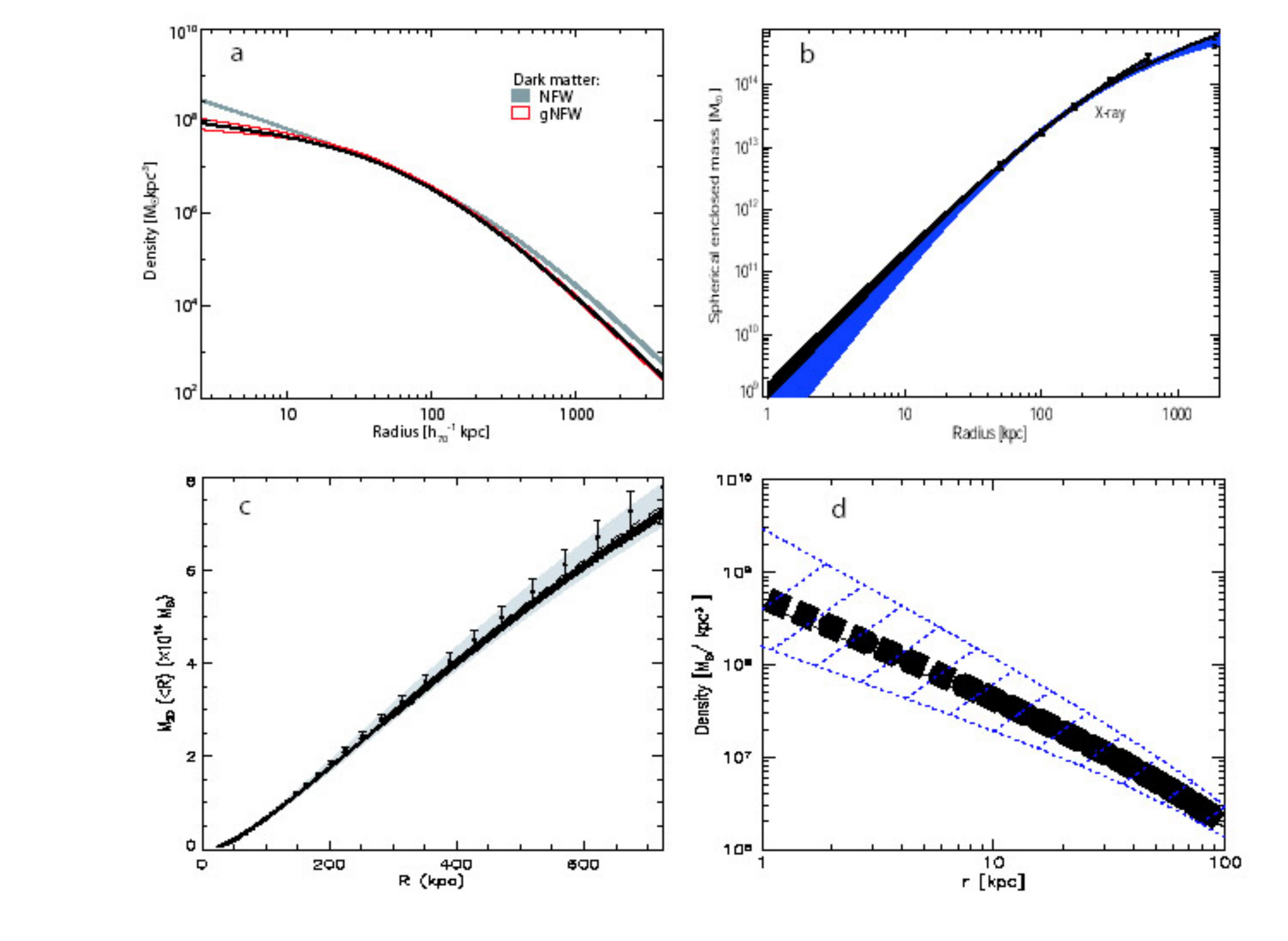}} 
\caption{Top left panel: density profile of A611. Grey solid line: NFW profile. Dashed lines: Newman et al. (2009)\cite{newman}
density profile. Black band: the result of Del Popolo (2012)\cite{delpopolo1}. Top right: mass profile of A383. Grey band: observational result of Newman et al. (2011)\cite{newman1}. Black band: mass distribution obtained in Del Popolo (2012)\cite{delpopolo1}. Bottom right panel: mass profile of MACS J1423. Grey band with error bars: observational result of \cite{morandi}. Black band: result obtained in Del Popolo (2012)\cite{delpopolo1}. Bottom right panel: density profile of RX J1133. Solid curve and the dotted-shaded region: Sand et al (2004)\cite{sand1} observational results. Black dashed band: the result of Del Popolo (2012)\cite{delpopolo1}.}
\label{spectra}
\end{figure}

As already reported, Governato et al. (2010)\cite{governato} simulated two dwarf galaxies using the SPH technique. 
In their simulation they took into account the role of supernova explosions and gas outflows from them. According to their 
simulation, gas outflows eliminate the gas having low angular momentum, and as a result the dwarf does not form a bulge and the density profile is cored and not cuspy. The idea at the base of Governato's result is not new at all. The process by them used was already proposed by several authors\cite{nv96,gelato,read,mashchenko}


In summary, the Cusp/Core problem can be solved in several ways, the challenge is to understand which solution is the correct one, and this is also valid for the others small scale problems.

\begin{figure}
\resizebox{10.4cm}{!}{\includegraphics{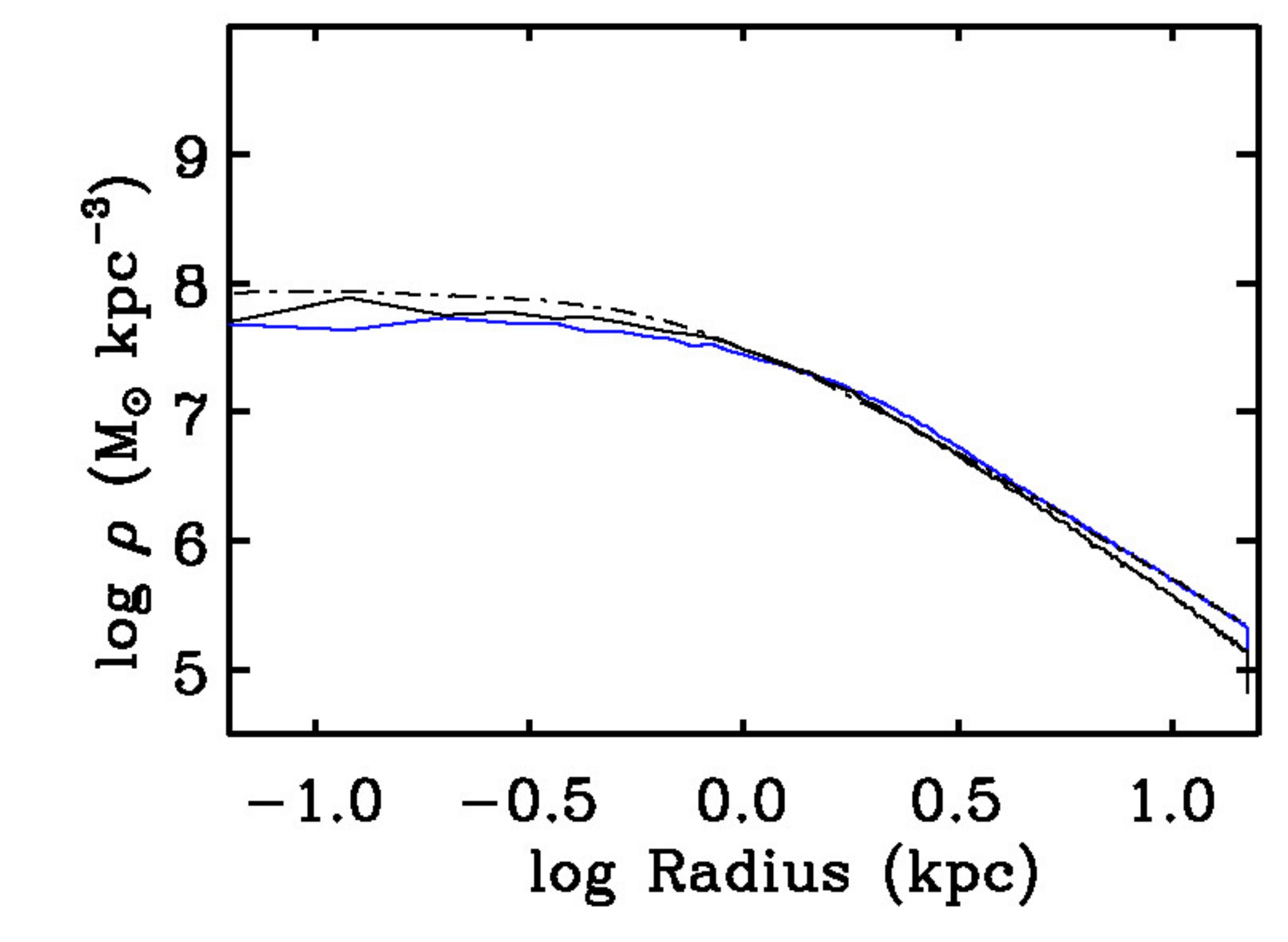}} 
\caption{Comparison of Governato et al. (2010) SPH simulation (continue lines) with the semianalytical result of 
Del Popolo (2009)\cite{delpopolo2}) (dashed line). }
\label{spectra}
\end{figure}

\subsection{A Unified baryonic solution to the $\Lambda$CDM small scale problems}

The quoted problems can be solved once we take into account the baryons physics, and connections among them are evident. For example, if cuspy density profiles are transformed into cored ones by SF feedback, or dynamical friction, this also has effects on the distribution and number of substructure/satellites of a large halo. In fact, cored profiles are more subject to be destroyed by tidal stripping. Ejection of low angular momentum gas by (e.g.) SF feedback can solve the angular momentum problem, as previously discussed. 
Recently, a series of papers have sketched a baryonic solution to the small scale problems of the $\Lambda$CDM model (Zolotov et al. (2012) \cite{zolotov}, and Brooks et al. (2013)\cite{brooks}). The quoted papers are based on the 
idea (\cite{nv96,gelato,read,mashchenko, governato}) that SF explosions removes angular momentum gas from the proto-galaxy with the result of a) flattening the density profile; b) giving rise to the correct angular momentum distribution in discs.  
Zolotov et al. (2012) \cite{zolotov} found a correction to the circular velocity in 1 kpc, $v_{1kpc}$, that applied to the results of large N-body simulations (e.g., Via Lactea) can noteworthy reduce the discrepancy of the observed and predicted number of satellites, and their too high density (see Brooks et al.v(2013)\cite{brooks}).

A similar result, is obtained in the "dynamical friction framework" (El-Zant et al. (2001, 2004)\cite{el1,el2}; Del Popolo (2009)\cite{delpopolo2}). In this framework, the cusp/core problem is resolved as discussed, and the angular momentum discrepancy is not present (Del Popolo (2009)\cite{delpopolo2}). Using Del Popolo (2009)\cite{delpopolo2} model, it is possible to obtain a correction to $v_{v_1kpc}$ similar to that of Zolotov et al. (2012) \cite{zolotov}. Applying the correction to the Via Lactea satellites, as in Brooks et al. (2013)\cite{brooks}, the missing satellite problem and the too-big-to-fail problem are solved (Del Popolo 2013\cite{del2013}.

\bibliographystyle{aipproc}   

\bibliography{sample}

\IfFileExists{\jobname.bbl}{}
 {\typeout{}
  \typeout{******************************************}
  \typeout{** Please run "bibtex \jobname" to optain}
  \typeout{** the bibliography and then re-run LaTeX}
  \typeout{** twice to fix the references!}
  \typeout{******************************************}
  \typeout{}
 }



\section{Summary and Outlook}

In this paper, I have discussed the many evidences for DM existence, how it is distributed in cosmic structures, its particle nature and the "zoo" of candidates that could be the constituents of the dark Universe. I have discussed the direct and indirect methods of detection, and finally some of the known problems of one of the CDM models that nowadays seem to be the favorite one from observations, namely the $\Lambda$CDM model often called "the concordance model" or "standard model of big bang cosmology".

Concerning the evidences of DM existence, in the last years, strong evidences come from galaxy clusters collisions, showing through weak lensing that clusters are made of a dissipationless component, not only gas. Cosmic shear, the weak lensing of the LSS is another strong evidence that Universe contains matter which deflect the light of remote objects. In 2012, was observed a weak lensing signature of a filament in the supercluster A222/A223, connecting the two clusters\cite{dietrich}
An important news, coming from particle physics is the absence of SUSY effects in LHC experiments, and if this will be confirmed in the next years, 
one of the most promising candidates of DM, the neutralino, should be substituted by other kind of DM. The great hope put on colliders
to reveal hints of the so called "new physics", has been up to this moment betrayed, but in 2015 LHC could give new and unexpected results. Colliders give a different point of view on DM with respect to astrophysical
experiments, and at the same time they could provide the needed information to reveal the physics at the base of the DM particles. At the same time,
colliders are not able to test its abundance in the universe or its cosmological stability. This is the reason colliders and direct and indirect searches must go hand in hand.  

Direct and indirect detection of DM has improved a lot in the last years. As early discussed, there was even a claim of axion detection, after disproved, and the DAMA experiment is claiming since a decade to have signal DM, even this never confirmed. The space telescope Fermi, has meanwhile studied the galactic center, MW dSphs, clusters of galaxies, the IGRB, finding possible evidences of DM existence but not any certainty. To disentangle the astrophysical signal from DM annihilation signal is a not easy task. 
The 511 MeV line observed several years ago by INTEGRAL, differently from other signals, is difficult to explain through astrophysics (SN Ia, Hipernovae, etc.), but MeV DM is difficult to explain. 

Indirect, experiments have put constraints to DM-nucleon cross-section which seems to be in the $1- 10 \simeq$ zb range.
An important improvement in direct search are the ton scale detectors (e.g., ArDM). These kind of detectors can test the most attractive DM models, including KK DM, that was before out of reach. Apart from the constraints to SUSY from LHC, these detectors could put strong constraints on SUSY and/or 
TeV scale physics. The next step, would be the detection of WIMPs, and the consequent precise measurements of its mass and interactions.  

\begin{theacknowledgments}

I would like to thank the organizer of the school for having invited me to their successful gathering, 
giving me also the opportunity to enjoy the wonderful location of the school and Puerto-Vallarta. 
This work was supported in part by FAPESP visiting research fellowship (Grant No. 2011/20688-1).
Many thanks go to the S\~ao Paulo university Astronomy Department for the facilities and hospitality. 

\end{theacknowledgments}

\end{document}